\documentclass[french]{amu_these}

\usepackage{amsmath,amssymb}
\usepackage{bm}

\usepackage{tikz}
\usetikzlibrary{shapes,arrows,arrows.meta,matrix,decorations.markings,calc,babel,quotes,angles,patterns}
\makeatletter \g@addto@macro\@floatboxreset\centering \makeatother
\usepackage{slashed,empheq}
\usepackage[shortlabels]{enumitem}
\renewcommand\bf{\bfseries}
\renewcommand\mathbf{\boldsymbol}

\newcommand\encircle[1]{%
  \tikz[baseline=(X.base)] 7
    \node (X) [draw, shape=circle, inner sep=0] {\strut #1};}

\DeclareMathOperator{\const}{const}
\DeclareMathOperator{\Vect}{Vect}
\DeclareMathOperator{\Diff}{Diff}
\DeclareMathOperator{\yamf}{\overset{o}{\Delta}_Y}
\DeclareMathOperator{\yam}{\Delta_Y}
\DeclareMathOperator{\diracf}{\slashed{D}^0}
\DeclareMathOperator{\dirac}{\slashed{D}}
\DeclareMathOperator{\Geom}{Geom}
\DeclareMathOperator{\Met}{Met}
\DeclareMathOperator{\Pf}{Pf}
\DeclareMathOperator{\Ima}{Im}
\newcommand{\id}{\mathbb{1}}

\newcommand{\R}{\mathbb{R}}
\newcommand{\Cc}{\mathbb{C}}

\newcommand{\T}{\mathbb{T}}

\newcommand{\angles}[1]{\left\langle #1 \right\rangle}

\newcommand{\Conf}{\mathrm{Conf}}

\newcommand{\Chr}{\mathrm{Chr}}

\newcommand{\Sch}{\mathrm{Sch}}
\newcommand{\Barg}{\mathrm{Barg}}
\newcommand{\Aut}{\mathrm{Aut}}
\newcommand{\Vol}{\mathrm{Vol(\rg)}}
\newcommand{\SL}{\mathrm{SL}}
\newcommand{\SO}{\mathrm{SO}}
\newcommand{\SE}{\mathrm{SE}}

\newcommand{\so}{\mathfrak{so}}
\newcommand{\volg}{\sqrt{|\rg|}\,}
\newcommand{\Ric}{\mathrm{Ric}}

\newcommand{\ie}{\textit{i.e.} }

\newcommand{\half}{\frac{1}{2}}
\newcommand{\tg}{\widetilde{g}}
\newcommand{\wh}{\widehat}

\newcommand{\rg}{\mathrm{g}}

\newcommand{\cC}{\mathcal{C}}
\newcommand{\cD}{\mathcal{D}}
\newcommand{\cE}{\mathcal{E}}
\newcommand{\cF}{\mathcal{F}}
\newcommand{\cG}{\mathcal{G}}
\newcommand{\cH}{\mathcal{H}}

\newcommand{\cJ}{\mathcal{J}}
\newcommand{\cK}{\mathcal{K}}
\newcommand{\cL}{\mathcal{L}}
\newcommand{\cM}{\mathcal{M}}
\newcommand{\cN}{\mathcal{N}}
\newcommand{\cO}{\mathcal{O}}
\newcommand{\cP}{\mathcal{P}}

\newcommand{\cS}{\mathcal{S}}
\newcommand{\cT}{\mathcal{T}}

\newcommand{\nvol}{|\mathrm{Vol}(\rg)|}
\newcommand{\bA}{\boldsymbol{A}}
\newcommand{\bB}{\boldsymbol{B}}
\newcommand{\bvarpi}{\boldsymbol{\varpi}}
\newcommand{\bomega}{\boldsymbol{\omega}}
\newcommand{\bOmega}{\boldsymbol{\Omega}}
\newcommand{\bpartial}{\boldsymbol{\partial}}
\newcommand{\bbeta}{\boldsymbol{\beta}}
\newcommand{\bgamma}{\boldsymbol{\gamma}}
\newcommand{\bsigma}{\boldsymbol{\sigma}}
\newcommand{\bnabla}{\boldsymbol{\nabla}}
\newcommand{\bb}{\boldsymbol{b}}
\newcommand{\Id}{\mathrm{Id}}
\newcommand{\bc}{\boldsymbol{c}}
\newcommand{\bzero}{\boldsymbol{0}}
\newcommand{\bS}{\boldsymbol{S}}
\newcommand{\diag}{\mathrm{diag}}
\newcommand{\Gal}{\mathrm{Gal}}
\newcommand{\GL}{\mathrm{GL}}
\newcommand{\eChr}{\widetilde{\mathrm{Chr}}}
\newcommand{\eSch}{\widetilde{\mathrm{Sch}}}

\newcommand{\bs}{{\boldsymbol{s}}}
\newcommand{\bst}{{\boldsymbol{s^\bot}}}
\newcommand{\st}{{s^\bot}}
\newcommand{\nst}{\Vert\bst\Vert}
\newcommand{\ns}{\Vert\bs\Vert}
\newcommand{\dS}{\dot{S}}

\newcommand{\sign}{\mathrm{sign}}
\newcommand{\dX}{\dot{X}}
\newcommand{\bx}{{\boldsymbol{x}}}

\newcommand{\bP}{{\boldsymbol P}}
\newcommand{\dP}{\dot{P}}
\newcommand{\barP}{\overline{P}}
\newcommand{\bp}{{\boldsymbol{p}}}
\newcommand{\bq}{{\mathbf{q}}}
\newcommand{\bbR}{\mathbb{R}}
\newcommand{\np}{\Vert\bp\Vert}
\newcommand{\xps}{\widetilde{\mathrm{xps}}}
\newcommand{\xl}{\widetilde{\mathrm{x\cL}}}
\newcommand{\vol}{\mathrm{vol}}
\newcommand{\bu}{{\mathbf{u}}}
\newcommand{\bv}{{\mathbf{v}}}
\newcommand{\bw}{{\mathbf{w}}}

\newcommand{\bcL}{\boldsymbol{\cL}}
\newcommand{\rO}{\mathrm{O}}

\newcommand{\dpp}{\vcentcolon}
\newcommand{\lb}{\left[}
\newcommand{\rb}{\right]}
\newcommand{\lp}{\left(}
\newcommand{\rp}{\right)}

\newcommand{\red}[1]{{\textcolor{red}{#1}}}
\newcommand{\blue}[1]{{\textcolor{blue}{#1}}}
\usepackage[normalem]{ulem} 

\newcommand\restr[2]{{
  \left.\kern-\nulldelimiterspace 
  #1 
  \vphantom{\big|} 
  \right|_{#2} 
  }}

\begin{document}

	\chead{}
	\pdfbookmark[0]{Page de titre}{titre}
	\thispagestyle{empty}
	\newgeometry{margin=2em}

\begin{center}
	\begin{minipage}[c]{.4\linewidth}
		\raggedright\includegraphics[height=7em]{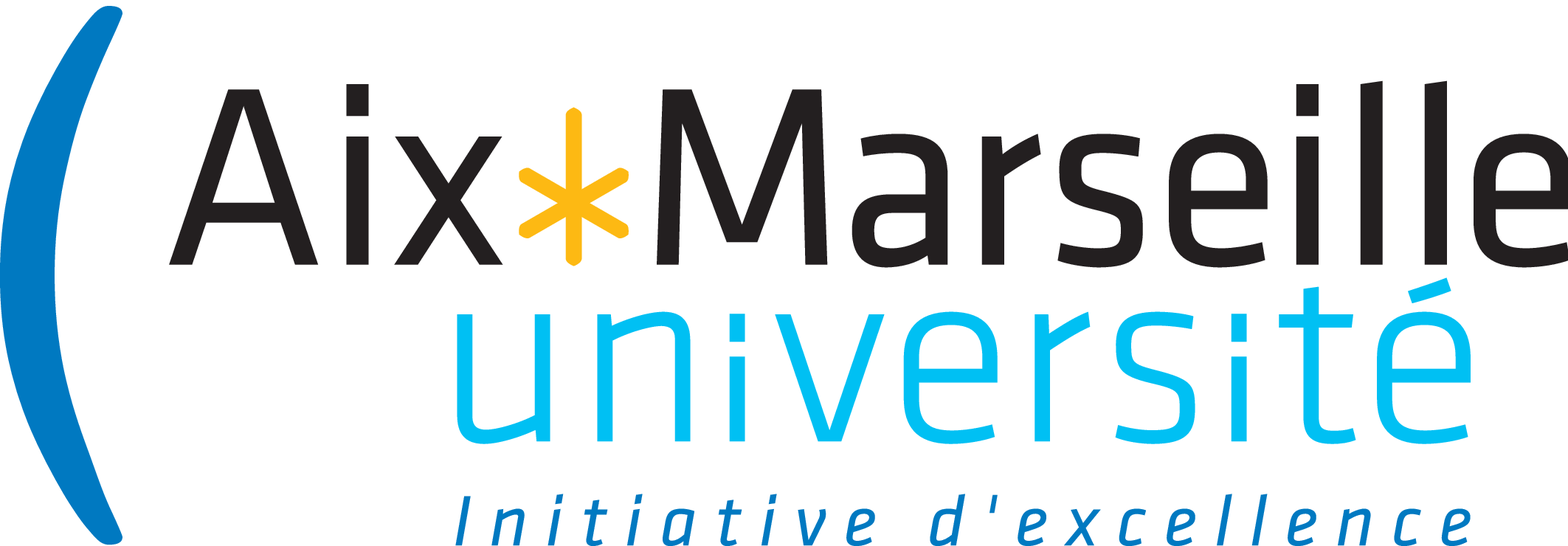}
	\end{minipage}\hfill
	\begin{minipage}[c]{.35\linewidth}
		\raggedleft\includegraphics[height=7em]{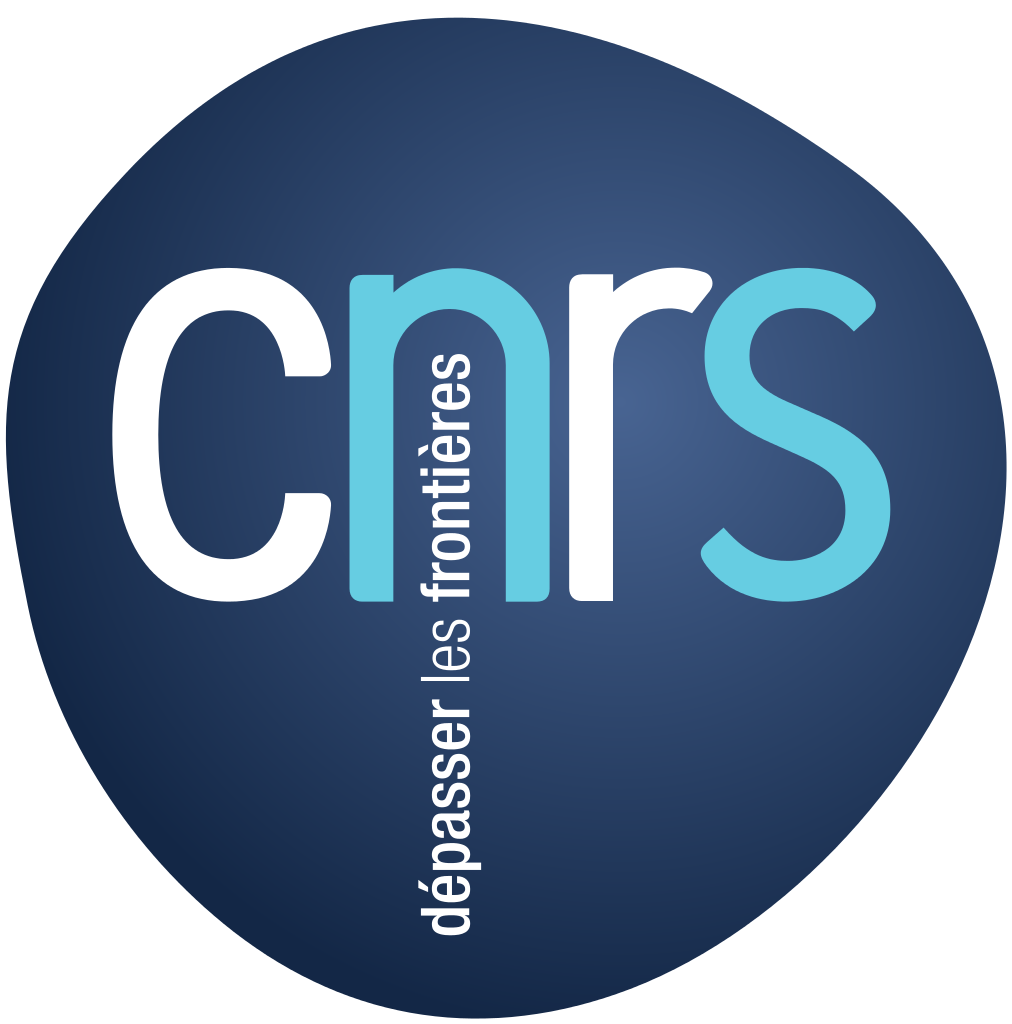}
	\end{minipage}\hfill
	\begin{minipage}[c]{.25\linewidth}
		\raggedleft\includegraphics[height=7em]{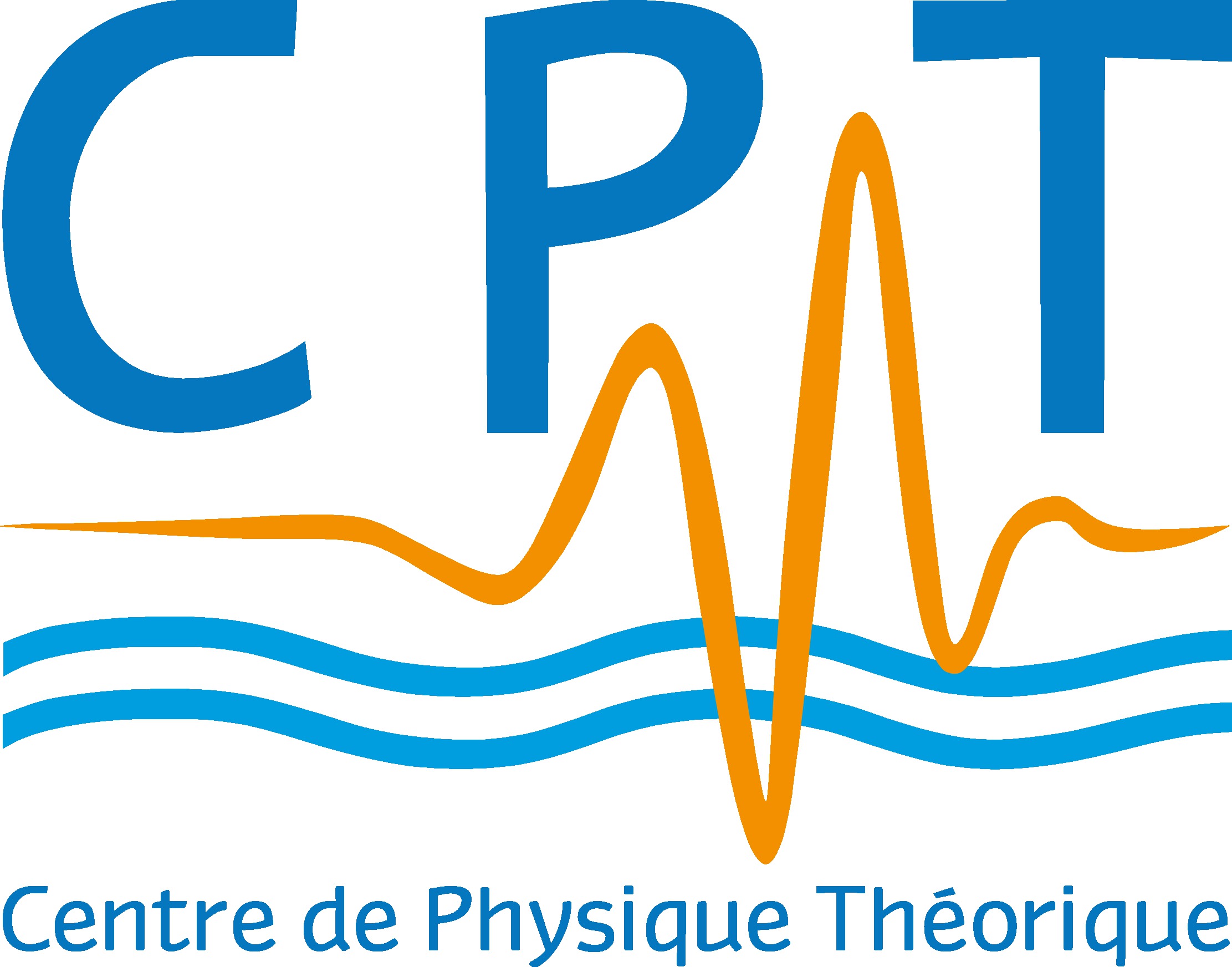} 
	\end{minipage}\hfill
\end{center}

\vspace{0.5em}

\begin{center}
	\begin{minipage}[c]{.61\linewidth}
		\dhorline{\textwidth}{4pt}
	\end{minipage}\hfill
	\begin{minipage}[c]{.37\linewidth}
		\raggedleft NNT/NL : 2020AIXM0401/045ED352
	\end{minipage}\hfill
\end{center}

\vspace{-2em}

\doublespacing
\begin{flushleft}
    École doctorale 352\\[-0.8em]
    Physique et Sciences de la Matière\\[-0.8em]
    	Physique Théorique et Mathématique\\[-0.4em]
    Centre de Physique Théorique UMR 7332
    \vspace{1em}
\end{flushleft}
\vspace{1em}
\begin{center}
    {\HUGE\textbf{\textcolor{cyanamu}{THÈSE DE DOCTORAT}}}\\
	{\Large Soutenue à Aix-Marseille Université}\\
	{\Large le 3 décembre 2020 par}\\
	\vspace{1em}
	{\Huge Lo\"ic \scshape{Marsot}}\\
    \vspace{2em}
    \hrulefill \\
    \vspace{-1.5em}
	{\Huge Geometric studies of the interplay between\\ spin and gravity}\\
    \hrulefill
\end{center}
\singlespacing

\vspace{2.5em}

\begin{center}
    	\textbf{Composition du jury \quad}
    	
	\begin{minipage}[t]{.48\linewidth}
	    \vspace{.5em}

	    \vspace{1em}
        \begin{tabular}{p{12em} p{9.5em}}
        Simone SPEZIALE (DR) & Président du jury \\
        	CNRS/INP \\
            \\
        Ruth DURRER (PR) & Rapporteure \\
        	Université de Genève \\
            \\
        Francesca VIDOTTO (AP) & Examinatrice \\
        	Western University \\
            \\
        	Charling TAO (DR) & Invitée \\
        	CNRS/IN2P3 \\
        \end{tabular}
	\end{minipage}\hfill
	\begin{minipage}[t]{.03\linewidth}
	    \dvertline{4pt}{-12.8em}
	\end{minipage}\hfill
	\begin{minipage}[t]{.49\linewidth}
	    \vspace{.5em}

	    \vspace{1em}
        \begin{tabular}{p{14.5em} p{9.5em}}
        Serge LAZZARINI (PR) & Directeur de thèse \\
        	Aix-Marseille Université \\
        	    \\
        	Francisco José HERRANZ (PR) & Rapporteur \\
        	Universidad de Burgos \\
            \\
        Thomas SCH\"UCKER (PR EM) & Examinateur \\
        	Aix-Marseille Université \\
        \end{tabular}
	\end{minipage}\hfill
\end{center}

\restoregeometry
	\newpage
	\pdfbookmark[0]{Affidavit}{affidavit}
	\thispagestyle{empty}
	\iftrue 
    Je soussigné, Lo\"ic Marsot, 
    déclare par la présente que le travail présenté dans ce manuscrit est mon propre travail, réalisé sous la direction scientifique de Serge Lazzarini, 
    dans le respect des principes d’honnêteté, d'intégrité et de responsabilité inhérents à la mission de recherche. Les travaux de recherche et la rédaction de ce manuscrit ont été réalisées dans le respect à la fois de la charte nationale de déontologie des métiers de la recherche et de la charte d’Aix-Marseille Université relative à la lutte contre le plagiat.
    
    Ce travail n'a pas été précédemment soumis en France ou à l'étranger dans une version identique ou similaire à un organisme examinateur.\\
    
    Fait à Marseille le 18/12/2020,
    
    \begin{flushright}\includegraphics[width=120px,height=40px]{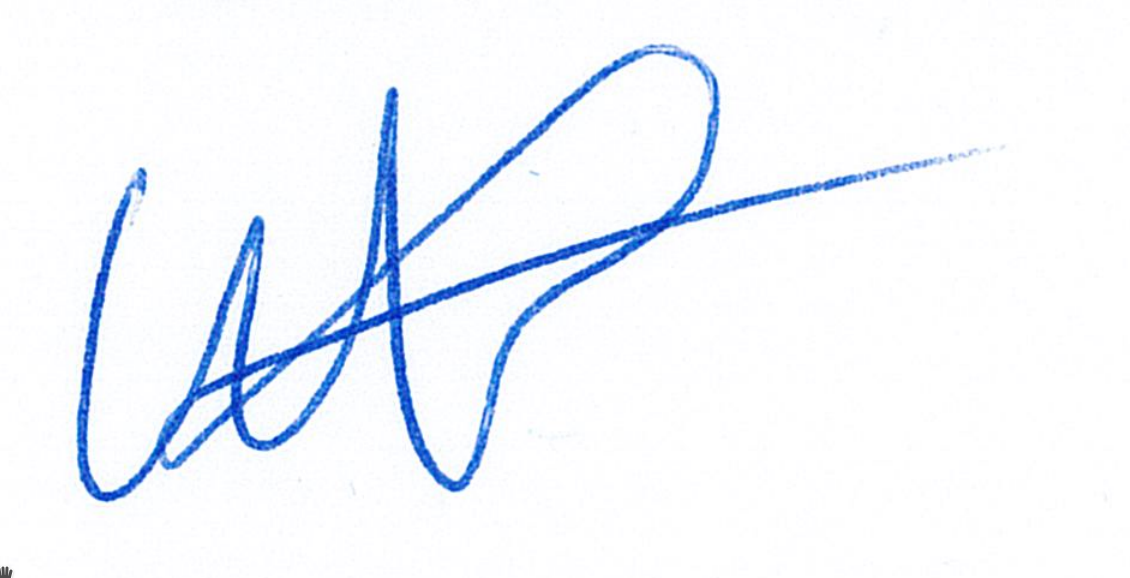}\end{flushright}
\fi

~\vfill
\begin{center}
	\begin{minipage}[c]{0.25\linewidth}
		\includegraphics[height=35px]{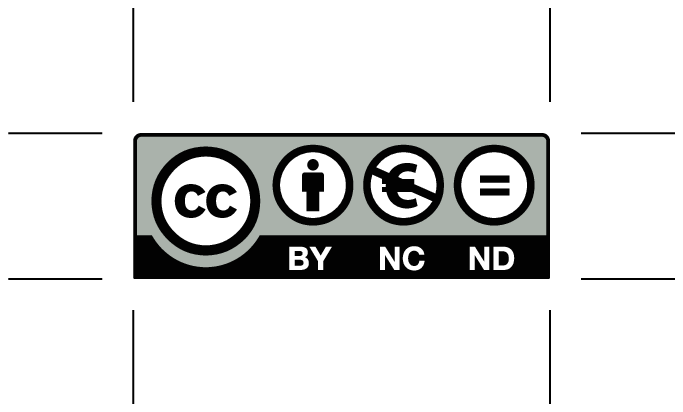}
	\end{minipage}\hfill
\end{center}

Cette \oe{}uvre est mise à disposition selon les termes de la \href{https://creativecommons.org/licenses/by-nc-nd/4.0/deed.fr}{Licence Creative Commons Attribution - Pas d’Utilisation Commerciale - Pas de Modification 4.0 International}. 

	\chapter*{Abstract}					
	\selectlanguage{english}

This thesis is the conclusion and summaries some of my works carried out at the Centre de Physique Th\'eorique, under the supervision of Serge Lazzarini. Two aspects are presented in the manuscript, both aiming at studying the effect of the spin of elementary particles on otherwise known theories. 

First is a study of the L\'evy-Leblond--Newton (LLN) equation, based on the works~\cite{LazzariniM20,Marsot17}. The LLN equation is used to describe the evolution of a quantum system with spin one half that is coupled to its own gravitational potential. After reviewing the (accidental) symmetries in non relativistic Quantum Mechanics, and how to geometrize them with the help of Bargmann structures, we recall what is the L\'evy-Leblond equation: it is to the Schr\"odinger equation what the Dirac equation is to the Klein--Gordon equation. Then, we recall some results of the Schr\"odinger--Newton (SN) equation, and write down the conserved quantities for this equation. The study of the LLN equation is aimed at describing this system in a fully covariant way, which is done through the help of Bargmann structures. This covariant formulation then helps to derive the dynamical symmetries of the equation, and its conserved quantities. The symmetry group of this equation turns out to be the Schr\"odinger--Newton group, that was derived to be the symmetry group of the SN equation in \cite{DuvalL15}. The conserved quantities of the LLN equation are computed, which are compared to the conserved quantities of the SN equation.  

The second part deals with the trajectory of particles with spin in General Relativity. Giving first an account on the extensive literature on the subject, especially highlighting Souriau's geometric method to obtain the Mathisson--Papapetrou--Dixon (MPD) equations, we discuss the different possible Spin Supplementary Conditions (SSC) that exist to close the system of MPD equations. We then recall how to derive the Souriau--Saturnini equations from the MPD equations, which describe the trajectory of photons in curved spacetime, assuming the Tulczyjew SSC holds. After reviewing a few applications, we present some works \cite{ChDLMTS,Marsot19}, where we have applied these equations to photons, respectively, in a Schwarzschild spacetime, and in a spacetime deformed by gravitational waves. In the first work \cite{ChDLMTS}, we looked for, and found, gravitational birefringence. That is, the trajectory, when taking the spin of the photon into account, deviates from the geodesic plane. This deviation depends on the helicity of the photon, and its wavelength. We shall also compare the predictions of \cite{ChDLMTS} to existing literature, and comment about possible experimental observations. The second example \cite{Marsot19}, a photon in gravitational wave background, consists in determining whether such spin effect could be observed in gravitational interferometry experiments. However, we found that the effect of the laser's polarisation on the interferometry pattern is many orders of magnitude lower than what we can detect with current technology. We shall also comment about the usefulness to consider the cosmological constant in these computations. 

\vspace{0.5cm}
Keywords: geometry, Bargmann structures, Lévy-Leblond--Newton equation, birefringence of light, Spin Hall Effect of Light, spin, helicity

\selectlanguage{english}

	\addcontentsline{toc}{chapter}{Abstract}
	
	\chapter*{Résumé}					
	\selectlanguage{french}

\vspace{0.5cm}
Cette thèse conclut et résume une partie de mes travaux au Centre de Physique Théorique, effectués sous la supervision de Serge Lazzarini. Deux thématiques sont abordées ici, toutes deux essayant de combler les lacunes de théories existantes, en incorporant les effets de spin, ou de polarisation, de particules élémentaires, qui sont souvent négligés.

En premier lieu, nous verrons une étude de l'équation de Lévy-Leblond--Newton (LLN) basée sur les travaux \cite{LazzariniM20,Marsot17}. Cette équation décrit l'évolution d'un système quantique consistant d'une particule élémentaire avec spin soumise à son propre potentiel gravitationnel. Après avoir revu les symétries (accidentelles) en mécanique quantique non relativiste, et comment les géométriser grâce aux structures de Bargmann, nous reverrons ce qu'est l'équation de Lévy-Leblond. Elle est à l'équation de Schr\"odinger ce que l'équation de Dirac est à l'équation de Klein--Gordon. Ensuite, nous reverrons quelques résultats à propos de l'équation de Schr\"odinger--Newton (SN), notamment ses symétries et les quantités conservées. Cette étude de l'équation de LLN a pour but de l'écrire d'une manière tout à fait covariante, ce qui est accompli en l'écrivant sur une structure de Bargmann. Cette formulation covariante a l'avantage de faciliter l'étude des symétries dynamiques de l'équation, et de ses quantités conservées. Le groupe de symétrie de cette équation se trouve être le groupe de Schr\"odinger--Newton, qui a été trouvé comme étant le groupe de symétrie de l'équation de SN \cite{DuvalL15}. Les quantités conservées de l'équation de LLN seront aussi déduites de cette analyse, et nous les comparerons aux quantités conservées de l'équation de SN.

La deuxième partie du manuscrit traite de la trajectoire des particules élémentaires en relativité générale lorsqu'on ne néglige pas leur spin. Tout d'abord nous reverrons la littérature existante sur ce sujet, notamment en soulignant la méthode géométrique de Souriau pour obtenir les équations de Mathisson--Papapetrou--Dixon (MPD). Ces équations n'étant pas fermées, nous discuterons aussi des différentes conditions supplémentaires sur le spin présentes dans la littérature qui permettent de les compléter. Ensuite, nous rappellerons comment obtenir les équations de Souriau--Saturnini à partir des équations de MPD, et en supposant l'équation supplémentaire de Tulczyjew pour décrire la trajectoire d'un photon avec son spin dans un espace-temps courbe. Après avoir rappelé quelques applications des équations de Souriau--Saturnini, nous présenterons deux résultats issues de \cite{ChDLMTS,Marsot19}, où nous avons appliqué ces équations dans, respectivement, un espace-temps de Schwarzshild, puis dans un espace temps déformé par une onde gravitationnelle. La première étude \cite{ChDLMTS} traite de la biréfringence gravitationnelle, c'est-à-dire que lorsqu'on prend la polarisation du photon en considération, sa trajectoire sort du plan géodésique usuel. Nous trouvons que le signe de l'angle que fait la trajectoire avec le plan dépend de l'hélicité du photon, et l'amplitude dépend de sa longueur d'onde et de la masse de l'étoile. La deuxième étude~\cite{Marsot19} essaye de déterminer si une onde gravitationnelle peut perturber la trajectoire d'un photon suffisamment pour être observable lors des expériences d'interférométries gravitationnelles. Bien que nous trouvions un effet, son ordre de magnitude est largement en dessous de ce que nous pouvons détecter avec la technologie actuelle. Nous commenterons aussi sur l'utilité de la constante cosmologique dans ces calculs.

\vspace{0.5cm}
Mots-clés : géometrie, structures de Bargmann, équation de Lévy-Leblond--Newton, biréfringence de la lumière, Spin Hall Effect of Light, spin, hélicité

\selectlanguage{english}
	\addcontentsline{toc}{chapter}{Résumé}

	\chapter*{Acknowledgements}			
	This thesis could not have happened without the numerous social links that were formed during these past years spent at the laboratory.

My first thoughts go to Christian Duval, who is sadly no longer with us to see the completion of this work. He introduced me to the world of research by accepting me for an internship during my Master studies, and thus welcoming me to the Centre de Physique Théorique. It is Christian who developed my interest for Mathematical Physics, through his teaching skills, his passion, and his exceptional knowledge. For me, he is the model of an accomplished physicist that I can only hope to follow.

Then, I would like to thank Serge Lazzarini, my PhD advisor, who accepted to supervise me in an internship, and then accepted me as a PhD student. I would like to thank him for his support, his interest in Science and his attention to details. With Serge, I was able to follow my interests in Mathematical Physics, and keep learning in the process. He was also of immense help during my PhD thesis, by answering my questions, whether they were related to Science or to administration. 

I am also very grateful to Thomas Sch\"ucker who welcomed me when I needed help. He suggested that we work together on the topic of birefringence of light, which makes up the second chapter of this thesis. I am also thankful for the countless hours we spent chatting about research or any other topic.

In general, I would like to thank the Centre de Physique Théorique for being especially welcoming in the wonderful setting that is the Campus de Luminy. In particular, I would like to thank Thierry Masson, Laurent Raymond, Thomas Krajewski, Alberto Verga, and Xavier Léoncini for always so stimulating discussions. I also do not forget my fellow PhD student friends.

A special mention to Jordan François, for discussions about what it takes to be a PhD student, and how to prepare properly to the future.

	\addcontentsline{toc}{chapter}{Acknowledgements}

    \microtypesetup{protrusion=false}	
	\tableofcontents					

    \microtypesetup{protrusion=true}	
	
	\chapter*{Résumé long}			
	\selectlanguage{french}
\newcounter{w}
\newcounter{p}
\newcounter{h}

\tikzstyle{hidden} = [dashed,line width=1.1pt]
\tikzstyle{lesser} = [line width=1.2pt]
\tikzstyle{normal} = [line width=0.8pt]
\tikzstyle{normalh} = [dashed,line width=0.8pt]
\tikzstyle{arrow} = [line width=0.9pt, draw, -latex']
\tikzstyle{labels} = [->]

\tikzset{middlearrow/.style={
        decoration={markings,
            mark= at position #1 with {\arrow{>}} ,
        },
        postaction={decorate}
    }
}

\addcontentsline{toc}{chapter}{Résumé long en français}
\section{Introduction}

Ce manuscrit regroupe deux thématiques apparaissant distinctes mais partageant néanmoins un point commun : l'étude de l'interaction d'une particule ou d'un système élémentaire avec un spin non nul et un champ de gravité. La première partie traite de l'équation de Lévy-Leblond--Newton (LLN) basée sur les travaux \cite{LazzariniM20,Marsot17}. Cette équation décrit l'évolution d'un système quantique formé d'une particule élémentaire avec spin$-\half$ soumise à son propre potentiel gravitationnel. Nous verrons comment calculer le groupe de symétrie dynamique de cette équation, ainsi que les quantités conservées associées à chaque degré de symétrie. La deuxième partie, quant à elle, traite de la trajectoire de particules élémentaires avec spin dans un espace-temps courbe, en particulier des photons. Pour ce faire, nous étudierons les équations de Souriau--Saturnini, qui sont des équations du mouvement pour des particules sans masse à spin 1 obéissant aux équations de Mathisson--Papapetrou--Dixon (MPD) et à la condition supplémentaire de spin de Tulczyjew. Nous verrons divers exemples d'application de ces équations, notamment dans un espace-temps de Schwarschild, basé sur \cite{ChDLMTS}, et dans un espace-temps plat déformé par une onde gravitationnelle, basé sur \cite{Marsot19}.

\section{Les symétries de l'équation de Lévy-Leblond--Newton}

Il existe une certaine incohérence entre la description offerte par la mécanique quantique et celle offerte par la mécanique Newtonienne, notamment le fait qu'en mécanique quantique un paquet d'onde s'étalera, ce qu'on n'observe pas expérimentalement pour certains objets macroscopiques qui pourraient y être susceptibles, comme de fines gouttelettes. 

Dans un effort pour répondre à cette incohérence, et pour introduire la gravité dans la mécanique quantique, Diosi a proposé en 1984 de considérer l'équation de Schr\"odinger--Newton (SN) pour décrire l'étalement de paquet d'onde d'objets macroscopiques \cite{Diosi84}. Le paquet d'onde est décrit par l'équation de Schr\"odinger avec potentiel, où le potentiel est donné par le champ gravitationnel du paquet lui même. Ce système d'équations s'écrit,
\begin{align*}
i\hbar \frac{\partial \psi}{\partial t}(\bx, t) & = \left(-\frac{\hbar^2}{2m}\Delta_{\R^n} + m U(\bx, t)\right) \psi(\bx, t), \\
\Delta_{\R^n} U(\bx, t) & = 4 \pi G m |\psi(\bx, t)|^2
\end{align*} 

\subsection{\'Equation de Lévy-Leblond--Newton}

Les expériences de mécanique quantique qui pourraient étudier ce type de phénomène impliquent typiquement des particules à spin $\half$. Il est légitime de considérer une description quantique en termes de spineurs non relativistes. Pour ce faire, nous posons l'équation de Lévy-Leblond--Newton \cite{LazzariniM20}, 
\begin{equation*}
\begin{array}{ll}
\displaystyle \left\lbrace\begin{array}{l}
\displaystyle \hbar \sigma(\bpartial) \varphi + 2 m \chi = 0 \\[0.5ex]
\displaystyle i \hbar \partial_t \varphi - m U \varphi - \hbar \sigma(\bpartial) \chi = 0
\end{array}\right. \\[2ex]
\displaystyle \Delta_{\R^n} U = 4 \pi G \rho \\
\displaystyle \, \rho = m  \, \varphi^\dagger \varphi
\end{array}
\end{equation*}

Au premier abord, \'etudier les symétries de ce système d'équations n'a pas l'air aisé. Nous allons donc le simplifier en l'écrivant sur une \emph{structure de Bargmann} \cite{Bargmann54,DuvalBKP85,Eisenhart28}. Les structures de Bargmann, dénotées par le triplet $(B, g, \xi)$, sont un moyen de rendre les calculs de symétries non relativistes plus commodes. En effet, bien qu'il existe une géométrisation de l'espace-temps non relativiste, connue sous le nom de structure de Newton--Cartan, les calculs sur ces structures sont pathologiques, étant donnée les dégénérescences intrinsèques sur ces dernières.

\medskip

\setcounter{w}{9}
\setcounter{p}{1}
\setcounter{h}{7}

\tikzstyle{hidden} = [dashed,line width=1.1pt]
\tikzstyle{lesser} = [line width=1.2pt]
\tikzstyle{normal} = [line width=0.8pt]
\tikzstyle{normalh} = [dashed,line width=0.8pt]
\tikzstyle{arrow} = [line width=0.9pt, draw, -latex']
\tikzstyle{cone} = [line width=0.7pt]
\tikzstyle{labels} = [->]
\tikzstyle{carr} = [black!50!blue]
\tikzstyle{line} = [draw, -latex']
\tikzstyle{nc} = [black!50!red]

\begin{minipage}[c]{0.45\textwidth}

Les structures de Bargmann sont construites comme un fibré au dessus de Newton--Cartan. La structure résultante est Lorentzienne, ce qui lève les dégénérescences de la structure de Newton--Cartan. La métrique $g$ sur $B$ est dite de Brinkmann, et la dimension supplémentaire est générée par un champ de vecteur $\xi$ covariant constant et isotrope, \ie tel que $g(\xi, \xi) = 0$ et $\nabla \xi = 0$.
\end{minipage}\hfill
\begin{minipage}[c]{0.53\textwidth}
{\Large
\resizebox{7.5cm}{!}{
\begin{tikzpicture}[line width=1.4pt]
  \draw [lesser] (0,0) -- (0.45 * \value{w},\value{p}) -- (\value{w},0);
  \draw (\value{w},0) -- (0.55 * \value{w},-\value{p}) -- (0,0);

  \draw [line width=1.4pt] (0,0) -- (0,-\value{h});
  \draw (0.55 * \value{w}, -\value{p}) -- (0.55 * \value{w}, -\value{p} - \value{h});
  \draw (\value{w}, 0) -- (\value{w}, -\value{h});
  \draw [hidden] (0.45 * \value{w},\value{p}) -- (0.45 * \value{w},\value{p} - \value{h});
  \draw [normal] (0.9 * \value{w}, - \value{p}) circle (0.6cm) node [scale=1.4]{$B$};

  \draw [hidden] (0, - \value{h}) -- (0.45 * \value{w},\value{p} - \value{h}) -- (\value{w}, - \value{h});
  \draw (\value{w}, - \value{h}) -- (0.55 * \value{w},-\value{p} - \value{h}) -- (0, - \value{h});
  
  \draw [lesser,nc] (0, - 1.4 * \value{h}) -- (0.45 * \value{w},\value{p} - 1.4 * \value{h}) -- (\value{w}, - 1.4 * \value{h});
  \draw [nc] (\value{w}, - 1.4 * \value{h}) coordinate (nc1) -- node[pos=1,above,scale=1.4]{$\cN$} (0.55 * \value{w},-\value{p} - 1.4 * \value{h}) coordinate (nc2) -- (0, -1.4 * \value{h}) coordinate (nc3);
  \draw [normal] pic["",draw=black,-,angle eccentricity=1.2,angle radius=0.85cm] {angle=nc1--nc2--nc3};

  \draw [normal,carr] (0.725 * \value{w}, 0.5*\value{p}) -- (0.275 * \value{w}, -0.5*\value{p}) coordinate (c1);
  \draw [normal,carr] (0.275 * \value{w}, -0.5*\value{p}) -- node[pos=0.92,right,scale=1.5]{$\widetilde{\Sigma}_t$} (0.275 * \value{w}, -1*\value{h} -0.5*\value{p});
  \draw [normalh,carr] (0.725 * \value{w}, 0.5*\value{p}) -- (0.725 * \value{w}, 0.5*\value{p} - \value{h}) coordinate (c3) -- (0.275 * \value{w}, -\value{h}-0.5*\value{p}) coordinate (c2);
  \draw [normal] pic["",draw=black,-,angle radius=1.25cm] {angle=c3--c2--c1};

  \draw [normal, middlearrow={0.82},black!60!green] (0.725 * \value{w}, 0.5 * \value{p} - 1.4 * \value{h}) -- node [left,pos=0.32,scale=1.2,black!50!red] {$(x, t)\qquad$} node[scale=3.5,pos=1]{.} (0.05 * \value{w}, -1.4 * \value{h} - 1 * \value{p});
  
  \draw [line] (-0.225 * \value{w}, -1.4 * \value{h} - 0.5* \value{p}) -- node [pos=1.18,below,scale=1.2] {$T \cong \R$ (time axis)} node[below, pos=0.58,scale=1.2]{$t\qquad$} (0.325 * \value{w}, -1.5 * \value{p} - 1.4 * \value{h});
  
  \draw [arrow] (0.80 * \value{w}, -0.65 * \value{h}) -- node [near end, left, scale=1.1] {$\xi$} (0.80 * \value{w}, -0.53 * \value{h});
  \draw [arrow] (0.40 * \value{w}, -0.4 * \value{h}) -- node[pos=0,scale=3]{.} node [left,pos=0,scale=1.2]{$(x,t,s)\;$} node [near end, right,scale=1.1] {$\xi$} (0.40 * \value{w}, -0.28 * \value{h});

  \draw [dashed,cone] (0.80 * \value{w}, -0.65 * \value{h}) -- (0.80 * \value{w}, -0.77 * \value{h});
  \draw [dashed,cone] (0.80 * \value{w}, -0.65 * \value{h}) -- (0.745 * \value{w}, -0.74 * \value{h});
  \draw [cone] (0.80 * \value{w}, -0.65 * \value{h}) -- (0.855 * \value{w}, -0.56 * \value{h});
  \draw [cone,rotate around={160:(0.827 * \value{w}, -0.55 * \value{h})}] (0.827 * \value{w}, -0.55 * \value{h}) ellipse (0.029 * \value{w} and 0.02 * \value{h});
  \draw [cone,dashed,rotate around={160:(0.773*\value{w}, -0.75 * \value{h})}] (0.773*\value{w}, -0.75 * \value{h}) ellipse (0.0295 * \value{w} and 0.02 * \value{h});
  
  \node[draw, align=center,scale=1.1] at (1.05*\value{w}, \value{p}+0.6) (barg) {Bargmann\\space-time-action\\$(B, \rg, \xi)$};
  \draw [labels] (barg) -- (0.83 * \value{w}, -0.1 * \value{p});
  
  \node[draw=black!50!blue, align=center,scale=1.1] at (-0.04 * \value{w}, \value{p}+0.6) (carr) {Carroll\\space-action\\$(\widetilde{\Sigma}_t, \Upsilon, \widetilde{\xi})$};
  \node[draw=none] at (0.4 * \value{w}, -0.5*\value{p}) (carr2) {};
  \draw [labels] (carr) to [out=0,in=100] (carr2);
  
  \node[draw=black!60!green, align=center,scale=1.1] at (1.1*\value{w}, -\value{h} - 1.5*\value{p}) (euclide) {Euclidean\\space\\$(\Sigma_t, h)$};
  \node[draw=none] at (0.55*\value{w}, -1*\value{h}-2.83*\value{p}) (euclide2) {};
  \draw [labels] (euclide) to [out=180,in=45] (euclide2);
  
  \node[draw=black!50!red, align=center,scale=1.1] at (1.03 * \value{w}, -\value{h} - 4.4 * \value{p}) (nc) {Newton-Cartan\\space-time\\$(\cN, h, \theta, \nabla^\cN)$};
  \node[draw=none] at (0.75 * \value{w}, -\value{h} - 2.8 * \value{p}) (nc2) {};
  \draw [labels] (nc) to [out=140,in=-20] (nc2);
  
  \draw [normalh, middlearrow=0.4] (0.40 * \value{w}, -0.4 * \value{h}) -- (0.40 * \value{w}, -\value{h} - 0.75 * \value{p});
  \draw [normal, middlearrow=0.36] (0.40 * \value{w}, -\value{h} - 0.75 * \value{p}) -- node [left,pos=0.32,scale=1.5]{$\pi$} node [pos=1,scale=3]{.} (0.40 * \value{w}, -1.12 * \value{h} - 2.22 * \value{p});
\end{tikzpicture}
}}
\end{minipage}\hfill

\smallskip

En effet, sur de telles structures, les équations de Lévy-Leblond--Newton s'écrivent simplement en termes d'opérateurs covariants,
\begin{center}
\begin{tabular}{lll}
$\begin{array}{ll}
\left\lbrace\begin{array}{l}
\hbar \sigma(\bpartial) \varphi + 2 m \chi = 0 \\
i \hbar \partial_t \varphi - m U \varphi - \hbar \sigma(\bpartial) \chi = 0
\end{array}\right. \\
\Delta_{\R^n} U = 4 \pi G \rho
\end{array}$
& 
$\Rightarrow$
&
$\begin{array}{l}
\dirac(g) \Psi = 0 \\
L_\xi \Psi = i \frac{m}{\hbar} \Psi \\
\Ric(g) = 4 \pi G \rho \vartheta \otimes \vartheta
\end{array}$
\end{tabular}
\end{center}
où $\Psi$ est un 4-spineur (en fait, une densité spinorielle) sur Bargmann. La deuxième équation, la relation d'équivariance, implique localement $\Psi = \left(\begin{array}{cc}
\varphi \\ \chi
\end{array}\right) e^\frac{ims}{\hbar}$.

Il sera donc plus simple de calculer les symétries du système d'équations sur les structures de Bargmann.

\subsection{Calcul des symétries}
Dire qu'on veut calculer le groupe de symétries du système d'équations de LLN implique trouver le groupe qui, en agissant sur une solution $\Psi$ du système, la transforme en une autre solution $\Phi^* \Psi$. Par exemple, si on prend la première équation du système, alors si $\dirac(g) \Psi = 0$, on veut $\dirac(g) \Phi^* \Psi = 0$.

Ayant des opérateurs covariants, ce calcul est simple. Les opérateurs des deux premières équations se trouvent être invariants sous transformations conformes, c'est-à-dire si,
\begin{align*}
\Phi^* g = \lambda \, g, \\
\Phi^* \xi = \nu \, \xi,
\end{align*}
avec $d \lambda \wedge \vartheta = 0$ et $d\nu = 0$.

Le calcul des symétries de la troisième équation montre qu'elle est préservée lorsque les deux facteurs conformes $\lambda$ et $\nu$ sont reliés par $\lambda^{2-\frac{n}{2}} \nu^3 = 1$.

Au final, le groupe de symétries de l'équation de Lévy-Leblond--Newton est donné par,
\begin{equation*}
\boxed{\mathrm{LLN}(B,\rg,\xi) = \left\lbrace\Phi \in \Diff(B) \vert \Phi^*\rg = \lambda \rg, \Phi^* \xi = \nu \xi, \lambda^{2-\frac{n}{2}} \nu^3 = 1\right\rbrace}
\end{equation*}

Si $n = 3$, ce groupe est de dimension 12.

\medskip

Plus concrètement, ce groupe agit sur les coordonnées de Bargmann $(\bx, t, s)$ comme,
\begin{equation*}
\left\lbrace\begin{array}{l}
\displaystyle \widehat{\bx} = \frac{A \bx + \bb t + \bc}{g}, \\
\displaystyle \widehat{t} = \frac{d t + e}{g}, \\
\displaystyle \widehat{s} = \frac{1}{\nu}\left( s - \langle \bb, A \bx \rangle - \frac{\Vert\bb\Vert^2}{2} t + h\right),
\end{array}\right.
\end{equation*}
où $A \in \SO(3), \bb, \bc \in \R^3, d, e, g, h\in \R$, et $d \, g = \nu$. L'action du groupe sur un spineur $\psi(\bx, t)$ est donnée par,
\begin{equation*}
\begin{aligned}
\left[\Phi^* \psi\right](\bx, t) = & \, \nu^{\frac{3(n+1)}{2(n-4)}} \exp\left(\frac{im}{\nu \hbar} \left(  - \langle\textbf{b}, A \textbf{x}\rangle - \frac{||\textbf{b}||^2 t}{2} + h\right)\right) \\
& \left(\begin{array}{cc}
d^{-1/2} \, a & 0 \\
-\frac{i}{2} d^{1/2} \sigma (\bb) a  & d^{1/2} \, a
\end{array} \right) \,
\psi\left(\frac{A \textbf{x} + \textbf{b} t + \textbf{c}}{g}, \frac{d t + e}{g} \right),
\end{aligned}
\end{equation*}
de telle sorte que si $\psi$ est une solution de l'équation de LLN, alors $\Phi^* \psi$ l'est aussi.

\subsection{Calcul des quantités conservées}
Ayant calculé les symétries dynamiques de l'équation de LLN, on peut maintenant déterminer les quantités conservées associées à chaque symétrie. Pour ce faire, on utilise un principe actionel, et on calcule les quantités conservées à la Noether. L'équation de LLN sur Bargmann est formée de 3 équations, mais les deux dernières, l'équivariance et l'équation de Poisson, sont intrinsèques aux structures de Bargmann. On considère donc seulement une action pour l'équation d'onde sur Bargmann, c'est-à-dire l'équation de Dirac sans masse,
\begin{equation*}
S_D[\psi,\rg] = i \hbar \int_B \overline \psi \diracf \psi \volg d^Nx, \quad N = 3 + 2.
\end{equation*}

Après calculs, on trouve les quantités conservées associées suivantes,
\begin{equation*}
\left\lbrace
\begin{array}{ll}
\displaystyle E = \int \varphi^\dagger H \varphi \, d^3 \bx \qquad \qquad & \text{énergie} \\[4mm]
\displaystyle \boldsymbol{P} \equiv \int \boldsymbol{\cP} \, d^3 \bx = \frac{i\hbar}{2} \int \left((\bnabla \varphi)^\dagger \varphi - \varphi^\dagger \bnabla \varphi\right) d^3 \bx & \text{impulsion} \\[4mm]
\displaystyle \boldsymbol{J} = \int \bx \times \boldsymbol{\cP} \, d^3 \bx + \frac{\hbar}{2} \int \varphi^\dagger \boldsymbol{\sigma} \varphi \, d^3\bx & \text{moment angulaire} \\[4mm]
\displaystyle M = m \int \varphi^\dagger \varphi \, d^3 \bx & \text{masse} \\[4mm]
\displaystyle \boldsymbol{G} = t \boldsymbol{P} - m \int \varphi^\dagger \varphi \, \bx \, d^3 \bx & \text{boost} \\[4mm]
\displaystyle D = \frac{n + 2}{n-4}\, t E + \frac{3}{n-4} \int \bx \cdot\! \boldsymbol{\cP} \, d^3 \bx & \text{dilatation}\ (n=3)
\end{array}
\right.
\end{equation*}
\subsection{Conclusions}

Les équations de Lévy-Leblond--Newton sont un autre exemple pour lequel les structures de Bargmann semblent indispensables. En effet, réécrire ces équations sur ces structures nous a permis de les étudier d'une manière tout à fait covariante, sans aucune dégénérescence, ce qui rend les calculs du groupe de symétrie dynamique bien plus aisés. Le groupe de symétrie de LLN se trouve être en fait le même groupe de symétries que les équations de SN \cite{DuvalL15}, ce qui n'est pas surprenant. Nous avons néanmoins pu calculer l'action de ce groupe sur des spineurs non relativistes comme représentation projective unitaire, ainsi que les quantités conservées. Non mentionné dans ce résumé court mais long, il est possible de généraliser les équations de LLN en utilisant la métrique la plus générale sur Bargmann, ce qui est fait dans le chapitre 1.

\section{La trajectoire de particules à spin en Relativité Générale}
\subsection{Le principe de covariance général (PCG) et équations de Souriau--Saturnini}
\begin{minipage}[c]{0.5 \textwidth}
\begin{figure}
\large
\resizebox{6.5cm}{!}{
\begin{tikzpicture}[line width=1.2pt,scale=1, every node/.style={transform shape}]
  \draw (0,0) to[out=5,in=175] coordinate[pos=0.5](midmid) (6,0) to[out=95,in=-95] node[pos=0.92,left,scale=1]{$\Met(M)$} (6,5) to[out=-175,in=-5] coordinate[pos=0.5](topmid) (0,5) to[out=-85,in=85] (0,0);
  \draw (-0.3,-1.85) to[out=13,in=167] node[pos=0.5,below,scale=1.1]{$[g]$} node[pos=0.2,below,align=center]{$\Geom(M) = \quad$ \\ $\Met(M)/\Diff_c(M) \quad$} coordinate[pos=0.5](botmid) (6.3,-1.85);
	  \draw [pattern=dots] (midmid) to[out=110,in=-110] node[pos=0.8,left,scale=1.1]{$\, \cO_g$} (topmid) to [out=-70,in=70] (midmid);
  \draw [->] (3,1.6) -- node[pos=0,scale=1.2] {$\bullet$} node[pos=0,scale=1.1,below] {$g$} node [pos=0.95, right,scale=1.1] {$\delta g$} (4.3,2.4);
  \draw [->] (3,1.6) to[out=130,in=-130] node [pos=0.95, above,scale=1.1] {$a^{\boldsymbol{*}} g$} (3,2.5);
  \draw [dashed] (midmid) -- node [pos=1.05,scale=1.2] {$\bullet$} (botmid);
  \draw (botmid) -- +(3,0) -- node[pos=0.85,above] {$T_{[g]}\Geom(M)$} +(-3,0);
  \draw [->] (botmid) -- node[pos=0.85,above] {$\delta \Gamma$} +(1.4,0);
\end{tikzpicture}
}
\end{figure}
\end{minipage}\hfill
\begin{minipage}{0.48\textwidth}
Pour Souriau \cite{Sou74}, l'invariance sous difféomorphismes en Relativité Générale implique que l'espace de toutes les métriques $\Met(M)$ d'une variété $M$ est trop grand pour faire de la physique. Il considère à la place ``l'espace des géométries'', défini par le quotient $\Geom(M) = \Met(M)/\Diff_c(M)$, où $\Diff_c(M)$ dénote les difféomorphismes à support compact.

L'information géométrique de l'Univers est encodée dans une distribution tensorielle $\cT \in T^*_{g} \Met(M)$ telle que,
\begin{equation}
\label{pgc_resume}
\boxed{\cT(L_\xi g) = 0, \; \forall \xi \in \Vect_c(M)}
\end{equation}
\end{minipage}

Pour utiliser ce principe, il faut faire l'hypothèse qu'une particule peut être décrite par une distribution multipolaire sur sa ligne d'univers $\cC$.

En considérant seulement le premier moment de la particule, qui est lié à sa masse, une telle distribution s'écrit,
\begin{equation*}
\cT_\cC(\delta g) = \half \int_\cC \theta^{\mu\nu}\delta g_{\mu\nu} \, ds.
\end{equation*}

De \eqref{pgc_resume}, on trouve que $\theta^{\mu\nu}$ est exprimée en fonction de la quadri-vitesse $\dot{X}$ et de la quadri-impulsion $P$ de la particule test, $\theta^{\mu\nu} = \dot{X}^\mu P^\nu$, et que cette particule suit une géodésique, \ie $\dot{P} = 0$ \& $\dot{X} \parallel P$ \footnote{Le point au dessus de $X$ symbolise la dérivée usuelle, alors que le point au dessus d'un tenseur, comme $P$ symbolise la dérivée covariante.}.

\smallskip
Ce calcul peut être généralisé en incluant le moment dipolaire de la particule test, qui va être relié à son moment angulaire (intrinsèque ou non), dans la distribution tensorielle,
\begin{equation*}
\cT_\cC(\delta g) = \half \int_\cC \big( \theta^{\mu\nu}\delta g_{\mu\nu} + \Phi^{\rho\mu\nu}\nabla_\rho \delta g_{\mu\nu}\big) \, ds.
\end{equation*}

Le PCG implique ensuite que la particule test suit les équations dites de Mathisson--Papapetrou--Dixon (MPD) \cite{Mat37,Pap51,Dix70},
\begin{align*}
\dot{P}^\mu & = - \half {R^\mu}_{\rho\alpha\beta} S^{\alpha \beta} \dot{X}^\rho, \\
\dot{S}^{\mu\nu} & = P^\mu \dot{X}^\nu - P^\nu \dot{X}^\mu.
\end{align*}

Ces équations sont largement acceptées dans la littérature, mais possèdent un défaut crucial : elles ne sont pas fermées. Une particule test possédant un moment dipolaire est décrite par sa position, son impulsion, et son tenseur de spin. Or, il n'y a ici que des équations différentielles sur l'impulsion et le tenseur de spin.

\smallskip

Pour résoudre cette dégénérescence, il faut postuler des conditions supplémentaires que la particule test doit satisfaire. Pour un photon, on postule la contrainte de Tulczyjew, ${S^\mu}_\nu P^\nu = 0$, et une masse conservée nulle $P_\mu P^\mu = 0$. Les équations de Souriau--Saturnini sont alors les équations de MPD auxquelles on adjoint ces deux conditions supplémentaires. Elles s'écrivent \cite{Sat76},
\begin{align}
\dot{X}^\mu & = P^\mu+\frac{2}{{R(S)^\lambda}_\sigma {S_\lambda}^\sigma}{S^\mu}_\nu {R(S)^\nu}_\rho P^\rho\,, \label{ss_resume1}\\
\dot{P}^\mu & = -s\,\frac{\Pf({R(S)^\mu}_\nu)}{{R(S)^\lambda}_\sigma {S_\lambda}^\sigma}\,P^\mu\,,\\
\dot{S}^{\mu\nu} & = P^\mu\dX^\nu-\dX^\mu P^\nu. \label{ss_resume2}
\end{align}
avec $-\half \Tr(S^2) = s^2, \, s = \pm \hbar$, ${R(S)^\mu}_\nu := {R^\mu}_{\nu\alpha\beta} S^{\alpha\beta}$, et ${R(S)^\lambda}_\sigma {S_\lambda}^\sigma \neq 0$.

\smallskip

Voici trois exemples d'application de ces équations. En premier lieu, dans un espace-temps de de Sitter, puis pour étudier les phénomènes de biréfringence de la lumière dans un espace-temps de Schwarzschild lors du ``weak lensing'', et enfin pour étudier l'interaction du spin d'un photon et d'une onde gravitationnelle. 

\subsection{Espace-temps de de Sitter}
Cet espace-temps étant maximalement symétrique, le tenseur de courbure peut être paramétrée par $R_{\mu\nu\lambda\rho} = \frac{\Lambda}{3}\left(g_{\mu\lambda} g_{\nu\rho} - g_{\mu\rho} g_{\nu\lambda}\right)$. Il suit,
\begin{equation*}
R(S)_{\mu\nu} := R_{\mu\nu\lambda\rho} S^{\lambda\rho} = \frac{2\Lambda}{3} S_{\mu\nu}.
\end{equation*}

Par définition, $P$ est dans le noyau de $S$, ce qui implique que les équations de Souriau--Saturnini \eqref{ss_resume1}--\eqref{ss_resume2} se réduisent aux équations géodésiques \cite{Sat76},
\begin{align*}
\dot{X}^\mu & = P^\mu,\\
\dot{P}^\mu & = 0,\\
\dot{S}^{\mu\nu} & = 0,
\end{align*}
où la dernière équation montre que le tenseur de spin est transporté parallèlement.

\subsection{Biréfringence gravitationnelle dans un espace-temps de Schwarzschild}

Les calculs dans une métrique de Schwarzschild étant plus complexes que dansl e cas précédent, il est commode d'utiliser des coordonnées dites isotropiques $(\bx,t)$, avec $\bx=(x^1,x^2,x^3)$, de telle sorte que la métrique s'écrive,
\begin{equation*}
\rg = -\lp\frac{r+a}{r}\rp^4\Vert d\bx\Vert^2+\lp\frac{r-a}{r+a}\rp^2\,dt^2,
\end{equation*}
avec $r := \Vert \bx \Vert = \sqrt{\bx\cdot\bx}$ et $0<a<r$, où $a=\half GM$ est le rayon de Schwarzschild.

On définit la quadri-impulsion à l'aide du vecteur impulsion $\bp = (p_1, p_2, p_3)$, par
\begin{equation*}
P:= (P^\mu) = \left(
\begin{array}{c}
\displaystyle
\frac{r^2}{(r+a)^2} \, \bp \\[10pt]
\displaystyle
 \frac{r+a}{r-a} \, \np
\end{array}\right),
\end{equation*}
telle que $P^2 = 0$. Le tenseur de spin est quant à lui donné par, avec $\bs = (s_1, s_2, s_3)$,
\begin{equation*}
S=({S^\mu }_\nu)=\left(
\begin{array}{cc}
j(\bs)&\displaystyle
-\frac{(\bs\times\bp)}{\np}\frac{r^2(r-a)}{(r+a)^3}\\[6pt]
\displaystyle
-\frac{(\bs\times\bp)^T}{\np}\frac{(r+a)^3}{r^2(r-a)}&0
\end{array}\right),
\end{equation*}
avec $SP = 0$, $j(\bs):\bp\mapsto\bs\times\bp$, et le spin longitudinal se trouvant être conservé,
\begin{equation*}
-\half\Tr(S^2) = \left(\frac{\bs\cdot\bp}{\np}\right)^2 = s^2.
\end{equation*}

Il y a de plus 4 quantités conservées associées aux symétries de l'espace-temps, l'énergie $\cE$ et les 3 composantes du moment angulaire $\cL$, définies par,
\begin{align*}
\cE & = \,\frac{r-a}{r+a}\,\np+\,\frac{2ar}{(r+a)^4\np}\lb\bx \times\bp\cdot\bs\rb\,, \\
\bcL & =\lp\frac{r+a}{r}\rp^2 \bx\times\bp+\,\frac{r-a}{r+a}\, \bs\,+\,\frac{2a}{r^2(r+a)}\,(\bs\cdot\bx)\, \bx\,,
\end{align*}


Les équations du mouvement pour le photon peuvent s'écrire,
\begin{align*}
\frac{d\bx}{dt}\,
 = & \,\frac{\blue{r^2(r-a)}}{\blue{(r+a)^3}D}
\Bigg\{
\blue{r^2 s \bp}\red{-3\np(\bcL\cdot\bx)\,\bx+3[\bx\times\bp\cdot\bs]\,\frac{\bx\times\bp}{\np}}\Bigg\}\,, \\[1.4ex]
\frac{d}{dt}\lp \frac{\bp}{\np} \rp
= & \, \frac{\blue{2 a}}{\blue{(r+a)^4} D} \Bigg\{\red{ 3(r-a) \frac{(\bx\cdot\bp)}{\np^2} \lb\bx\times\bp\cdot\bs\rb \, \bx\times\bp} \, + \nonumber \\ 
& \quad +\Big(\red{3r(\bcL\cdot\bx) (\bx\cdot\bp)}\blue{-(2r-a) s \np r^2\Big)  \left(\bx-\frac{(\bx\cdot\bp)\, \bp}{\np^2}\right)}\Bigg\}.
\end{align*}
avec $D := \blue{r^2 s \np} \red{-3(\bp\cdot\bx)(\bcL\cdot\bx)}$, et,
\begin{equation*}
\bx\times\bp\cdot\bs = \frac{r+a}{r-a}\lp \bx\times\bp\cdot\bcL-\lp\frac{r+a}{r}\rp^2\lp r^2\np^2-(\bx\cdot\bp)^2\rp\rp.
\end{equation*}

Le code couleur est tel que les termes en bleu correspondent aux termes géodésiques, et les termes rouges ne sont  présents seulement lorsqu'on considère le moment dipolaire de la particule test.

\smallskip

Ces équations du mouvement étant compliquées, considérons pour simplifier l'exemple du cas radial, où l'impulsion initiale du photon $\bp_0$ est radiale. Dans ce cas, les équations se réduisent aux équations des géodésiques, plus le transport parallèle du spin.

\begin{minipage}[c]{0.45\textwidth}
\begin{tikzpicture}[line width=1pt,scale=0.78]
  \node [draw,circle,minimum size = 15mm, label=below:Star] at (0,0) {};
  \draw [dashed] (0,0) to node[pos=0.6,scale=1]{$\bullet$} +(5,0);
  \draw [->] (3,0) to node[pos=0.7,below,scale=1]{$\bp_0$} +(1,0);
\end{tikzpicture}
\end{minipage}\hfill
\begin{minipage}{0.55\textwidth}
\vskip -1ex
\begin{align*}
\frac{d\bx}{dt} & = \frac{r^2(r-a)}{(r+a)^3} \frac{\bp}{\np}, \\[1.3ex]
\frac{d\bp}{dt} & = - \frac{2\,a\, r^2}{(r+a)^4} \, \bp, \\[1.3ex]
\frac{d\bs^\perp}{dt} & = - \frac{2\,a\, r^2}{(r+a)^4} \, \bst
\end{align*}
\end{minipage}\hfill

\vskip 1ex
Même si le cas radial est intéressant de par l'effet Einstein, et de part sa simplicité, nous allons maintenant considérer le ``weak lensing''. Pour rappel, dans le cas sans spin, tout ce passe dans le plan géodésique,

\begin{center}
\begin{tikzpicture}[line width=1pt,scale=0.78]
  \node [draw,circle,minimum size = 12mm, label=below:Gravitational lens] at (5.5,-0.5) {};
  \draw (0,0.7) to[out=10,in=170] (11,0.7) coordinate (enddash);
  \node [draw,circle,minimum size = 7mm, label=below:Emitting star] at (-1,0.5) {};
  \node [draw,circle,minimum size = 5mm, label=below:Earth] at (12,0.5) {};
  \draw [dashed] (enddash) -- +(171:11.1) coordinate (fakestar);
  \node [draw,circle,minimum size = 7mm,dashed, label=below:fake position] at ($(fakestar)+(-1.04,0)$) {};
  \coordinate (repere) at (9,1.5);
  \draw [->] (repere) to node[pos=1,above,scale=1]{$x_1$} +(0,1);
  \draw [->] (repere) to node[pos=1,right,scale=1]{$x_2$} +(1,0);
  \draw [dashed] (5.5,-0.5) to node[pos=0.6,right,scale=1]{$r_0$} +(0,1.8);
\end{tikzpicture}
\end{center}

Considérons maintenant le weak lensing tel que décrit par les équations de Souriau--Saturnini. Pour se faire, simulons numériquement trois photons : l'un dont la trajectoire est décrite par une géodésique isotrope, et deux autres suivant les équations de Souriau--Saturnini, avec hécilité $+1$ et $-1$. Les conditions initiales sont prises identiques pour les trois photons au périhélion de la trajectoire géodésique,
\begin{equation*}
\bx_0 = \lp
\begin{array}{c}
r_0\\
0\\
0
\end{array}\rp,\qquad
\bp_0 = \lp
\begin{array}{c}
0\\
p_0\\
0
\end{array}\rp,\qquad
\bs_0 = \lp
\begin{array}{c}
0\\
s\\
0
\end{array}\rp,
\end{equation*}

\begin{minipage}[c]{.5\linewidth}
\includegraphics[scale=0.47]{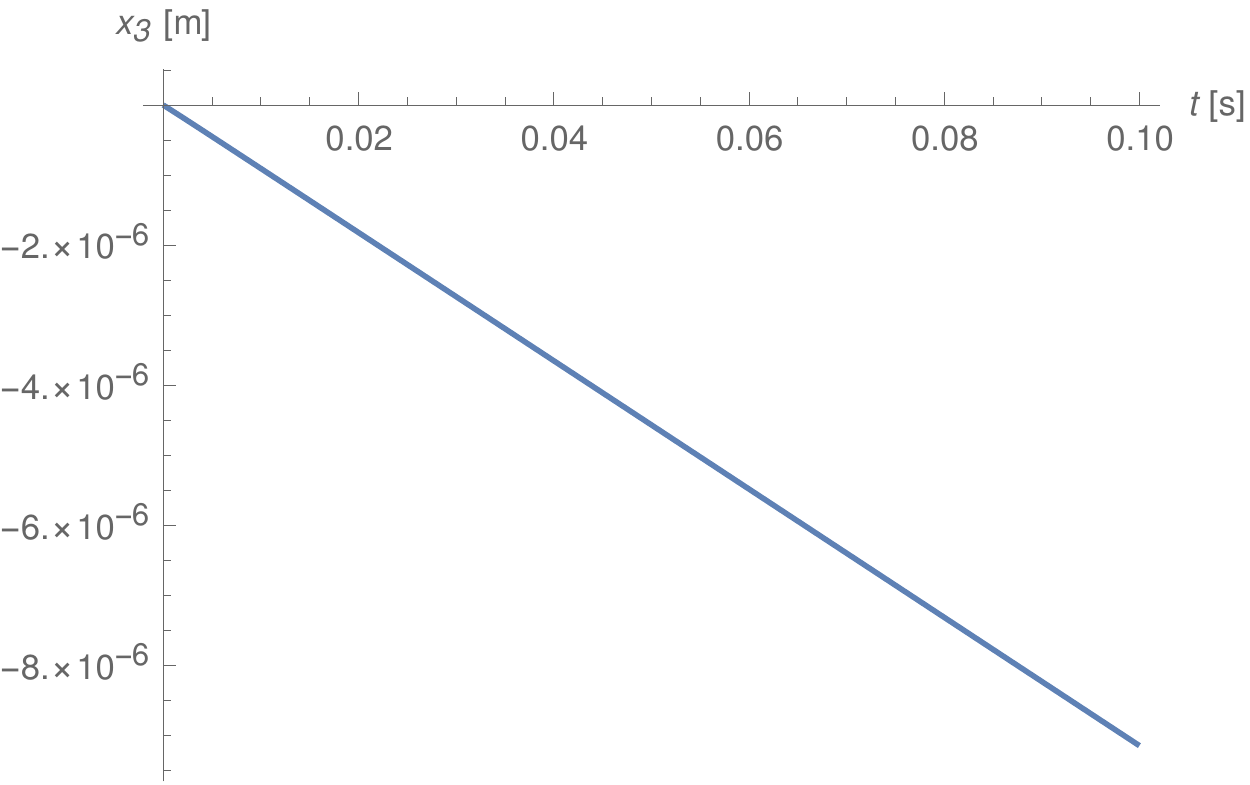}
\end{minipage}\hfill
\begin{minipage}[c]{.5\linewidth}
\includegraphics[scale=0.47]{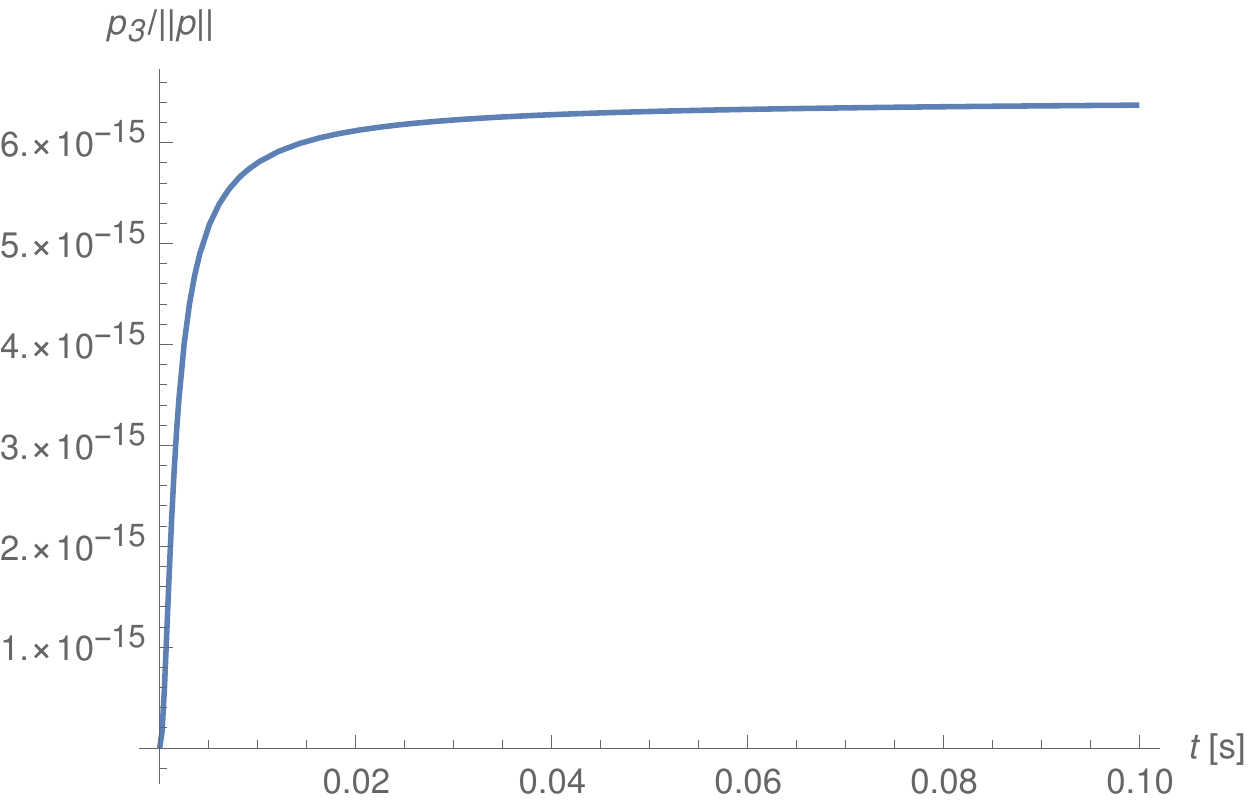}
\end{minipage}\hfill

\vskip 1ex
Numériquement, on trouve que les trois photons ont des trajectoires qui coïncident presque parfaitement dans le plan géodésique. Néanmoins, les photons décrits par l'équation de Souriau--Saturnini ont chacun une composante de leur trajectoire hors de ce plan, avec une direction dépendant de leur hélicité.

Pour y voir plus clair, il est possible d'obtenir une solution perturbative de ces équations avec de telles conditions initiales. Pour ce faire, on utilise deux petits paramètres,
\begin{equation*}
\alpha = \frac{a}{r_0} \qquad \& \qquad \epsilon = \frac{\hbar}{r_0\, p_0} = \frac{\lambda_0}{2\pi \, r_0},
\end{equation*}

On a vu numériquement que la trajectoire d'un photon avec spin est proche d'une géodésique isotrope. De plus, on peut s'attendre à ce que les effets de spin disparaissent lorsque le paramètre $\epsilon$ tend vers 0. Il y a donc du sens à chercher une solution perturbative d'ordre $\epsilon$ autour de la géodésique isotrope pour les équations du mouvement. 

On trouve que les trajectoires coïncident au premier ordre dans le plan géodésique, mais qu'il y a une déviation hors du plan. En effet, on trouve,
\begin{equation*}
\qquad x_3 = - \epsilon \, \chi \,t \qquad\& \qquad
p_3 = 2 \, \epsilon \, \alpha \, \chi \, p_0 \left(1- \frac{r_0}{\sqrt{r_0^2+t^2}}\right).
\end{equation*}

Ces expressions correspondent à un très bon niveau de précision aux résultats numériques précédents. On retrouve dans ces expressions que la trajectoire dépend de l'état d'hélicité  $\chi$ du photon. Il y a néanmoins un problème apparent : les signes de la trajectoire et de l'impulsion hors plan ne coïncident pas. Il semblerait donc que le photon ne suive pas sa propre impulsion à l'infini. Pour étudier ce phénomène, on définit deux angles $\beta$, $\gamma$,
\begin{equation*}
\beta \sim -(1-4\alpha) \frac{\chi \, \lambda_0}{2\pi \, r_0} \qquad\& \qquad \gamma \sim \chi\frac{a \, \lambda_0}{\pi \, r_0^2},
\end{equation*}
qui correspondent à l'angle que fait la trajectoire du photon à spin avec le plan géodésique dans le cas de $\beta$, et pour $\gamma$ c'est l'angle que ferait la trajectoire du photon avec le plan s'il suivait son impulsion à l'infini.

En utilisant des valeurs réalistes, pour le Soleil et $\lambda_0 = 600$nm, on a $2 |\beta| \sim 5 \cdot 10^{-11}\, \arcsec$ et $2 |\gamma| \sim 5 \cdot 10^{-16}\, \arcsec$ : ces angles sont extrêmement faibles.

Pour tester si cette différence de signe est une erreur provenant du fait que les équations de Souriau--Saturnini sont mal définies dans le cas plat (l'espace-temps de Schwarzschild est asymptotiquement plat), il est possible de rajouter la constante cosmologique dans les calculs, et ainsi considérer ces équations dans l'espace-temps de Kottler (ou Schwarzschild--de Sitter). Les équations du mouvement résultantes sont extrêmement compliquées, mais on y voit que ces signes se régularisent.

{\Large
\begin{figure}[ht]
\begin{tikzpicture}[line width=1pt,scale=0.8, every node/.style={transform shape}]
  \node [draw,circle,minimum size = 17mm, label=below:Lens] at (0,0) {};
  \draw [dashed] (12,0) coordinate (allright) -- node[pos=0.35,above,scale=1]{$\gamma$} (8,0) -- node[pos=0.8,below,scale=0.9]{Spinless} (2,0) -- node[pos=0.08,above,scale=1]{$\beta$} (0,0);
  \draw (8,0) -- +(-15:4);
  \draw (8,0) coordinate (center) -- +(15:4) coordinate (topleft);
  \draw (8,0) to[out=165,in=35] node[pos=0.58,above,scale=1]{$\chi = -1$} (0,0) coordinate (allleft);
  \draw (8,0) to[out=-165,in=-35] node[pos=0.58,below,scale=1]{$\chi = +1$} (0,0);
  \draw [line width=0.6pt] pic["",draw=black,-,angle radius=2.3cm] {angle=allright--center--topleft};
  \coordinate (abovecenter) at ($(0,0)+(28:2)$);
  \draw [line width=0.6pt] pic["",draw=black,-,angle radius=1.6cm] {angle=allright--allleft--abovecenter};
\end{tikzpicture}
\end{figure}
} 

Localement autour de l'étoile, il semble y avoir une sorte de phénomène de spin-orbite, qui pousse le photon dans une direction et son impulsion dans l'autre, et en s'éloignant de l'étoile on voit qu'elle perd de l'influence sur le photon, pour qu'à la fin celui-ci suive exactement son impulsion. On trouve donc une figure ressemblant à celle ci-dessus.

\subsection{Interaction entre le spin d'un photon et une onde gravitationnelle}

Les expériences de détection d'ondes gravitationnelles étant des plus précises au monde, et impliquant tant des photons qu'un champ gravitationnel non homogène, il est pertinent de se demander si elles pourraient être utiliser pour détecter une interaction spin-gravité. Pour ce faire, nous allons calculer ici la variation du temps de vol d'un photon dans le bras d'un interféromètre lorsqu'une onde gravitationnelle passe.

Pour rappel, une onde gravitationnel peut être décrite perturbativement autour de l'espace-temps de Minkowski grâce à sa faible amplitude $\epsilon$,
\begin{equation*}
g = dt^2 - \Big(1-\epsilon \cos(\omega(t-z))\Big) dx_1^2 - \Big(1+\epsilon \cos(\omega(t-z))\Big) dx_2^2 - dx_3^2.
\end{equation*}

Il est tout de suite évident, que des problèmes surviendront ici à cause de la perturbation autour de Minkowski, les équations étant mal définies pour le cas plat, ce qui n'apparaît que pour les particules de masse nulle. 

Nous allons donc tenter de régulariser ces équations en introduisant une masse  $m_\gamma$ pour le photon. Une première justification est qu'expérimentallement, nous n'avons qu'une limite haute sur sa masse,
\begin{equation*}
{m_\gamma}_{\mathrm{exp}} < 10^{-54} \mathrm{ kg}.
\end{equation*}

Pour justifier plus amplement cette méthode de régularisation, écrivons les équations de MPD avec $P^2 = {m_\gamma}^2$ et toujours $SP = 0$. On trouve,
\begin{align*}
\dot{X} & = P - \,\frac{2 \, S R(S) P}{4 \, P^2 - R(S)(S)},\\
\dot{P} & = - \half R(S) \dot{X},\\
\dot{S} & = P\overline{\dX}-\dX\barP.
\end{align*}
Il est tout à fait notable, et c'est même un point nécessaire pour utiliser cette régularisation, que ces équations se réduisent aux équations de Souriau--Saturnini dans la limite $m_\gamma \rightarrow 0$. 

De plus, si on considère un photon tel que son impulsion initiale est dans une seule direction, par exemple $\bp_0 = (0, p_2, 0)$, nous avons une suite d'ordres de grandeur,
\begin{equation*}
{p_2}^2 \gg {m_\gamma}^2 \gg R(S)(S)
\end{equation*}

Le terme de masse $P^2 = {m_\gamma}^2$ va régulariser le dénominateur pour la première équation, étant plus grand que $R(S)(S)$. Mais vu que l'on considère seulement la direction de propagation principale du photon, pour calculer son temps de vol, les termes de masse sont négligés devant l'énergie ${p_2}^2$ du photon. Nous devrions donc avoir des équations du mouvement régularisées, mais ne dépendant pas d'une masse arbitraire.

On trouve l'équation du mouvement dans la direction principale de propagation,
\begin{equation*}
\frac{dx_2}{dt} = \underbrace{1-\frac{\epsilon}{2} \cos(\omega(t-z))}_{\textrm{null geodesic}} \underbrace{-\frac{\epsilon}{2}  \frac{\lambda_\gamma^2}{\lambda_{\mathrm{GW}}^2} \frac{s_1^2-s_3(s_2+s_3))}{\hbar^2} \cos(\omega(t-z))}_{\textrm{Spin-GW interaction}}.
\end{equation*}

Nous voyons un terme supplémentaire par rapport à la géodésique isotrope, qui décrit l'interaction entre le spin du photon et l'onde gravitationnelle. Il s'agit d'un terme oscillant d'amplitude
\begin{equation*}
\frac{\epsilon}{2} \frac{\lambda_\gamma^2}{\lambda_{\mathrm{GW}}^2} \sim 10^{-46},
\end{equation*}
en prenant les valeurs expérimentales pour l'amplitude de l'onde gravitationnelle et des longueurs d'onde.

Ces équations prédisent donc un effet d'interaction entre le spin du photon et l'onde gravitationnelle, mais cet effet est extraordinairement faible. Il est d'ailleurs même plus faible que les effets géodésiques de seconds ordres, en $\epsilon^2$.

\subsection{Conclusions}

Les équations de Souriau--Saturnini sont des équations du mouvement pour un photon avec spin 1. Nous avons vu trois exemples d'application de ces équations. Dans un espace-temps de de Sitter, elles se réduisent à l'équation des géodésiques, en prédisant que le tenseur de spin est transporté parallèlement, donc n'est pas exactement conservé. Dans un espace-temps de Schwarzschild nous avons vu que ces équations prédisent une biréfringence gravitationnelle du photon lors du weak lensing. Enfin, dans le contexte de détection d'ondes gravitationnelles, ces équations prédisent un effet du spin sur le temps de vol du photon, mais beaucoup trop faible pour être observé. 

Ces équations présentent néanmoins des inconvénients. En premier lieu, bien que le spin longitudinal soit une constante du système, il n'en n'est rien de la norme du vecteur de spin. On ne peut pas s'attendre à ce que cette norme soit quantifiée dans une théorie purement classique, mais cette non conservation a l'air d'entrainer quelques instabilités dans les calculs. Plus contraignant encore, dans les deux exemples non triviaux présentés ici, les espace-temps de Schwarzschild et d'onde gravitationnelle, il a été nécessaire de recourir à une méthode de régularisation des équations. Soit en considérant la constante cosmologique, soit en incluant une \textit{dummy mass} pour le photon. Ces équations semblent donc souffrir d'être purement classiques. Une théorie décrivant la trajectoire de particules à spin en espace-temps courbe gagnerait à être au moins une limite classique d'une description quantique. Quelques pistes existent dans ce style, comme \cite{Gos06,OanceaJDRPA20}.

	\ohead{\leftmark\ifstr{\rightmark}{\leftmark}{}{ -- \rightmark}}	

	\chapter*{Introduction}
	It has been shown by Wigner in 1939 \cite{Wigner39} that elementary particles can be classified according to the different irreducible unitary representations of the coadjoint orbits of the Poincar\'e group. This classification depends on two quantities, called Casimir invariants, which are labelled as the mass squared $m^2$ and $s$ which is either called spin, or helicity, depending on the value of the mass. Wigner showed that massive and massless particles are fundamentally different within this classification. The physical elementary particles are classified as follow. First, one need to know if the mass squared is positive or vanishes, then study the so-called \emph{little group}, which is the subgroup of the Poincaré group which stabilizes the 4-momentum $P$ of the particle considered. In the case where the mass squared is strictly positive we have massive particles. The little group of such particles is the rotation group $\SO(3)$. This little group further classifies elementary particles depending on the representation of $\SO(3)$ that they belong to. The representations are labelled by the spin $s$, which can only take half-integer, or integer, values. Now, in the case where the mass squared vanishes, we have massless particles. The little group is the the special Euclidean group $\SE(2)$, which again discriminates massless particles depending on the representation of $\SE(2)$ they belong to, where such representations are labelled by the helicity $s$, which can only take half-integer, or integer, values. Other classifications are possible, such that as negative mass squared, or with continuous spin representations, but these have not been observed in Nature.

Hence, according to this classification, it turns out that the notion of spin of an elementary particle is already fundamental in classical Mechanics \cite{Sou70,ChD82}. 

Moreover, according to Lévy-Leblond \cite{LevyLeblond67,LevyLeblond74}, the notion of spin appears already in non relativistic (or Galilean relativistic as Lévy-Leblond would emphasis, see, \textit{e.g.}~\cite{LevyLeblond74}\footnote{Since some of the work presented in this thesis in based on Lévy-Leblond's contributions, it seems fitting to employ his (justified) vocabulary. We shall therefore refer to ``non relativistic'' mechanics as Galilean mechanics.}) Quantum Mechanics. While it is often taught that spin can be brought down to the level of non relativistic Quantum Mechanics as a limit of the Dirac equation, which then yields the Schr\"odinger--Pauli equation, Lévy-Leblond showed that one can naturally define a first order differential equation acting on spinors to describe non relativistic quantum systems. This equation is called the Lévy-Leblond equation, and it is to the Schr\"odinger equation what the Dirac equation is to the Klein--Gordon equation. It is also worth mentioning that this equation yields the correct Landé g-factor of 2 for the magnetic moment of a spin one half particle.

This thesis will focus around the notion of spin with two main topics, each in a domain that is often slightly put aside: particles with spin in non relativistic quantum mechanics, and particles with spin in classical mechanics. The first subject, developed in chapter 1, will be the study of the Lévy-Leblond--Newton equation, which aims at describing the evolution of a quantum system of spin one half that is coupled to its own gravitational potential. The second subject, developed in chapter 2, is about the equations of motion of spinning particles in General Relativity. Indeed, the geodesic equation ignores the spin, or more generally angular momentum, of the trajectory of test particles it describes. We will focus on the Souriau--Saturnini equations, which aim at describing the trajectory of photons in curved spacetime, without neglecting their helicity. 

\bigskip

This thesis is based on three published works,

S.~Lazzarini, L.~Marsot, ``On the Lévy-Leblond--Newton equation and its
  symmetries: a geometric view'',
  \href{https://dx.doi.org/10.1088/1361-6382/ab6998 }{ Class. Quant.
  Grav. {\bf 37}, p.~055008, (2020)}, arXiv:
  \href{https://arxiv.org/abs/1911.03099}{1911.03099}.
  
\medskip  
  
C.~Duval, L.~Marsot, T.~Sch{\"u}cker, ``{Gravitational birefringence of light
  in Schwarzschild spacetime}'',
  \href{https://dx.doi.org/10.1103/PhysRevD.99.124037 }{ Phys. Rev. D {\bf 99},
  p.~124037, (2019)}, arXiv:
  \href{https://arxiv.org/abs/1812.03014}{1812.03014}.
  
\medskip  
  
L.~Marsot, ``{How does the photon's spin affect Gravitational Wave
  measurements?}'', \href{https://dx.doi.org/10.1103/PhysRevD.100.064050 }{ Phys. Rev. D {\bf 100}, p.~064050, (2019)}, arXiv:
  \href{https://arxiv.org/abs/1904.09260}{1904.09260}.
	\addcontentsline{toc}{chapter}{Introduction}

	\chapter{Symmetries of the L\'evy-Leblond--Newton equation}
	\chaptertoc{}

\section{Introduction to the subject}
\label{sec-Intro}

\hfuzz=\maxdimen

This chapter is devoted to the study of the L\'evy-Leblond--Newton (LLN) equation. This equation describes the behavior of non-relativistic fermions (or L\'evy-Leblond fermions \cite{LevyLeblond67}) when coupling the evolution of the fermion's wave packet with its own Newtonian gravitational potential. It is closely linked to the Schr\"odinger--Newton (SN) equation, originally introduced by Di\`osi in \cite{Diosi84}. The Schr\"odinger--Newton equation has then been proposed \cite{GiuliniG11, GiuliniG13} to have an effect on the spreading of wave packets, and could support Penrose's idea of ``Gravitization of Quantum Mechanics'' \cite{Penrose96,Penrose14}. This introduction will review some aspects of Quantum Mechanics, symmetries in particular for both the Schr\"odinger--Newton equation and the L\'evy-Leblond--Newton equation.

\subsection{Open questions in Quantum Mechanics}

\subsubsection{Inhibited spreading of the wave packet through gravity}

In the world of macro objects, we have two contradicting principles. First is the Newtonian principle, which states that if we have an object upon which no external force is acting, then it stays at rest, \ie it is stationary. Then, Quantum Mechanics tells us that this macro object is in fact described at the quantum level by a wave packet, with a characteristic size. In Quantum Mechanics, this wave packet spreads, becomes larger, as time goes by, as if we were losing precision on the object. The macro object would thus not be stationary, and hence not appear as localized as what we seem to observe, according to Di\`osi \cite{Diosi84}.

This spread of the wave packet depends on its characteristic size: the larger the wave packet, the slower the spread. If the characteristic size is large enough, for example on the order of $10^{-8}$ cm, then the spreading is so slow that we would not see it. But Di\`osi argues \cite{Diosi84} that the characteristic size of atomic wave packets is much smaller than that, around $10^{-12}$ cm, meaning that the quantum spread would be much faster, and we should be able to see it experimentally.

A soliton is a wave packet, solution to a wave equation, that keeps its shape through time, just like a wave on the ocean that would never change as it moves forward, which are called solitary waves. The problem in standard Quantum Mechanics, described by the Schr\"odinger equation, is that, as discussed above, wave packets slowly spread out, rendering soliton-like solutions impossible. It should be clear then, that to make these wave packets keep their shape, there should be something holding them in place. According to Di\`osi \cite{Diosi84}, this can be gravity. He shows that the Schr\"odinger--Newton equation, which is the Schr\"odinger equation to which we add the Newtonnian gravitational potential of the system, thus creating a self gravitating process, possesses soliton like solutions, which would solve the above-mentioned problem. He also gives the order of the characteristic width $a_0$ of the ground state wave packet for a pointlike macro object,
\begin{equation}
\label{Diosi_width}
a_0 \approx \frac{\hbar^2}{G M^3},
\end{equation}
with $M$ the mass of the macro object. 

As for the width of the ground state of an extended macro object of radius $R$, as opposed to a pointlike object, he suggests,
\begin{equation}
a \approx a_0^{1/4} \, R^{3/4}.
\end{equation}

In \cite{GiuliniG11}, instead of considering solitons, the authors explicitly analyzed numerically wave packets of different masses and compare both evolutions using the free Schr\"odinger equation, and the Schr\"odinger--Newton system. Using the atomic mass unit, $1 \, \mathrm{u} = 1.66 \cdot 10^{-27} \, \mathrm{kg}$, they found that for an initial Gaussian distribution of width $0.5 \, \mathrm{\mu m}$ and a mass less than $6 \cdot 10^9 \, \mathrm{u}$, the wave packet is still spreading, just like the case of the free Schr\"odinger equation, but at a reduced rate. For masses greater than $7 \cdot 10^9 \, \mathrm{u}$, the wave packet collapses, making macroscopic objects localized. For the width considered here, $0.5 \, \mathrm{\mu m}$, the soliton-like solution would be one with a fine tuned mass between 6 and $7 \cdot 10^{-9}$ u, so that the spread and the gravitational collapse are at equilibrium. This numerical result is in accordance with the formula given by Di\`osi (\ref{Diosi_width}).

\subsubsection{Measure problem, reduction of the wave packet}

Two major theories emerged in the 20th century, General Relativity and Quantum Mechanics, which are both undefeated in their domain of application, but whose base principles are incompatible. Because of its wider applications on our scale, Quantum Mechanics is considered by many as more fundamental than General Relativity. This led to a movement of people trying to bring General Relativity in line with Quantum Mechanics. Penrose argues \cite{Penrose96,Penrose14} that this should not be the case, and that it is worth investigating the opposite way of thinking, bringing Quantum Mechanics in line with General Relativity principles.

An important open question in Quantum Mechanics is about the measuring process, or the reduction of the wave function. The evolution of the quantum world seems unitary and linear.
Problems arise when considering the superposition principle. A wavefunction made of the superposition of two states can be described with the Schr\"odinger equation, with its linear unitary evolution, yet when observing the wavefunction, it collapses into one state or the other. 
A measurement seems to lead to a discontinuous jump of the wave function, as is depicted in Schr\"odinger's cat thought-experiment: we see the cat either dead or alive when we open the box, we do not see the superposition. Several interpretations, Copenhagen, Bohm, Many Worlds, etc, came to life through the years to explain this apparent inconsistency of Quantum Mechanics, suggesting a \textit{reduction} process to explain this discontinuous jump.

Penrose suggests that the consideration of gravity and GR principles in Quantum Mechanics can lead to a solution to this measurement problem. Indeed, consider an experiment with macro objects, much like Schr\"odinger's cat, where we have, in a linear order, a photon emitter, a beam splitter, and a photon detector that moves a massive object if a photon is detected. If the photon goes through the beam splitter, the massive object is moved, but if the photon is reflected, then nothing happens. In both cases, after the photon was emitted, the resulting configurations would be stationary. If we consider Quantum Mechanics, before measuring this system, it would be a superposition of the two states: \textit{object moved} and \textit{object untouched}. When we introduce gravity in this system, including the principles of General Relativity, we get a superposition of two spacetimes. We encounter two problems with this:
\begin{itemize}
\item The superposed state would not be stationary, even if each state is independently stationary, because of gravity ;
\item The principle of general covariance forbids the identification of points between two spacetimes, making the superposition of the two spacetimes in this experiment, that differ by a translation, ill-defined.
\end{itemize}

Penrose argues \cite{Penrose96,Penrose14} that a measure of this ill-definiteness is the gravitational self-energy $E_G$ of the \emph{difference} between the gravitational fields of the two spacetimes. Indeed, the ``closer'' the two states are, the smaller this energy is. Since this ill-definiteness can be related to an energy, we can call for an analogy with particle physics, and define a characteristic time of instability, through the Heisenberg uncertainty. We get a lifetime $\tau$ of the order of
\begin{equation}
\label{lifetime_penrose}
\tau \approx \frac{\hbar}{E_G}.
\end{equation}

\begin{center}
\begin{figure}[ht]
\includegraphics[scale=0.3]{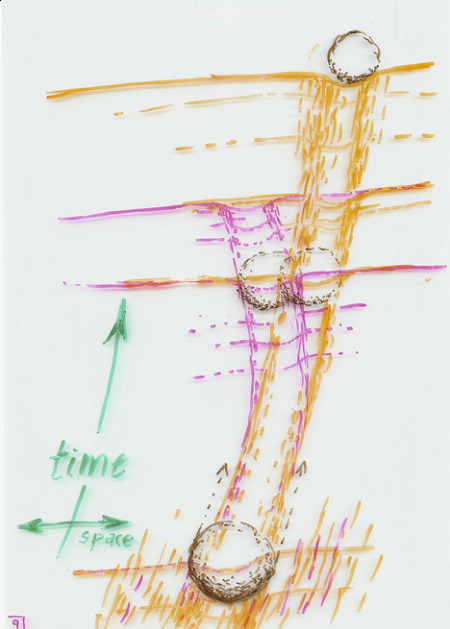}
\caption{Image from Penrose, 2014. A macroscopic object, pictured at the bottom, leads to the superposition of two spacetimes after a quantum event, one in yellow, one in pink. After a characteristic time, the superposition decays into one: the yellow spacetime.}
\label{decay_spacetime}
\end{figure}
\end{center}

According to Penrose, in this previous experiment, the resulting state being a superposition of two spacetimes, is inherently unstable. What happens, according to Penrose, is that after a lifetime of the order of (\ref{lifetime_penrose}), the superposed state naturally decays into one of the two states, without any measurement. This is depicted by Penrose in figure~\ref{decay_spacetime}. This might explain why we do not see superposition of states on our macro world.

The value of the lifetime (\ref{lifetime_penrose}) is certainly interesting, because while Plank's constant is very small, the gravitational self energy of the \emph{difference} of the two gravitational fields is also very small. As a result, the decay time for macro object ought to be accessible through experiments. Penrose computed \cite{Penrose96}, that doing this experiment with a water drop of $10^{-5}$~cm of radius would result a in superposition state with a lifetime of the other of one hour, while for a water drop of radius of $10^{-3}$~cm it would be about $10^{-6}$~s. Several experiments are planned \cite{MarshallSPB03,Kaltenbaek12,Kaltenbaek15} to study this. One of them is the experience MAQRO \cite{Kaltenbaek12,Kaltenbaek15}, which stands for macroscopic quantum resonators. It is a proposed experiment to the European Spatial Agency, embarked in a satellite, aimed at testing quantum experiments in microgravity to see the relevance of alternative theories for macro objects, such as Schr\"odinger--Newton. Indeed, the SN equation is a natural candidate to describe the Quantum Mechanics processes behind Penrose's ideas.
 
\subsection{Notations in this chapter}

The spatial dimension will be denoted $n$, and we will often use $N = n + 2$ which will be the dimension of a Bargmann structure associated to a $n+1$ dimensional Newton--Cartan spacetime.

Bold characters will reference vectors of dimension $n$. For instance the vector $(x^1, \ldots, x^n)$ will be denoted $\bx$. The transposed of a vector will be written with a superscript $T$, for instance $\bx^T$.

\subsection{The Schr\"odinger--Newton equation}

The Schr\"odinger--Newton equation can be derived on two assumptions \cite{Diosi84,BahramiGDB14}. First, that gravity is a classical theory, and second that its coupling to matter is described by the semi classical\footnote{By semi classical, it is understood that the matter fields are quantized, but gravity remains classical.} Einstein equations \cite{Moller62,Rosenfeld63},
\begin{equation}
R_{\mu\nu} - \half R g_{\mu\nu} = \frac{8\pi G}{c^4} \langle \psi \vert \hat{T}_{\mu\nu} \vert \psi \rangle,
\end{equation}
where on the right hand side we have the expectation value of the energy momentum tensor operator in a quantum state $\psi$. 

Then, in linearized gravity \cite{MisnerTW73}, where the metric is expanded around Minkowski spacetime, as $g_{\mu\nu} = \eta_{\mu\nu} + h_{\mu\nu}$ with $\eta$ the Minkowski metric, and in the Newtonian limit where the dominant component of the right hand side is $\langle \psi \vert \hat{T}_{00} \vert \psi \rangle$, one finds a Hamiltonian describing the interaction between matter and gravity \cite{BahramiGDB14},
\begin{equation}
\hat{H}_{\text{int}} = - G \int d^3 \bx d^3\bx' \frac{\langle \psi \vert \hat{\rho}(\bx')\vert \psi\rangle}{\vert \bx-\bx'\vert}\hat{\rho}(\bx),
\end{equation}
with $\hat{\rho}(\bx)$ the mass density operator, arising from $\hat{T}_{00} = c^2 \hat{\rho}$ in the Newtonian limit.

Finally, when considering only a one-particle system, the evolution of the quantum state $\psi$ is described by the Schr\"odinger--Newton equation,
\begin{equation}
i\hbar \frac{\partial \psi}{\partial t}(\bx, t) = - \frac{\hbar^2}{2m}\Delta_{\R^3} \psi(\bx, t) - G m^2 \left( \int \frac{|\psi(\bx', t)|^2}{|\bx'-\bx|} d^3 \bx'\right) \psi(\bx, t)
\end{equation}

It can be generalized to arbitrary spatial dimension $n$ as,
\begin{equation}
\label{sch_spacetime}
i\hbar \frac{\partial \psi}{\partial t}(\bx, t) = \left(-\frac{\hbar^2}{2m}\Delta_{\R^n} + m U(\bx, t)\right) \psi(\bx, t),
\end{equation}
with a potential $m U(x, t)$ which we identify as the Newtonian gravitational potential. It is a solution of the Poisson equation,
\begin{equation}
\label{poissonintro}
\Delta_{\R^n} U(\bx, t) = 4 \pi G \rho(\bx, t),
\end{equation}
with the mass density $\rho$ here related to the quantum probability density,
\begin{equation}
\label{densite_fctonde}
\rho(\bx, t) = m |\psi(\bx, t)|^2.
\end{equation}

This equation has been extensively studied, for instance numerically \cite{MorozPT98,MelkoM00,vanMeter11,GiuliniG11,GiuliniG13}, as finding its scaling symmetries \cite{GiuliniG11}, or complete symmetries \cite{DuvalL15}.

\subsection{The L\'evy-Leblond--Newton equation: overview of the chapter}

In this chapter, which is largely based on \cite{LazzariniM20}, we shall study the symmetries of what we call the L\'evy-Leblond--Newton equation as describing L\'evy-Leblond fermions coupled to Newton-Cartan (NC) geometry through their gravitational self-interaction. To some extent, one may consider the LLN equation as the ``square root'' of the SN equation. As such, it is rather natural to ask oneself whether the L\'evy-Leblond fermions can also be treated in the Bargmann framework. Indeed, such a framework has been developed~\cite{Bargmann54,DuvalBKP85,Eisenhart28} as a way to introduce the powerful covariant relativistic geometric tools to study NC structures. Previous indications in that direction were shown in~\cite{KunzleD86,Duval85}. In particular, one may wonder which scale laws L\'evy-Leblond fermions are subject to, and which dynamical exponent in any spatial dimension characterizes them along the seminal idea given in~\cite{GiuliniG11} for the SN equation. As a major result, the latter turns out to be the same as for the SN case, as computed in~\cite{DuvalL15}. 

In this respect, we mainly follow the line given in a previous work \cite{DuvalL15} in which most of the Bargmann study for the SN equation has been introduced. The reader will often be referred to the latter. The following sections are organized as follows. 
In Section \ref{sec-Prelim}, a quick review is made about the Bargmann geometry over a Newton-Cartan spacetime. Some delicacies are required in dealing with spinorial densities in order to have a correct geometrical description for the Dirac operator, the covariant derivative and the infinitesimal transformation (Lie derivative) of spinors.
Next, Section \ref{sec-LLN} is devoted to the LLN coupled system along the line given in \cite{DuvalL15}. In particular, the generalized LLN equation is discussed in relation with gauge transformations.
Section \ref{sec-LLN-sym} treats the symmetries of the LLN equation collected in the SN group with in addition the corresponding spinorial representation. 
Explicit representations of this group will be given for spatially flat Bargmann structures. Of course, the corresponding projective unitary representation on LL spinors which is of importance at the quantum level is given. Also, conserved quantities of the LLN equation are exhibited.
Conclusions and some remarks are gathered in Section \ref{sec-conclusion}.

\section{Introduction to symmetries and geometric tools}
\subsection{Why study symmetries?}

Differential equations can be complicated to solve. Studying their symmetries is one way to obtain information about these equations, without solving them. Once the symmetries are known, one can transform a specific solution of the differential equation to a whole family of solutions. Another valuable knowledge is the set of quantities that are conserved along the evolution of the system. Conserved quantities can help to solve a differential equation both analytically and numerically. For instance, one check for the accuracy of a numerical integration is that conserved quantities should indeed be conserved, up to some error margin.

Let us see some examples of computing symmetries, namely the symmetries of the Klein--Gordon equation, and of the Schr\"odinger equation. Then, we will recall the symmetries of the Schr\"odinger--Newton equation.

\subsubsection{The simple example of the Klein-Gordon equation}

The free Klein-Gorden equation is given by, with $c = \hbar = 1$,
\begin{equation}
\label{eq_kg}
\square \psi - m^2 \psi = 0.
\end{equation}

We assume the spacetime to be 4-dimensional, and the metric has signature $(p,q) = (3,1) = (+++,-)$.
We want to look for all diffeomorphisms of Minkowski spacetime which send a solution of the Klein-Gordon equation to another solution of the same equation. In other words, we are looking for the group which permutes the solutions of the equation. We can proceed in a simple way, by writing $\widetilde{\psi}(x) := \psi(a(x)) \exp(i S(x))$\footnote{We do not seek to look for a global dilation term in this transformation, as it necessarily has to be a constant, which we fix to 1 to keep the same energy eigenvalue.} with $a \in \Diff(\R^{3,1})$, and $S \in C^\infty(\R)$ to be determined. We want $\widetilde{\psi}$ to be again solution of \eqref{eq_kg}, hence $\square \widetilde{\psi}(\widetilde{x}) - m^2 \widetilde{\psi}(\widetilde{x}) = 0$, with $\widetilde{x} := a(x)$.

One can show that to satisfy the Klein--Gordon equation, the function $S$ must be a constant, that we will denote $h$. Then, one finds two sets of conditions on $a(x)$,
\begin{equation}
g^{\mu\nu} \frac{\partial \widetilde{x}^\alpha}{\partial x^\mu} \frac{\partial \widetilde{x}^\beta}{\partial x^\nu} = g^{\alpha\beta}, \quad  \mathrm{and} \quad g^{\mu\nu} \frac{\partial^2 \widetilde{x}^\alpha}{\partial x^\mu \partial x^\nu} = 0.
\end{equation}

The second set dictates that we have an affine transformation, \ie $\widetilde{x} = L x + c$, with $L$ a $4 \times 4$ matrix, and $c \in \R^4$. Then, the first set of conditions becomes $g^{\mu\nu} L^\alpha_\mu L^\beta_\nu = g^{\alpha\beta}$. This is the well-known defining relation of Lorentz transformations. Hence, the 4-dimensional Klein-Gordon equation is invariant under the 11-dimensional trivial central extension of the Poincar\'e group, $\SE_+(3,1) \times \R$, with $h \in \R$ as the central extension parameter.

The central extension parameter $h$ is often forgotten here since the extension is just a phase factor, and hence trivial here. However, in some cases, such as for the Schr\"odinger equation, it needs to be taken into consideration.

\subsubsection{The symmetries of the Schr\"odinger equation}
\label{ss:sym_schr}
One can obtain the Schr\"odinger equation from the Klein-Gordon equation in the non-relativistic limit. However, while the Galilean limit of the Poincar\'e group is the Galilei group, the Schr\"odinger equation is invariant under a larger group than the central extension of the Galilei group (also named the Bargmann group). 

To exemplify this fact, let us compute the maximal symmetry group of the 1+1 dimensional free Schr\"odinger equation,
\begin{equation}\label{eq_schro}
\frac{\hbar}{2mi} \frac{\partial^2 \psi(x, t)}{\partial x^2} + \frac{\partial \psi(x, t)}{\partial t} = 0.
\end{equation}

We want to find real functions $F, G, R, S$ such that,
\begin{equation}\label{expr_psi}
\widetilde{\psi}(x, t) := \psi\left(F(x, t), G(x, t)\right) R(x, t) \exp(i S(x, t))
\end{equation}
is again a solution of \eqref{eq_schro}. Similarly to the Klein-Gordon equation in the previous section, we find a set of conditions,

{\setstretch{2.2}\begin{equation}\label{syst_eq_diff_sch}
\left\lbrace\begin{array}{l}
\displaystyle \frac{\partial G}{\partial x} = 0, \\
\displaystyle \frac{\partial G}{\partial t} = \left( \frac{\partial F}{\partial x} \right)^2, \\
\displaystyle \frac{\partial^2 F}{\partial x^2} R + 2 \frac{\partial F}{\partial x} \frac{\partial R}{\partial x} = 0, \\

\displaystyle \frac{\partial F}{\partial x} \frac{\partial S}{\partial x} + \frac{m}{\hbar} \frac{\partial F}{\partial t} = 0, \\

\displaystyle \frac{\partial^2 R}{\partial x^2} - \left(\frac{\partial S}{\partial x}\right)^2 R - \frac{2m}{\hbar} \frac{\partial S}{\partial t} R = 0, \\
\displaystyle 2 \frac{\partial R}{\partial x} \frac{\partial S}{\partial x} + R \frac{\partial^2 S}{\partial x^2} + \frac{2m}{\hbar} \frac{\partial R}{\partial t} = 0.
\end{array}\right.
\end{equation}
}

We easily see that $G$  is a function of time only. We can safely assume $\partial_x F \neq 0$. It follows that $\partial^2_x F = 0$, and we find that $R$ is also a function of time only. Then, it is interesting to find that the Schwarzian derivative of $G$, with respect to time, is zero, \ie $G'''/G' - 3/2 (G''/G')^2 = 0$. This implies that $G$ is an homography. Finally, the other functions are easily found. The results can immediately be generalized to the $n+1$ dimensional equation. We find the functions to be,

\begin{align}
& F(\textbf{x}, t) = \frac{A \textbf{x} + \textbf{b} t + \textbf{c}}{f t + g} \\
& G(t) = \frac{d t + e}{f t + g} \\
& R(t) = \frac{1}{(ft + g)^{n/2}} \\
& S(\textbf{x}, t) = \frac{m}{\hbar} \left(\frac{f}{2} \frac{||A\textbf{x}+\textbf{b}t+\textbf{c}||^2}{ft+g} - \langle\textbf{b}, A \textbf{x}\rangle - \frac{||\textbf{b}||^2 t}{2} + h\right) \label{sym_phase_schr}
\end{align}
with $A \in \mathrm{O}(n), \textbf{b}, \textbf{c} \in \R^n, d, e, f, g, h \in \R$, respectively a rotation, boosts, spatial translations, time dilation, time translation, inversion, spatial dilation, and finally the parameter of the central extension, here non trivial (see the transformation law of the wavefunction in \eqref{repr_sn_neq4} down below). We also have the compatibility condition $dg-ef = 1$. This symmetry group, of the free Sch\"odinger equation, is called the (extended) Schr\"odinger group \cite{Niederer72}\footnote{There are two ``Schr\"odinger'' groups in the literature: a 12-dimensional one, and its central extension that is 13-dimensional. Often, the first group is called the Schr\"odinger group, while the second one is called the extended Schr\"odinger group. Niederer finds the 12-dimensional group in his paper, but to obtain all the non trivial symmetries of the Schr\"odinger equation, one needs the 13-dimensional group. The additional symmetry corresponds to the phase transformation.}. Note that the normalization of the wavefunction is preserved.

Remarkably, for $n = 3$, the Schr\"odinger group has 13 dimensions. We recover, as a subgroup, the Bargmann group, of 11 dimensions, (it is recalled that this is the central extension of the Galilei group) with $f = 0$ and $d = g = 1$.

Hence, free non-relativistic Quantum Mechanics, as described by the Schr\"odinger equation, has more symmetries than relativistic Quantum Mechanics. These new symmetries are sometimes called accidental symmetries.

\subsubsection{Symmetries of the SN equation}

The study of the symmetries of the SN equation can be found in \cite{DuvalL15}, and the conserved quantities in \cite{Marsot17}.

Obtaining the symmetries of the SN equation turns out to be more complicated than the process of getting the symmetries for the Schr\"odinger equation. In \cite{DuvalL15}, they were obtained by recasting the SN system on a Bargmann structure\footnote{See section \ref{ss_barg} for an overview of Bargmann structures.} in a fully covariant way, which amounts to deducing them directly.

The full symmetry group of the Schr\"odinger--Newton equation turns out to be larger than one can expect at first glance. There are two cases, depending on the spatial dimension $n$. If $n \neq 4$, then the group is called the Schr\"odinger--Newton group \cite{DuvalL15}. For $n = 4$, the group of symmetry turns out to be the full extended Schr\"odinger group, as derived in the previous section. Interestingly, the Schr\"odinger--Newton group turns out to have a matrix representation,
\begin{equation}
\label{rep_matrix_sn}
\left(\begin{array}{cccc}
A & \bb & 0 & \bc \\
\bzero^T & d & 0 & e \\
\displaystyle - \frac{\bb^T A}{d} & \displaystyle - \frac{\Vert \bb \Vert^2}{2d} & \displaystyle \frac{1}{d} & \displaystyle \frac{h}{d} \\
\bzero^T & 0 & 0 & g
\end{array}\right),
\end{equation}
with $A \in O(n)$, $\bb, \bc \in \R^n$, $d, e, g, h \in \R$, respectively a rotation, spatial boosts, spatial translations, time dilations, spatial dilations, and central extension translation\footnote{This corresponds to translations in the fiber above Newton--Cartan spacetime in Bargmann structures.}. The two dilations are not independant, they are linked by $d = \nu^{\frac{n-1}{n-4}}$ and $g = \nu^{-\frac{3}{n-4}}$, for $\nu \in \R^{*+}$. 

These groups act on wavefunctions such that they map solutions to the SN equation to solutions. For $n \neq 4$, the action of the projective unitary representation on the wavefunctions reads \cite{DuvalL15},
\begin{equation}
\label{repr_sn}
\widetilde{\psi}(\bx, t) := \, g^{-\frac{n}{2}} \exp\left(\frac{im}{\nu \hbar} \left(  - \langle\textbf{b}, A \textbf{x}\rangle - \frac{||\textbf{b}||^2 t}{2} + h\right)\right) \psi\left(\frac{A \textbf{x} + \textbf{b} t + \textbf{c}}{g}, \frac{d t + e}{g} \right)
\end{equation}
Note that \eqref{repr_sn} can be obtained by the transformation law of $\psi$ in the previous section of the symmetries of the Schr\"odinger equation, by eliminating the inversions, \ie $f = 0$, and requiring $dg = \nu$ instead of $dg = 1$.

\medskip

Once we have the symmetry group, one can also deduce the conserved quantities by considering an action principle of the wave equation on Bargmann manifold, and using a procedure adapted to these manifolds \cite{DuvalHP94}. They read, in the case where $n \neq 4$ \cite{Marsot17},

{\setstretch{2}\begin{equation}
\label{conserved_n_neq4_sn}
\left\lbrace
\begin{array}{ll}
\displaystyle \cH = \int \left[ \frac{-\hbar^2}{2m} \vec{\nabla} \psi \cdot \vec{\nabla} \bar\psi - m U |\psi|^2 \right] d^n \vec{x} \qquad \qquad & \mathrm{energy} \\
\displaystyle \vec{\cP} = \int \vec{P} d^n \vec{x} = \frac{i \hbar}{2} \int (\psi \vec{\nabla} \bar \psi -  \bar \psi \vec{\nabla} \psi) d^n \vec{x} & \mathrm{linear\; momentum} \\
\displaystyle \vec{\cJ} = \int \vec{x} \wedge \vec{P} \, d^n \vec{x} & \mathrm{angular\; momentum} \\
\displaystyle \cM = m \int |\psi|^2 \, d^n \vec{x} & \mathrm{mass} \\
\displaystyle \vec{\cG} = t \vec{\cP} - m \int |\psi|^2 \, \vec{x} \, d^n \vec{x} & \mathrm{boost} \\
\displaystyle \cD = \frac{n + 2}{n-4} t \cH + \frac{3}{n-4} \int \vec{x} \cdot \vec{P} \, d^n \vec{x} & \mathrm{dilation}
\end{array}
\right.
\end{equation}
}
with $\displaystyle U(x, t) = \Delta_{\R^n}^{-1} 4 \pi G m |\psi|^2$.

In the case where $n = 4$, the symmetry group is the full Schr\"odinger group, \ie the same symmetry group as that of the free Schr\"odinger equation, that we review in section~\ref{ss:sym_schr}. Hence, the wavefunction transforms as,
\begin{align}
\label{repr_sn_neq4}
\widetilde{\psi}(\bx, t) = \, & \frac{1}{(ft + g)^{n/2}} \exp \left(\frac{im}{\hbar} \left(\frac{f}{2} \frac{||A\textbf{x}+\textbf{b}t+\textbf{c}||^2}{ft+g} - \langle\textbf{b}, A \textbf{x}\rangle - \frac{||\textbf{b}||^2 t}{2} + h \right) \right) \times \nonumber \\
& \psi\left(\frac{A \textbf{x} + \textbf{b} t + \textbf{c}}{f t + g}, \frac{d t + e}{f t + g}\right),
\end{align}
with $f \in \R$ an inversion, and $dg-ef = 1$. The conserved quantities are then given by,

{\setstretch{2}\begin{equation}
\label{conserved_n_eq4_sn}
\left\lbrace
\begin{array}{ll}
\displaystyle \cH = \int \left[ \frac{-\hbar^2}{2m} \vec{\nabla} \psi \cdot \vec{\nabla} \bar\psi - m U |\psi|^2 \right] d^4 \vec{x} \qquad \qquad & \mathrm{energy} \\
\displaystyle \vec{\cP} = \int \vec{P} d^4 \vec{x} = \frac{i \hbar}{2} \int (\psi \vec{\nabla} \bar \psi -  \bar \psi \vec{\nabla} \psi) d^4 \vec{x} & \mathrm{linear\;momentum} \\
\displaystyle \vec{\cJ} = \int \vec{x} \wedge \vec{P} \, d^4 \vec{x} & \mathrm{angular\; momentum} \\
\displaystyle \cM = m \int |\psi|^2 \, d^4 \vec{x} & \mathrm{mass} \\
\displaystyle \vec{\cG} = t \vec{\cP} - m \int |\psi|^2 \, \vec{x} \, d^4 \vec{x} & \mathrm{boost} \\
\displaystyle \cD = 2 t \cH + \int \vec{x} \cdot \vec{P} \, d^4 \vec{x} & \mathrm{dilation} \\
\displaystyle \cK = t \cD - t^2 \cH - \frac{m}{2} \int |\psi|^2 r^2 d^4 \vec{x} & \mathrm{inversion}
\end{array}
\right.
\end{equation}
}

It is remarkable that $n = 4$ is such a different case in that it displays the full symmetries of the free Schr\"odinger equation.

\subsection{Newton--Cartan structures: the geometry of Galilean relativistic spacetime}

\subsubsection{Definition of a Newton--Cartan structure}
\label{ss:def_nc}

Historically, the Newton--Cartan (NC) structures \cite{Cartan23,Trautman63,Havas64,Kunzle72} were introduced as a way to study Galilean relativistic theories in a geometrical way. Just as it is possible to obtain Newton's equations of motion for gravity from Einstein's equations in the Galilean limit, it is possible to obtain the structure of non relativistic space time, also called Newton--Cartan structures, as the Galilean limit of a Lorentzian manifold.

Let us take the example of a flat spacetime, where in the relativistic case the structure is given by Minkowski's spacetime which, with coordinates $(\bx, t)$, is described by a (covariant) metric $\eta = \delta_{ij} dx^i \otimes dx^j - c^2 dt \otimes dt$. Since the Newtonian limit is computed by taking the speed of light $c$ to be infinite, one should consider the inverse metric, or contravariant metric, $g^{-1} = \delta^{ij} \partial_i \otimes \partial_j - c^{-2} \partial_t \otimes \partial_t$. In the Newtonian limit $h = g^{-1}\vert_{c\rightarrow\infty}$, the temporal part of the metric vanishes, and we are left with a contravariant degenerate ``metric'' $h = \delta^{ij} \partial_i \otimes \partial_j$\footnote{This ``metric'' $h$ must not be confused with the central extension parameter appearing in \eqref{sym_phase_schr}}. Its kernel is generated by the ``clock'' $\theta = dt$, \ie $h(\theta) = 0$. It is also possible to take the Newtonian limit of connection on the Lorentzian spacetime to obtain the Newton--Cartan connection $\nabla^\cN$, provided the compatibility condition $d\theta = 0$, which is trivially realized in our example.

It is also possible to define a Newton--Cartan structure without taking the Newtonian limit. Indeed, a Newton--Cartan structure $(\cN, h, \theta, \nabla^\cN)$ is the data of a manifold $\cN$ endowed with a degenerate contravariant 2-tensor $h$ together with a 1-form $\theta$, called the ``clock'', such that $h(\theta) = 0$, and a connection $\nabla^\cN$ compatible with $(h, \theta)$, \ie $\nabla^\cN h = 0$ and $\nabla^\cN \theta = 0$. The last relation automatically implies that $\theta$ is a closed form, $d\theta = 0$. Unlike on Lorentzian manifolds with the Levi-Civita connection, a connection compatible with both $h$ and $\theta$ on an NC structure is not unique. Indeed, if $\Gamma^\alpha_{\beta\gamma}$ are the coefficients of a connection $\nabla^\cN$ compatible with $h$ and $\theta$, then the connection $\widetilde{\nabla}^\cN$ is also compatible with coefficients $\widetilde{\Gamma}^\lambda_{\mu\nu} = \Gamma^\lambda_{\mu\nu} + 2 \theta_{(\mu} \alpha_{\nu)\rho} h^{\rho\lambda}$, with $\alpha$ an arbitrary 2-form \cite{Kunzle72}.

A Newton--Cartan spacetime also has interesting structure with respect to the time axis. Indeed, consider a vector field $X \in \ker\theta$. Since $\theta$ is a closed form, we have $L_X \theta = 0$, meaning that $\ker\theta$ is an integrable distribution, and that there exists a foliation on $\cN$, with each leaf $\Sigma_t$ being n-dimensional and parametrized by $t$, which we call the time. The set of leaves $\Sigma_t$ is the time axis $\T = \cN / \ker \theta$. Hence, we have a projection $\pi$ from the Newton--Cartan spacetime to the time axis, as depicted in figure~\ref{f:nc}. This echoes Newton's idea of an absolute time.

\setcounter{w}{6}
\setcounter{p}{1}
\setcounter{h}{4}

\tikzstyle{hidden} = [dashed,line width=1.1pt]
\tikzstyle{lesser} = [line width=1.2pt]
\tikzstyle{normal} = [line width=0.8pt]
\tikzstyle{normalh} = [dashed,line width=0.8pt]
\tikzstyle{arrow} = [line width=0.9pt, draw, -latex']
\tikzstyle{labels} = [->]

\tikzset{middlearrow/.style={
        decoration={markings,
            mark= at position #1 with {\arrow{>}} ,
        },
        postaction={decorate}
    }
}

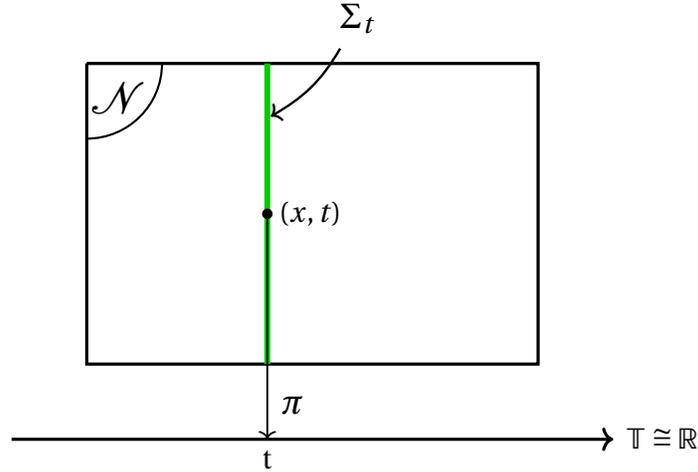
\begin{figure}[ht!]
\begin{tikzpicture}[line width=1.4pt]
  \draw [lesser] (0, 0) coordinate (n2) -- node [pos=0.07,scale=1.5,below]{$\cN$} (\value{w}, 0) coordinate (n3) -- (\value{w}, -\value{h}) -- (0, -\value{h}) coordinate (n1) -- (0, 0);
  \draw [normal] pic["",draw=black,-,angle eccentricity=1.2,angle radius=1cm] {angle=n1--n2--n3};

  \draw [lesser,->] (-\value{p}, -\value{h}-\value{p}) -- node [pos=1,right]{$\T \cong \R$} (\value{w} + \value{p}, -\value{h}-\value{p});

  \draw [line width=2.2pt,black!20!green] (0.4 * \value{w}, 0) -- (0.4 * \value{w}, -\value{h});
  \draw [line width=0.7pt,->] (0.4 * \value{w}, -0.5 * \value{h}) -- node [pos=0,scale=3]{.} node[pos=0,right]{$(x,t)$} node[pos=0.85,right,scale=1.2]{$\pi$} node[pos=1,below]{t} (0.4 * \value{w}, -\value{h} - \value{p});
  
  \node [draw=none,scale=1.2] at (0.6 * \value{w}, 0.6 * \value{p}) (surf) {$\Sigma_t$};
  \node [draw=none] at (0.38 * \value{w}, -0.2 * \value{h}) (surf2) {};
  \draw [labels,line width=0.9] (surf) to [out=240,in=30] (surf2);
  
\end{tikzpicture}
\caption{A Newton--Cartan structure projects onto the time axis.}
\label{f:nc}
\end{figure}

Newton--Cartan structures are greatly suited to study Galilean relativistic gravitational problems \cite{Cartan23,Trautman63,Trautman67}. Newton's field equation for gravity is realized through $\Ric(h) = 4 \pi G \rho \, \theta \otimes \theta$, with $\Ric(h)$ the Ricci tensor associated to the degenerate ``metric'' $h$, and $\rho$ the density of matter on $\cN$.

\subsubsection{Structural symmetries}
\label{ss:nc_sym}

Galilean relativistic symmetries can also be defined through Newton--Cartan structures. We will review here three of the most common symmetry groups. The first one is of course the Galilei group, and the other two are Galilean relativistic conformal groups, namely the Schr\"odinger group \cite{Niederer72} and the Chronoprojective group \cite{Duval82,DuvalH09}.

\paragraph{The Galilean group} is one of the first Lie group people are introduced to, as it describes the symmetries of our seemingly Galilean-relativistic world. While it may be defined as a group contraction of the Poincar\'e group \cite{InonuW53}, just like the Newton--Cartan structure can be defined as the Galilean limit of a Lorentzian manifold, it is also possible to define it naturally as a symmetry group of Newton--Cartan structures. Given a flat Newton--Cartan structure $(\cN, h, \theta, \nabla)$, the Galilean group $\Gal(n)$ is simply its group of automorphisms, 
\begin{equation}
\label{def_gal}
\Gal(n) = \Aut(\cN, h, \theta, \nabla^\cN) = \left\lbrace\Phi \in \Diff(\cN) \vert \Phi^* h = h, \Phi^* \theta = \theta, \Phi^* \nabla = \nabla\right\rbrace.
\end{equation}
Note that we do have to preserve the connection $\nabla$ to obtain the Galilean group. Indeed, due to the degeneracy of $h$, the group of automorphisms of $(\cN, h, \theta)$ is infinite dimensional\footnote{This group, $C^\infty(\R,\SE(n))\times \R$ is sometimes called the Coriolis group \cite{Duval93}}.

This group has the well-known matrix representation with its elements $a$ being of the form, see \textit{e.g.} \cite{Sou70,LevyLeblond72},
\begin{equation}
a = \left(\begin{array}{ccc}
A & \bb & \bc \\
0 & 1 & e \\
0 & 0 & 1
\end{array}\right),
\end{equation}
with $A \in O(n)$ a rotation, $\bb, \bc \in \R^n$ Galilean boosts and spatial translations, and $e \in \R$ a time translation. Its action on spacetime is given by,
\begin{equation}
\left\lbrace \begin{array}{l}
\bx' = A \bx +\bb t + \bc, \\
t' \; = t + e,
\end{array}\right.
\end{equation}

Its Lie algebra has $n(n-1)/2 + 2n + 1$ generators, $(J_i), (K_i), (P_i), E$ for, respectively, infinitesimal rotations, boosts, spatial translations and time translations, with non trivial commutators,
\begin{equation}
\label{alg_gal}
[J_i, J_j] = {\epsilon_{ij}}^k J_k, \quad [J_i, K_j] = {\epsilon_{ij}}^k K_k, \quad [J_i, P_j] = {\epsilon_{ij}}^k P_k, \quad [K_i, E] = P_i,
\end{equation}
with ${\epsilon_{ij}}^k$ the structure constants of $\so(n)$.

\paragraph{The Chronoprojective group} is a way to define conformal symmetries on Galilean relativistic structures. See \cite{DuvalH09} for an overview. Much like in the relativistic case where conformal symmetries means to the metric up to a (strictly positive) rescaling, Galilean relativistic symmetries will ask to preserve $h$ and $\theta$ up to rescalings, \ie $\Phi^* h = f h$ and $\Phi^* \theta = g \theta$ with $f, g$ two strictly positive real functions of $\cN$. Now, just like before, due to the degeneracy of $h$, the resulting group is infinite dimensional\footnote{If, in addition to $\Phi^* h = f h$ and $\Phi^* \theta = g \theta$, one asks for $\Phi^*(\gamma \otimes \theta) = \gamma \otimes \theta$, the resulting infinite dimensional group is the Schr\"odinger--Virasoro group.}. An additional structure one can ask for is to preserve the projective character of the connection, \ie we want to permute the geodesics, without necessarily keeping the affine parametrisation. The resulting group is called the Chronoprojective group \cite{ChD82,Duval82}, $\Chr(n)$. It can be realized as a matrix groups, with elements $a$ of the form,
\begin{equation}
\label{element_chr}
a = \left(
\begin{array}{ccc}
A & \bb & \bc \\
0 & d & e \\
0 & f & g
\end{array}
\right),
\end{equation}
with $A \in O(n)$ a rotation, $\bb, \bc \in \R^n$ Galilean boosts and spatial translations, and $d,e,f,g \in \R$ respectively a time dilation, time translation, inversion, and spatial dilation. Interestingly, the chronoprojective group can be found to be isomorphic to the ``orthonormal group'' of the degenerate metric $\Sigma = \diag(1, \ldots, 1, 0, 0)$, \ie $\Chr(n) = O(n,0) := \{a \in \GL(n+2,\R) / a\Sigma \bar{a} = \Sigma\}$. In the physical case $n = 3$, this group is 13 dimensional. This group acts projectively on $\cN$,
\begin{equation}
\left\lbrace \begin{array}{l}
\displaystyle \bx' = \frac{A \bx +\bb t + \bc}{ft + g}, \\
\displaystyle t' \; = \frac{d\, t + e}{f \, t + g},
\end{array}\right.,
\end{equation}
and its algebra is generated by $(J_i), (K_i), (P_i), E, S, T, I$ where we have added $S, T, I$ compared to the Galilean algebra, which are generators of infinitesimal spatial dilations, time dilations, and inversions. Their non trivial commutators are given by \cite{ChD82},
\begin{equation}
\label{alg_chr}
\begin{array}{llll}
[J_i, J_j] = {\epsilon_{ij}}^k J_k, & [J_i, K_j] = {\epsilon_{ij}}^k K_k, & [J_i, P_j] = {\epsilon_{ij}}^k P_k, & [K_i, E] = P_i, \\ \relax
[P_i, I] = - K_i, & [E, I] = -S-2T, & [I, T] = I,& [K_i, T] = K_i, \\ \relax
[E, T] = - E, & [K_i, S] = - K_i, & [P_i, S] = -P_i. & 
\end{array}
\end{equation}

\paragraph{The Schr\"odinger group} is a contraction of the Chronoprojective group, it appears as we link space and time dilations. It can also be represented as the group of matrix of the form \eqref{element_chr}, but instead of having $\displaystyle \left(\begin{array}{cc}
d & e \\
f & g
\end{array}\right) \in \GL(2,\R)$, we have $\displaystyle \left(\begin{array}{cc}
d & e \\
f & g
\end{array}\right) \in \SL(2,\R)$, \ie with $dg-ef = 1$. It is denoted as $\Sch(n)$. At the algebraic level, it means to define a generator $D = S + 2 T$, so that the algebra is now given by the non trivial commutators \cite{ChD82},
\begin{equation}
\label{alg_sch}
\begin{array}{llll}
[J_i, J_j] = {\epsilon_{ij}}^k J_k, & [J_i, K_j] = {\epsilon_{ij}}^k K_k, & [J_i, P_j] = {\epsilon_{ij}}^k P_k, & [K_i, E] = P_i, \\ \relax
[P_i, I] = - K_i, & [E, I] = -D, & [I, D] = 2 I, & [K_i, D] = K_i, \\ \relax
[P_i, D] = -P_i, & [E, D] = - 2 E. & &
\end{array}
\end{equation}

Note that this means time is dilated twice as much as space, or that the dynamical exponent is 2.

While this group is called the Schr\"odinger group, it is not the maximal symmetry group of the Schr\"odinger equation. Instead, its central extension, which will correspond to phase transformation, is. This group is called the extended Schr\"odinger group, and we will see its definition later on.

\subsection{Bargmann structures: a Lorentzian tool to study Galilean relativistic symmetries}

\label{sec-Prelim}


\subsubsection{Bargmann structure and its link to Newton-Cartan}
\label{ss_barg}

While Newton--Cartan structures are certainly interesting and useful to study gravitational problems, they are somewhat lacking for quantum problems. For instance, to study the symmetries of an equation, one could hope to recast this equation using only geometric objects of the structure. This should allow for a much easier time computing the symmetries. However, the Schr\"odinger equation does not have a simple geometric formulation on Newton--Cartan structures. Instead, it is usually convenient to write down non-relativistic systems in the formalism of what is called a Bargmann structure \cite{Bargmann54,DuvalBKP85,Eisenhart28}. This is a Lorentzian structure, which which is a principal bundle Newton--Cartan spacetime, where the group can be taken to be $(\R, +)$ or $U(1)$. It possesses geometrical tools which make the study of non-relativistic systems more geometrical and much easier to handle.

A Bargmann structure is defined as a manifold $M$ endowed with a Lorentzian metric $\rg$ and a light-like vector field $\xi$, nowhere vanishing, with $\rg(\xi, \xi) = 0$. It is also equipped with the usual Levi-Civita connection, compatible with $g$ and $\xi$, such that $\nabla \rg = 0$ and $\nabla \xi = 0$. A Bargmann structure will then be denoted by the triple $(M, \rg, \xi)$.

\setcounter{w}{9}
\setcounter{p}{1}
\setcounter{h}{7}

\tikzstyle{hidden} = [dashed,line width=1.1pt]
\tikzstyle{lesser} = [line width=1.2pt]
\tikzstyle{normal} = [line width=0.8pt]
\tikzstyle{normalh} = [dashed,line width=0.8pt]
\tikzstyle{arrow} = [line width=0.9pt, draw, -latex']
\tikzstyle{cone} = [line width=0.7pt]
\tikzstyle{labels} = [->]
\tikzstyle{carr} = [black!50!blue]
\tikzstyle{line} = [draw, -latex']
\tikzstyle{nc} = [black!50!red]

\tikzset{middlearrow/.style={
        decoration={markings,
            mark= at position #1 with {\arrow{{>}[scale=1.5]}} ,
        },
        postaction={decorate}
    }
}

\begin{figure}[ht]
\begin{tikzpicture}[line width=1.4pt,scale=0.85, every node/.style={transform shape}]
  \draw [lesser] (0,0) -- (0.45 * \value{w},\value{p}) -- (\value{w},0);
  \draw (\value{w},0) -- (0.55 * \value{w},-\value{p}) -- (0,0);

  \draw [line width=1.4pt] (0,0) -- (0,-\value{h});
  \draw (0.55 * \value{w}, -\value{p}) -- (0.55 * \value{w}, -\value{p} - \value{h});
  \draw (\value{w}, 0) -- (\value{w}, -\value{h});
  \draw [hidden] (0.45 * \value{w},\value{p}) -- (0.45 * \value{w},\value{p} - \value{h});
  \draw [normal] (0.9 * \value{w}, - \value{p}) circle (0.6cm) node [scale=1.5]{$M$};

  \draw [hidden] (0, - \value{h}) -- (0.45 * \value{w},\value{p} - \value{h}) -- (\value{w}, - \value{h});
  \draw (\value{w}, - \value{h}) -- (0.55 * \value{w},-\value{p} - \value{h}) -- (0, - \value{h});
  
  \draw [lesser,nc] (0, - 1.4 * \value{h}) -- (0.45 * \value{w},\value{p} - 1.4 * \value{h}) -- (\value{w}, - 1.4 * \value{h});
  \draw [nc] (\value{w}, - 1.4 * \value{h}) coordinate (nc1) -- node[pos=1,above,scale=1.5]{$\cN$} (0.55 * \value{w},-\value{p} - 1.4 * \value{h}) coordinate (nc2) -- (0, -1.4 * \value{h}) coordinate (nc3);
  \draw [normal] pic["",draw=black,-,angle eccentricity=1.2,angle radius=0.85cm] {angle=nc1--nc2--nc3};

  \draw [normal,carr] (0.725 * \value{w}, 0.5*\value{p}) -- (0.275 * \value{w}, -0.5*\value{p}) coordinate (c1);
  \draw [normal,carr] (0.275 * \value{w}, -0.5*\value{p}) -- node[pos=0.92,right,scale=1.5]{$\widetilde{\Sigma}_t$} (0.275 * \value{w}, -1*\value{h} -0.5*\value{p});
  \draw [normalh,carr] (0.725 * \value{w}, 0.5*\value{p}) -- (0.725 * \value{w}, 0.5*\value{p} - \value{h}) coordinate (c3) -- (0.275 * \value{w}, -\value{h}-0.5*\value{p}) coordinate (c2);
  \draw [normal] pic["",draw=black,-,angle radius=1.25cm] {angle=c3--c2--c1};

  \draw [normal, middlearrow={0.82},black!60!green] (0.725 * \value{w}, 0.5 * \value{p} - 1.4 * \value{h}) -- node [left,pos=0.32,scale=1.1] {$(x, t)\qquad$} node[scale=3.5,pos=1]{.} (0.05 * \value{w}, -1.4 * \value{h} - 1 * \value{p});
  
  \draw [line] (-0.225 * \value{w}, -1.4 * \value{h} - 0.5* \value{p}) -- node [pos=1.18,below,scale=1.1] {$T \cong \R$ (time axis)} node[below, pos=0.58,scale=1.1]{$t = \const \qquad$} (0.325 * \value{w}, -1.5 * \value{p} - 1.4 * \value{h});
  
  \draw [arrow] (0.1 * \value{w}, -0.5 * \value{h}) -- node [near end, right,scale=1.1] {$\xi$} (0.1 * \value{w}, -0.38 * \value{h});
  \draw [arrow] (0.80 * \value{w}, -0.65 * \value{h}) -- node [near end, left, scale=1.1] {$\xi$} (0.80 * \value{w}, -0.53 * \value{h});
  \draw [arrow] (0.40 * \value{w}, -0.4 * \value{h}) -- node[pos=0,scale=3]{.} node [left,pos=0,scale=1.1]{$(x,t,s)$} node [near end, right,scale=1.1] {$\xi$} (0.40 * \value{w}, -0.28 * \value{h});

  \draw [dashed,cone] (0.80 * \value{w}, -0.65 * \value{h}) -- (0.80 * \value{w}, -0.77 * \value{h});
  \draw [dashed,cone] (0.80 * \value{w}, -0.65 * \value{h}) -- (0.745 * \value{w}, -0.74 * \value{h});
  \draw [cone] (0.80 * \value{w}, -0.65 * \value{h}) -- (0.855 * \value{w}, -0.56 * \value{h});
  \draw [cone,rotate around={160:(0.827 * \value{w}, -0.55 * \value{h})}] (0.827 * \value{w}, -0.55 * \value{h}) ellipse (0.029 * \value{w} and 0.02 * \value{h});
  \draw [cone,dashed,rotate around={160:(0.773*\value{w}, -0.75 * \value{h})}] (0.773*\value{w}, -0.75 * \value{h}) ellipse (0.0295 * \value{w} and 0.02 * \value{h});
  
  \node[draw, align=center,scale=1.1] at (1.3*\value{w}, \value{p}) (barg) {Extended\\Bargmann\\space-time-action\\$(M, \rg, \xi)$};
  \draw [labels] (barg) -- (0.83 * \value{w}, -0.1 * \value{p});
  
  \node[draw=black!50!blue, align=center,scale=1.1] at (-0.20 * \value{w}, \value{p}) (carr) {Carroll\\space-action\\$(\widetilde{\Sigma}_t, \Upsilon, \widetilde{\xi})$};
  \node[draw=none] at (0.4 * \value{w}, -0.5*\value{p}) (carr2) {};
  \draw [labels] (carr) to [out=0,in=100] (carr2);
  
  \node[draw=black!60!green, align=center,scale=1.1] at (1*\value{w}, -\value{h} - 1.5*\value{p}) (euclide) {Euclidean\\space\\$(\Sigma_t, h)$};
  \node[draw=none] at (0.55*\value{w}, -1*\value{h}-2.83*\value{p}) (euclide2) {};
  \draw [labels] (euclide) to [out=180,in=45] (euclide2);
  
  \node[draw=black!50!red, align=center,scale=1.1] at (1.1 * \value{w}, -\value{h} - 4.1 * \value{p}) (nc) {Newton-Cartan\\space-time\\$(\cN, h, \theta, \nabla^\cN)$};
  \node[draw=none] at (0.8 * \value{w}, -\value{h} - 2.95 * \value{p}) (nc2) {};
  \draw [labels] (nc) to [out=140,in=25] (nc2);
  
  \draw [normalh, middlearrow=0.4] (0.40 * \value{w}, -0.4 * \value{h}) -- (0.40 * \value{w}, -\value{h} - 0.75 * \value{p});
  \draw [normal, middlearrow=0.36] (0.40 * \value{w}, -\value{h} - 0.75 * \value{p}) -- node [left,pos=0.32,scale=1.5]{$\pi$} node [pos=1,scale=3]{.} (0.40 * \value{w}, -1.12 * \value{h} - 2.22 * \value{p});
\end{tikzpicture}
\caption{Visualization of a 1+2 dimensional Bargmann structure, and its link to Newton-Cartan and Carroll structures \cite{Marsot17}.}
\label{f:barg}
\end{figure}
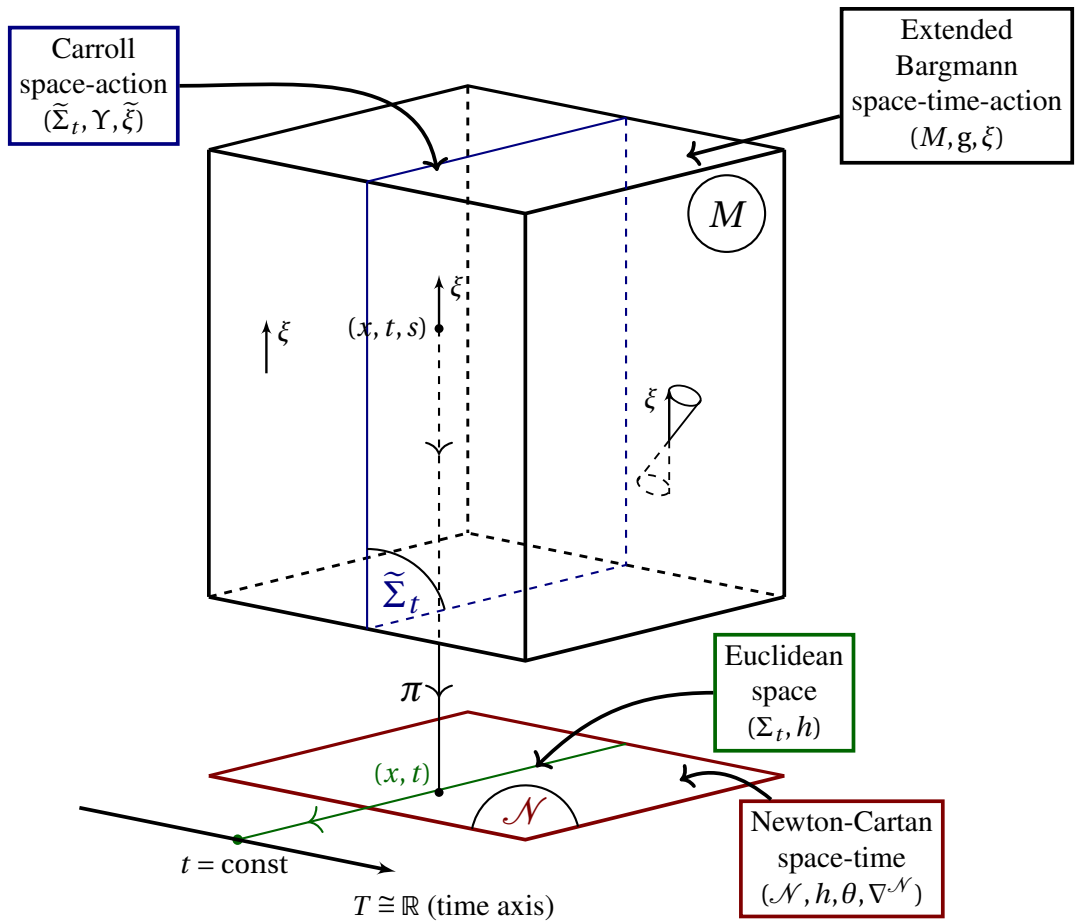

By definition of such structures, the quotient $\cN = M/\R\xi$ is endowed with a Newton-Cartan structure \cite{DuvalBKP85}, or non-relativistic spacetime, see section \ref{ss:def_nc} for an overview. In other words, there is a projection $\pi$ from the Bargmann structure along the direction given by the vector $\xi$ yielding a Newton--Cartan structure. The projection yields a degenerate contravariant ``metric'' $h = \pi^* g^{-1}$, and a ``clock'' $\theta = \pi^*g(\xi)$, generating the kernel of $h$. Finally, the connection $\nabla^\cN$ on the Newton--Cartan structure is also given by the projection of the one living in Bargmann spacetime. The Newton--Cartan structure thus defined is denoted $(\cN, h, \theta, \nabla^\cN)$. Taking a section $t = \const$, called $\Sigma_t$, of this Galilean relativistic spacetime naturally gives a Euclidean manifold $(\Sigma_t, h)$.

It is worthwhile to note that another kind of structures can be recovered from Bargmann spacetime: those of Carroll \footnote{The name, coined by L\'evy-Leblond, refers to Lewis Carroll.} \cite{LevyLeblond65,Dautcourt98,DuvalGH14,DuvalGH91,DuvalGHZ14}. Each slice of constant time $\widetilde{\Sigma}_t$ in the Bargmann structure is endowed with a Carrollean structure. Since $\xi$ on Bargmann is light-like, the ``metric'' $\Upsilon$ induced on these Carroll space-actions is again degenerate because the induced vector field $\widetilde{\xi}$  generates the kernel of the metric. A Carroll structure is denoted by the triple $(\widetilde{\Sigma}_t, \Upsilon, \widetilde{\xi})$. All these geometries are sketched in Fig.\ref{f:barg}.

\medskip
Let us go back to a Bargmann space. It can be equipped with local coordinates $(x^1, \ldots, x^n, t, s)$, where $s$ has the dimension of an action (per mass). Locally, the metric $\rg$ and the vector field $\xi$ can be written as what is known as a Brinkmann metric \cite{Brinkmann25}:
\begin{equation}
\label{def_g_barg}
\rg = \rg_{{}_{\Sigma_t}} + dt \otimes \omega + \omega \otimes dt, \qquad \omega = \varpi_i(x, t) dx^i - U(x, t) dt + ds, \qquad \xi = \partial_s
\end{equation}
with $\rg_{{}_{\Sigma_t}} = \rg_{ij}(x, t) \, dx^i \otimes dx^j$ with $x = (x^1, \ldots, x^n)$ the metric on each slice $\Sigma_t$, and where $\omega$ is a connection form on the principal $(\mathbb{R},+)$-bundle $\pi:M\to{}\cN$, with coefficients not depending on $s$. As explained in \cite{DuvalL15}, the spacetime function~$U$ (for example, the profile of the gravitational wave whose wave-vector $\xi$ is null and parallel) is interpreted in the present context as the Newtonian gravitational potential on NC spacetime, $\cN$.
The functions $\varpi_i$ can be interpreted as some kind of Coriolis potential \cite{Ehlers97,CostaN15}, or thought of as a gravitational magnetic moment. They have physical dimension $[\varpi] = L T^{-1}$, and the Coriolis curvature $\Omega = d_{\Sigma_t} \varpi$ that will appear below in the Christoffel symbols \eqref{christo} has dimension $[\Omega] =  T^{-1}$. The Coriolis curvature $\Omega$ was shown to be relevant in various physical situations \cite{CostaN15}, and it also appears in the Newtonian limit of the Taub-NUT spacetime \cite{Ehlers97}\footnote{\label{fnnut}In this limit, one obtains the curvature $\Omega$ as $\Omega = \star d_{\R^3}\frac{a}{r}$, with $a$ the Taub-NUT parameter. We can recover $\varpi$ by analogy with a magnetic monopole. Indeed, solving for $\varpi$ is the same equation as solving for the potential vector in the case where the magnetic field is given by a magnetic monopole. One solution for the corresponding connection is therefore given by $\varpi_\mp = \frac{y dx - x dy}{r(z\pm r)}$ \cite{GockelerS87}.}.
While one could think that $\varpi$ could appear in the Kerr spacetime, this spacetime does not have a physically acceptable Newtonian limit, unless one wants to add the concept of negative mass \cite{Keres67,Ehlers97}.

\smallskip
For example, the simple flat Bargmann structure is given by the metric $g_0$ such that:
\begin{equation}
\label{barg_flat}
\rg_0 = \delta_{ij} dx^i \otimes dx^j + dt \otimes ds + ds \otimes dt\ .
\end{equation}
By projection \cite{DuvalBKP85}, it induces the flat Newton-Cartan spacetime, given by $\cN=T\times\mathbb{R}^n$
with $(t,x^1,\ldots,x^n)$ as local coordinates, $h=\delta^{ij}\,\partial_i\otimes\partial_j$ and $\theta=dt$.

\smallskip
It is noteworthy to study what happens when the coordinate $s$ is transformed while preserving the fiber characterized by $\xi = \partial_s$. A general transformation of this kind is of the form
\begin{equation}
\label{trsf_s}
s \mapsto s + f(x,t),
\end{equation}
where $f$ is some function on $\cN$. Under this transformation, only the connection form $\omega$ from \eqref{def_g_barg} is modified in $\widehat{\omega}$ according to
\begin{equation}
\omega \mapsto \widehat{\omega} = \big(\varpi_i(x, t) - \partial_i f(x,t) \big) dx^i - \big(U(x, t) + \partial_t f(x,t)\big) dt + ds.
\end{equation}
Hence, in the general case, one is thus free to kill either \emph{one}  of the $n$ functions $\varpi_i$ or $U$ with such transformations.

In the particular case where $\varpi$ is exact, \ie $\varpi = d_{\Sigma_t}\vartheta$, with $\vartheta \in C^\infty(\cN,\R)$, one can turn off the terms $\varpi_i$ with such transformations~\eqref{trsf_s}\footnote{One could impose this form for $\varpi$ by postulating an additional field equation, due to Trautman \cite{Trautman63}, namely~${R^{\mu\nu}}_{\lambda\rho} = 0$, and thus ignore $\varpi$ \cite{Ehlers97}.}. Pushing this particular case further, if $\varpi = d_{\Sigma_t}\vartheta$ and $U = \partial_t \vartheta$, we can then turn off all gravitational and Coriolis potentials. 
We shall come back to this point in section~\ref{subsec-LLNonNC} in relation with gauge transformations.

\subsubsection{The Yamabe operator}
\label{ss_yam_and_densities}

Now, to be able to write the lifted differential equations from Newton--Cartan spacetime to Bargmann structures, one needs suitable differential operators on such Bargmann structures. A suitable operator for the lift of the Schr\"odinger equation is the Yamabe operator. 

The Yamabe operator, denoted by $\Delta_Y(g)$, stands for the conformally invariant version of the Laplacian, when acting on densities on a manifold $M$. To understand this operator, start by considering the action of the related operator on complex valued functions $\psi \in C^\infty(M, \Cc)$. We denote this operator by $\yamf(g)$, and it is defined as
\begin{equation}
\label{yamf}
\yamf(g) = g^{\mu\nu} \nabla_\mu \nabla_\nu - \frac{N-2}{4(N-1)} R(g),
\end{equation} 
with $\nabla_\mu$ the covariant derivative, $R(g)$ the curvature scalar, and $N = \dim M$.

Since the Yamabe operator is supposed to be conformally invariant, consider now a conformal rescaling of the above definition. That is, dilating the metric by a non-vanishing positive function $\Omega$ such that the new metric $\tg$ is related to the old one by $\tg = \Omega g$, with $\Omega \in C^\infty(M,\R_{>0})$. Under such rescaling, the curvature scalar $R(g)$ transforms as
\begin{equation}
\label{rescale_R}
R(\tg) = \frac{R(g)}{\Omega} - (N - 1) \left(\frac{\Delta \Omega}{\Omega^2} + \frac{N-6}{4} \frac{|d\Omega|^2}{\Omega^3}\right),
\end{equation}
with $\Delta \Omega = g^{\mu\nu} \nabla_\mu \partial_\nu \Omega$, and $|d \Omega|^2 = g^{\mu\nu} \partial_\mu \Omega \partial_\nu \Omega$.

Using the definition (\ref{yamf}) and the transformation of $R(g)$ above, we can compute the transformation of the operator $\yamf$. We get,
\begin{equation}
\label{rescale_yamf}
\yamf(\tg) = \Omega^{- \frac{N+2}{4}} \circ \yamf(g) \circ \, \Omega^{\frac{N-2}{4}}.
\end{equation}

We clearly see that this operator is not conformally invariant, though this is not the true Yamabe operator. The Yamabe operator is defined to act on densities $\Psi$. Recall that densities can be seen as functions proportional to some power, called \emph{weight}, of the norm of the volume element. Here, on a Riemannian manifold, the volume element is taken to be $\Vol = \sqrt{|g|} \, dx^1 \wedge \cdots \wedge dx^N$ in the local coordinate system $(x^1, \ldots, x^N)$, with $|g| = |\det(g_{\mu\nu})|$. The space of complex-valued densities of weight $\omega$ is written $\cF_{\omega}(M, \Cc)$. A $\omega$-density $\Psi \in \cF_{\omega}(M, \Cc)$ is thus locally written as,
\begin{equation}
\label{def_density}
\Psi = \psi(x) \, \nvol^\omega,
\end{equation}
with $\psi \in C^\infty(M, \Cc)$ a function. Then, going from $C^\infty(M, \Cc)$ to $\cF_\omega(M, \Cc)$ is done by composing functions with $\nvol^\omega$. Notice that (true) functions are in fact $0$-densities.

\begin{figure}[ht]
\begin{tikzpicture}[line width=1.4pt,scale=1.1, every node/.style={transform shape}]
  \node [draw=white,align=center] at (0, 0) (fl) {$\cF_{\lambda=\frac{N-2}{2N}}(M, \Cc)$};
  \node [draw=white,align=center] at (5, 0) (fm) {$\cF_{\mu=\frac{N+2}{2N}}(M, \Cc)$};
  \node [draw=white,align=center] at (0, -4) (cinf1) {$C^\infty(M, \Cc)$};
  \node [draw=white,align=center] at (5, -4) (cinf2) {$C^\infty(M, \Cc)$};

  \draw [arrow] (fl) to node [pos=0.5,above] {$\yam(g)$} (fm);
  \draw [arrow] (cinf1) to node [pos=0.5,left] {$\nvol^\lambda$} (fl);
  \draw [arrow] (cinf2) to node [pos=0.5,right] {$\nvol^\mu$} (fm);
  \draw [arrow] (cinf1) to node [pos=0.5,below] {$\yamf(g)$} (cinf2);
\end{tikzpicture}
\caption{Relations between the spaces of functions, densities, and both Yamabe operators}
\label{f:link_yams}
\end{figure}

We can see the relation between the Yamabe operator acting on functions, and the one acting on densities of weight $\lambda$, sending them to densities of weight $\mu$, on the diagram \ref{f:link_yams}. The Yamabe operator is then related to $\yamf(g)$ by,
\begin{equation}
\label{rel_yams}
\yamf(g) = \nvol^{-\mu} \circ \yam(g) \circ \nvol^\lambda.
\end{equation}

We want the Yamabe operator, $\yam(g)$, to be conformally invariant. The only degrees of freedom to get this invariance are the weights $\lambda$ and $\mu$ of the densities. Rescaling $\yamf(g)$ with $\tg = \Omega g$ in (\ref{rel_yams}) and using (\ref{rescale_yamf}), we get that
\begin{equation}
\begin{split}
& \Omega^{- \frac{N+2}{4}} \circ \nvol^{-\mu} \circ \yam(g) \circ \nvol^\lambda \circ \Omega^{\frac{N-2}{4}} = \\
& \qquad \qquad \qquad = \Omega^{-\frac{N \mu}{2}} \circ \nvol^{-\mu} \circ \yam(\tg) \circ \nvol^\lambda \circ \Omega^{\frac{N \lambda}{2}}
\end{split}
\end{equation}

From this equation, we see that to impose the invariance of the Yamabe operator $\yam(g)$ under conformal rescaling, \ie to have $\yam(\tg) = \yam(g)$, we need to have 
\begin{equation}
\lambda = \frac{N-2}{2N} \qquad \& \qquad \mu = \frac{N+2}{2N},
\end{equation}
that is, the Yamabe operator is a map from $\cF_{\frac{N-2}{2N}}(M, \Cc)$ to $\cF_{\frac{N+2}{2N}}(M, \Cc)$.

With such weights, the Yamabe operator has the nice property of being conformally invariant, \ie
\begin{equation}
\label{yam_inv}
\Delta_Y(\tg) \Psi = \Delta_Y(g) \Psi,
\end{equation}
for all $\tg = \Omega g$, with $\Omega \in C^\infty(M)$, and $\Psi \in \cF_{\frac{N-2}{2N}}$.

The action of $\yam(g)$ on densities of weight $\frac{N-2}{2N}$ can be written in term of $\yamf(g)$ using the relation (\ref{rel_yams}) :
\begin{equation}
\label{action_yam_yamf}
\yam(g) \Psi = \left(\yamf(g) \, \psi(x)\right) \nvol^\frac{N+2}{2N},
\end{equation}
with $\Psi = \psi(x) \nvol^\frac{N-2}{2N}$. In particular, when $\Psi$ is a Yamabe harmonic, \ie when $\yam(g) \Psi = 0$, we have the equivalence $\yam(g) \, \Psi = 0 \Leftrightarrow \yamf(g) \, \psi(x) = 0$.

\subsubsection{Lie derivatives and covariant derivatives of densities}

We want to consider now the action of a vector field $v$ on a density. Its action on the volume element $\Vol$, through the Lie derivative $\cL_v$, is $\displaystyle \cL_v \volg = \frac{\partial \volg}{\partial g_{\mu\nu}} \cL_v g_{\mu\nu}$. Then, since $\displaystyle \frac{\partial |g|}{\partial g_{\mu\nu}} = |g| g^{\mu\nu}$, and $\cL_v g_{\mu\nu} = 2 \nabla_{(\mu} v_{\nu)}$, we have
\begin{equation}
\label{lie_volume}
\cL_v \Vol = \nabla_\mu v^\mu \, \Vol.
\end{equation}

In the next sections, we will mainly be interested in the case where the Lie derivative in (\ref{lie_volume}) is taken along $\xi$, the null vector belonging to a Bargmann structure $(M, g, \xi)$. By definition, $\xi$ is covariant constant on such structure, $\nabla_\mu \xi^\nu = 0$. In such a case, we immediately get $\cL_\xi \volg = 0$, and thus the Lie derivative along $\xi$ of a density is simply given by,
\begin{equation}
\label{lie_density_bargmann}
\cL_\xi \Psi = \left(\cL_\xi \psi\right) \nvol^\omega
\end{equation}
with $\Psi = \psi \nvol^\omega$, and $\xi$ the null vector belonging to the Bargmann structure $(M, g, \xi)$.

For any covariant derivative $\nabla_\mu$ compatible with the metric, we have $\nabla_\mu g_{\nu\rho} = 0$, which implies $\nabla_\mu \volg = 0$. Thus, a covariant derivative acting on a density $\Psi = \psi \nvol^\omega$, when $\nvol = \volg$, only sees the function $\psi$, and we have \begin{equation}
\label{d_density}
\nabla_\mu \Psi = (\nabla_\mu \psi) \Vol^\omega.
\end{equation}

In conclusion, working with densities is almost transparent:
\begin{itemize}
\item The action of the Yamabe operator on densities is the action of the differential operator on the associated function, with a change of weight \eqref{action_yam_yamf}. Moreover, this change of weight is irrelevant when dealing with Yamabe-harmonic densities, which will be the case in this chapter.
\item The action of the Lie derivative along the null and divergent free vector $\xi$ in a Bargmann structure on a density is merely the action of the Lie derivative on the associated function \eqref{lie_density_bargmann}.
\item The action of a covariant derivative on a density is again just the action of this derivative on the associated function \eqref{d_density}.
\end{itemize}

Hence, in most cases in this chapter, we can treat densities as functions that happen to transform in a special way under dilations, while keeping in mind the change of weight after acting with the Yamabe operator.

\subsubsection{Symmetries of Bargmann structures}
\label{ss:sym_barg}

Now that we have defined a Bargmann structure as a geometric triple $(M, g, \xi)$, let us look at some of its structural (conformal) symmetries. We shall review here three symmetry groups: the Bargmann group, the extended Schr\"odinger group, and the extended chronoprojective group.

\paragraph{The Bargmann group}
It is the group of strict automophisms of a flat Bargmann structure, $\Barg(n) = \Aut(M, g, \xi)$. This group is isomorphic to a matrix group where an element $u(A, \bb, \bc, e, h) \in \Barg(n)$ is mapped to \cite{LevyLeblond74,DuvalBKP85},
\begin{equation}
u(A, \bb, \bc, e, h) = \left( \begin{array}{cccc}
A & \bb & \bzero & \bc \\
\bzero^T & 1 & 0 & e \\
- \bb^T A & - \Vert \bb \Vert^2/2 & 1 & h \\
\bzero^T & 0 & 0 & 1
\end{array}\right),
\end{equation}
with $A \in O(n)$ a rotation, $\bb, \bc \in \R^n$ spatial boosts and spatial translations, $e, h \in \R$ time translations, and fiber translations. The identity is $u_0(\Id_n, \bzero, \bzero, 0, 0) = \Id_{n+3}$. This group maps Bargmann coordinates $(\bx, t, s)$ to $(\bx', t', s')$ with,
\begin{equation}
\left\lbrace \begin{array}{l}
\bx' = A \bx +\bb t + \bc, \\[1ex]
t' \; = t + e, \\
\displaystyle s' \, = s - \langle \bb, A\bx\rangle - \frac{t}{2} \Vert \bb \Vert^2 + h.
\end{array}\right.
\end{equation}

Note that the set of elements $u(\Id_n, \bzero, \bzero, 0, h)$ forms a subgroup of the Bargmann group. It is relevant, as the quotient of the Bargmann group by this subgroup yields the Galilei group \eqref{def_gal}. As it turns out, the Bargmann group is a non trivial central extension of the Galilei group. This is of course reminiscent of the fact that the quotient of a Bargmann structure by the fiber generated by the vector field $\xi$ yield a Newton--Cartan structure.

At the infinitesimal level, the generators of the Lie algebra associated to the Bargmann group are $(J_i), (K_i), (P_i), E, M$ for infinitesimal rotations, boosts, spatial translations, time translation and fiber translation. The non trivial commutators are \cite{LevyLeblond74},
\begin{equation}
\label{alg_barg}
\begin{array}{lll}
[J_i, J_j] = {\epsilon_{ij}}^k J_k, & [J_i, K_j] = {\epsilon_{ij}}^k K_k, & [J_i, P_j] = {\epsilon_{ij}}^k P_k, \\ \relax
[K_i, E] = P_i, & [P_i, K_j] = M \delta_{ij}. & 
\end{array}
\end{equation}

We clearly see that the Bargmann algebra is the central extension of the Galilei group, as the sub algebra generated by $M$ lies in the center of the Bargmann algebra.

\medskip

\paragraph{The extended Schr\"odinger group} is the group that preserves the flat metric comformally, while the vector field is exactly preserved, and also preserving the Bargmann structure \ie with $\nabla_{\widehat{g}} \xi = 0$,
\begin{equation}
\label{def_esch}
\eSch(n) = \Aut(M, [g], \xi) = \lbrace\Phi\in \Diff(M) \vert \Phi^*g = \lambda g, \Phi^*\xi = \xi\rbrace,
\end{equation}
with $[g]$ the conformal class of metrics. The condition that the vector field must stay covariant constant is equivalent to ask for $d\lambda \wedge \vartheta = 0$, with $\vartheta = g(\xi)$ \cite{BurdetDP85}. With an appropriate choice of coordinates $(\bx, t, s)$, this means in practice that $\lambda$ must be a function of time.

This group acts projectively on the coordinates by,
\begin{equation}
\left\lbrace \begin{array}{l}
\displaystyle \bx' = \frac{A \bx +\bb t + \bc}{ft + g}, \\
\displaystyle t' \; = \frac{d\, t + e}{f \, t + g}, \\
\displaystyle s' \, = s + \frac{f}{2} \frac{\Vert A\bx + \bb t + \bc\Vert^2}{ft+g}- \langle \bb, A\bx\rangle - \frac{t}{2} \Vert \bb \Vert^2 + h,
\end{array}\right.
\end{equation}
where we have three additional parameters compared to the Bargmann group, namely, $d, f, g \in \R$ which are respectively a time dilation, an inversion, and a spatial dilation, such that $\displaystyle \left(\begin{array}{cc}
d & e \\
f & g
\end{array}\right) \in \SL(2,\R)$, \ie with $dg-ef = 1$. This group is called the \emph{extended} Schr\"odinger group, because just like in the case of the Bargmann and the Galilei group, quotienting out the extended Schr\"odinger group by the subgroup of fiber translations yield the Schr\"odinger group, that we have seen previously in section \ref{ss:sym_schr}. Note that once again, the extended Schr\"odinger group is a non trivial central extension of the Schr\"odinger group. Also, we recover the Bargmann group as a subgroup with $d = g = 1$ and $f = 0$.

The generators of the Lie algebra associated to the extended Schr\"odinger group are $(J_i), (K_i), (P_i), E, I, D, M$, with the same generators as the Bargmann algebra and with $I$ the generator of inversions and $D$ the generator of dilations. The non trivial commutators are,
\begin{equation}
\label{alg_esch}
\begin{array}{llll}
[J_i, J_j] = {\epsilon_{ij}}^k J_k, & [J_i, K_j] = {\epsilon_{ij}}^k K_k, & [J_i, P_j] = {\epsilon_{ij}}^k P_k, & [K_i, E] = P_i, \\ \relax
[P_i, K_j] = M \delta_{ij}, & [P_i, I] = - K_i, & [E, I] = -D, & [I, D] = 2 I, \\ \relax
[K_i, D] = K_i, & [P_i, D] = -P_i, & [E, D] = - 2 E. & 
\end{array}
\end{equation}

\paragraph{The extended Chronoprojective group}\footnote{In the literature, the extended chronoprojective group as defined here is often called the chronoprojective group, indistinguishably from the chronoprojective group defined on Newton--Cartan structures, leading to some confusions. We choose here to add \emph{extended} to its name to match the nomenclature of the (extended) Schr\"odinger group.} is the group preserving conformally both the metric and the vector field, 
\begin{equation}
\label{def_echr}
\eChr(n) = \Aut(M, [g], [\xi]) = \lbrace\Phi\in \Diff(M) \vert \Phi^*g = \lambda g, \Phi^*\xi = \nu \xi\rangle,
\end{equation}
with the requirement that $(M, \Phi^*g, \Phi^*\xi)$ be Bargmann implying $d\lambda \wedge \vartheta$ and $d\nu = 0$. Its projective action on $M$ is, \cite{DuvalL15}
\begin{equation}
\left\lbrace \begin{array}{l}
\displaystyle \bx' = \frac{A \bx +\bb t + \bc}{ft + g}, \\
\displaystyle t' \; = \frac{d\, t + e}{f \, t + g}, \\
\displaystyle s' \, = \frac{1}{\nu} \left(s + \frac{f}{2} \frac{\Vert A\bx + \bb t + \bc\Vert^2}{ft+g}- \langle \bb, A\bx\rangle - \frac{t}{2} \Vert \bb \Vert^2 + h\right).
\end{array}\right.
\end{equation}

The generators of the Lie algebra associated to the extended Chronoprojective group are $(J_i), (K_i), (P_i), E, I, S, T, M$, with the same generators as the Bargmann algebra and with $I$ the generator of inversions, and $S$ and $T$ the generators of spatial dilations and time dilations. The non trivial commutators are,
\begin{equation}
\label{alg_echr}
\begin{array}{llll}
[J_i, J_j] = {\epsilon_{ij}}^k J_k, & [J_i, K_j] = {\epsilon_{ij}}^k K_k, & [J_i, P_j] = {\epsilon_{ij}}^k P_k, & [K_i, E] = P_i, \\ \relax
[P_i, K_j] = M \delta_{ij}, & [P_i, I] = - K_i, & [E, I] = -S-2T, & [I, T] = I, \\ \relax
[K_i, T] = K_i, & [E, T] = - E, & [K_i, S] = - K_i, & [P_i, S] = -P_i, \\ \relax
[S, M] = 2M, & [T, M] = -M & &
\end{array}
\end{equation}

Just like how one recovers the Schr\"odinger group from the Chronoprojective group by contraction, see section \ref{ss:nc_sym}, it is possible to recover the extended Schr\"odinger group from this extended Chronoprojective group. Indeed, we see that from this algebra \eqref{alg_echr}, one recovers the extended Schr\"odinger algebra \eqref{alg_esch} with the contraction $D = 2 T + S$. While in the extended Chronoprojective algebra time and space dilations are independent, in the extended Schr\"odinger algebra time is dilated twice as much as space. It is said that the dynamical exponent $z$ is 2. Another remarkable fact is that the extended Chronoprojective group is an extension of the Chronoprojective group, which we defined in section \ref{ss:nc_sym}, but it is not central. This can clearly be seen at the algebraic level in~\eqref{alg_echr}, since $S$ and $T$ do not commute with $M$. However the contraction to the extended Schr\"odinger algebra yields indeed a central extension since $[D,M] = 2 [T, M] + [S, M] = 0$.

\subsubsection{The lift of the Schr\"odinger equation}

We have seen how to compute the symmetries of the Schr\"odinger equation~\eqref{eq_schro} in section \ref{ss:sym_schr} when it is written in its usual form. We will now see how writing the Schr\"odinger equation on a Bargmann structure renders the computation of its symmetries almost trivial.

The Schr\"odinger equation on a Bargmann structure $(M,g,\xi)$ is given by,
\begin{subequations} \label{schro_barg}
\begin{empheq}[left=\empheqlbrace]{align}
& \yam(g) \Psi = 0, \label{schro_barg_yam}\\
& L_\xi \Psi = i \frac{m}{\hbar} \Psi, \label{schro_barg_lxi}
\end{empheq}
\end{subequations}
with $\Psi \in \cF_\frac{N-2}{2N}$ a density and $\yam(g)$ the Yamabe operator associated to the metric $g$, see section \ref{ss_yam_and_densities} for an overview of densities and the Yamabe operator. As we have seen before when mentioning densities \eqref{def_density}, and together with the second equation \eqref{schro_barg_lxi}, in view of \eqref{lie_density_bargmann}, we can write locally $\Psi$, with suitable coordinates $(\bx, t, s)$ where $\xi = \partial_s$, as
\begin{equation}
\label{decomp_psi_schro}
\Psi = \psi(\bx, t) \exp\left(i \frac{m}{\hbar} s\right) \vert \Vol\vert^\frac{N-2}{2N}.
\end{equation}

Now, take the Bargmann metric to be $g = \Vert d\bx \Vert^2 + 2 dt \, ds - 2 U(\bx, t) dt^2$, and let us look at the corresponding Yamabe operator. As we have discussed in section \ref{ss_yam_and_densities}, and in view of \eqref{decomp_psi_schro}, equation \eqref{schro_barg_yam} is equivalent to $\yamf(g) \left(\psi(\bx, t) \exp\left(i m s/\hbar\right)\right) = 0$. Now, the expression of the differential operator on functions $\yamf$ \eqref{yamf} simplifies since here the Ricci scalar $R(g)$ vanishes. Moreover, when specifying this operator to the metric $g$ defined in this paragraph, one finds $\yamf(g) = \Delta_{\R^n} + 2 \partial_t \partial_s - 2 U (\partial_s)^2$, with $\Delta_{\R^n}$ the flat Laplacian on $\R^n$. Finally, the system of equations \eqref{schro_barg} with the above mentioned metric reduces to,
\begin{equation}
\left(\Delta_{\R^n} + \frac{2 i m}{\hbar} \partial_t + 2 \frac{m^2}{\hbar^2} U(\bx, t) \right) \psi(\bx, t) = 0,
\end{equation}
which is nothing but the Schr\"odinger equation with arbitrary potential $U(\bx, t)$. Note that we did not use the most general Bargmann metric~\eqref{def_g_barg}. It is indeed possible to generalize the Schr\"odinger equation such that the spatial metric is not flat, and more interestingly it is possible to include the Coriolis terms $\varpi$ that we have seen in section~\ref{ss_barg}.

\medskip

Let us now look at computing the symmetries of the \emph{free} Schr\"odinger equation in this framework, where it is recast into a system of two equations \eqref{schro_barg}. We want to find the group of transformations $\Phi$ such that if $\Psi$ is a solution to the Schr\"odinger equation \eqref{schro_barg}, then so is $\Phi^* \Psi$. From the first equation, we have $0 = \Phi^*\left(\yam(g) \Psi\right) = \yam(\Phi^* g) \left(\Phi^* \Psi\right)$ due to the naturality relation of the Yamabe operator \cite{MichelRS14}. Hence, $\Phi^* \Psi$ is a solution to \eqref{schro_barg_yam} if $\yam(\Phi^* g) = \yam(g)$. Now, by construction, the Yamabe operator is conformally invariant when acting on $\frac{N-2}{2N}$-densities, see \eqref{yam_inv}. This means that the transformations $\Phi$ belong to the conformal group $\Conf(M, g)$.

The symmetry of the second equation \eqref{schro_barg_lxi} is even more straightforward. Indeed, we have $\Phi^*\left(L_\xi \Psi\right) = \Phi^*\left(i \frac{m}{\hbar} \Psi\right)$, which implies that $L_{\Phi^* \xi} \left(\Phi^* \Psi\right) = i \frac{m}{\hbar} \left(\Phi^* \Psi\right)$. Hence, $\Phi^* \Psi$ is a solution to \eqref{schro_barg_lxi} as long as $\Phi^* \xi = \xi$. 

In the end, the symmetry group of the free Schr\"odinger equation written on a Bargmann structure is the group which preserves conformally the flat metric and keeps invariant the vector field $\xi$. This is exactly the definition of the extended Schr\"odinger group that we have seen in the section \ref{ss:sym_barg}.


\subsection{Including spinors in the theory}
\subsubsection{Justifications}

While the Schr\"odinger--Newton equation aims at describing the evolution of the wavefunction of Galilean relativistic massive spinless particles, some experiments are planned to take place with spin $1/2$ massive particles. L\'evy-Leblond fermions~\cite{LevyLeblond67} turn out to be the natural candidates. For instance, they ought to be studied as slow neutrons in an ultra cold neutron beam at the Institut Laue-Langevin (ILL-Grenoble, France) along the line as suggested by \cite{ColellaOW75}. Other experiments like those as proposed in \cite{Kaltenbaek12,Kaltenbaek15,MAQRO}, could also be supported by experiments at ILL in order to reveal a wave packet reduction process with a major change in the spreading of wave packets around and above a critical mass of a system composed by L\'evy-Leblond fermions. Moreover, this type of particles has been receiving some attention in different contexts, see for instance~\cite{DuvalHP96,CarigliaG18}.

\subsubsection{L\'evy-Leblond equation}

To properly study Galilean relativistic spinors, one needs to introduce the L\'evy-Leblond equation \cite{LevyLeblond67}. This is a wave equation for particles of spin $\half$\footnote{There also exist L\'evy-Leblond equations for any spin.}. It is similar to the Dirac equation in the sense that we have a 4-spinor $\Psi = \left(\begin{array}{l}\varphi \\ \chi\end{array}\right)$ on which first order differential operators act. It can be constructed in a similar fashion too: by seeking a linearization of the Schr\"odinger equation. Electromagnetism can also be taken into account, using the minimal coupling scheme. The L\'evy-Leblond equation is given by the system of coupled equations,

\begin{equation}
\label{ll_eq_em}
\left\lbrace\begin{array}{l}
(E - q V) \varphi + \sigma(\bp- q \bA) \, \chi = 0, \\[1ex]
 \sigma(\bp- q \bA) \, \varphi + 2 \, m \, \chi = 0,
\end{array}\right.
\end{equation}
where $\varphi$ and $\chi$ are two bispinors, and $E$ and $\bp$ respectively the energy and the momentum. The probability density is defined as $\rho = \varphi^\dagger \varphi$, so that the norm is,
\begin{equation}
\Vert \Psi \Vert^2 = \int \rho \, d^3 \bx.
\end{equation}
Note that the probability is not degenerate: $\varphi = 0$ implies $\chi = 0$ through the second equation of \eqref{ll_eq_em}. Also, as mentioned by L\'evy-Leblond \cite{LevyLeblond67}, $\rho$ has the correct transformation law under Galilean symmetries, which makes it a Galilean scalar.

It is possible to eliminate one of the two bispinor from \eqref{ll_eq_em} to recover the well-known Schr\"odinger--Pauli equation. For example, by eliminating $\chi$, we have,
\begin{equation}
\label{schro_pauli}
\left(E - q V - \frac{1}{2m} \left(\bp - q \bA\right)^2 - \frac{q}{2m} \sigma(\bB)\right) \varphi = 0,
\end{equation}
with $\bB = \mathrm{curl} \, \bA$.

As emphasised by L\'evy-Leblond, while the Schr\"odinger--Pauli equation can be derived from a limit of the Dirac equation, or is postulated, here in this completely Galilean relativistic theory, without taking any limit, the spinor degrees of freedom appear naturally, as well as the correct Land\'e g-factor of 2 for the magnetic moment $\mu = \frac{q}{2m} B$ of a spin-$\half$ particle.

\subsubsection{Dirac operator and spinor densities}

Now, we are going to want to lift the L\'evy-Leblond equation on a Bargmann structure. To this end, much like in the Schr\"odinger case in a previous section, we will need a suitable differential operator, preferably conformally invariant. While for the Schr\"odinger equation this was the Yamabe operator, see section \ref{ss_yam_and_densities}, for the L\'evy-Leblond equation we need the conformally invariant Dirac operator.

Let $(M, \rg, \xi)$ be a N-dimensional Bargmann manifold, with $N = n + 2$. The Dirac operator was originally introduced by Dirac to describe relativistic Quantum Mechanics with spin. This operator, that we denote $\diracf(\rg) = \gamma^\mu \nabla_\mu$ acts on spinors  $\psi \in \cS(M) = L^2(M)\otimes \Cc^k$, with $N = 2 k$ if $N$ is even, and $N = 2 k + 1$ if $N$ is odd, and with $(\gamma^\mu)$ the set of gamma matrices belonging to the Clifford algebra associated to the Bargmann space, such that $\gamma^\mu \gamma^\nu + \gamma^\nu \gamma^\mu = - 2 \rg^{\mu\nu}$ \footnote{Mind the sign convention.\label{sign-footnote}}.

This Dirac operator $\diracf(\rg)$ transforms non trivially under conformal dilations (or Weyl rescalings) of the metric $\rg \rightarrow \wh \rg = \lambda \rg$, with $\lambda \in C^\infty(M, \R_+^*)$. Yet, we would like to have a conformally invariant operator. We will write $\dirac(\rg)$ an operator satisfying
\begin{equation}
\label{inv_conf_dirac}
\dirac(\wh \rg) = \dirac(\rg)\ .
\end{equation}

It turns out that such a conformally invariant operator $\dirac(\rg)$ can be constructed from the usual Dirac operator $\diracf(\rg)$ in the following way grounded on geometry, see \cite{MichelRS14}
\begin{equation}
\label{rel_dirac}
\dirac(\rg) = \nvol^{\frac{N+1}{2N}} \circ \diracf(\rg) \circ \nvol^{- \frac{N-1}{2N}},
\end{equation}
where $\nvol$ is, as before, the canonical volume element of $M$. These volume forms cancel out the non trivial dilation terms coming from $\diracf(\rg)$. However, definition \eqref{rel_dirac} has a cost, now our operator $\dirac(\rg)$ does not act on spinors anymore, it rather acts on spinor densities which will be denoted by $\Psi \in \cS_w(M) = \cS(M) \otimes \cF_w(M)$, where $\cF_w(M)$ stands for the space of densities of weight $w$. This means that locally, spinor densities are written as
\begin{equation}
\label{spinor_densities}
\Psi = \psi \, \nvol^w,
\end{equation}
where $\psi \in \cS(M)$. Notice that usual spinors $\psi$ are merely 0-density spinors.

\tikzstyle{hidden} = [dashed,line width=1.1pt]
\tikzstyle{lesser} = [line width=1.2pt]
\tikzstyle{normal} = [line width=0.8pt]
\tikzstyle{normalh} = [dashed,line width=0.8pt]
\tikzstyle{arrow} = [line width=0.9pt, draw, -latex']
\tikzstyle{cone} = [line width=0.7pt]
\tikzstyle{labels} = [->]
\tikzstyle{carr} = [black!50!blue]
\tikzstyle{line} = [draw, -latex']
\tikzstyle{nc} = [black!50!red]

\tikzset{middlearrow/.style={
        decoration={markings,
            mark= at position #1 with {\arrow{{>}[scale=1.5]}} ,
        },
        postaction={decorate}
    }
}
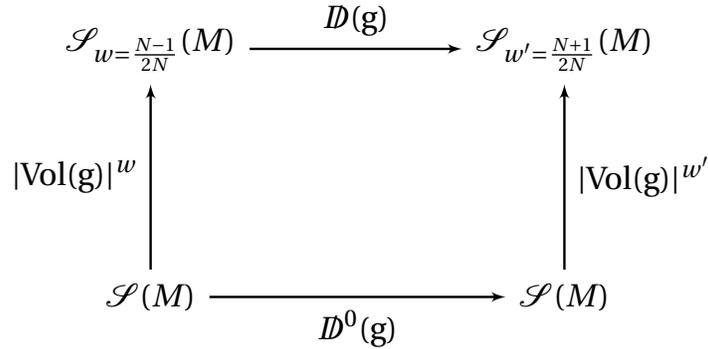
\begin{figure}[ht]
\begin{tikzpicture}[line width=1.4pt,scale=1.1, every node/.style={transform shape}]
  \node [draw=white,align=center] at (0, 0) (fl) {$\cS_{w=\frac{N-1}{2N}}(M)$};
  \node [draw=white,align=center] at (5, 0) (fm) {$\cS_{w'=\frac{N+1}{2N}}(M)$};
  \node [draw=white,align=center] at (0, -3) (cinf1) {$\cS(M)$};
  \node [draw=white,align=center] at (5, -3) (cinf2) {$\cS(M)$};

  \draw [arrow] (fl) to node [pos=0.5,above] {$\slashed{D}(\rg)$} (fm);
  \draw [arrow] (cinf1) to node [pos=0.5,left] {$\nvol^w$} (fl);
  \draw [arrow] (cinf2) to node [pos=0.5,right] {$\nvol^{w'}$} (fm);
  \draw [arrow] (cinf1) to node [pos=0.5,below] {$\slashed{D}^0(\rg)$} (cinf2);
\end{tikzpicture}
\caption{Relations between the spaces of spinors, spinor densities, and both Dirac operators}
\label{f:link_diracs}
\end{figure}

With the definitions \eqref{rel_dirac} and \eqref{spinor_densities}, and the fact that the spinor densities are of weight $w = \frac{N-1}{2N}$ here, it is easy to see the action of the conformally invariant Dirac operator on spinor densities,
\begin{equation}
\label{dirac_spinor}
\dirac(\rg) \Psi = \left(\diracf(\rg) \psi\right) \nvol^\frac{N+1}{2N},
\end{equation}
meaning that the operator $\dirac(\rg)$ sends $\frac{N-1}{2N}$-densities into $\frac{N+1}{2N}$-densities, while the action on the spinorial part is just like the usual Dirac operator, as summed up in the diagram in Fig.\ref{f:link_diracs}.

\medskip
Let us now consider the action of covariant derivatives on spinor densities. Since the connection used here is the usual Levi-Civita connection, it is compatible with the metric and thus the covariant derivative only sees the spinorial part, and not the volume. Its action on a spinor $\psi \in \cS(M)$ is defined as \cite{Kosmann71,Souriau64}
\begin{equation}
\label{der_spinors}
\nabla_X \psi = X^\mu \partial_\mu \psi - \frac{1}{8} X^\mu \left[\gamma^\rho, \partial_\mu \gamma_\rho - \Gamma^\sigma_{\mu\rho} \gamma_\sigma\right] \psi,
\end{equation}
with $\gamma_\mu = \rg_{\mu\nu} \gamma^\nu$, and is such that $\nabla \gamma = 0$.

\medskip
The action of a Lie derivative along a vector field $X$ on a spinor can also be defined \cite{Kosmann71},
\begin{equation}
\label{def_lie_spinor}
L_X \psi = \nabla_X \psi - \frac{1}{4} \gamma^\mu \gamma^\nu \nabla_{[\mu} X_{\nu]} \, \psi.
\end{equation}
In order to obtain the action of a Lie derivative on a spinor density, one must combine \eqref{def_lie_spinor} and the action of a Lie derivative on the volume element \eqref{lie_volume} with \eqref{spinor_densities}, to get
\begin{equation}
\label{lie_spinor_density}
L_X \Psi = \left( \nabla_X \psi \right)\nvol^\frac{N-1}{2N} - \frac{1}{4} \gamma^\mu \gamma^\nu \nabla_{[\mu} X_{\nu]} \Psi + \frac{N-1}{2N} \left(\nabla_\mu X^\mu\right) \Psi \ .
\end{equation}

Notice for later use, that in the case of a Lie derivative along $\xi$, the covariantly constant null vector field entering in the definition of a Bargmann structure, we get that $L_\xi \Psi = \left(L_\xi \psi \right) \nvol^\frac{N-1}{2N}$.

In conclusion, much like in the Schr\"odinger case with the Yamabe operator, working with spinor densities is almost transparent, as most operators used here act on spinor densities just like they do on spinors. We still have to be careful when considering dilations and conformal transformations, as the densities will play an important role there.

\section{The L\'evy-Leblond--Newton system}
\label{sec-LLN}

It is recalled that the L\'evy-Leblond equation \cite{LevyLeblond67} is an equation describing Galilean relativistic fermions in 3-dimensional space. While one could have worked with the Schr\"odinger-Pauli equation, L\'evy-Leblond showed that the Schr\"odinger equation can be factorized into a system of first order partial differential equations, in analogy with the derivation of the Dirac equation from the Klein-Gordon's one. The free L\'evy-Leblond system of PDE's is given by
\begin{equation}
\label{ll_eq}
\left\lbrace\begin{array}{l}
\hbar \, \sigma(\bpartial) \, \varphi + 2 \, m \, \chi = 0 \\[1ex]
i \, \hbar \, \partial_t \varphi - \hbar \, \sigma(\bpartial) \, \chi = 0
\end{array}\right.
\end{equation}
with two bispinors $\varphi$ and $\chi$, where $\sigma$ denotes the set of the three Pauli matrices, and\footnote{Throughout this section, the notation $\sigma(\boldsymbol{a}) = \sigma_k a^k = \vec{\sigma}\!\cdot\!\vec{a}$, where bold letters $\boldsymbol{a}=\vec{a}$ for vectors in $\mathbb{R}^3$, will be used. A slight abuse of notation yields $\boldsymbol{\sigma}$ as well to compactly denote the three Pauli matrices.} 
$\sigma(\bpartial) = \sigma^i \partial_i$. It is worthwhile to notice that the second bispinor $\chi$ is non-dynamical, unlike in the Dirac equation. This is to be expected since the Schr\"odinger equation is of first order in time. The two bispinors fit into a 4-spinor $\psi = \left(\begin{array}{c}\varphi \\ \chi \end{array}\right)$.

The L\'evy-Leblond--Newton equation, or LLN for short, is when we add a gravitational potential in the L\'evy-Leblond equation, whose source is the probability density of the 4-spinor $\psi$. This is in the same spirit of the Schr\"odinger-Newton equation \cite{Diosi84,DuvalL15}, which is the Schr\"odinger equation with a gravitational potential whose source is the probability density of the wavefunction. The system thus becomes,
\begin{equation}
\label{lln}
\left\lbrace\begin{array}{l}
\hbar \, \sigma(\bpartial) \, \varphi + 2 \, m \, \chi = 0 \\[1ex]
i \, \hbar \, \partial_t \varphi - m U \varphi - \hbar \, \sigma(\bpartial) \, \chi = 0
\end{array}\right.
\end{equation}
together with the Poisson equation for the potential $U$ and mass density $\rho$
\begin{equation}
\label{poisson}
\Delta U(x, t) = 4 \pi G \rho, \qquad \rho = m \varphi^\dagger \varphi\, .
\end{equation}

While one could work with this system directly, we will see that writing the LLN equation in the formalism of a Bargmann structure will make the symmetries apparent and the general study of this system more transparent.

\subsection{Lifting LLN on the Bargmann space}

Motivated by the previous considerations, let us call the L\'evy-Leblond--Newton system on Bargmann the set of coupled equations\footnote{Notice also that the covariant form of the LL equation was first provided in~\cite{KunzleD86}.}
\begin{subequations} \label{lln_barg}
\begin{empheq}[left=\empheqlbrace]{align}
& \gamma^\mu \gamma^\nu + \gamma^\nu \gamma^\mu = - 2 \rg^{\mu\nu} \label{clialg} \\[1ex]
& \dirac \Psi = 0 \label{diracmnul} \\[1ex]
& L_\xi \Psi = i \frac{m}{\hbar}\, \Psi \label{equivariance} \\[1ex]
& \Ric(\rg) = 4 \pi G \rho \, \theta \otimes \theta \label{poissonbarg} \\[1ex]
& \rho = m \overline\Psi^\sharp \gamma(\xi) \Psi^\sharp \label{density}
\end{empheq}
\end{subequations}
with $\Psi$ a spinor $\frac{N-1}{2N}$-density, that we can locally decompose as $\Psi = \psi \, \nvol^\frac{N-1}{2N}$, where here $\psi = \left(\begin{array}{c}\varphi \\ \chi \end{array}\right)$, with $\varphi$ and $\chi$ two bispinors. Then, $m$ is a mass, and $\overline \Psi = \Psi^\dagger G$, with $G$ such that $\overline \gamma_\mu = G^{-1} \gamma^\dagger_\mu G = \gamma_\mu$, and $G^\dagger = G$.
Some comments are in order. 

\smallskip
The reader's attention is drawn to 
the sign in the relation \eqref{clialg} defining the Clifford algebra, (see footnote~\ref{sign-footnote}). This comes from the signature of the metric, chosen to be $(+, \ldots, +, -)$ on the Bargmann space, so that we recover a positive metric when projecting onto the non-relativistic Newton--Cartan spacetime. Also, while a seemingly arbitrary dimension $N$, or $n$, appears in the relations, the reader must keep in mind\footnote{We shall generically work in space dimension $n$, going back to $n=3$ when required.}
that this work is focused on the $n = 3$ case, namely $N=5$.

In the probability density definition \eqref{density}, the notation $\Psi^\sharp$ corresponds to a \emph{normalized} spinor which is defined as
\begin{equation}
\label{norm_psi}
\Psi^\sharp = \frac{\Psi}{||\Psi||_\rg} \qquad \& \qquad ||\Psi||^2_\rg = \int_{\Sigma_t} \overline \psi \gamma(\xi) \psi \, \mathrm{vol}( h)
\end{equation}
so that $||\Psi^\sharp||^2_\rg = 1$. 
Here $\mathrm{vol}(h)$ stands for the canonical volume form of $\Sigma_t$,\footnote{This volume form can be defined intrinsically. Indeed, call $\eta=\rg^{-1}(\omega)$ the vector field associated with the connection form $\omega$ given by \eqref{def_g_barg}; one checks that $\eta$ is null and $\omega$-horizontal. Then $\mathrm{vol}(\rg)(\xi,\eta)$ flows down to NC spacetime, $\cN$; once pulled-back to $\Sigma_t$, it canonically defines the volume $n$-form $\mathrm{vol}(h)$. 
The latter admits the following local expression, namely $\mathrm{vol}(h)=\sqrt{\det(\rg_{ij}(x,t))}\,dx^1\wedge\cdots\wedge dx^n$, where $h=h^{ij}(x,t)\,\partial/\partial{x^i}\otimes\partial/\partial{x^j}$ and $(h^{ij})=(\rg_{ij})^{-1}$.
}
and $\psi(\bx,t)=\psi_t(\bx)$ --- with $\psi_t\in{}L^2\left(\Sigma_t,\mathrm{vol}(h)\right)\otimes \mathbb{C}^k$ --- is as in \eqref{spinor_densities}.

\medskip
The intent to write this system on a Bargmann space and not directly on the usual Galilean relativistic space time, Newton--Cartan, is that on Bargmann the system is written in a completely covariant and geometrical way, which makes it easier to compute its symmetries. Note that on Bargmann, this system is written with a Dirac equation for a null mass~\eqref{diracmnul}.

\subsection{Recovering the LLN system on Newton--Cartan \label{subsec-LLNonNC}}

The L\'evy-Leblond equation \eqref{ll_eq}, as originally written \cite{LevyLeblond67}, was on flat space of dimension $n = 3$. In order to recover the LLN equation from the system \eqref{lln_barg}, we first put ourselves in this case, with a spatially flat metric on Bargmann space,
\begin{equation}
\label{barg_metric}
\rg = \Vert d\bx\Vert^2 + 2 \, dt \, ds - 2 \,U(\bx, t) dt^2 + 2 \,\bvarpi(\bx, t)\!\cdot\! d\bx \, dt
\end{equation}
with $\bx\in\mathbb{R}^3$, $U(\bx, t)$ a scalar potential, and $\bvarpi(\bx, t)$ a covariant Coriolis vector potential.

We are now going to see what each of the relations in the system \eqref{lln_barg} becomes when we specify the metric to \eqref{barg_metric}.

\paragraph{Clifford algebra}
In order to satisfy the Clifford algebra \eqref{clialg} for the Bargmann metric \eqref{barg_metric} whose matrix reads
\[
\rg = (\rg_{\mu\nu}) = \begin{pmatrix}
I_3 &  \bvarpi & \boldsymbol{0} \\
\bvarpi^{\text{T}} & -2U & 1\\
\boldsymbol{0}^{\text{T}} & 1 & 0
\end{pmatrix}, \qquad
\text{and} \quad 
\rg^{-1} = (\rg^{\mu\nu}) = \begin{pmatrix}
I_3 & \boldsymbol{0} & -\bvarpi \\
\boldsymbol{0}^{\text{T}} & 0 & 1\\
- \bvarpi^{\text{T}} & 1 & 2U+ \bvarpi^2
\end{pmatrix},
\]
the set of gamma matrices is computed to be 
\begin{equation}
\gamma^t = \left(\begin{array}{cc}
0 & 0 \\
1 & 0
\end{array}\right), \qquad
\gamma^j = \left(\begin{array}{cc}
- i \sigma^j & 0 \\
0 & i \sigma^j
\end{array}\right), \qquad
\gamma^s = \left(\begin{array}{cc}
i \sigma(\bvarpi) & -2 \\
U & -i \sigma(\bvarpi)
\end{array}\right)
\end{equation}
where the $\sigma^j$ are the Pauli matrices, $U = U(\bx, t)$, $\bvarpi = \bvarpi(\bx, t)$, and $\sigma(\bvarpi) = \sigma^i \varpi_i$. Note that since the metric is spatially flat, we have $\varpi^i = \varpi_i$, and likewise for $\boldsymbol{\sigma}$. We have $\gamma_\mu = \rg_{\mu\nu} \gamma^\nu$, which becomes,
\begin{equation}
\gamma_t = \left(\begin{array}{cc}
0 & -2 \\
- U & 0
\end{array}\right), \qquad
\gamma_j = \left(\begin{array}{cc}
- i \sigma_j & 0 \\
\varpi_j & i \sigma_j
\end{array}\right), \qquad
\gamma_s = \left(\begin{array}{cc}
0 & 0 \\
1 & 0
\end{array}\right),
\end{equation}
such that we also have $\gamma_\mu \gamma_\nu + \gamma_\nu \gamma_\mu = - 2 \rg_{\mu\nu}$.

\paragraph{Equivariance relation}
To compute the equivariance relation \eqref{equivariance}, we need to compute the Christoffel symbols associated to the Bargmann metric \eqref{barg_metric}. The non-zero ones are, (see \cite{DuvalL15}):
\begin{align} \label{christo}
\Gamma^i_{tt} &= \partial_i U + \partial_t \varpi_i; \quad \Gamma^s_{it} = - \partial_i U - \half \Omega_{ij} \varpi^j; \notag\\[-3mm]
\\[-3mm]
\Gamma^s_{tt} &= - \partial_t U - \varpi^i(\partial_i U + \partial_t \varpi_i); \quad \Gamma^s_{ij} = \partial_{(i}\varpi_{j)}; \quad \Gamma^i_{jt} = - \half \Omega_{ij} \notag
\end{align}
with $\Omega = d_{\Sigma_t} \bvarpi$ the Coriolis curvature.

\smallskip
Since $\xi(U) = 0$, and $\xi^\mu \Gamma^\rho_{\mu\nu} = 0$, applying the definition of the Lie derivative on a spinor density \eqref{lie_spinor_density} gives $L_\xi \Psi = \partial_\xi \Psi = i \frac{m}{\hbar} \Psi$, exactly like in the free case where the potential $U$ and the Coriolis vector potential $\bvarpi$ vanish.

Note that this equivariance relation together with the density character of the spinors \eqref{spinor_densities} imply the following decomposition of $\Psi$,
\begin{equation}
\label{decomp_psi}
\Psi(\bx, t, s) = e^\frac{ims}{\hbar} \, \psi(\bx, t) \, \nvol^\frac{n+1}{2(n+2)}.
\end{equation}

\paragraph{Poisson equation}
From the metric  \eqref{barg_metric} used here, the Ricci tensor gives constraints on $U$ and $\omega$, so that the gravitation equation \eqref{poissonbarg} takes the form
\begin{equation}
\delta \Omega = 0 \qquad \& \qquad \Delta_{\R^n} U + \frac{\partial}{\partial t} \delta \bvarpi + \half \Vert\Omega\Vert^2 = 4 \pi G \rho
\end{equation}
with $\delta$ the codifferential acting on differential forms on the Euclidean space~$\Sigma_t\simeq\mathbb{R}^n$ and $\Vert\Omega\Vert^2 = \half \delta^{ik}\delta^{jl} \Omega_{ij}\Omega_{kl}$. Note that in the case $\bvarpi = 0$, we recover the usual Poisson equation \eqref{poisson}.\footnote{The Galilean limit of Taub-NUT spacetimes, see footnote \ref{fnnut}, yields $\delta \Omega = 0$, $\delta \varpi = 0$, and $\Vert\Omega\Vert^2~=~4 a^2/r^4$.}

The density \eqref{density} is such that $\rho = m \overline\Psi \gamma(\xi) \Psi = m \Psi^\dagger G \gamma_s \Psi = m \varphi^\dagger \varphi$, with $G = \left(\begin{array}{cc} 0 & 1\\1 & 0
\end{array}\right)$. Note that the probability density only involves the first bispinor $\varphi$, as was remarked by L\'evy-Leblond in~\cite{LevyLeblond67}. This is not a problem, as we will see later on.

\paragraph{The massless Dirac equation}
We are now left with the massless Dirac equation on Bargmann \eqref{diracmnul}, $\dirac(\rg) \Psi = 0$.

The second term in the expression of the covariant derivative of spinors \eqref{der_spinors} can be split into two parts: $\gamma^\mu \left[\gamma^\rho, \partial_\mu \gamma_\rho\right]$ and $- \gamma^\mu \left[\gamma^\rho, \Gamma^\sigma_{\mu\rho} \gamma_\sigma\right]$. We have for the former
\begin{equation}
\label{eq-former}
\gamma^\mu \left[\gamma^\rho, \partial_\mu \gamma_\rho\right] = - 2 \sigma^i \sigma^j \partial_i \varpi_j \left(\begin{array}{cc}
0 & 0 \\
1 & 0
\end{array}\right),
\end{equation}
while the latter becomes
\begin{align}
- \left[\gamma^\rho, \Gamma^\sigma_{\mu\rho} \gamma_\sigma\right] & = \Gamma^i_{\mu t} [\gamma_i, \gamma^t] + \Gamma^s_{\mu t} [\gamma_s, \gamma^t] + \Gamma^s_{\mu i} [\gamma_s, \gamma^i] + \Gamma^i_{\mu j} [\gamma_i, \gamma^j] \notag\\[1mm]
& = 2 \, i \, \sigma^j \, \left(\Gamma^j_{\mu t} - \Gamma^s_{\mu j}\right) \left(\begin{array}{cc}
0 & 0 \\
1 & 0
\end{array}\right) - 2 \, i \, \Gamma^j_{t k} \left(\begin{array}{cc}
\epsilon_{j k l} \sigma^l & 0 \\
\sigma^k \varpi^j & \epsilon_{j k l} \sigma^l
\end{array}\right),
\end{align}
with $\epsilon_{j k l}$ the fully skewsymmetric Levi-Civita tensor, and $\epsilon_{1 2 3} = 1$. The non zero components are for $\mu = t$ and $\mu = j$; they read
\begin{align}
- \left[\gamma^\rho, \Gamma^\sigma_{t\rho} \gamma_\sigma\right] & = 2 \,i\, \sigma^k \left(2 \partial_k U + \partial_t \varpi_k\right) \left(\begin{array}{cc}0 & 0\\1 & 0\end{array}\right) + 2 \, i \, \epsilon_{k l m} \sigma^m \partial_k \varpi_l \left(\begin{array}{cc}1 & 0 \\ 0 & 1\end{array}\right),\\
- \left[\gamma^\rho, \Gamma^\sigma_{j\rho} \gamma_\sigma\right] & = - 2 \, i \, \sigma^k \partial_k \varpi_j \left(\begin{array}{cc}0 & 0\\1 & 0\end{array}\right).
\end{align}
Upon contracting with $\gamma^\mu$, we get,
\begin{equation}
- \gamma^\mu \left[\gamma^\rho, \Gamma^\sigma_{\mu\rho} \gamma_\sigma\right] = 2 \, \delta^i_j \, \partial_i \varpi^j \left(\begin{array}{cc}
0 & 0 \\
1 & 0
\end{array}\right),
\end{equation}
which combined with \eqref{eq-former} yields,
\begin{equation}
\gamma^\mu \left[\gamma^\rho, \partial_\mu \gamma_\rho - \Gamma^\sigma_{\mu\rho} \gamma_\sigma\right] = - 2 \, i \, \sigma\left(\bpartial\times\bvarpi\right) \left(\begin{array}{cc}
0 & 0 \\
1 & 0
\end{array}\right).
\end{equation}
The massless Dirac equation \eqref{diracmnul} on Bargmann can thus be developed as
\begin{align}
\left[ \left(\begin{array}{cc}
0 & 0 \\
\partial_t & 0
\end{array}\right)
\right. & +
\left(\begin{array}{cc}
- i \sigma(\bpartial) & 0 \\
0 & i \sigma(\bpartial)
\end{array}\right)  \notag \\[3mm]
& \left. +\,
\frac{i \, m}{\hbar} \left(\begin{array}{cc}
i \sigma(\bvarpi) & -2 \\
U & -i \sigma(\bvarpi)
\end{array}\right)
+
\frac{1}{4}\left(\begin{array}{cc}
0 & 0 \\
i \sigma(\bpartial\times\bvarpi) & 0
\end{array}\right)
\right]
\left(\begin{array}{c}
\varphi \\
\chi
\end{array}\right) = 0,
\end{align}
which generalizes the original LL equation since it equivalently reads in bispinor components as
\begin{equation}
\label{generalized_lln_eq}
\left\lbrace\begin{array}{l}
\hbar \, \sigma(\bpartial) \, \varphi + 2 \, m \, \chi - i \, m \, \sigma(\bvarpi)\, \varphi = 0 \\[1ex]
i \, \hbar \, \partial_t \varphi - m U \varphi - \hbar \, \sigma(\bpartial) \, \chi + i \, m \, \sigma(\bvarpi) \, \chi - \frac{1}{4} \hbar \, \sigma(\bpartial\times\bvarpi) \, \varphi = 0.
\end{array}\right.
\end{equation}
The first equation can be recast to show the 1st order relation between the two bispinors,
\begin{equation}
\label{rel_two_bispinors}
\chi = -\frac{\hbar}{2m} \sigma(\bpartial) \, \varphi + \frac{i}{2} \sigma(\bvarpi) \, \varphi \, ,
\end{equation}
and gives us the opportunity to write the system \eqref{generalized_lln_eq} solely in terms of the principal bispinor $\varphi$. This is the reason why writing the probability density  only in terms of $\varphi$ is not a problem, the second bispinor is somewhat redundant in the LL model. We thus recover a second order differential equation, akin to the Schr\"odinger equation, for a bispinor $\varphi$ with a gravitational potential $U$, and the Coriolis (co)vector potential $\bvarpi$,
\begin{equation} \label{eq-H}
\left(-\frac{\hbar^2}{2m} \Delta + \frac{i \hbar}{2}\left[\sigma(\bpartial)\circ\sigma(\bvarpi) + \sigma(\bvarpi)\circ\sigma(\bpartial)\right] + m \!\left(U + \frac{\Vert\bvarpi\Vert^2}{2}\right) + \frac{1}{4} \hbar \sigma(\bpartial\times\bvarpi) \right) \varphi = i\hbar \, \partial_t\, \varphi.
\end{equation}
It is worthwhile to notice at this stage that the (self-adjoint) Hamiltonian in the l.h.s. of \eqref{eq-H} fulfills the most general form dictated by the Galilean relativity principle as stated in \cite{Jauch68,Piron76} and refreshed in a modern language in \cite[§ 8.4 Galilean invariance]{LeBellac}. This principle provides a way to justify the minimal coupling form through the strong link between translation in momentum and the action of Galilean boosts. 
According to \cite{Piron76} the most general form for a Hamiltonian acting on a bispinor is thus given by 
\[
H = \frac{1}{2m} \big( \bP I_2 - \boldsymbol{A}_\mu(\bx,t) \sigma^\mu \big)^2 + V_\mu(\bx,t) \sigma^\mu
\]
where $\mathbf{P}$ is the momentum operator, for $\mu=0,1,2,3,4$, $\mathbf{A}_\mu$ gives four vector fields, $V_\mu$ stands for four scalar fields and $(\sigma^\mu) = (I_2,\bsigma)$ is a basis for $2\times 2$ complex matrices. After some algebra, a direct comparison yields (dropping the unit matrix) the equivalent expression\footnote{The Hamiltonian occurring in the generalized SN equation \cite[Eq.(3.9)]{DuvalL15} is readily seen to be recast into the canonical form as
$H =  \frac{1}{2m} \big( \mathbf{P} - m \bvarpi \big)^2 + m U$.} for the Hamiltonian obtained in \eqref{eq-H}\footnote{Since Galilean boosts form an abelian subgroup of the SN group, such a canonical form for the Hamiltonian was expected.}
\begin{equation}
\label{hamiltonien}
H = \frac{1}{2m} \big( \mathbf{P} - m \bvarpi \big)^2 + m U - \frac{1}{4}\, \hbar \,  \sigma(\bpartial\times\bvarpi)
\end{equation}
for $\mathbf{A}_0 =  m \bvarpi$, $\mathbf{A}_k \equiv 0$, $V_0 = m U$ and 
$\boldsymbol{V} = (V_1,V_2,V_3) = - \frac{1}{4}\, \hbar\, (\bpartial\times\bvarpi)$. The last term is reminiscent of the Pauli coupling term $\sigma(\boldsymbol{B})$ for spin $1/2$. Note that the (pseudo) vector $\bOmega = \bpartial\times\bvarpi$ is linked to the curvature 2-form $\Omega$ by $\bOmega = \star \Omega$. It remains to interpret the coupling upon setting $\mathbf{S} = \hbar\,\boldsymbol{\sigma}/2$ for the spin operator
\[
- \frac{1}{4}\, \hbar \,  \sigma(\bpartial\times\bvarpi) = -\half \, \bS\!\cdot\! \bOmega
\]
where $\bOmega = \bpartial\times\bvarpi$ is very similar to $\boldsymbol{B} = \bpartial\times \boldsymbol{A}$ for $\boldsymbol{A}$ the usual Maxwell vector potential. 

In order to complete the analogy with electromagnetism, we can look at spin precession due to this Coriolis term. Computing the usual time evolution of the operator through $d\bS/dt = \frac{i}{\hbar} \left[H, \bS\right]$ and the Hamiltonian \eqref{hamiltonien}, we obtain,
\begin{equation}
\frac{d\bS}{dt} = \half\, \bS \times \bOmega,
\end{equation}
in accordance with \cite{CostaN15}.

\smallskip
On the other hand, thanks to the canonical form of the Hamiltonian $H$ given in \eqref{hamiltonien}, it is well-known that a $U(1)$-gauge transformation of the wave-function corresponds to a gauge transformation of the potentials, see \textit{e.g.} \cite[§13-5]{Jauch68}.
In light of these observations, one may wonder whether by a phase change on the bispinor $\varphi(\bx,t) \mapsto (\Theta\varphi)(\bx,t) = e^{\frac{im}{\hbar} \vartheta (\bx,t)} \varphi(\bx,t)$ the Coriolis potential could be put to zero. Mimicking \cite[§13-5]{Jauch68}, for $\varphi$ subject to the Schrödinger equation $ i\hbar \, \partial_t\, \varphi= H\varphi$, one gets
\begin{align*}
&\Theta \mathbf{P} \Theta^{-1} = \mathbf{P} - m \bpartial \vartheta \quad\Rightarrow \quad \Theta \big( \mathbf{P} - m \bvarpi \big)^2 \Theta^{-1} = \big( \mathbf{P} - m (\bvarpi + \bpartial \vartheta) \big)^2 \\
& H' = \Theta H \Theta^{-1} + i\hbar (\partial_t \Theta) \Theta^{-1} = \frac{1}{2m}\big( \mathbf{P} - m (\bvarpi + \bpartial \vartheta) \big)^2 + m (U - \partial_t \vartheta) - \frac{1}{4}\, \hbar \,  \sigma(\bpartial\times\bvarpi).
\end{align*}
If $\bvarpi = - \bpartial \vartheta$, (namely, the Coriolis curvature $\Omega = d \bvarpi = i_{\bpartial\times\bvarpi}\mathrm{vol}(h)= 0$) and hence $\bpartial\times\bvarpi \equiv 0$. Remember that $\delta \Omega = \delta d \omega = \bpartial \times (\bpartial\times\bvarpi) \!\cdot\! d\bx = (\bpartial (\bpartial\!\cdot\!\bvarpi) - \Delta_{\mathbb{R}^3} \bvarpi)\!\cdot\! d\bx$.
Moreover, the self-gravitating coupling is at least modified, or if moreover $U - \partial_t \vartheta=0$ then the Newton potential can be turned off allowing the recovering the free LL equation. This makes contact with the general discussion given at the end of section~\ref{ss_barg}. In particular, the meaning of the gauge transformation on the bispinor $\varphi$ correponds to a translation $s \mapsto s + \vartheta (\bx,t)$ in the $s$ variable in the Bargmann space.

\medskip
In the usual case where we have $\bvarpi = 0$, we recover the original L\'evy-Leblond equations \cite{LevyLeblond67} with a scalar potential $U$, \eqref{lln} which forms, with the Poisson equation \eqref{poisson}, the L\'evy-Leblond--Newton system projected onto Newton--Cartan spacetime.

In this case, the relation between the two bispinors \eqref{rel_two_bispinors} becomes,
\begin{equation}
\label{rel_two_bispinors_varpi0}
\chi = -\frac{\hbar}{2m} \sigma(\bpartial) \, \varphi,
\end{equation}
and on replacing $\chi$ in \eqref{lln} by \eqref{rel_two_bispinors}, we recover the usual Schr\"odinger equation, for the bispinor~$\varphi$,
\begin{equation}
-\frac{\hbar^2}{2m} \Delta \varphi + m \, U \varphi = i\hbar \, \partial_t \varphi.
\end{equation}

\smallskip

\subsection{Current and chirality}

Let us first investigate the current associated to the LLN equation.
Recall that the Bargmann structure is a relativistic structure, and for this reason, we can write the Dirac equation, although for the massless case here. We can thus define a Dirac current
\begin{equation}
j^\mu = \overline \Psi \gamma^\mu \Psi
\end{equation}
that is naturally conserved, \ie $\nabla_\mu j^\mu = 0$. What we want though, is a current on the Newton--Cartan non-relativistic spacetime. First, note that $j^0 = j^\mu \xi_\mu$, in the spatially flat case, is coherent with the definition of the mass density in \eqref{density}. Then, since $\xi$ is covariantly constant by definition, and taking into account the equivariance relation \eqref{equivariance}, we have $\nabla_s j^s = 0$. This current $(j^\mu)$ on Bargmann thus projects onto a current $(J^\alpha)$ on Newton--Cartan, which is again conserved, $\nabla_\alpha J^\alpha = 0$, with components\footnote{$\varrho$ must not be confused with $\rho = m \varphi^\dagger \varphi = m \varrho$ introduced in \eqref{poisson}.}
\begin{equation}
\varrho = \varphi^\dagger \varphi \qquad \& \qquad \boldsymbol{J} = i \left(\varphi^\dagger \bsigma \chi - \chi^\dagger \bsigma \varphi \right) \in \R^3\, ;
\end{equation}
an alternative expression of $\boldsymbol{J}$, only in terms of the principal bispinor $\varphi$, reads
\begin{equation}
\boldsymbol{J} = \frac{\hbar}{2mi} \left[ \varphi^\dagger \left(\bpartial \varphi\right) - \left(\bpartial \varphi\right)^\dagger \varphi \right] + \frac{\hbar}{2m} \bpartial \times \left(\varphi^\dagger \bsigma \varphi \right) - i \bvarpi \times \left(\varphi^\dagger \bsigma \varphi\right).
\end{equation}
We clearly notice that the first part of this current has the same general expression as the usual Schr\"odinger current, and the second part accounts for the spinorial aspect.

\medskip
Let us now turn to the study of the chirality by considering
the chiral operator $\Gamma$ acting on spinors on Bargmann space of $N=3+2$ dimensions. Since the Brinkmann metrics on Bargmann space are non diagonal, the general definition of the chiral operator has to be used,
\begin{equation}
\label{chiral}
\Gamma = - \frac{\sqrt{-g}}{5!} \epsilon_{\mu\nu\rho\lambda\sigma} \gamma^\mu \gamma^\nu \gamma^\rho \gamma^\lambda \gamma^\sigma
\end{equation}
(with the convention $\epsilon_{123ts}=+1$), which, in our case, simply gives
\begin{equation}
\label{chiral_flat}
\Gamma = I_4 \, .
\end{equation}
The triviality of the chirality operator comes from the odd dimension of Bargmann space (here $N=5$). Indeed, according to the Clifford algebra, in odd dimensions, $\Gamma$ commutes with all $\gamma^\mu$, and hence, by Schur's lemma, has to be a multiple of the identity. This is in accordance with \cite{Duval85} where the chiral operator does not seem to be relevant in non-relativistic dynamics within a space of spatial dimension 3.

\section{LLN symmetries}
\label{sec-LLN-sym}

With the formulation of the LLN equations on a Lorentzian Bargmann spacetime, we are in position to investigate their symmetries, in particular, the maximal symmetry group.

\subsection{Spacetime symmetries}

Finding the symmetries of the system of equations \eqref{lln_barg} is to find the transformations $\Phi$ such that if $\Psi$ is a solution of \eqref{lln_barg} then so is $\Phi^* \Psi$. 
In the following, while we explicitly show the dependence in $n$, we assume the physical case $n = 3$.

As a prerequisite, note the naturality relationship \cite{KolarMS93} for the Dirac operator,
\begin{equation}
\label{nat_dirac}
\Phi^*(\dirac(\rg)) = \dirac(\Phi^*\rg)
\end{equation}
for all $\Phi \in \Diff(M, \rg)$, together with the naturality of the Ricci tensor \cite{Besse87},
\begin{equation}
\label{nat_ricci}
\Phi^*(\Ric(\rg)) = \Ric(\Phi^*\rg)
\end{equation}
and of the equivariance operator,
\begin{equation}
\label{nat_lie}
\Phi^*(L_\xi) = L_{\Phi^*\xi}\ .
\end{equation}

From the massless Dirac equation \eqref{diracmnul}, for any transformation $\Phi$, we have $\Phi^* \left(\dirac(\rg) \Psi\right) = 0$. Introducing the naturality relationship \eqref{nat_dirac}, we have $\dirac(\Phi^* \rg) \Phi^* \Psi = 0$. To obtain the desired result, namely
\begin{equation}
\dirac(\rg) \Phi^* \Psi = 0,
\end{equation}
we need to restrict the transformations $\Phi$ to those preserving the Dirac operator, which are transformations preserving the metric up to a conformal factor, as seen with \eqref{inv_conf_dirac}. This means the $\Phi$s are such that
\begin{equation}
\label{conf_g}
\Phi^* \rg = \lambda \rg,
\end{equation}
for $\lambda$ a strictly positive valued function of $M$. Since we want the transformations to be expressed on the Newton-Cartan spacetime, the direction of the fiber generated by $\xi$ should also be preserved, hence the restriction,
\begin{equation}
\label{conf_xi}
\Phi^* \xi = \nu \xi,
\end{equation}
with $\nu$ another function of $M$.

If we want to preserve the Bargmann structure, $\Phi^* \xi$ needs to be compatible with the connection built from the transformed metric $\Phi^* \rg$. This gives the following conditions on $\lambda$ and $\nu$,
\begin{equation}
\label{cond_conn}
d \lambda \wedge \theta = 0 \qquad \mathrm{\&} \qquad d\nu = 0.
\end{equation}
In practice, $\lambda$ turns out to be a positive non-vanishing function of time $\lambda(t)$, and $\nu \in \R$.

So far, the conditions on $\Phi$ we have worked out, namely \eqref{conf_xi}, \eqref{conf_g} and \eqref{cond_conn}, are exactly those of the extended Chronoprojective group that we have review in section~\ref{ss:sym_barg}, see definition \eqref{def_echr}.

Let us now look at the Clifford algebra. From the equation \eqref{clialg}, we immediately get,
\begin{equation}
\Phi^* \gamma_\mu = \lambda^{\half} \gamma_\mu.
\end{equation}

From the equivariance equation \eqref{equivariance}, we have $\Phi^*\left(L_\xi \Psi\right) = \Phi^*\left(\frac{i \, m}{\hbar} \Psi\right)$. Or, with \eqref{nat_lie}, \eqref{conf_xi} and by definition of a Lie derivative, $\nu L_\xi \Phi^* \Psi = \frac{i}{\hbar} \left(\Phi^* m\right) \left(\Phi^* \Psi\right)$. If we impose the dilation of the mass parameter $m$ under these transformations, $\Phi^* m = \nu \, m$, we recover the equivariance equation for $\Phi^* \Psi$,
\begin{equation}
L_\xi \left(\Phi^* \Psi\right) = \frac{i}{\hbar} m \, \left(\Phi^* \Psi\right).
\end{equation}

To check the symmetries of the gravitation equation \eqref{poissonbarg}, we first need to learn how the density $\rho$ transforms in \eqref{density}. From the definition, $\Phi^* \rho = \Phi^*\left(m \overline\Psi^\sharp \gamma(\xi) \Psi^\sharp\right)$, we see with the help of \eqref{norm_psi} and the dilation of the mass in the paragraph above, that
\begin{equation}
\label{nat_prob_density}
\Phi^* \rho = \lambda^{-\frac{n}{2}} \nu \,m \, \overline{\left(\Phi^*\Psi\right)}^\sharp \gamma(\xi) \left(\Phi^* \Psi\right)^\sharp \ .
\end{equation}

Moving on to the last equation of the set \eqref{lln_barg} to study, \ie \eqref{poissonbarg}, we have, with \eqref{nat_ricci} $\Ric(\Phi^*\rg) = 4\pi G \left(\Phi^* \rho\right) \left(\Phi^* \theta\right) \otimes \left(\Phi^* \theta\right)$. The Ricci tensor is to be rescaled here with the conformal factor $\lambda(t)$. If we write $\lambda(t) = \phi'(t)$, then the conformal transformation law of the Ricci tensor can be put into the remarkable form \cite{DuvalL15},
\begin{equation}
\label{trsf_ricci}
\Ric\left(\phi' \, \rg\right) = \Ric(\rg) - \half (N-2) S(\phi)\, \theta \otimes \theta,
\end{equation}
where $\displaystyle S(\phi) = \frac{\phi'''}{\phi'} - \frac{3}{2}\left(\frac{\phi''}{\phi'}\right)^2 = \big(\ln\lambda \big)'' - \frac{1}{2} \big( (\ln\lambda)'\big)^2$, is the well-known Schwarzian derivative.

Upon combining the transformation law \eqref{trsf_ricci} together with the transformation of the probability density \eqref{nat_prob_density} and since $\theta = \rg(\xi)$, we obtain,
\begin{equation}
\Ric(\rg) = 4 \pi G \,m \, \nu^3 \lambda^{2-\frac{n}{2}} \, \overline{\left(\Phi^*\Psi\right)}^\sharp \gamma(\xi) \left(\Phi^* \Psi\right)^\sharp + \half (N-2) S(\phi)\, \theta \otimes \theta\ .
\end{equation}
Hence, the gravitation equation is preserved for $\Phi^* \Psi$ as long as
\begin{equation}
\label{contrainte}
\lambda^{2-\frac{n}{2}} \nu^3 = 1,
\end{equation}
(hence, $\lambda$ and $\nu$ are constant functions) and
\begin{equation}
\label{schwarzien_nul}
S(\phi) = 0.
\end{equation}
As detailed in \cite[§ 4.4 and \textit{ff.}]{DuvalL15}, this constraint which characterizes homographic transformations in time, reduces to affine time transformation as given below in \eqref{action_sn}.

\smallskip
At the end, we find that the transformations preserving the LLN system, are
\begin{equation}
\label{def_group_sym}
\mathrm{LLN}(M,\rg,\xi) = \{\Phi \in \Diff(M) \vert \Phi^*\rg = \lambda \rg, \Phi^* \xi = \nu \xi, \lambda^{2-\frac{n}{2}} \nu^3 = 1\}.
\end{equation}
The symmetrygroup of the L\'evy-Leblond--Newton equation turns out to be isomorphic to the symmetry group of the Schr\"odinger--Newton equation \cite{DuvalL15}. Thus, its action on the coordinates is given by \cite[§ 5.4.2]{DuvalL15}, for $n = 3$,
\begin{subequations}
\label{action_sn}
\begin{empheq}[left=\empheqlbrace]{align}
& \widehat{\bx} = \frac{A \bx + \bb t + \bc}{g} \\
& \widehat{t} = \frac{d t + e}{g} \\
& \widehat{s} = \frac{1}{\nu}\left( s - \langle \bb, A \bx \rangle - \frac{\Vert\bb\Vert^2}{2} t + h\right),
\end{empheq}
\end{subequations}
with $A \in \SO(3), \bb, \bc \in \R^3, d, e, g, h\in \R$, and $d \, g = \nu$.

\smallskip
Infinitesimally, this corresponds to the Lie algebra of vector fields $X$ which can be written as,
\begin{equation}
\label{conf_sn}
\left(X^\mu\right) = 
\left(
\begin{array}{l}
\displaystyle \omega \bx + t \bbeta + \bgamma + \frac{3}{n-4} \delta \bx \\
\displaystyle \frac{n+2}{n-4} \delta t + \epsilon \\
\displaystyle - \bbeta \cdot \bx - \delta s + \eta 
\end{array}
\right)
\end{equation}
with $\omega \in \so(n), \bbeta, \bgamma \in \R^n, \epsilon, \delta, \eta \in \R$ which are, respectively, generators of rotations, boosts, spatial translations, time translations, dilations, and ``vertical'' translations. For the case $n = 3$, we have $\omega \bx = \epsilon_{ijk} \omega^i x^j \boldsymbol{e}^k \equiv j(\bomega) \bx$, where $j(\bomega)$ is a skew-symmmetric matrix parametrized by $\bomega$.

\subsection{Infinitesimal actions of the LLN group}

We want to find the representation of the group action \eqref{action_sn} acting on the spinors which are solutions of the LLN equation. To this end, we will first compute the action of a Lie derivative acting on a spinor along the vector field \eqref{conf_sn} generating the Lie algebra.

To define the effect of the group action \eqref{action_sn} on objects of interests such as the gravitational potential $U$ and the Coriolis vector potential $\bvarpi$, remember that these transformations act conformally on the metric \eqref{conf_g}. We want $\wh \rg = \Phi^* \rg = \lambda \, \rg$, and since $U$ and $\bvarpi$ appear in the metric, we readily find the transformation laws \cite{DuvalL15},
\begin{equation}
\label{trsf_u_w}
\wh U(\wh \bx, \wh t) = \lambda^{-1} \nu^{-2} \left(U(\bx, t) + \bvarpi(\bx,t) \cdot A^{-1}\bb\right) \quad \mathrm{\&} \quad \wh \bvarpi(\wh \bx, \wh t) = \lambda^{-\half} \nu^{-1} \bvarpi(\bx, t) \cdot A^{-1}.
\end{equation}

Infinitesimally, the conformal condition is written as $L_X \rg = \frac{2}{N} \left(\nabla_\mu X^\mu\right) \rg$, with $X$ the vector field as in \eqref{conf_sn}. Using the general expression for the metric $\rg = \rg_0 - 2 U(\bx,t) \, dt\otimes dt + \varpi_i(\bx,t) \,dx^i\otimes dt + \varpi_i(\bx,t) \, dt\otimes dx^i$ with $\rg_0$ the flat Bargmann metric as in \eqref{barg_flat}, we obtain the Lie derivative acting on $U$ and the $\varpi_i$. Since $U$ and the $\varpi_i$ are functions, we obtain the useful relations,
\begin{equation}
\label{trsf_inf_u_w}
X^\mu \partial_\mu U = -2\, \tfrac{n-1}{n-4}\, \delta \, U + \bvarpi \cdot \bbeta \quad \mathrm{\&} \quad (X^\mu \partial_\mu \varpi_i)\boldsymbol{e}^i  = - \tfrac{n-1}{n-4}\, \delta \, \bvarpi + \bomega \times \bvarpi \, ,
\end{equation}
where $\boldsymbol{e}^i, i=1,2,3$ is the canonical basis of $\mathbb{R}^3$.
We are now ready to compute the action of a Lie derivative of a spinor density along a conformal vector field $X$. Developing the expression of a Lie derivative of a spinor density \eqref{lie_spinor_density} in terms of partial derivatives, we get
\begin{equation}
\begin{split}
L_X \Psi = \, & X^\mu \partial_\mu \Psi - \frac{1}{8} X^\mu \left[\gamma^\rho, \partial_\mu \gamma_\rho\right] \Psi + \frac{1}{8} X^\mu \left[\gamma^\rho, \Gamma^\sigma_{\mu\rho} \gamma_\sigma\right] \Psi - \frac{1}{8} \left[\gamma^\mu, \gamma^\nu\right] \partial_\mu X_\nu \Psi + \\
&  + \frac{N-1}{2N} \, \partial_\mu X^\mu \Psi + \, \frac{N-1}{2N} \Gamma^\mu_{\mu\lambda} X^\lambda \Psi
\end{split}
\end{equation}
for any conformal Killing vector field $X$.

Computing all these terms for the expression of the vector field \eqref{conf_sn}, for $n = 3$, and in view of \eqref{contrainte} and \eqref{trsf_inf_u_w} we find the expression,
\begin{equation}
\label{inf_rep}
L_X \Psi = \underbrace{X^\mu \partial_\mu}_\textrm{\encircle{1}} \Psi + 
\underbrace{\left(
\begin{array}{cc}
-\frac{n-1}{2(n-4)} \, \delta + \frac{i}{2} \sigma(\bomega) & 0 \\[2ex]
\frac{i}{2} \sigma\left(\bbeta\right) & \frac{n-1}{2(n-4)} \, \delta + \frac{i}{2} \sigma(\bomega)
\end{array}
\right)}_\textrm{\encircle{2}}  \Psi
+ \underbrace{\frac{3(n+1)}{2(n-4)} \, \delta}_\textrm{\encircle{3}} \, \Psi \ .
\end{equation}

\smallskip
These conformal transformations thus act in three parts on our spinors:
\begin{enumerate}[1)]
\item
The first part is the coordinate transformation, \ie $\Psi(x, t, s) \rightarrow \Psi(\wh x, \wh t, \wh s)$.
\item
The second part of the transformation shows how the two bispinors behave under rotations, the fact that the two bispinors are dilated separately under these transformations, and that boosts mix of two bispinors.
\item
The last part of the transformation comes from the dilation of the volume of the densities. This is a global factor encompassing the two bispinors.
\end{enumerate}

\subsection{Integration to group representation}

To obtain a representation of the LLN group through \eqref{inf_rep}, is to find $\rho(\Phi) \Psi = (\Phi^{-1})^* \Psi$, such that if $\Psi$ is a solution of the LLN system \eqref{lln_barg}, then $\rho(\Phi) \Psi$ is again a solution.

\smallskip
The first step is thus to find the reverse action of \eqref{action_sn}, \ie $(\wh \bx, \wh t, \wh s) = \Phi^{-1}(\bx, t, s)$, for $\Phi = (a, \bb, \bc, d, e, g, h)$ belonging to the LLN group, where $a \in \mathrm{SU}(2)$ is such that $a \, \sigma( \bx) \, a^{-1} = \sigma  (A \bx)$. We get \cite{DuvalL15},
\begin{subequations}
\label{rev_action_sn}
\begin{empheq}[left=\empheqlbrace]{align}
& \wh \bx = A^{-1} \left[g \bx - \frac{g t - e}{d} \bb - \bc\right] \\[2mm]
& \wh t = \frac{gt - e}{d} \\[2mm]
& \wh s = \nu s + g \left\langle\bb, \bx\right\rangle - \frac{g}{2d} \Vert\bb\Vert^2 t + \frac{e}{2d} \Vert\bb\Vert^2 - \left\langle \bb, \bc \right\rangle - h
\end{empheq}
\end{subequations}

\smallskip\noindent
with $d = \nu^{\frac{n-1}{n-4}}$ and $g = \nu^{-\frac{3}{n-4}}$.

\smallskip
First, if we restrict ourselves to the subgroup of dilations, we have, using \eqref{inf_rep} and \eqref{decomp_psi},

\begin{equation}
\label{rep_dil2}
\left[\rho(u_\nu) \psi\right](\bx, t) = \nu^{-\frac{3(n+1)}{2(n-4)}} \, \left(
\begin{array}{cc}
\nu^{\frac{n-1}{2(n-4)}} & 0 \\
0 & \nu^{-\frac{n-1}{2(n-4)}}
\end{array} \right) \,
\psi\left(\nu^{-\frac{3}{n-4}} \bx, \nu^{-\frac{n+2}{n-4}} t\right)
\end{equation}
where we find again the three elements of the conformal transformations. From left to right: the global factor coming from the dilation of the volume element; then the matrix transforming the two bispinors, which can also be put in the remarkable form $\left(\begin{array}{cc}d^\half & 0 \\ 0 & d^{-\half}\end{array}\right)$; then the action on the coordinate variables. Hence the dynamical exponent of this model,
\begin{equation}
\label{dyn_exp}
z = \frac{N}{3} = \frac{n+2}{3},
\end{equation}
which is the same as in the Schr\"odinger--Newton case as found in \cite{DuvalL15}. This ought to be expected as we can recover the same form of the (generalized) Schr\"odinger--Newton equation \eqref{eq-H}, though for a bispinor and with a spin contribution.

In the special case of $n = 3$, we get the representation for the dilation subgroup, $\left[\rho(u_\nu) \psi\right](\bx, t)~=~\nu^{6} \left(\begin{array}{cc} \nu^{-1} & 0 \\ 0 & \nu\end{array}\right) \psi(\nu^3 \bx, \nu^5 t)$, and thus $z = 5/3$.

Let us now consider a general element of the LLN group of the form $u(a, \bb, \bc, d, e, g, h)$. We can extract the dilations, acting with $d$ and $g$, using the decomposition,
\begin{equation}
\label{decomposition_u}
u(a, d^{-1} \bb, g^{-1} \bc, 1, g^{-1} e, 1, (dg)^{-1} h) \cdot u_\nu(1, 0, 0, d, 0, g, 0) = u(a, \bb, \bc, d, e, g, h).
\end{equation}

The left element above, without dilations, belongs to the Bargmann subgroup, which is the group of isometries of a Bargmann structure $(M, \rg, \xi)$. For such element of the form $u_B(a, \bb, \bc, 1, e, 1, h)$, we have the known representation \cite{LevyLeblond67},
\begin{equation}
\label{rep_barg}
\left[\rho(u_B) \psi\right](\bx, t) = \exp\left( \frac{im}{\hbar} \left(\angles{\bb, \bx-\bc} - \frac{\Vert\bb\Vert^2 }{2}( t-e) - h\right) \right)
\left(
\begin{array}{cc}
a^{-1} & 0 \\
\frac{i}{2} a^{-1} \sigma (\bb)  & a^{-1}
\end{array} \right)
\psi\left(\wh \bx, \wh t\right).
\end{equation}
It is worthwhile to notice that the transformation (\ref{rev_action_sn}c) yields the phase factor.

When combining the two representations \eqref{rep_barg} and \eqref{rep_dil2} by using the decomposition \eqref{decomposition_u} we then get for the full action of the LLN group on bispinor. For a general element $u(a, \bb, \bc, d, e, g, h)$ of the LLN group, one has the following projective unitary (anti-)representation,
\begin{equation}
\label{full_rep}
\boxed{
\begin{aligned}
\left[\rho(u) \, \psi\right](\bx, t) = & \, \nu^{-\frac{3(n+1)}{2(n-4)}} \, \exp\left( \frac{im}{\nu \hbar} \left(g \angles{\bb, \bx} - \frac{g}{2d} \Vert\bb\Vert^2 t + \frac{e}{2d} \Vert\bb\Vert^2 - \angles{\bb, \bc} - h\right) \right)\\[2mm] 
& \left(
\begin{array}{cc}
d^{1/2} \, a^{-1} & 0 \\
\frac{i}{2} d^{1/2} a^{-1} \sigma (\bb)  & d^{-1/2} \, a^{-1}
\end{array} \right) \,
\psi\left(A^{-1} \left(g \bx - \frac{gt-e}{d}\, \bb - \bc \right) , \, \frac{gt-e}{d}\right)
\end{aligned}
}
\end{equation}
once again with $d = \nu^{\frac{n-1}{n-4}}$ and $g = \nu^{-\frac{3}{n-4}}$ (with $n=3$).
The inverse is given by,
\begin{equation}
\label{repr_lln}
\begin{aligned}
\left[\rho(u)^{-1} \, \psi\right](\bx, t) = & \, \nu^{\frac{3(n+1)}{2(n-4)}} \exp\left(\frac{im}{\nu \hbar} \left(  - \langle\textbf{b}, A \textbf{x}\rangle - \frac{||\textbf{b}||^2 t}{2} + h\right)\right) \\
& \left(\begin{array}{cc}
d^{-1/2} \, a & 0 \\
-\frac{i}{2} d^{1/2} \sigma (\bb) a  & d^{1/2} \, a
\end{array} \right) \,
\psi\left(\frac{A \textbf{x} + \textbf{b} t + \textbf{c}}{g}, \frac{d t + e}{g} \right).
\end{aligned}
\end{equation}

It can be verified that $\rho(u) \psi$ is indeed a solution of the generalized LLN equation \eqref{generalized_lln_eq} if $\psi$
is.\footnote{\samepage This can be seen at the infinitesimal level with the Lie derivative $L_X$ on spinor densities \eqref{lie_spinor_density} along a conformal Killing vector field $X$, \ie such that $\nabla_{(\mu} X_{\nu)} = \frac{1}{N} (\nabla_\rho X^\rho) \rg_{\mu\nu}$, and of the Dirac operator $\dirac(\rg)$ \eqref{dirac_spinor} on a spinor density, we find the commutator,
\begin{equation*}
\label{lxdirac}
\left[ L_X, \dirac(\rg)\right] \Psi = \frac{N-1}{2N} (\nabla_\mu X^\mu) \, \dirac(\rg) \Psi.
\end{equation*}
This means that whenever $\Psi$ is a solution of $\dirac(\rg) \Psi = 0$, then so is $\Psi_\epsilon \equiv \Psi + \epsilon L_X \Psi + \ldots$, for any conformal Killing vector field $X$.
}
This is also true for the LLN equation \eqref{lln} without the Coriolis vector potential.

\subsection{Action, energy-momentum tensor and conserved quantities}

To obtain the symmetries of the system, one way to proceed is through an action principle. Having succeeded in adapting the LLN system \eqref{lln_barg} to a Bargmann structure, it is natural to define the action principle on the Bargmann manifold $M$. Since the wave equation \eqref{diracmnul} is what determines the time evolution of the system, we will consider its action $S_D$, while the other equations, notably the gravitational equation and the equivariance relation are postulated without deriving them from an action principle. A justification for this could be that both the gravitational equation and the equivariance are inherent to the Bargmann structure, in the sense that they stem from its geometry.

\smallskip
Thus, for the massless Dirac equation \eqref{diracmnul}, we have the usual action of the Dirac equation in curved spacetime, with here zero mass,
\begin{equation}
\label{action_dirac}
S_D[\psi,\rg] = i \hbar \int_M \overline \psi \diracf \psi \volg d^Nx.
\end{equation}

Variations of the fields lead to the massless Dirac equation, and variations of the metric gives us the energy-momentum tensor (EMT), which we will use to compute conserved quantities. Recall its definition,
\begin{equation}
\label{def_emt}
T_{\mu\nu} = - \frac{2}{\volg} \frac{\delta S_D}{\delta \rg^{\mu\nu}}.
\end{equation}
The EMT obtained from \eqref{action_dirac} is much simpler than the one for SN. After symmetrization, it is given by
\begin{equation}
\label{emt}
T_{\mu\nu} = \frac{i \hbar}{4} \left(\overline \psi \gamma_\mu \nabla_\nu \psi + \overline \psi \gamma_\nu \nabla_\mu \psi - \nabla_\mu \overline \psi \gamma_ \nu \psi - \nabla_\nu \overline \psi \gamma_\mu \psi\right).
\end{equation}
This expression of the EMT for spinors already appears in \cite{Weldon01}.

\medskip
The next step is now to compute the conserved currents and quantities associated to the EMT (\ref{emt}) and the conformal symmetries \eqref{conf_sn}. To build these, a method similar to Souriau's is used~\cite{Sou74}. See also \cite{DuvalHP94,Duval76,DuvalHP96}.

Diffeomorphisms act infinitesimally on the Lagrangian $\cL_D$, which is defined as $S_D = \int_M \cL_D \volg d^Nx$, associated to the action functional \eqref{action_dirac}, by
\begin{equation}
L_X \left(\cL_D \volg\right) = (\nabla_\mu X^\mu) \cL_D \volg + \left(L_X \cL_D\right) \volg\, ,
\end{equation}
with $X \in \Vect(M)$.

On the equations of motion, we have $\cL_D = 0$. Then, since $\cL_D$ is represented by a closed N-form, we have by Cartan's formula that $L_X \cL_D = d\left(i_X \cL_D\right)$. Hence, on the equations of motion, $L_X \left(\cL_D \volg\right) = d\left(i_X \cL_D\right) \volg$. Thus,
\begin{equation}
L_X S_D = 0\, .
\end{equation}

With an action invariant under diffeomorphisms, the EMT is automatically divergence free. Indeed, from the definition of the EMT \eqref{def_emt}, we have, $0 = L_X S_D = \half \int_M T^{\mu\nu} (L_X \rg)_{\mu\nu} \volg d^Nx$. From the definition of a Lie derivative, the EMT being symmetric, and an integration by parts, we have $0 = \int_M \left(\nabla_\mu T^{\mu\nu}\right) X_\nu \volg d^Nx, \, \forall X \in \Vect(M)$. Hence the well known result for the EMT of the Dirac equations,
\begin{equation}
\nabla_\mu T^{\mu\nu} = 0.
\end{equation}
This can also be computed directly with the help of the field equation, and the various symmetries of the Riemann tensor. Also, through the field equations, we clearly have that the energy-momentum tensor is traceless, or $\rg^{\mu\nu} T_{\mu\nu} = 0$.

We now have all the ingredients to build up conserved charges. We want to build currents $k = (k^\mu)$ that are conserved, \ie $\nabla_\mu k^\mu = 0$. Two objects are of particular interest here: the EMT \eqref{emt}, which is divergence-free and traceless, and the conformal Killing vector field $X^\nu$ associated to the conformal symmetries of our system. Now, a current built as:
\begin{equation}
\label{k_barg}
k^\mu = T^{\mu\nu} X_\nu
\end{equation}
is conserved. Indeed, by taking the divergence of this expression, and using the fact that $\nabla_{(\mu} X_{\nu)} = \cL_X \rg_{\mu\nu} = \lambda \rg_{\mu\nu}$ for a conformal Killing field, and the properties that the EMT is traceless, symmetric, and divergent free, we have,
\begin{equation}
\label{cons_curr}
\nabla_\mu k^\mu = 0.
\end{equation}

However, for now, $k^\mu$ lives in Bargmann space, of dimension $N = n + 2$, but we would like conserved currents on the Galilean relativistic spacetime. Notice that the action does not depend on $s$, it is $\xi$-invariant. The same goes for the EMT, but unlike \cite{DuvalHP96}, here we have $\nabla_s X^s \neq 0$ , because of the dilations. Thus, the current $k^\mu$ does not project onto spacetime here. However, to get a charge living on NC spacetime, we can integrate the current on $\widetilde{\Sigma}_t$, \ie on both space and the fiber of Bargmann spacetime, instead of only space $\Sigma_t$.

Since $k^0 = \xi_\mu k^\mu$, the conserved charges read as:
\begin{equation}
\label{cons_charges}
Q_X = \frac{1}{2 \pi} \int_{\widetilde{\Sigma}_t} T_{\mu\nu} X^\nu \xi^\mu \sqrt{\rg_{{}_{\Sigma_t}}}\, \mu(s) \, d^n x \, ds,
\end{equation}
with $\mu(s)$ the integration measure of the variable $s$. Indeed, this is a time like dimension, and we can choose the fiber to be $S^1$ instead of $\R$, so that $\widetilde{\Sigma}_t = \Sigma_t \times S^1$, to get convergent integrals. If $\theta \in (-\pi,+\pi)$ is the angular coordinate on $S^1$, then $s = 2 \tan(\theta/2)$ is an affine coordinate. The integration measure is thus $\mu(s) = \frac{1}{1+s^2/4}$. Note that $\int^\infty_{-\infty} \mu(s) ds = 2 \pi$, and $\int^\infty_{-\infty} s \mu(s) ds = 0$. Most charges do not depend on $s$, and thus only get a $2 \pi$ factor. The only exception is for the charge associated to dilations, for $n \neq 4$, in which case the contribution linear in $s$ in the integrand will disappear after integration.

Altogether, this is the formulation of the Noether theorem applied to Bargmann structures.

\smallskip
Since there is one conserved quantity for each generator of the Lie algebra of the LLN group, one can write $Q_X$ as
\begin{equation}
Q_X = \boldsymbol{J}\cdot\boldsymbol{\omega}  + \boldsymbol{P} \cdot \bgamma + \boldsymbol{G} \cdot \bbeta + H \, \epsilon + D \, \chi + M \, \eta.
\end{equation}

Computing \eqref{cons_charges}, we find the following conserved charges, for $n = 3$ and in the flat case, but with Coriolis forces,

\begin{equation}
\label{conserved_n_neq4}
\left\lbrace
\begin{array}{ll}
\displaystyle E = \int \varphi^\dagger H \varphi \, d^3 \bx \qquad \qquad & \mathrm{energy} \\[4mm]
\displaystyle \boldsymbol{P} \equiv \int \boldsymbol{\cP} \, d^3 \bx = \frac{i\hbar}{2} \int \left((\bnabla \varphi)^\dagger \varphi - \varphi^\dagger \bnabla \varphi - i \, m \, \bvarpi\times (\varphi^\dagger \boldsymbol{\sigma} \varphi)\right) d^3 \bx & \mathrm{linear\; momentum} \\[4mm]
\displaystyle \boldsymbol{J} = \int \bx \times \boldsymbol{\cP} \, d^3 \bx + \frac{\hbar}{2} \int \varphi^\dagger \boldsymbol{\sigma} \varphi \, d^3\bx & \mathrm{angular\; momentum} \\[4mm]
\displaystyle M = m \int \varphi^\dagger \varphi \, d^3 \bx & \mathrm{mass} \\[4mm]
\displaystyle \boldsymbol{G} = t \boldsymbol{P} - m \int \varphi^\dagger \varphi \, \bx \, d^3 \bx & \mathrm{boost} \\[4mm]
\displaystyle D = \frac{n + 2}{n-4}\, t E + \frac{3}{n-4} \int \bx \cdot\! \boldsymbol{\cP} \, d^3 \bx & \mathrm{dilation}\ (n=3)
\end{array}
\right.
\end{equation}
with $H$ the Hamiltonian given in \eqref{hamiltonien}. Notice that in the conserved quantity $D$ with $n=3$, the dynamical exponent $z=5/3$ is split into $-5$ for the time part and $-3$ for the space part.

\smallskip\noindent
These conserved quantities are qualitatively the same as for Schr\"odinger--Newton \cite{Marsot17}, with two slight differences. We now have a bispinor $\varphi$ instead of a scalar wave-function, and we have a new contribution to the angular momentum due to the spin.
Here, once again, we note that the second bispinor plays no role, only the first one, $\varphi$, is important. These conserved quantities must also be compared with those obtained in~\cite{DuvalHP94}.

\section{Conclusion}
\label{sec-conclusion}

Di\`osi introduced the Schr\"odinger--Newton (SN) equation as a way to answer some open questions in Quantum Mechanics, notably the lack of spread of the wave packet for macro objects. This happens to be in the vein of Penrose's suggestion of gravitazing Quantum Mechanics. While the SN equation has been extensively studied, experiments that are planned typically involve particles or molecules with spin. To include spinors in Galilean relativistic Quantum Mechanics, it is necessary to replace the Schr\"odinger equation with the Lévy-Leblond equation. We thus study the Lévy-Leblond--Newton equation in this chapter, notably its symmetries.

To study symmetries of Galilean relativistic theories, one could potentially write them on Galilean relativistic spacetime, or Newton--Cartan (NC) structures. We have given a short account on some of the structural symmetries of such structures, namely the Galilean group, the Schr\"odinger group, and the Chronoprojective group. However, NC structures are somewhat pathological because of their degenerate ``metric'', which renders the computation of symmetries somewhat involved. We have then reviewed how to lift these NC structures to Bargmann structures, which are Lorentzian manifolds, hence allowing us to use the usual geometric tools to study symmetries. We also reviewed some structural symmetries of Bargmann structures, namely the Bargmann group, the extended Schr\"odinger group, and the extended Chronoprojective group, which all turn out to be extensions of the Galilean relativistic groups we had obtained before. We then saw how easy it is to compute the symmetries of the Schr\"odinger equation, when recast on Bargmann structures.

This justifies the recasting of the Lévy-Leblond--Newton equation on Bargmann structures in a completely covariant formulation as to study its symmetries, which we have done later on. In addition, this geometrical framework yields a natural generalization of the L\'evy-Leblond--Newton equations, where Coriolis forces can be taken into account.Despite the self-coupling of the spinor with itself by gravity, and the Coriolis forces, the second bispinor remains non-dynamical, in accordance with L\'evy-Leblond's remarks~\cite{LevyLeblond67}. This is to be expected since the Schr\"odinger equation is first order in time and the LL equation is morally its ``square root''. To some extent, the physical interpretation of this second bispinor in the non-relativistic framework deserves to be better understood.

Thanks to the geometrical framework of Bargmann structures and the covariant rewriting of the LLN equations, we were able to find the maximal symmetry group of this system which turns out to be the same as that of the Schr\"odinger--Newton equations, namely the SN group~\cite{DuvalL15}. This group is of dimension 12 in 3+1 dimensional space-time. The action of this group on 4-component spinors was computed, and of particular interest is the scaling law of the theory: in 3+1 dimensions, the dynamical exponent turns out to be $z = 5/3$. This is the same unusual dynamical exponent as in the Schr\"odinger--Newton case which also occurs in \cite{StichelZ10}. It is a curiosity that the dynamical exponent $z=N/3$ obtained in \eqref{dyn_exp} (with $n\neq 4$) keeps a trace of the $N$-dimensional Bargmann space. Finally, we computed the conserved quantities associated to the symmetries of the generalized LLN system driven by the SN group. They depend on the main dynamical bispinor.

	\chapter{Motion of spinning particles in General Relativity}
	\chaptertoc{}

\section{Introduction}
\subsection{An equation of motion for test particles in General Relativity}

The geodesic equation describes the trajectory followed by spinless test particles in a given (curved) spacetime. By test particles, it is understood that they are small enough so that they do not alter the background spacetime in any appreciable way. This equation can be written,
\begin{equation}
\label{geod_eq}
\frac{d^2 x^\mu}{ds^2} + \Gamma^{\mu}_{\lambda\rho} \frac{dx^\lambda}{ds} \frac{dx^\rho}{ds} = 0,
\end{equation}
where $s$ is a suitable parameter indexing the trajectory. If the test particle described is massive, then this parameter may be its proper time, often denoted $\tau$. In this case, the geodesic is said to be time like. If the test particle is massless, then one has to find another parameter, as the proper time of a massless particle is zero, and the geodesic is said to be light like.

The timelike geodesic equation was originally derived by means of an action principle, by maximizing the line element $ds$ between two timelike points of spacetime \cite{Einstein16}. The action is given by,
\begin{equation}
S = \int_A^B ds,
\end{equation}
with $ds^2 = g_{\mu\nu}dx^\mu dx^\nu$. The usual Euler--Lagrange equations then lead to the geodesic equation~\eqref{geod_eq}. Intuitively, the test particle goes from point $A$ to point $B$ in the least amount of its proper time. 

The geodesic equation can also be derived from the Equivalence Principle \cite{Weinberg72}.

It is also possible to derive the geodesic equation through diffeomorphism invariance, or the Principle of General Covariance \cite{Sou74}. Now, this last derivation, which we will get back to later on, allows for interesting generalization. In this framework, it is possible to introduce the effect of angular momentum of a test particle on its trajectory on curved spacetime. This is a legitimate wish, for example if we wonder if the angular momentum of the Earth has an influence on its trajectory around the Sun. Note that in general, such a framework allows for equations of motion that take into consideration a multipole expansion of the test particle, up to an arbitrary order. Truncating the expansion after the monopole moment yields the geodesic equation, and keeping the dipole moment, which is linked to angular momentum, produces the so-called Mathisson--Papapetrou--Dixon equations \cite{Mat37,Pap51,Dix70} (MPD equations for short).

Now, if we have equations of motion that describe a test particle with angular momentum, it is legitimate to wonder if such a framework could describe the trajectory of elementary particles, including their intrinsic angular momentum, spin. 

\subsection{Overview of the chapter}

This chapter will be devoted to the study of the Souriau--Saturnini equations, which aim at describing the trajectory of massless particles with spin, \textit{e.g.} photons.

As we have discussed above, the geodesic equation does not take into account the spin of a test body. (For massless particles, one prefers to say helicity rather than spin.) Hence, we will first see how to include the spin of test particles in equations of motions, through the MPD equations in section \ref{ss:eom}. There are different ways to derive these equations, and we choose to highlight Souriau's elegant derivation, which has the advantage of being completely geometric. While these equations are widely accepted in the literature, they are not sufficient to describe the trajectory of spinning particles. Indeed, they are not closed: there are more degrees of freedom than equations. Hence, supplementary conditions are required, and we will review some of these conditions. Unfortunately, there does not seem to be a canonical way to single out one of these conditions. We will justify choosing the Tulczyjew condition, and derive the ensuing closed set of equations, the Souriau--Saturnini equations.

These equations are much more complicated than the (light-like) geodesic equation. To get a feeling of how they work and hope to find a general scheme for solving them, we are going to see four examples. Firstly, we will write these equations in the case of the simple de Sitter spacetime in section \ref{ss:desitter}. This was done originally by Saturnini in his thesis \cite{Sat76}. Then, we will quickly review in section \ref{ss:flrw} an article of Duval and Sch\"ucker \cite{ChDTSRW} who applied these equations in the case of cosmology, in a Friedmann--Lema\^itre--Robertson--Walker (FLRW) background. Then, a more tricky example will be to study in \ref{ss:schw} the trajectory of a spinning photon as it passses by a Schwarzschild body. This is based on the work \cite{ChDLMTS}, and will feature additional comments about the cosmological constant. The last example, in \ref{ss:gw}, will be the study of a photon in a gravitational wave background, based on the work \cite{Marsot19}.

Thanks to these four examples, each technically different, we will try to summarize, in the last section \ref{ss:conclusions}, the different techniques that were used to solve these equations of motion. We will also discuss some open problems resulting from the models still ignoring quantum properties. Finally, we mention possible experimental tests of the model.

\section{Deriving equations of motions for spinning massless test particles}
\label{ss:eom}
\subsection{A bit of history}

There have been multiple attempts to define equations of motions for particles with spin in an electromagnetic field or in a gravitational field \cite{Fre26,Mat37,Pap51,Pir56,Tul59}, with different approaches. However, it is Dixon who finally provided a treatment of extended test particles in a fully covariant manner \cite{Dix70}. His approach is based on a general multipole expansion of extended test particles along a certain worldline, \textit{a posteriori} identified as representing the history of the body. Such an expansion makes sense when the length scales associated to the body are much shorter than the curvature length scale.

In this chapter we focus on the first two multipoles, hence the name of ``pole-dipole'' approximation which may arise. The first multipole is associated to the mass of the (extended) test particle, while the dipole moment is associated to the rotation of the test particle. Dixon shows that we can define equations, which are now called the Mathisson--Papapetrou--Dixon (MPD) equations, for extended test particles with angular momentum,
\begin{align}
\dot{P}^\mu & = - \half {R^\mu}_{\rho\alpha\beta} S^{\alpha \beta} \dot{X}^\rho, \label{mpd_pdot}\\
\dot{S}^{\mu\nu} & = P^\mu \dot{X}^\nu - P^\nu \dot{X}^\mu, \label{mpd_sdot}
\end{align}
with $X$ the position, $P$ the momentum, and $S$ a 2-tensor representing the angular momentum of the test particle. The dot on $X$ denotes the usual derivative with respect to a parameter $s$, while on other vectors and tensors, such as $P$ or $S$, it denotes the covariant derivative with respect to $s$.

These equations are fairly straightforward to obtain, and are thus widely accepted in the literature to describe the behaviour of extended test particles with spin. However, this system of equation has more degrees of freedom than equations as we will see. 

Souriau also obtained these equations \cite{Sou74}, but in a geometric way, which we will describe below.

\subsection{Souriau's model}
\label{s:souriau_fram}
\subsubsection{Geometrical framework of the Principe of General Covariance}

Souriau revisited the diffeomorphism invariance of General Relativity (GR) in what he calls the Principle of General Covariance \cite{Sou74}. In GR (potentially modified), Nature is described with the help of a pseudo-Riemannian metric, which belongs to the ``space'' of all metrics. The group of diffeomorphisms naturally acts on this ``space'', and the diffeomorphism invariance of GR states that the action of this group is unobservable. Intuitively, this means that the space of all metrics is ``too big'' to represent the physical information of the Universe. Indeed, we can build equivalence classes, where all metrics in a class are physically equivalent, by considering the orbit of a representative element by the group of diffeomorphisms. Souriau states that the space formed by the set of equivalence classes thus forms the right space to encode the physical information of the Universe.

Hence the following geometrical construction of this space. Let $M$ be a pseudo-Riemannian manifold, $\Diff(M)$ the group of diffeomorphisms acting on $M$, and $\Met(M)$ the space of all metrics of $M$. Souriau calls the space representing the physical information the ``hyperspace'', or space of geometries, which we denote $\Geom(M)$. It is the quotient of $\Met(M)$ by the group of diffeomorphisms $\Diff(M)$. Due to some topological difficulties, one usually restricts this definition to the quotient by the subgroup of diffeomorphisms with compact support $\Diff_c(M)$. This can have consequences in some pathological cases, but not in the applications which will be presented in this chapter. Of course, we need $M$ to be non compact. Hence the definition,
\begin{equation}
\label{def_geom}
\Geom(M) = \Met(M) / \Diff_c(M).
\end{equation}

\begin{figure}[ht]
\centering
\begin{tikzpicture}[line width=1.2pt,scale=1, every node/.style={transform shape}]
  \draw (0,0) to[out=5,in=175] coordinate[pos=0.5](midmid) (6,0) to[out=95,in=-95] node[pos=0.92,left,scale=1]{$\Met(M)$} (6,5) to[out=-175,in=-5] coordinate[pos=0.5](topmid) (0,5) to[out=-85,in=85] (0,0);
  \draw (-1,-1.85) to[out=15,in=165] node[pos=0.5,below,scale=1.1]{$[g]$} node[pos=0.85,below,align=center]{$\Geom(M) =$ \\ $\Met(M)/\Diff_c(M)$} coordinate[pos=0.5](botmid) (7,-1.85);
	  \draw [pattern=dots] (midmid) to[out=110,in=-110] node[pos=0.8,left,scale=1.1]{$\, \cO_g$} (topmid) to [out=-70,in=70] (midmid);
  \draw [->] (3,1.6) -- node[pos=0,scale=1.2] {$\bullet$} node[pos=0,scale=1.1,below] {$g$} node [pos=0.95, right,scale=1.1] {$\delta g$} (4.3,2.4);
  \draw [->] (3,1.6) to[out=130,in=-130] node [pos=0.95, above,scale=1.1] {$a^* g$} (3,2.5);
  \draw [dashed] (midmid) -- node [pos=1.05,scale=1.2] {$\bullet$} (botmid);
  \draw (botmid) -- +(3,0) -- node[pos=0.90,above] {$T_{[g]}\Geom(M)$} +(-3,0);
  \draw [->] (botmid) -- node[pos=0.85,above] {$\delta \Gamma$} +(1.4,0);
\end{tikzpicture}
\caption{The geometry of the Principle of General Covariance}
\label{f:pgc}
\end{figure}
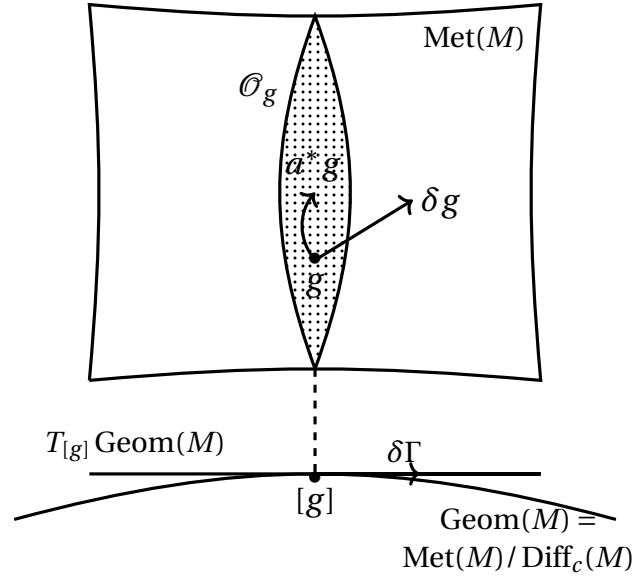

Now, the space of geometries $\Geom(M)$ is complicated, but we do not need to define a manifold structure on it. We only need to characterize its (co)tangent vector space. Let us look at the local variation $\delta \Gamma \in T_{[g]} \Geom(M)$ of the geometry induced by an infinitesimal, arbitrary, variation $\delta g$ of a metric $g$ in $\Met(M)$. To characterize the projection $\delta g \rightarrow \delta \Gamma$, note that by definition, a variation of $g$ by a diffeomorphism leaves the geometry invariant. Thus, if we take $\delta g$ to be a ``vertical'' infinitesimal variation, we have $\delta \Gamma = 0$. Since here a vertical variation is a diffeomorphism, infinitesimally a local vertical variation is given by $\delta g = L_\xi g \in T_g \Met(M)$, for $\xi \in \Vect_c(M)$. Hence the definition of the tangent vector space $T_{[g]} \Geom(M)$ as a quotient,
\begin{equation}
T_{[g]} \Geom(M) = T_g \Met(M) / T_g \cO_g,
\end{equation}
where $\cO_g$ is the orbit of the metric $g$ by the group of diffeomorphisms with compact support.

By duality, we can also characterize the cotangent vector space $T^*_{[g]} \Geom(M)$ by projections. We have,
\begin{equation}
T^*_{[g]} \Geom(M) = \left\lbrace\cT \in T^*_g \Met(M) \mid \cT(\delta g) = 0, \forall \delta g \mathrm{\,``vertical''}\right\rbrace.
\end{equation}

The Principle of General Covariance is thus stated as follows. The geometric information of the Universe is represented by a distribution tensor $\cT \in T^*_{[g]} \Geom(M)$, such that,
\begin{equation}
\label{PGC}
\boxed{\cT(L_\xi g) = 0, \quad \forall \xi \in \Vect_c(M)}
\end{equation}

\bigskip
\subsubsection{Continuous matter distribution}

If we want to describe the continuous matter content on $M$ by its energy momentum tensor $T^{\mu\nu}$, the distribution $\cT$ may look like,

\begin{equation}
\label{continuous_case}
\cT(\delta g) = \half \int_M T^{\mu\nu} \delta g_{\mu\nu} \vol_g.
\end{equation}

As $T^{\mu\nu}$ is a symmetric tensor, and using the identity $(L_\xi g)_{\mu\nu} = \nabla_\mu \xi_\nu + \nabla_\nu \xi_\mu$, \eqref{continuous_case} and \eqref{PGC} imply $\int_M T^{\mu\nu} \nabla_\mu \xi_\nu \vol_g = 0$. Since $\xi$ is a vector field with compact support, integrating by parts gives $\int_M \left(\nabla_\mu T^{\mu\nu}\right) \xi_\nu \vol_g = 0, \, \forall \xi \in \Vect_c(M)$, hence,
\begin{equation}
\label{divT}
\nabla_\mu T^{\mu\nu} = 0.
\end{equation}

In the case of a continuous matter distribution, the Principle of General Covariance implies the covariant conservation of the energy momentum tensor.

\bigskip
\subsubsection{Conservation laws}
\label{ss:conslaws}

Souriau's framework is also suitable to derive conservation laws. 
Indeed, given a tensor distribution $\cT$, we have $\cT(L_Z g) = 0$ if $L_Z g = 0$, \ie if $Z$ is a Killing vector field. Now, if we take the same distribution representing the continuous matter case \eqref{continuous_case} as an example, we have $0 = \int_M T^{\mu\nu} \nabla_\mu Z_\nu \vol_g$. Now integrating by part, but keeping in mind that now $Z$ is not with compact support in general, and with the result \eqref{divT}, we get
\begin{equation}
\label{interm_noether}
\int_M \nabla_\mu\left(T^{\mu\nu}Z_\nu\right) \vol_g \equiv \int_M \mathrm{div}(T\cdot Z) \vol_g = 0.
\end{equation}

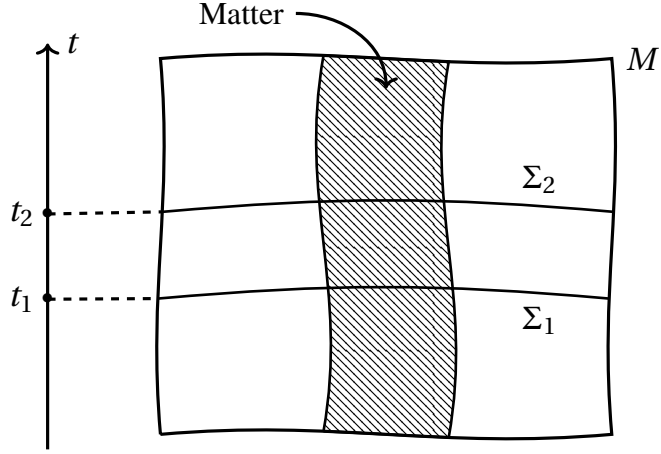
\begin{figure}[ht]
\centering
\begin{tikzpicture}[line width=1.2pt,scale=1]
  \draw (0,0) to[out=5,in=-175] coordinate[pos=0.35](b1) coordinate[pos=0.65](b2) (6,0) to[out=95,in=-85] coordinate[pos=0.35](d1) coordinate[pos=0.6](d2) node[pos=1,right,scale=1.1]{$M$} (6,5) to[out=-175,in=5] coordinate[pos=0.35](h2) coordinate[pos=0.65](h1) (0,5) to[out=-85,in=95] coordinate[pos=0.4](g2) coordinate[pos=0.65](g1) (0,0);
  \draw[->] (-1.5,-0.2) -- coordinate[pos=0.37](t1) coordinate[pos=0.58](t2) node[pos=1,right,scale=1.1] {$\,t$} (-1.5,5.2);
  \draw[line width=1pt,pattern=north west lines] (b1) to[out=80,in=-100] (h1) -- (h2) to[out=-100,in=80] (b2) -- (b1);
  \draw[line width=1pt] (g1) to[out=5,in=175] node[pos=0.86,below,scale=1.1]{$\Sigma_1$} (d1);
  \draw[line width=1pt] (g2) to[out=5,in=175] node[pos=0.85,above,scale=1.1]{$\Sigma_2$} (d2);
  \draw[dashed] (t1) -- node[pos=0,left,scale=1.1]{$t_1$} node[pos=0]{$\bullet$} (g1);
  \draw[dashed] (t2) -- node[pos=0,left,scale=1.1]{$t_2$} node[pos=0]{$\bullet$} (g2);
  \draw[->] (1.8,5.6) to[out=0,in=90] node[pos=0,left]{Matter} (3,4.6);
\end{tikzpicture}
\caption{Geometry of the conservation law}
\label{f:cons}
\end{figure}

Now, assuming that the topology of $M$ permits the definition of a coherent time variable $t$, and that the distribution of matter is bounded in space, we can consider two Cauchy surfaces $\Sigma_1$ and $\Sigma_2$ at times $t_1$ and $t_2$ respectively, and apply Stokes' theorem to obtain from \eqref{interm_noether}, $0 = \int_{\Sigma_1} \vol_{\Sigma_1}(T\cdot Z)-\int_{\Sigma_2} \vol_{\Sigma_2}(T\cdot Z)$. See figure \ref{f:cons}. Since this is true for any Cauchy surface $\Sigma$ at a time $t$, we have,
\begin{equation}
\label{cons_law}
\int_{\Sigma} \vol_{\Sigma}(T\cdot Z) = \mathrm{const.},
\end{equation}
if $Z$ is a Killing vector field.

\subsubsection{Test particle without spin localized on a worldline}

We now consider a test particle localized on a worldline $\cC$, which is parametrized by $s$, and characterized only by its monopole moment $\theta^{\mu\nu}$, which we assume to be nowhere vanishing. Its distribution $\cT_\cC$ along the wordline $\cC$ reads,
\begin{equation}
\label{t_particle}
\cT_\cC(\delta g) = \half \int_\cC \theta^{\mu\nu}\delta g_{\mu\nu} \, ds.
\end{equation}

We will now show that the Principle of General Covariance applied to the tensor distribution of a localized particle \eqref{t_particle} leads to the geodesic equation of motion for that particle.

To apply the Principle of General Covariance, consider a test function $\delta g$ as the Lie derivative of the metric along a vector field of the form $\alpha \xi$ with $\xi \in \Vect_c(M)$ and $\alpha \in C^\infty(M,\R)$, such that $\restr{\alpha}{\cC} = 0$. Writing down \eqref{PGC} and doing the usual integration by parts, keeping in mind that $\restr{\alpha}{\cC} = 0$, we get $0 = \int_{-\infty}^{\infty} \left(\theta^{\mu\nu}\partial_\mu \alpha\right)\xi_\nu \, ds, \forall \xi \in \Vect_c(M)$. Hence,
\begin{equation}
\theta^{\mu\nu} \partial_\mu \alpha = 0.
\end{equation}

The contraction of the tensor $\theta^{\mu\nu}$ with any vector orthogonal to the worldline vanishes. Thus, $\theta^{\mu\nu}$ can be decomposed in full generality as 
\begin{equation}
\label{expr_theta}
\theta^{\mu\nu} = P^\mu {\dot X}^\nu,
\end{equation}
for some vector $P \in T_x M$. Recall that dot on $X$ denotes the ordinary derivative with respect to the parameter $s$, while on other vectors and tensors, such as $P$ or $\xi$, it denotes the covariant derivative with respect to $s$. Since, by construction, $\theta$ is a symmetric tensor, we also have $P \parallel \dot X$.

Putting \eqref{expr_theta} into \eqref{t_particle}, and with $\delta g = L_\xi g$, we have $0 = \int_{-\infty}^\infty P^\mu \dot{\xi}_\mu \, ds$. Hence, if the worldline leaves any compact, an integration by part leaves us with $0 = - \int_{-\infty}^\infty \dot{P}^\mu \xi_\mu ds$. It is true for all $\xi$ with compact support, and thus comes the final result,
\begin{equation}
\label{geod}
\dot{P}^\mu = 0 \quad \& \quad P \parallel \dot X.
\end{equation}

The worldline of the particle without dipole moment is described by the well-known geodesic equation.

\subsubsection{Spinning test particle localized on a worldline}

In the case of a localized test particle with angular momentum or spin, we have to also consider the dipole moment $\Phi^{\rho\mu\nu}$ with the distribution,
\begin{equation}
\label{t_spin}
\cT_\cC(\delta g) = \half \int_\cC \big( \theta^{\mu\nu}\delta g_{\mu\nu} + \Phi^{\rho\mu\nu}\nabla_\rho \delta g_{\mu\nu}\big) \, ds.
\end{equation}

The computation is similar to the case of a spinless particle. With the help of (\ref{PGC}), we obtain in full generality $\Phi^{\rho\mu\nu} = S^{\rho\mu} \dot{X}^\nu$, with $S$ a skew-symmetric tensor, and so in the end we end up with the Mathisson-Papapetrou-Dixon equations \eqref{mpd_pdot}--\eqref{mpd_sdot},
\begin{align}
\dot{P}^\mu & = - \half {R^\mu}_{\rho\alpha\beta} S^{\alpha \beta} \dot{X}^\rho,\\
\dot{S}^{\mu\nu} & = P^\mu \dot{X}^\nu - P^\nu \dot{X}^\mu.
\end{align}

Souriau also shows that this framework can be used to obtained the conserved quantities associated to each Killing vector. 
Now, let us apply the conservation laws to the case of a spinning particle. To obtain the expression of the conserved quantities for the MPD equations \eqref{mpd_pdot}--\eqref{mpd_sdot}, one considers the distribution describing the particle on its worldpath \eqref{t_spin} with $\Phi^{\rho\mu\nu} = S^{\rho\mu} \dot{X}^\nu$. Then, we need to use the idea of the previous paragraph on conservation laws, namely that $\cT_\cC(L_Z g) = 0$ if $Z$ is Killing, even though $Z$ is not with compact support. A similar computation to the conservation laws paragraph \ref{ss:conslaws} leads to,
\begin{equation}
\cT_\cC(L_Z g) = \int_\cC d\left(\Psi(Z)\right),
\end{equation}
where $\Psi$ is such that,
\begin{equation}
\Psi(Z) = P_\mu Z^\mu + \half S^{\mu\nu}\nabla_\mu Z_\nu.
\end{equation}
The quantity $\Psi(Z) = \mathrm{const.}$ is a first integral of the MPD equations \eqref{mpd_pdot}--\eqref{mpd_sdot}.

While the geodesic equation of motion \eqref{geod} is deterministic, meaning there are as many unknowns as equations, one can clearly see that the MPD equations \eqref{mpd_pdot}--\eqref{mpd_sdot} are not: an equation is missing for $\dot{X}$, and we will need to impose additional constraint to the system, as we will see later. However, we can still construct all the conserved quantities, even before specifying any constraint.

\subsubsection{Adding electromagnetism}

Souriau shows that a background electromagnetic field can be added in the description of a charged spinning test particle. To accomplish this, the geometric description (see figure \ref{f:pgc}) has to be generalized. Souriau shows this can be done by not only considering the space of all metrics, but the space of all potentials $(g, A)$. Also, while previously we were considering the action of $\Diff(M)$ on $g$, now we have to consider the action of the semi direct product of the group of gauge transformations on $A$, by the group of diffeomorphisms on $(g,A)$. Now, a ``vertical'' variation is given with the help of a vector field $\xi \in \Vect_c(M)$ with compact support and a function $\alpha \in C^\infty(M,\R)$ such that,
\begin{equation}
\label{deltaga}
\delta(g, A) = (L_\xi g, L_\xi A + d\alpha).
\end{equation}

Then, to describe a charged spinning test particle, Souriau considers the distribution,
\begin{equation}
\label{t_charge_spin}
\cT_\cC(\delta g, \delta A) = \int_\cC \left(\half \theta^{\mu\nu}\delta g_{\mu\nu} + \half \Phi^{\rho\mu\nu}\nabla_\rho \delta g_{\mu\nu} + \psi^\mu \delta A_\mu + \Omega^{\mu\nu} \nabla_\mu \delta A_\nu\right) ds.
\end{equation}

The Principle of General Covariance leads to the definitions, $\psi^\mu = q \dot{X}^\mu$, where $q$ will be interpreted as the charge of the particle, $\Omega^{\mu\nu} = \dot{X}^\mu B^\nu + \cM^{\mu\nu}$ for some vector $B$ and skewsymmetric tensor $\cM$ which will be interpreted as the electromagnetic moment of the particle, and $2 \theta^{\mu\nu} = P^\mu \dot{X}^\nu + P^\nu \dot{X}^\mu + \cM^{\mu\rho} {F_\rho}^\nu + \cM^{\nu\rho}{F_\rho}^\mu$ where $F = dA$. This then leads to the MPD equations for charged spinning particles in a background gravitational and electromagnetic fields,

\begin{align}
\frac{dq}{ds} & = 0, \label{cons_charge}\\
\dot{P}^\mu & = q {F^\mu}_\nu \dot{X}^\nu + \half \cM^{\rho\sigma} \nabla^\mu F_{\rho\sigma} - \half {R^\mu}_{\rho\alpha\beta} S^{\alpha \beta} \dot{X}^\rho, \label{pdot_charge}\\
\dot{S}^{\mu\nu} & = P^\mu \dot{X}^\nu - P^\nu \dot{X}^\mu - \cM^{\mu\rho} {F_\rho}^\nu + \cM^{\nu\rho}{F_\rho}^\mu.\label{sdot_charge}
\end{align}

In this case, the expression
\begin{equation}
\label{cons_a}
\half S^{\mu\nu}\nabla_\mu Z_\nu + P^\mu V_\mu + q A^\mu V_\mu + qu
\end{equation}
is a first integral of the above equations \eqref{cons_charge}--\eqref{sdot_charge}, for any $Z$ and $u$ such that $L_Z g = 0$ and $L_Z A + du = 0$.

\medskip

Again, the system \eqref{cons_charge}--\eqref{sdot_charge} is not deterministic. This is due to adding new unknowns in the form of the spin and electromagnetic moment of the test particle, without new equations. We thus need to impose phenomenological equations, associated to the kind of test particle we want to describe. Several such equations exist, for instance, $P^2 = m^2 + \frac{q g}{2} F_{\mu\nu}S^{\mu\nu}$, $\cM \parallel S$, ${S^\mu}_\nu P^\nu = 0$ or ${S^\mu}_\nu \dot{X}^\nu = 0$, etc. We will discuss the later two constraints in the next section.

\subsection{Spin Supplementary Conditions}
\label{s:sscs}

While the Mathisson--Papapetrou--Dixon(--Souriau) equations provide a solid starting point to equations of motion for test particles with spin, they do not determine the evolution uniquely: we lack an equation for $\dot{X}$ (the latter needs not be parallel to $P$). This reflects an ambiguity in the selection of the worldline $X(s)$ representing the particle history. One thus has to impose certain constraints to close the system. In the case of a gravitational field, \ie without electromagnetism, these can be written in the form ${S^\mu}_\nu V^\nu=0$, where $V^\mu$ is a suitable vector. These constraints are usually called spin supplementary conditions (SSCs). The vector $V^\mu$ may in principle be chosen freely, though there are several obvious ``intrinsic'' options, provided by the geometry of the problem itself. In the literature, two of such possibilities are mainly studied. The first one being the Mathisson--Pirani (or Frenkel--Pirani) SSC, where $V \parallel \dot{X}$ \cite{Mat37,Pir56,Fre26}, and the second one being the Tulczyjew SSC, where $V \parallel P$ \cite{Tul59,Dix70}. 

The lack of constitutive laws which determine how the body responds to gravitational and inertial strains leads to the freedom which the different supplementary conditions fix, each in a different way. In particular, they lead to different trajectories. It is not possible yet to say which of the conditions is the ``correct'' one. See \cite{KyrianS07,KunstLLGS15,HarmsLGBN16,LGHBN17,WitzanySLG18} for comparison of different SSCs. In the case of extended massive test particles, the choice of SSC seems to relate to the choice of worldline of the center of mass of the test particle which is used to define how the spin tensor is defined \cite{KozamehNQ19}.

In the past, the Mathisson--Pirani SSC was sometimes deemed unsatisfactory due to there not being a unique representative worldline, depending on the choice of initial conditions \cite{Mat372,Wey47,Moller49}. This issue has been clarified recently in \cite{CostaNZ12,CostaHNZ11,CostaLGS17}, in connection with discovering the momentum-velocity relation for that SSC. The Tulczyjew SSC, on the other hand, does provide a unique worldline, irrespectively of how the initial conditions are prescribed \cite{Dix70}.

Since $\dot{X}$ needs not be parallel to $P$ anymore, the theory naturally offers different definitions of the body's ``mass'', $m=\sqrt{P^\mu P_\mu}$, $\tilde{m}=\dot{X}^\mu P_\mu$, and possibly $V^\mu P_\mu$. The MPD equations by themselves do not ensure that any of the above masses remains constant, not even that the vectors $P$ and $\dot{X}$ are, or remain, timelike. However, we obtain more information with the help of the chosen SSC. For the Mathisson-Pirani SSC ${S^\mu}_\nu\dot{X}^\nu=0$, it is $\tilde{m}$ that is conserved. For the Tulczyjew SSC ${S^\mu}_\nu P^\nu=0$, it is $m$ that is conserved. Interestingly, in the massless case, the Mathisson--Pirani SSC leads to a 4-momentum that may be spacelike \cite{Mas75,BailynR81,BiniCAJ06,Semerak15}, while the Tulczyjew SSC leads to a 4-velocity which may be spacelike \cite{Sat76}.

Let us add that the MPD equations ensure, independently of the SSC, the conservation, along the representative worldline, of the spin-tensor invariant $2s^2=S_{\alpha\beta}S^{\alpha\beta}$. This scalar is sometimes called the longitudinal spin and for photons it equals $\pm\hbar$, with sign$(s)$ called helicity or handedness. By fixing the conserved mass and spin, this completes the description of a classical elementary particle as belonging to one of the coadjoint representations of the Poincar\'e group.

In the massless case, the choice of the SSC is even more subtle than in the massive one. Two main arguments have been given in favor of the Mathisson-Pirani SSC: i) Maxwell equations minimally coupled to gravity yield null geodesics in the geometric-optic limit \cite{Laue20}, like do the MPD equations together with this SSC \cite{Mas75,BailynR81,BiniCAJ06} (with just one type of counterexample given in \cite{BailynR81}). (ii) Imposing conformal invariance of the theory, in particular the tracelessness of the energy-momentum tensor, implies (a slight generalization of) the Mathisson-Pirani constraint \cite{Duv78,BailynR77,Semerak15}. Less satisfactorily, the MPD equations supplemented with that constraint do not behave well in the $\tilde{m}\rightarrow 0$ limit, the massless problem is actually unrelated to the massive one \cite{Wey47,Obu11}. On the other hand, Tulczyjew’s SSC has often been considered inappropriate because, as already mentioned, it generally leads to a spacelike motion, which is more serious than the spacelike momentum yielded by the Mathisson-Pirani SSC. It also leads to a certain degeneracy of the massless problem in flat spacetime: rather than a localized particle, it yields a plane traveling at the speed of light.

Recently, however, the Tulczyjew SSC has been revisited in connection with phenomena observed in spinoptics. As already predicted by Fedorov and Imbert \cite{Fed55,Imb72}, the wave packet of spinning light should perform an ``instantaneous'' transverse shift when being reflected at an interface. This effect can be described theoretically using the symplectic mechanics in a 3-dimensional manifold \cite{Duv06,Duv07,Duv08,Duv13} similar to the symplectic representation of Souriau’s spinning-particle model involving the Tulczyjew SSC \cite{Sou74}. The effect, also called Spin Hall Effect of light, was confirmed experimentally in 2008 \cite{Hos08,Bli08}. Recall that Fermat’s principle can be rephrased to say that light rays follow null geodesics in a 3-dimensional Riemannian space conformally related to the Euclidean one by a scale factor represented by the local refractive index squared. One can then summarize the 2008's experiments as follows: the spinning light rays deviate from null geodesics in the above space. More specifically, the speed of spinning light can locally become higher than the speed of spinless light, without violating causality over distances larger than the wavelength of the photon.

Also in favor of the Tulczyjew SSC, one can mention the presence of the Berry phase in quantum mechanics, which is in general connected with a deviation from geodesics as well. In specific examples, the treatment of the problem with the help of a Berry phase and the treatment with the MPD equations with the Tulczyjew SSC, or their symplectic description, agree with each other. See, for instance \cite{StoneDZ14,DuvalH14} for the treatment of chiral fermions, and \cite{Gos06,ChDLMTS} for birefringence of a photon in a Schwarzschild spacetime. Still another support for the Tulczyjew SSC was provided by Souriau who showed \cite{Sou70} that geometric quantization of the symplectic system which derives the MPD equations with this SSC, when considered with a flat background, leads to the Maxwell equations.

To summarize, the MPD equations with the Tulczyjew SSC may provide an effective, semi-classical description of phenomena tied to the photon spin and involving the occurrence of superluminal speeds. Note that if causality is not violated over distances larger than the wavelength of the photon, it should not imply any problem, since the pole-dipole approximation as such only holds if the length scales tied to the particle (here the wavelength of the photon) are much smaller than the curvature length scale. Indeed, in papers where the Tulczyjew SSC was employed, e.g., to study photons in the Schwarzschild, de Sitter or FLRW backgrounds \cite{Sat76,ChDTSRW,ArmazaHKZ16,ChDLMTS,Marsot19}, causality has not been found to be violated over meaningful distances.

Note that these SSC also appear outside the study of test particles in a gravitational field. Indeed, it is possible to include the electromagnetic field in the MPD equations, as we have seen in the previous section, but yet again the equations are not completely determined, and one needs to choose an SSC to close them. For example, for massive and charged elementary particles, both the Mathisson--Pirani and the Tulczyjew SSC recover the spin precession equation~$\dot{S}$ of the Bargmann-Michel-Telegdi~(BMT) equations~\cite{BMT59} from the MPD equations, in the weak field limit \cite{Sou74}. Recall that the BMT equations describe the spin precession of an electron in a constant and weak electromagnetic field. They underly the interpretation of the very precise experimental measurements of the gyromagnetic moment of the particle \cite{Jeg09}. While both SSC lead to the same spin precession equation, they feature (in general) non vanishing anomalous velocities~$\dot{X}$ \cite{Sou74,Duv16,Del95,Der14} (not necessarily the same for the two different SSC), not present in the original BMT equations. Hence, precise experiments of electrons in an electromagnetic field could give an hint about the choice of SSC for elementary particles.

While we will concentrate on using the Tulczyjew SSC in the rest of this chapter, let us review in the next section an argument for the Mathisson--Pirani SSC by Duval and Fliche \cite{Duv78}, which uses Souriau's framework that we highlighted in the previous section.

\subsubsection{Localization and conformal invariance: the case for the MP SSC}

To obtain a closed system of equations describing an elementary particle, one may want to impose, alongside localization on its worldline, conformal invariance. Considering that conformal invariance is the relevant symmetry when dealing with massless (and spinless) particles, see for instance the Maxwell equations, it may be legitimate to wonder if a theory, in the framework of the MPD equations, describing massless photons with spin 1 should manifest conformal invariance. This has been studied by Duval and Fliche in \cite{Duv78}. They have shown that, when using Souriau's framework (see the previous section \ref{s:souriau_fram}) to obtain the MPD equations, and when imposing that the distribution $\cT$ associated to the massless spinning particle is conformally invariant, the Mathisson--Pirani SSC is recovered. Let us outline the procedure.

Since they are using Souriau's framework, the spinning test particle is represented by the matter distribution \eqref{t_spin} on a worldline $\cC$, written explicitly,
\begin{equation}
\label{t_spin_2}
\cT_\cC(\delta g) = \half \int_\cC \big( P^\mu\dot{X}^\nu\delta g_{\mu\nu} + S^{\rho\mu} \dot{X}^\nu \nabla_\rho \delta g_{\mu\nu}\big) \, ds.
\end{equation}

As we have seen in the previous section, from this distribution, the Principle of General Covariance \eqref{PGC} leads to the MPD equations \eqref{mpd_pdot}--\eqref{mpd_sdot}. 

One can now ask for conformal invariance. In \cite{Duv78}, it is implemented by requiring,
\begin{equation}
\label{req_inv_conf}
\cT_\cC(\lambda g) = 0, \forall \lambda \in C_c^\infty(M).
\end{equation}

From \eqref{t_spin_2} and \eqref{req_inv_conf}, one immediately finds that ${S^\mu}_\rho \dot{X}^\rho = \alpha \dot{X}^\mu$ and $P^\rho\dot{X}_\rho = \dot{\alpha}$, with $\alpha \in C^\infty(\cC, \R)$ a function on the test particle's worldline. From their study, and the standard assumption that $S$ is of rank 2, or $\det(S) = 0$, it then follows that $\alpha = 0$, hence recovering the Mathisson--Pirani constraint, and that $\dot{X}^\rho \dot{X}_\rho = 0$, $\Tr(S^2) = \const$, and $\ddot{X} \parallel \dot{X}$.

The main result from \cite{Duv78} is that when imposing conformal invariance together with the MPD equations to describe a spinning massless test particle, the Mathisson--Pirani constraint appears naturally, and the particle travels on a null geodesic.

However, two caveats are worth mentioning. First, as noted in \cite{Duv78}, the 4-momentum $P$ of the particle considered here is spacelike. This is also mentionned by Mashhoon in \cite{Mas75} who argues that the canonical momentum of massless spinning particles is not restricted to be timelike. Second caveat, which appears in the work of Duval and Fliche \cite{Duv78} but is not discussed, is that when studying the symplectic structure of their conformally invariant model for the massless spinning photon, its evolution space has dimension 11. It is recalled in the work of Duval and Sch\"ucker \cite{ChDTSRW} in a footnote p.~7, mentioning that the dimension of this evolution space leads to, in the flat spacetime limit, degrees of freedom with unclear physical interpretation. These extra degrees of freedom may be linked to the analysis of \cite{CostaNZ12,CostaHNZ11,CostaLGS17}, where they mention that the Mathisson--Pirani SSC leaves some ``residual gauge freedom'', and they argue that these are not a problem.

\subsection{Equation of motions for photons with Tulczyjew SSC}
\label{s:eom_ss}

\subsubsection{Notations}
\label{s:notations}
First, let us introduce the notations that will be used in the rest of this chapter. The metric has signature $(-, -, -, +)$. The components of the Riemann curvature tensor are defined by the convention ${R^\mu}_{\nu\alpha\beta} = \partial_\alpha \Gamma^\mu_{\beta\nu} - \partial_\beta \Gamma^\mu_{\alpha\nu} + \cdots$. In this paper, we often suppress indices by considering linear maps instead of 2-tensors. For instance, we use the linear map $S = ({S^\mu}_\nu)$ and likewise for the shorthand notation $R(S)$, with ${R(S)^\mu}_\nu = {R^\mu}_{\nu\alpha\beta} S^{\alpha\beta}$. In the same vein, we write $P$ for the vector $P$ and $\barP = (\barP_\mu)$ for the associated covector $P_\mu$, where indices are lowered with the metric. Another shorthand notation will be $R(S)(S)=R_{\mu\nu\alpha\beta}S^{\mu\nu}S^{\alpha\beta}$. 

For a skew-symmetric linear map $F$, the operator $\Pf$ gives its Pfaffian $\Pf(F)$. With the fully skew-symmetric Levi-Civita tensor $\epsilon_{\mu\nu\rho\sigma}$, with $\epsilon_{1234} = 1$, we have the expression $\Pf(F) = - \frac{1}{8} \sqrt{-\det(g_{\alpha\beta})} \epsilon_{\mu\nu\rho\sigma} F^{\mu\nu}F^{\rho\sigma}$. We have the relation $\Pf(F)^2 = \det(F)$. Indeed, the determinant of a skew-symmetric matrix can always be written as a perfect square.

\subsubsection{Spinning photons}

Let us now derive the Souriau--Saturnini equations. This section \ref{s:eom_ss} is based on the studies \cite{Sou74,Sat76,ChDTSRW}. 

As a starting point, we have the MPD equations \eqref{mpd_pdot}--\eqref{mpd_sdot}, which read in our notations,
\begin{align}
\dot{P} & = - \half R(S) \dot{X}, \label{mpdp_notations}\\
\dot{S} & = P \overline{\dot{X}} - \dot{X} \overline{P} . \label{mpds_notations} 
\end{align}

Since these equations are not closed, we need to consider additional conditions. See section \ref{s:sscs} for a discussion on this subject. We consider here two conditions, for $P \neq 0$,
\begin{align}
P^2 & = 0, \label{cond_p2_eq0} \\
{S^\mu}_\nu P^\nu & =: SP = 0. \label{cond_tulczyjew}
\end{align}

\subsubsection{Conservation of the mass}

Let us first show that the Tulczyjew SSC $SP = 0$ leads to $P^2$ being a constant of the system, justifying condition \eqref{cond_p2_eq0}, for a massless particle. Differentiating the Tulczyjew condition \eqref{cond_tulczyjew} leads to $\dot{S}P + S\dot{P} = 0$. Contracting this relation with $\dot{P}$, remembering that $S$ is skew-symmetric, using the equations \eqref{mpdp_notations} and \eqref{mpds_notations}, and the fact that $R(S)$ is also skew-symmetric, one obtains $\overline{P}\dot{X} \overline{P} \dot{P} = 0$. We also obtain, if we replace directly $\dot{S}$ and $\dot{P}$ using \eqref{mpdp_notations} and \eqref{mpds_notations},
\begin{equation}
\label{rel_diff_sp}
P(\overline{P}\dot{X}) - \dot{X} P^2 - \half SR(S)\dot{X} = 0.
\end{equation}

The former relation, can be rewritten $\overline{P} \dot{X} d P^2/ds = 0$, which leads to two possibilities. Either $P^2 = \const$, and we are done, or $\overline{P}\dot{X} = 0$. With the latter possibility, the relation \eqref{rel_diff_sp} becomes $\dot{X} P^2 + \half SR(S)\dot{X} = 0$. Contracting by $R(S)$ and using the useful relation that, for any skew-symmetric linear operators $F$ and $\Omega$, one has
\begin{equation}
\label{magic_formula}
\Omega F \Omega = \Pf(\Omega) \star(F) + \half \Tr(\Omega F)\Omega,
\end{equation}
we find that $\left(P^2 + \frac{1}{4} \Tr(SR(S))\right) R(S) \dot{X} + \half \Pf(R(S)) \star(S) \dot{X} = 0$.
Since $\Pf(S) = \pm \sqrt{\det(S)} = 0$
and since $\Pf(S) \id = S \star (S)$,
multiplying by $S$ on the left the previous relation leads to
$\left(P^2 + \frac{1}{4} \Tr(SR(S))\right) SR(S) \dot{X} = 0$. Hence, if we assume,
\begin{equation}
\label{rss_neq_0}
P^2 + \frac{1}{4} \Tr(SR(S)) \neq 0,
\end{equation}
we have $SR(S) \dot{X} = 0$, and thus we obtain $P^2 = 0$ from \eqref{rel_diff_sp} since we suppose $\dot{X} \neq 0$ for a massless particle.

The MPD equations, with the Tulczyjew condition \eqref{cond_tulczyjew} thus lead to the conservation of the quantity $P^2 = 0$. We wish to call this quantity the mass squared of our particle, and since we want to describe a massless particle, we write the condition,
\begin{equation}
P^2 = 0.
\end{equation}

\subsubsection{Conservation of the longitudinal spin}

It is well known that a classical elementary particle, described as belonging to the coadjoint representation of the Poincar\'e group, should have two invariants to qualify it: its mass and its (scalar) spin. We have just seen how to define the invariant mass of a photon with the MPD equations and the Tulczyjew condition. Let us now look at the definition of the spin.

Since $S$ is a skew-symmetric matrix, its rank must be even. We know that $P$ is in the kernel of $S$, by \eqref{cond_tulczyjew}, hence, since we assume that $S$ does not vanish, it must have rank 2. Since $\Pf(S) \id = S \star (S) = \star(S) S = 0$, and since $S$ is skew-symmetric of rank 2, we have $\Ima(\star S) = \ker(S)$. We can choose a vector $J \in \ker(S)$, not parallel to $J$, such that $\{P, J\}$ is a basis of $\ker(S)$. Then, we can write $\star (S) P = \lambda P + \mu J$, for $\lambda, \mu \in \R$. Since $S$ is skew-symmetric, and $P^2 = 0$ and $J$ is such that $\overline{P}J \neq 0$, we find that $\mu$ necessarily vanishes. Then, we write $\lambda = s$. We thus have $\star(S) P = s P$, and we call $\vert s \vert$ the spin and $\sign(s)$ the helicity of the photon. A further computation shows, with the help of $\Tr(S^2) = - \Tr((\star(S))^2)$, that $\Tr(S^2) = -2 s^2$. One then shows, with \eqref{mpds_notations} and \eqref{cond_tulczyjew}, that the spin $s$ of the particle is a constant of this system of equations.

For the description of the photon, we will have $s = \pm \hbar$.

\subsubsection{Equations of motion}
\label{ss:eom2}

We have seen that a basis of $\ker(S)$ is $\{P,J\}$, with $\overline{P}J \neq 0$. Now, the relation \eqref{rel_diff_sp} with $P^2 = 0$ is $P(\overline{P}\dot{X}) - \half SR(S)\dot{X} = 0$. Since $SJ = 0$ and $\overline{P}J \neq 0$, contracting this relation with $J$ leads to
\begin{equation}
\label{pxdot_eq0}
\overline{P}\dot{X} = 0.
\end{equation}

Then, plugging this back into the relation \eqref{rel_diff_sp}, we get,
\begin{equation}
\label{srsdx_eq0}
SR(S)\dot{X} = 0.
\end{equation}

We can decompose $\dot{X}$ in full generality as $\dot{X} = \lambda P + \mu J + SK$, for some $\lambda, \mu \in C^\infty(M,\R)$ and some vector $K$. From \eqref{pxdot_eq0}, \eqref{cond_p2_eq0} and \eqref{cond_tulczyjew}, we immediately get that $\mu = 0$, and thus $\dot{X} = \lambda P + SK$. Contracting this relation on the left by $SR(S)$, and in the view of \eqref{srsdx_eq0}, we obtain $0 = \lambda SR(S)P + SR(S)SK$. Using again the useful relation \eqref{magic_formula} for $SR(S)S$, and with $\Pf(S) = 0$, one gets $0 = \lambda SR(S)P - \half R(S)(S) \, SK$, where $R(S)(S) = - \Tr(SR(S))$. Thus, $SK = 2 \lambda \frac{SR(S)P}{R(S)(S)}$ if $R(S)(S) \neq 0$. Notice that $R(S)(S) \neq 0$ is the condition \eqref{rss_neq_0} with $P^2 = 0$. We thus have the expression for the velocity,
\begin{equation}
\dot{X} = \lambda\left(P + 2 \frac{SR(S)P}{R(S)(S)}\right).
\end{equation}

Then, injecting $\dot{X}$ into the MPD equation for $\dot{P}$ \eqref{mpdp_notations}, we get $\dot{P} = -\half \lambda R(S) P - \lambda \frac{R(S)SR(S) P}{R(S)(S)}$. Since $R(S)SR(S) = \Pf(R(S)) \star (S) - \half R(S)(S) \, R(S)$ by \eqref{magic_formula}, and since $\star(S) P = s P$, we have, $\dot{P} = - \lambda s \frac{\Pf(R(S))}{R(S)(S)} P$. The complete equations of motion for a photon, which we call the Souriau-Saturnini equations, are thus, with a suitable worldline parameter $\tau$ such that $\lambda = 1$,
\begin{align}
\dot{X} & = P + 2 \frac{SR(S)P}{R(S)(S)}, \label{ss_xdot}\\
\dot{P} & = - s \frac{\Pf(R(S))}{R(S)(S)} P, \label{ss_pdot}\\
\dot{S} & = P \overline{\dot{X}} - \dot{X} \overline{P}. \label{ss_sdot}
\end{align}

Note that, as was already mentioned by Saturnini \cite{Sat76}, these equations do not depend on the value of $\hbar$. Indeed, one could redefine the spin tensor $S$ as $\Omega = S/s$ such that $\Omega$ is dimensionless. Then, the system of equations, together with the redefinitions $P \rightarrow P/s$ and $\tau \rightarrow \tau s$, is the same as the Souriau--Saturnini equations above, with dimensionless $\Omega$ instead of $S$. This means that at this point in the derivation of the equations, there is no hope of recovering the (light-like) geodesic equation in the limit $\hbar \rightarrow 0$.

\subsubsection{Comment on the vanishing curvature limit}

To derive the Souriau-Saturnini equations in the form \eqref{ss_xdot}--\eqref{ss_sdot}, one needs the assumption that $R(S)(S) \neq 0$ \eqref{rss_neq_0}, hence that curvature does not vanish. Let us now look at the case when curvature does vanish. We have the MPD equations,
\begin{align}
\dot{P} & = 0, \\
\dot{S} & = P \overline{\dot{X}} - \dot{X} \overline{P},
\end{align}
and the constraints $P^2 = 0$ and $SP = 0$. The linear operator $S$ has once again rank 2, with a kernel generated by the vectors $\{P, J\}$ for some $J$ such that $\overline{P} J \neq 0$. Decomposing in full generality the velocity, as in the previous section, we have $\dot{X} = \lambda P + \mu J + SK$, for some vector $K$ and $\lambda, \mu \in \R$. Just as before, contracting with $P$ leads to $\mu = 0$. Then, one can choose the parameter $\tau$ and redefine $K$ such that $\lambda = 1$. We get
\begin{equation}
\dot{X} = P + S K. 
\end{equation}
This is as far as we can go, there is no way to completely determine $K$ with the equations and constraints of this system. 

This is noted in \cite{Kun72,Sou74,Sat76}. When writing this system in a symplectic formalism, the dimension of the kernel of the symplectic form is not constant: it is of dimension 1 in the curved case, and of dimension 3 in the vanishing curvature limit. This means that 2 degrees of freedom are not constrained anymore when we go to the vanishing curvature limit, and we get a degenerate system. This degenerate system was studied in \cite{Sou70}, and in the spin 1 case, it turns out to lead to the Maxwell equations after geometric quantization. This seems to imply that in the presence of gravity, the spinning photon can be localized in this theory, while in flat spacetime, this theory leads to the wave equation for the photon with spin. 

Thus, one needs to be careful when dealing with vanishing curvature and the Souriau-Saturnini equations \eqref{ss_xdot}--\eqref{ss_sdot}.

%

\section{The simple example of de Sitter spacetime}
\label{ss:desitter}

\subsection{Direct computation}

The results of this section can be found in Saturnini's thesis \cite{Sat76}. The Souriau--Saturnini equations of motion for photons with spin \eqref{ss_xdot}--\eqref{ss_sdot} are particularly simple to write in the case of a de Sitter spacetime. Indeed, de Sitter spacetime being maximally symmetric, we have for the Riemann tensor,
\begin{equation}
R_{\mu\nu\lambda\rho} = \frac{\Lambda}{3}\left(g_{\mu\lambda} g_{\nu\rho} - g_{\mu\rho} g_{\nu\lambda}\right), 
\end{equation}
with $\Lambda$ the cosmological constant, and $g$ the de Sitter metric. We can immediately compute $R(S)$,
\begin{equation}
\label{rs_ds}
R(S)_{\mu\nu} := R_{\mu\nu\lambda\rho} S^{\lambda\rho} = \frac{2\Lambda}{3} S_{\mu\nu}.
\end{equation}

Recalling the Tulczyjew condition that $P$ lies in the kernel of $S$, see \eqref{cond_tulczyjew}, the previous result implies that $R(S)P = 0$. Since $\Pf(S) = 0$, we also have $\Pf(R(S)) = 0$, and we can compute $R(S)(S) := - \Tr(SR(S)) = R(S)_{\mu\nu} S^{\mu\nu} = \frac{4 \Lambda}{3} s^2$, with $s = \pm \hbar$. The Souriau-Saturnini equations \eqref{ss_xdot}--\eqref{ss_sdot} thus take the following simple form,
\begin{align}
\dot{X} & = P, \label{ssl_xdot} \\
\dot{P} & = 0, \label{ssl_pdot} \\
\dot{S} & = 0, \label{ssl_sdot}
\end{align}
with the two conditions $P^2 = 0$ \eqref{cond_p2_eq0} and $SP = 0$ \eqref{cond_tulczyjew}. Recall the notations: the dot over $X$ means the usual derivative with respect to the affine parameter describing the trajectory, while the dot over $P$ and $S$ denotes the covariant derivative with respect to that same parameter. 

The main take away from the above equations is that according to the Souriau--Saturnini equations, in a de Sitter universe, a spinning photon follows the same null geodesic as that of the usual ``spinless'' photon. Now, some comments are in order.

First, note that the spin tensor is parallel transported \eqref{ssl_sdot}, just as the 4-momentum is \eqref{ssl_pdot}. Parallel transport of the 4-momentum implies that the energy of the photon varies through time, according to the curvature of the spacetime. This is called redshift. Now for the spin of the photon, while the longitudinal spin $s$ is a constant of the system, nothing fixes the norm of the transverse spin $\st$. In particular here, parallel transport of the spin tensor implies that the transverse spin also follows the same redshift as the energy of the photon does. This is a good place to remind the reader that these equations of motions are for \emph{classical} spinning particles. Quantum effects are not taken into considerations here, as the notion of the spin of an elementary particle already appears at the classical level. One can expect this oddity to disappear in a complete quantum theory of gravitation, which sadly still seems out of reach for now.

Second comment, notice that the equations \eqref{ssl_xdot}--\eqref{ssl_sdot} do not depend on the value of $\Lambda$ -- other than through the covariant derivative. In a space of constant curvature, which can be arbitrarily close to 0, the spinning photon follows a null geodesic. Yet, remember that the condition to derive the Souriau--Saturnini equations is that $R(S)(S) \neq 0$, and hence here $\Lambda \neq 0$.

\subsection{Explicit computation and general remarks about coordinates}
\label{ss:procedure}

\subsubsection{Writing the metric in isotropic coordinates}
\label{ss:iso_coords}

When computing the Souriau--Saturnini equations, it is convenient to use isotropic coordinates whenever. These coordinates have the benefit to put the three space coordinates $\bx = (x^1, x^2, x^3)$ on an equal footing, allowing for easy computations of cross products and Euclidean norms. Consider metrics of the form,
\begin{equation}
A^2(t, \bx) \, dt^2 - B^2(t, \bx) \, \Vert d\bx \Vert^2,
\end{equation}
with $\Vert d\bx \Vert$ the norm of the vector $d\bx$. In a way, isotropic coordinates put the \emph{spatial} part of the metric in a form that is conformally flat.

Let us look at the case where we have a metric of the form,
\begin{equation}
g = D(\rho) dt^2 - C(\rho) d\rho^2 - \rho^2 d\Omega^2,
\end{equation}
with $d\Omega^2 = d\theta^2 + \sin^2 \theta d\varphi^2$. We need to find a change of coordinate $\rho \rightarrow r$ so that we have $C(\rho) d\rho^2 + \rho^2 d\Omega^2 = B^2(r) \left(dr^2 + r^2 d\Omega^2\right)$, for some function $B(r)$. If $\rho = \rho(r)$, we need to solve $\displaystyle C(\rho) \left(\frac{d\rho}{dr}\right)^2 = \frac{\rho^2}{r^2}$ for $\rho(r)$, or, if $C(\rho)$ is strictly positive, $\displaystyle \frac{d\rho}{dr} = \frac{1}{\sqrt{C(\rho)}} \frac{\rho}{r}$. Then, $B(r)$ will simply be $\displaystyle B^2(r) = \frac{\rho^2(r)}{r^2}$, and define $A^2(r)$ so that $A^2(r) := D(\rho(r))$, to have the metric $\displaystyle g = A^2(r) dt^2 - B^2(r) \left(dr^2 + r^2 d\Omega^2\right)$ with isotropic coordinates.

As an example, consider the de Sitter metric, as usually written in spherical coordinates $(t, \rho, \theta, \varphi)$,
\begin{equation}
g = \left(1 - \frac{\rho^2 \Lambda}{3}\right) dt^2 - \left(1 - \frac{\rho^2 \Lambda}{3}\right)^{-1} d\rho^2 - \rho^2 d\Omega^2,
\end{equation}
for $0 \leq \rho < \sqrt{3/\Lambda}$.

Following the previous procedure, we have here $\displaystyle C(\rho) = \left(1 - \frac{\rho^2 \Lambda}{3}\right)^{-1}$. Hence, one needs to solve the differential equation $\displaystyle \frac{d\rho}{dr} = \frac{\rho}{r}\sqrt{1-\frac{\rho^2\Lambda}{3}}$ for $\rho(r)$. We find a solution to be $\displaystyle \rho(r) = \frac{r}{1 + r^2 \Lambda/12}$. We then find the de Sitter metric to be, in isotropic coordinates, $g = \left(\frac{12 - r^2 \Lambda}{12 + r^2\Lambda}\right)^2 dt^2 - \left(1 + \frac{r^2 \Lambda}{12}\right)^{-2} \left(dr^2 + r^2 d\Omega^2\right)$, for $0 \leq r < 2\sqrt{3/\Lambda}$. It is then usually common to replace the spherical coordinates $(r, \theta, \varphi)$ by cartesian coordinates $(x^1, x^2, x^3)$, so that we have the metric,
\begin{equation}
g = A^2(r) dt^2 - B^2(r) \Vert d\bx \Vert^2,
\end{equation}
with the shorthands $A(r) = \frac{12 - r^2 \Lambda}{12 + r^2\Lambda}$ and $B(r) = \left(1 + \frac{r^2 \Lambda}{12}\right)^{-1}$, and $r = \sqrt{\bx\cdot\bx}$. In later computations, we will simplify notations and write, \textit{e.g.} $A = A(r)$.

\subsubsection{Definitions and writing the Souriau--Saturnini equations}

Once one has the metric, the different objects appearing in the Souriau--Saturnini equations \eqref{ss_xdot}--\eqref{ss_sdot} should be defined and computed. The first definition is the 4-momentum of the photon. We write it here as,
\begin{equation}
P=\left(
\begin{array}{c}
\displaystyle
\frac{\bp}{B}\\[10pt]
\displaystyle
 \frac{\np}{A}
\end{array}\right)
\label{Pds}
\end{equation}
with $\bp\in\R^3\setminus\{0\}$, the spatial linear momentum, and $\np\dpp=\sqrt{\bp\cdot\bp}$. The 4-momentum is light-like, $P^2=0$.

The second object to define is the map $S$, which is skewsymmetric with respect to the metric, $\rg(SV,W)=-\rg(V,SW)$ for all 4-vectors $V$ and $W$. The spin tensor is also required to obey the Tulczyjew constraint $SP=0$. A solution to these constraints is the tensor,
\begin{equation}
S=({S^\mu }_\nu)=\left(
\begin{array}{cc}
j(\bs)&\displaystyle
-\frac{(\bs\times\bp)}{\np}\frac{A}{B}\\[6pt]
\displaystyle
-\frac{(\bs\times\bp)^T}{\np}\frac{B}{A}&0
\end{array}\right)
\label{Sds}
\end{equation}
with the spin vector $\bs\in\R^3\setminus\{0\}$. The vector-product is with respect to the Euclidean metric and we define the linear map $j(\bs):\bp\mapsto\bs\times\bp$. We have,
\begin{equation}
-\half\Tr(S^2)=s^2\,,
\end{equation}
with the scalar spin, or longitudinal spin,
\begin{equation}
s\dpp=\frac{\bs\cdot\bp}{\np},
\label{scalarspinds}
\end{equation}
which we can show is a constant of the system.
The scalar spin $s$ is not to be confused with the norm
$\Vert\bs\Vert$ of the spin vector.
The helicity  or handedness of the photon is $\sign(s)$.

Next would be the computation of the quantities involving the Riemann tensor, such as $R(S)$, $R(S)(S)$, or $\Pf(R(S))$, however we already have computed such quantities in the previous section, see for example \eqref{rs_ds}.

From the first Souriau--Saturnini equation \eqref{ss_xdot}, we find 
\begin{equation}
\frac{dt}{d\tau} = P_4 + 2 \frac{SR(S)P_4}{R(S)(S)} = \frac{12+r^2 \Lambda}{12-r^2 \Lambda} \np,
\end{equation}
and therefore the equations of motion read,
\begin{align}
\frac{d\bx}{dt} & = \left(1 - \frac{r^2 \Lambda}{12}\right) \frac{\bp}{\np}, \label{ds_dxdt}\\
\frac{d\bp}{dt} & = \Lambda \frac{\np^2 \left(12+r^2 \Lambda\right) \bx + (\bx \cdot \bp) \left(12-r^2\Lambda\right) \bp}{6 \np \left(12+ r^2 \Lambda\right)}, \label{ds_dpdt}\\
\frac{d\bs}{dt} & = \Lambda \frac{24 (\bp \cdot \bx) \bs-\left(12-r^2 \Lambda\right)(\bs\cdot\bp) \bx - \left(12+r^2 \Lambda\right)(\bs\cdot\bx) \bp}{6 \np \left(12+ r^2 \Lambda\right)},\label{ds_dsdt}
\end{align}
which are the equations of motion \eqref{ssl_xdot}--\eqref{ssl_sdot} we found already in the previous section, but here explicitly in terms of the time coordinate $t$ and without covariant derivatives.

\section{A spinning photon in a flat FLRW background}
\label{ss:flrw}

An application of the Souriau--Saturnini equations to derive the trajectory of spinning photons in a Friedmann--Lema\^itre--Robertson--Walker (FLRW) spacetime can be found in \cite{ChDTSRW}. In this section, we outline the main results.

The background metric, with no spatial curvature ($K = 0$) and in coordinates $(t, \bx)$,
\begin{equation}
g = dt^2 - a(t) \, \Vert d\bx \Vert^2,
\end{equation}
is characterized by the scale factor	$a$, a strictly positive function of cosmic time $t$.

The authors then define the 4-momentum $P$ and the spin tensor $S$,
\begin{equation}
P = \left(\begin{array}{c}
\displaystyle \frac{\bp}{a} \\
\Vert \bp \Vert
\end{array}\right)
\qquad \& \qquad
S = \left(\begin{array}{cc}
j(\bs) & \displaystyle - \frac{\bs \times \bp}{a \Vert \bp \Vert} \\
\displaystyle - \frac{(\bs \times \bp)^T}{\Vert \bp \Vert} a & 0
\end{array}\right),
\end{equation}
with $\bp$ the spatial linear momentum of our test particle, and $\bs$ its spin vector, such that the Tulczyjew constraint \eqref{cond_tulczyjew} and the null 4-momentum condition \eqref{cond_p2_eq0} are satisfied. We also have the conserved longitudinal spin $s$ through $\Tr(S^2) = -2s^2$, with here $\displaystyle s = \frac{\bs \cdot \bp}{\Vert \bp \Vert}$. Another useful definition is that of the transverse spin $\displaystyle \bst = \bs - s \frac{\bp}{\Vert\bp\Vert}$.

With these definitions in mind, they compute the Souriau--Saturnini equations \eqref{ss_xdot}--\eqref{ss_sdot} for a spinning photon, in terms of cosmic time $t$,
\begin{align}
\frac{d\bx}{dt} & = \frac{a''}{a'^2} \frac{\bp}{\Vert\bp\Vert} + \frac{1}{a}\left(1 - \frac{aa''}{a'^2}\right) \frac{\bs}{s}, \label{rw_xdot}\\
\frac{d\bp}{dt} & = - \frac{a'}{a}\left(\frac{aa''}{a'^2} \bp + \Vert\bp\Vert\left(1 - \frac{aa''}{a'^2}\right) \frac{\bs}{s}\right), \label{rw_pdot}\\
\frac{d\bs}{dt} & = - \left(1-\frac{aa''}{a'^2}\right) \frac{\bs}{s} \times \bp - \frac{a'}{a}\bs + \frac{a'}{a}\left(\frac{\Vert\bs\Vert^2}{s}\left(1-\frac{aa''}{a'^2}\right) + s \frac{aa''}{a'^2}\right) \frac{\bp}{\Vert\bp\Vert}. \label{rw_sdot}
\end{align}

The first interesting result they observe, is that in the case of ``enslaved'' spin, \ie when $\bst = 0$, the photon travels on a light-like geodesic. 

\begin{figure}
\centering
\includegraphics[scale=0.7]{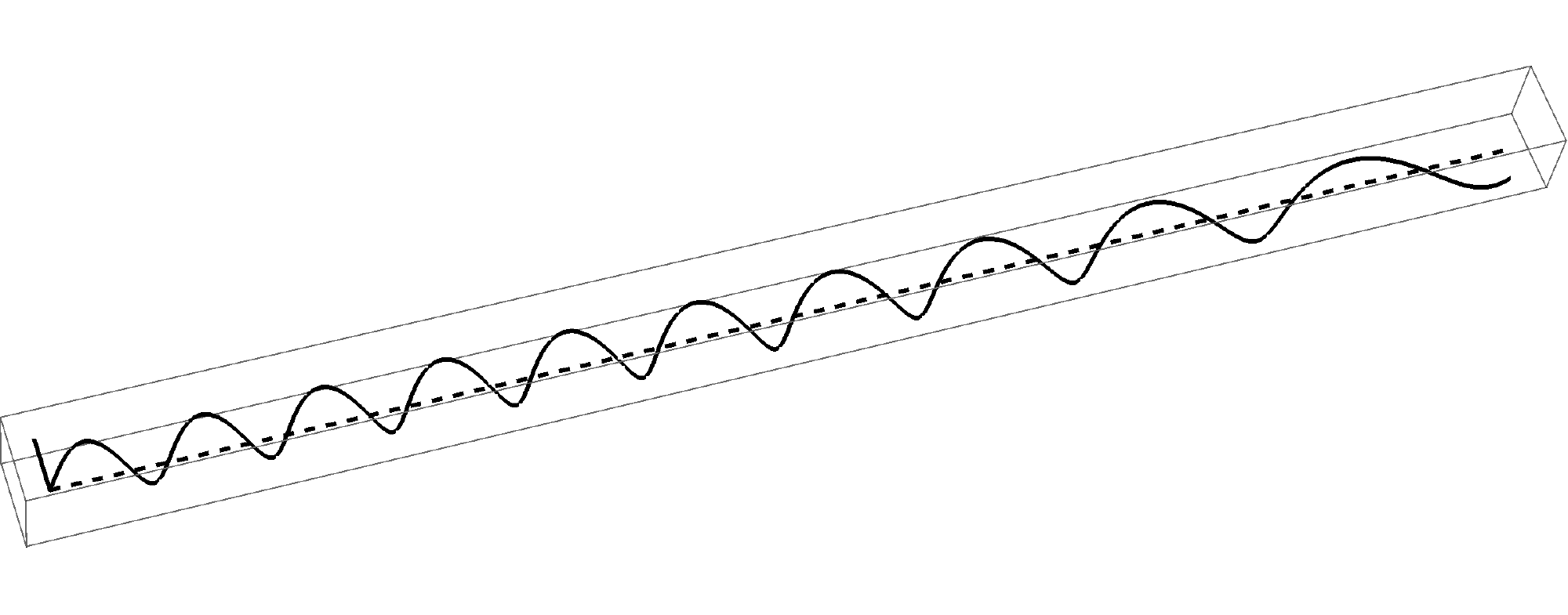}
\caption{Numerical simulation from \cite{ChDTSRW} of the trajectory of a spinning photon (black line) according to the Souriau--Saturnini equations in a FLRW background. The dashed line is the (geodesic) trajectory of a spinless photon with identical initial conditions, other than the spin.}
\label{fig_helix}
\end{figure}

The authors of \cite{ChDTSRW} then perform a numerical simulation of a spinning photon. In ``comoving coordinates'' $\bx$, the trajectory computed is displayed in figure \ref{fig_helix}. This trajectory is most interesting. Indeed, the spinning photon seems to have a helical trajectory around the usually computed lightlike geodesic. Moreover, these two test particles, spinning and spinless photon, travel on the same front. This means that the spinning photon must have a ``superluminal'' velocity at any point of its trajectory. While it initially seems to break causality, the radius of the helix remains of the order of the wavelength of the photon. Hence, information cannot propagate outside of this tube of small radius. This is reminiscent of the Spin Hall Effect of Light, which was observed experimentally, where an ``instantaneous'' transverse shift of the photon's trajectory, of the order of the wavelength, happens when it is reflected on a surface. Two more comments about the helical motion are in order. First, the direction of rotation depends on the helicity $\sign(s)$ of the photon. Second, the period of rotation seems to increase with cosmic time $t$.

A perturbative solution of the equations of motions of a spinning photon is also provided in \cite{ChDTSRW}. To this end, the authors use two small parameters related to, respectively, the longitudinal spin and the transverse spin. They are defined as $\eta = s / \cE = \pm T_e / 2 \pi a_e$ and $\epsilon = s_e^\bot / \cE$ with $T_e, a_e, s_e^\bot$ respectively the atomic period, the scale factor $a$, and the norm of the transverse spin at emission time, and $\cE = a \np$ the conserved ``energy'' associated to the photon. The idea is that while both numbers are typically very small, $\epsilon = 0$ implies a pure longitudinal spin, and hence a light-like geodesic trajectory.

Rewriting the equations of motion \eqref{rw_xdot}--\eqref{rw_sdot} in linear order in terms of $\eta$ and $\epsilon$, they are able to find analytical perturbative solutions for the helix period $T_{helix}(t)$ and the ``comoving transverse coordinate'' $x^\bot(t)$, defined as $x^\bot(t) = \sqrt{x^2(t)^2 + x^3(t)^2}$, with $x^1(t)$ the axis along which the spinless photon propagates.

The helix period is given by,
\begin{equation}
T_{helix}(t) \approx \frac{a(t)}{a_e} \frac{1}{1+q(t)} T_e,
\end{equation}
with the deceleration parameter $q(t) = - a(t) a''(t) / a'(t)^2$. The transverse coordinate by,
\begin{equation}
x^\bot(t) \approx \vert\eta\vert \sqrt{1-2 \cos \frac{x^1(t)+1/a'(t) - 1/a_e'}{\vert\eta\vert}\left(1-a_e' x^1(t)\right)+\left(1-a_e' x^1(t)\right)^2}.
\end{equation}

Note that in their work, the transverse coordinate doesn't match with the radius of the helix. Indeed, due to initial conditions of the spinning photon being on the geodesic axis, it undergoes an immediate shift out of the axis, and the resulting helix is not centered on the geodesic axis.

\section{Birefringence of light around a Schwarzschild body}
\label{ss:schw}
Birefringence of light is a well known phenomenon in anisotropic matter like quartz or calcite. Thanks to the Souriau--Saturnini equations, we have a framework to study the effect of the spin of a photon on its trajectory in a gravitational background. In the previous section, we saw that photons in a FLRW background show birefringence. We may wonder whether this is also true in a Schwarzschild spacetime. This section is based on the article \cite{ChDLMTS}, with additional comments about the cosmological constant.

\subsection{Spinning massless particles}
\label{s:2}

\subsubsection{Schwarzschild metric in isotropic coordinates}

Let us follow the procedure of section \ref{ss:procedure} to write down the Souriau--Saturnini equations for the Schwarzschild metric. The first step will be to find the Schwarzschild metric with isotropic coordinates. The metric in Schwarzschild coordinates $(t, \rho, \theta, \varphi)$ is given by,
\begin{equation}
g = \left(1- \frac{2 GM}{\rho}\right) dt^2 - \left(1 - \frac{2GM}{\rho}\right)^{-1} d\rho^2 - \rho^2 d\Omega^2,
\end{equation}
for $2GM < \rho$.

This is the same kind of metric as we have studied in the section \ref{ss:procedure}, with $C(\rho) = - \left(1 - \frac{2GM}{\rho}\right)^{-1}$. Hence, the differential equation to solve to obtain the isotropic coordinates is $\displaystyle \frac{d\rho}{dr} = \frac{\rho}{r}\sqrt{1 - \frac{2GM}{\rho}}$ for $\rho(r)$. One finds,
\begin{equation}
\label{schw_trsf}
\rho=r\left(1+\frac{GM}{2r}\right)^2
\qquad
\text{or}
\qquad
r=\half\left(\rho-GM+\sqrt{\rho(\rho-2GM)}\right).
\end{equation}

The Schwarzschild metric can thus be expressed in an isotropic coordinate patch $(X^\mu)=(\bx,t)$ by,
\begin{equation}
\label{gbis}
g = A^2(r)\,dt^2-B^2(r)\Vert d\bx\Vert^2
\end{equation}
with
\begin{equation}
A(r)\dpp=\,\frac{r-a}{r+a}\,, \qquad B(r)\dpp=\lp\frac{r+a}{r}\rp^2,\qquad r\dpp=\sqrt{\bx\cdot\bx}\,,\qquad 0<a<r\,.
\end{equation}
where $\bx=(x^1,x^2,x^3)$, and $a=\half GM$ is the Schwarzschild radius in isotropic coordinates.

We compute the following Christoffel symbols,
\begin{equation}
{\Gamma^j}_{ii}=-{\Gamma^i}_{ji}=-{\Gamma^j}_{jj}=\frac{2a\, x^j}{r^2(r+a)}\,,
\quad
{\Gamma^j}_{44}=\frac{2ar^3(r-a)\, x^j}{(r+a)^7}\,,
\qquad
{\Gamma^4}_{4j}=\frac{2a\, x^j}{r\,(r+a)(r-a)}\,,
\label{Gamma}
\end{equation}
for all $i\not=j=1,2,3$, no summation over repeated indices. 

For the Riemann tensor ${R^\mu}_{\nu\alpha\beta}=\partial_{\alpha}{\Gamma^\mu}_{\beta\nu}-\partial_{\beta}{\Gamma^\mu}_{\alpha\nu}+\cdots$ with $i,j$ and $k$ all different, we have
\begin{align}
{R^i}_{jij}&=\,\frac{2a\,[2(x^k)^2-(x^i)^2-(x^j)^2]}{r^3(r+a)^2}\,,&
 {R^j}_{iki}&=-\,\frac{6a\,x^jx^k}{r^3(r+a)^2}\,,\\
 {R^4}_{i4i}&=\,\frac{2a\,[2(x^i)^2-(x^j)^2-(x^k)^2]}{r^3(r+a)^2}\,,&
{R^4}_{i4j}&=\ \,\frac{6a\,x^ix^j}{r^3(r+a)^2}\,.
\end{align}
The Ricci tensor vanishes.

\subsubsection{Momentum and spin}

In the above coordinate system, the (future pointing)  4-momentum of the photon is written as
\begin{equation}
P=\left(
\begin{array}{c}
\displaystyle
\frac{\bp}{B}\\[10pt]
\displaystyle
 \frac{\np}{A}
\end{array}\right)
\label{Pbis}
\end{equation}
with $\bp\in\R^3\setminus\{0\}$, the spatial linear momentum, and $\np\dpp=\sqrt{\bp\cdot\bp}$ (Euclidean scalar product). We suppose positive energy, $\np>0$. The 4-momentum is light-like, $P^2=0$.

The map $S$ is skewsymmetric with respect to the metric: $\rg(SV,W)=-\rg(V,SW)$ for all vectors $V$ and $W$. Accordingly, the spin tensor is also defined by the Tulczyjew constraint $SP=0$.

A solution to these constraints is the tensor,
\begin{equation}
S=({S^\mu }_\nu)=\left(
\begin{array}{cc}
j(\bs)&\displaystyle
-\frac{(\bs\times\bp)}{\np}\frac{A}{B}\\[6pt]
\displaystyle
-\frac{(\bs\times\bp)^T}{\np}\frac{B}{A}&0
\end{array}\right)
\label{Sbis}
\end{equation}
with the spin vector $\bs\in\R^3\setminus\{0\}$. The vector-product is with respect to the Euclidean metric and we define the linear map $j(\bs):\bp\mapsto\bs\times\bp$. We have,
\begin{equation}
-\half\Tr(S^2)=s^2\,,
\end{equation}
with the scalar spin, or longitudinal spin,
\begin{equation}
s\dpp=\frac{\bs\cdot\bp}{\np},
\label{scalarspin}
\end{equation}
which we have shown is a constant of the system.
The scalar spin $s$ is not to be confused with the norm
$\Vert\bs\Vert$ of the spin vector.
The helicity  or handedness of the photon is $\sign(s)$.

In the Schwarzschild metric we obtain, with the notations (\ref{s:notations}),
\begin{align}
\Pf(R(S))&=\,\frac{48\, a^2r^4}{(r+a)^{12}\np}\lb\bx \times\bp\cdot\bs\rb(\bs\cdot\bx),\\[2mm]
R(S)(S)&=\,\frac{8\, ar}{(r+a)^6}\, \lb 3\lb\bx \times\bp\cdot\bs\rb^2/\np^2-3 (\bs\cdot\bx)^2+s^2r^2\rb,\\[2mm]
S\,R(S)\,P&=\dpp
\begin{pmatrix}\bc\\d
\end{pmatrix}\qquad\text{ with} \\[2mm]
\nonumber
\bc &=\,\frac{12\,a\,r^3}{(r+a)^8}\,\Bigg[(\bs\cdot\bx)^2{\bp}\,-\np s\,(\bs\cdot\bx )\,\bx-\lb\bx \times\bp\cdot\bs\rb\,\bs\times\bx\\[2mm]
&\qquad\qquad\qquad\qquad\qquad\qquad\qquad\qquad+\lb\bx \times\bp\cdot\bs\rb\lp\bx\cdot\frac{\bp}{\np}\rp\bs \times\frac{\bp}{\np}\Bigg]\,,\\[2mm]
d&=\,\frac{12\,a\,r}{(r-a)(r+a)^5}\,\Big[ s\,(\bs\cdot\bx)\lp\bx\cdot{\bp}\rp-\ns^2/\np\,\lp\bx\cdot{\bp}\rp^2
\nonumber
\\[2mm]
&\qquad\qquad\qquad\qquad\qquad\qquad
+\np\,(\ns^2-s^2)\,r^2-2\lb\bx \times\bp\cdot\bs\rb^2/\np\Big] .
\end{align}
The following vector identity will be useful:
\begin{eqnarray}
\lb\bu \times\bv\cdot\bw\rb^2&=&\Vert\bu\Vert^2\Vert\bv\Vert^2\Vert\bw\Vert^2
+
2\,(\bu\cdot\bv)(\bu\cdot\bw)(\bv\cdot\bw)\nonumber\\
&&-\Vert\bu\Vert^2(\bv\cdot\bw)^2
-\Vert\bv\Vert^2(\bu\cdot\bw)^2
-\Vert\bw\Vert^2(\bu\cdot\bv)^2\,.
\end{eqnarray}

\subsubsection{Conservation laws}

The group of isometries of Schwarzschild spacetime is $\rO(3)\times\R$, its generators are the Killing vector fields of the metric~(\ref{gbis}),
$Z=\varepsilon^i_{\,jk}\,\omega^jx^k\,\partial/\partial x^i+\epsilon \,\partial/\partial t $,
where $\bomega\in\R^3$ and $\epsilon \in\R$ stand for infinitesimal rotations and time translations, respectively; the $\varepsilon^i_{\,jk}$ are the structure constants of $\so(3)$.
Using the general expression \cite{Sou70}
\begin{equation}
\Psi(Z)=P_\mu{}Z^\mu+\half{}S^{\mu\nu}\nabla_\mu{}Z_\nu
\label{Psi}
\end{equation}
 of the ``moment map'', $\Psi$, associated with a Killing vector field, $Z$, together with the expressions (\ref{Pbis}) and (\ref{Sbis}) for $P$ and $S$, we find in a straightforward fashion
$
\Psi(Z)=-\bcL\cdot\bomega+\cE\,\epsilon
$
where 
\begin{equation}
\cE=
\,\frac{r-a}{r+a}\,\np+\,\frac{2ar}{(r+a)^4\np}\lb\bx \times\bp\cdot\bs\rb\,,  
\label{cE}
\end{equation}
is the conserved energy and
\begin{equation}
\bcL=\lp\frac{r+a}{r}\rp^2 \bx\times\bp+\,\frac{r-a}{r+a}\, \bs\,+\,\frac{2a}{r^2(r+a)}\,(\bs\cdot\bx)\, \bx,
\label{cL}
\end{equation}
 the conserved angular momentum featuring both an extra spin contribution. The latter equation defines an affine map between spin and angular momentum. We will use its inverse:
 \begin{equation}
 \bs\,=\,\frac{r+a}{r-a}\, \lb \bcL-\lp\frac{r+a}{r}\rp^2\,\bx\times\bp-\,\frac{2a}{r^2(r+a)}\, (\bcL\cdot\bx)\,\bx\rb. \label{bs}
 \end{equation} 

\subsubsection{Specifying the Souriau-Saturnini equations}

Now that we have defined the metric together with the objects appearing in the equations and the conserved quantities, we are ready to specialize the Souriau-Saturnini equations \eqref{ss_xdot}--\eqref{ss_sdot} to the case of Schwarzschild spacetime.

Let us introduce the shorthand,
\begin{equation}
\label{def_d}
D\dpp=r^2(\bs\cdot\bp)-3(\bp\cdot\bx)(\bs\cdot\bx).
\end{equation}

To obtain the equations of motion in $3$-space, we trade the affine parameter $\tau$ for the coordinate time $t$ using \eqref{ss_xdot},

\begin{equation}
\label{dtdtau}
\frac{dt}{d\tau}= \frac{r+a}{r-a}\,\np
\!\left[
\frac{s\,D \, \np}{s^2r^2\np^2-3(\bs\cdot\bx)^2\np^2+3[\bx\times\bp\cdot\bs]^2}
\right]\,,
\end{equation}
which we assume non-vanishing. 
By abuse of notation we write $\tau(t)$ for the inverse function of~$t(\tau)$ and we do not distinguish $\bx=\bx(t)=\bx(\tau(t))$ and likewise for $\bp$ and $\bs$. Then we have, from the Souriau-Saturnini equations \eqref{ss_xdot}--\eqref{ss_sdot} and (\ref{dtdtau}):
\begin{align}
\,\frac{d\bx}{dt}\,
& =\,\frac{r^2(r-a)}{\np\,(r+a)^3D}
&\hspace{-10mm}\Big\{&
r^2(\bs\cdot\bp)\,\bp-3\np^2(\bs\cdot\bx)\,\bx+3[\bx\times\bp\cdot\bs]\,\bx\times\bp\Big\}\,, \label{dx_spin}\\
\,
\frac{d\bp}{dt}\,
& =\,\frac{2\,a}{\np\,(r+a)^4D}
&\hspace{-10mm}\Bigg\{&
r^2(r-a)\left[(\bs\cdot\bp)(\bp\cdot\bx)-\frac{3r}{(r+a)^3}\,(\bs\cdot\bx)[\bx\times\bp\cdot\bs]\right]\,\bp  \qquad\qquad\nonumber\\
&&&\qquad-r\,\np^2\big[D+r\,(r-a)\,(\bs\cdot\bp)\big]\,\bx  \nonumber\\
&&&\qquad\qquad\qquad+3\,(r-a)[\bx\times\bp\cdot\bs]\,(\bp\cdot\bx)\,\bx\times\bp\Bigg\},\label{dp_spin}\\
\,\frac{d\bs}{dt}\,
& =\,\frac{1}{\np\,(r+a)^4D}
&\hspace{-10mm}\Big\{&
3(r-a)(r+a)^3\big[\lp-r^2 \np^2+(\bx\cdot\bp)^2\rp \bs\times\bp\nonumber\\
&&&\hspace{-20mm}+\lp 2 \np^2(\bx\cdot\bs)-(\bx\cdot\bp)(\bs\cdot\bp)\rp \bx\times\bp\big]+ 2 a r D\lp(\bx\cdot\bs)\bp-(\bx\cdot\bp)\bs\rp\nonumber\\
&&&+ 2 a(r-a)\big(-r^2 (\bs\cdot\bp)^2\bx-3[\bx\times\bp\cdot\bs]^2 \bx\nonumber\\
&&&\qquad\qquad+r^2(\bx\cdot\bs)(\bs\cdot\bp) \bp+3[\bx\times\bp\cdot\bs](\bx\cdot\bs) \bx\times\bp\big)\Big\}.\label{ds_spin}
\end{align}

With the equations above, we can verify that the conserved quantities, namely the scalar spin (\ref{scalarspin}), the energy (\ref{cE}) and the total angular momentum (\ref{cL}) are conserved. We have indeed $d\cE/dt = d\bcL/dt = ds/dt = 0$.

We can simplify the system by only considering the equations of position and momentum (\ref{dx_spin},\ref{dp_spin}) and by eliminating $[\bx\times\bp\cdot\bs]$ and $(\bs\cdot\bx)$ in favour of  the conserved angular momentum $\bcL$ using equation (\ref{bs}) and by eliminating $(\bs\cdot\bp)$ in favour of the conserved scalar spin $s$ using equation (\ref{scalarspin}). We use the following relations
\begin{eqnarray}
\bx\times\bp\cdot\bs &=& \frac{r+a}{r-a}\lp \bx\times\bp\cdot\bcL-\lp\frac{r+a}{r}\rp^2\lp r^2\np^2-(\bx\cdot\bp)^2\rp\rp
\label{xps}
,\\[6pt]
\bs\cdot\bx&=&\bcL\cdot\bx,\\[6pt]
\bs\cdot\bp&=&s\np. \label{sp}
\end{eqnarray}
We are thus left with six equations for six unknown functions of $t$, which will be spelled out later, (\ref{pert_dx},\,\ref{pert_dp}).

We also have a formula for the norm of $\bp$ from the conserved quantities (\ref{cE}) and (\ref{cL}),
\begin{equation}
\np = \frac{r-a}{r+a} \; \frac{(r+a)^2 \cE - \frac{2 a r}{(r^2-a^2) \np} \lp \bx\times\bp\cdot\bcL\rp}{(r-a)^2 \; - \; \frac{2a}{r \np^2} \Vert\bx\times\bp\Vert^2}
\label{redshift}
\end{equation}
and 
\begin{align}
\frac{d}{dt}\lp \frac{\bp}{\np} \rp
= \, \frac{2 a}{(r+a)^4 D} \Bigg\{ &\Big(3r(\bs\cdot\bx) (\bx\cdot\bp)-(2r-a) (\bs\cdot\bp)r^2\Big)  \left(\bx-\frac{(\bx\cdot\bp)\, \bp}{\np^2}\right) \nonumber \\ 
& \qquad +3(r-a) \frac{(\bx\cdot\bp)}{\np^2} \lb\bx\times\bp\cdot\bs\rb \, \bx\times\bp \Bigg\}.
\end{align}
Noticing that this last equation and the three equations for position only depend on $\bp/\np$ our system effectively reduces to five equations.

The results above can already by found in Saturnini's thesis \cite{Sat76} of 1976.

 
\subsubsection{Radial case}
The first observation is that in the radial case, \textit{i.e.} with an initial momentum parallel to the initial position, the equations of motion (\ref{dx_spin}-\ref{ds_spin}) reduce to those of the radial geodesics,
\begin{align}
\frac{d\bx}{dt} & = \frac{r^2(r-a)}{(r+a)^3} \frac{\bp}{\np}, \label{radial_dx} \\
\frac{d\bp}{dt} & = - \frac{2\,a\, r^2}{(r+a)^4} \, \bp, \label{radial_dp} \\
\frac{d\bs^\perp}{dt} & = - \frac{2\,a\, r^2}{(r+a)^4} \, \bst \label{radial_ds}.
\end{align}

While the differential equation (\ref{radial_dp}) displays the well known redshift effect of light, it is striking that we have the same expression (\ref{radial_ds}) for the evolution of the transverse spin. This can be expected when looking at the Souriau--Saturnini equations \eqref{ss_xdot}--\eqref{ss_sdot} and noticing that the redshift terms in (\ref{radial_dp}) and (\ref{radial_ds}) come from the covariant derivative. Indeed, when the photon is following the geodesic trajectory, the Souriau-Saturnini equations reduce to the geodesic equations \textit{i.e.} $\dX = P$ and $\dP = \dS = 0$, meaning that both $P$ and $S$ are parallel transported.

We also take the opportunity to note that equation (\ref{cE}) tells us that the conserved energy $\cE$ is modified by the transverse spin in general, but not in the radial case.
 
\subsection{Null geodesics \& spinless gravitational lensing}
\label{s:3}

In this section, we first show that we recover the known spinless massless case, albeit in a slightly different form than the usual geodesic equations, by putting $\bs = 0$ in our equations. Then we show an unusual derivation of the well known deviation angle $\Delta \varphi$.

\subsubsection{Some preliminaries}

If we put $\bs=0$ in (\ref{cE}) and (\ref{cL}), the Noether quantities are of the form
\begin{equation}
\cE=
\frac{r-a}{r+a}\,\np\,,
\qquad
\&
\qquad  
\bcL=\lp\frac{r+a}{r}\rp^2 \bx\times\bp\,.
\label{cEcLs=0}
\end{equation}
From $dX/d\tau=P$, and equation (\ref{Pbis}), we find
\begin{equation}
\bp=\left(\frac{r+a}{r}\right)^2\frac{d\bx}{d\tau}\,.
\label{bps=0}
\end{equation}
For null geodesics, $P^2=0$, we have
\begin{equation}
\left\Vert\frac{d\bx}{d\tau}\right\Vert=\frac{\cE\,r^2}{r^2-a^2}\,.
\label{normdxdtaus=0}
\end{equation}
Taking advantage of the conservation of total angular momentum, $\bcL$, we compute $\bx\times\bcL$ and end up with
\begin{equation}
\frac{d\bx}{d\tau}=-\frac{r^2}{(r+a)^4}\,\bx\times\bcL+\lambda\,\bx
\label{dxdtau}
\end{equation}
where the function $\lambda$ satisfies (using (\ref{bps=0}),  (\ref{dxdtau}) and (\ref{normdxdtaus=0}))
\begin{equation}
\lambda =\frac{\bx\cdot\bp}{(r+a)^2}\qquad\text{and}\qquad
\lambda^2=\frac{\cE^2\,r^2}{(r^2-a^2)^2}-\frac{\cL^2\,r^4}{(r+a)^8}
\end{equation}
with $\cL=\Vert\bcL\Vert$. We note that $\lambda^2\ge0$ implies a condition on $\cE,\cL$ and $r$. By taking the scalar product on both sides of equation (\ref{dxdtau}) with $\bx$, we obtain the simple expression
\begin{equation}
\frac{dr}{d\tau}=\lambda\,r\,.
\label{drdtau}
\end{equation}
We record for further use that
\begin{equation}
\frac{d\lambda}{d\tau}=-r^2\left[\frac{\cE^2(r^2+a^2)}{(r^2-a^2)^3}-\frac{2\cL^2r^2(r-a)}{(r+a)^9}\right].
\label{dlambdadtau}
\end{equation}
Let us stress that the latter equations lead precisely to the equations of null geodesics given in terms of the Christoffel symbols (\ref{Gamma}). Here we used, instead, the conservation laws, including a number of computational tricks, to obtain the velocity (\ref{dxdtau}). Note that the time-component of the geodesic equation yields (up to a global sign):
\begin{equation}
\frac{dt}{d\tau}=\cE\left(\frac{r+a}{r-a}\right)^2
\label{dtdtaus=0}
\end{equation}
which is clearly non-vanishing. Comparison with the general equation (\ref{dtdtau}), which is ill-defined in the limit $s\to0$, shows a striking similarity with equation (\ref{dtdtaus=0}), namely the latter  is identical with the the former provided we ignore the spin-dependent factor on the RHS.

To make the link with equations (\ref{dx_spin}) and (\ref{dp_spin}), let us write down the equations of motion of the null geodesic in the form:
\begin{align}
\,\frac{d\bx}{dt}\,
& =\,\frac{r^2(r-a)}{(r+a)^3\np}\,\bp\,,\label{dxdtgeo}
\\[3mm]
\,\frac{d\bp}{dt}\,
& =\,\frac{2a}{(r+a)^4\np}
\Big\{(r-a)(\bp\cdot\bx)\,\bp \, -\,(2r-a)\,\np^2\,\bx  
\Big\}.\label{dpdtgeo}
\end{align}

\subsubsection{Lensing in weak fields}
\label{sec_deltaphi}
We restrict our analysis to geodesics remaining in regions of space where the gravitational field is weak, \textit{i.e.} where all distances $r(t)$ remain much larger than the Schwarzschild radius $a$,  
\begin{equation} \alpha (t)\dpp=\,\frac{a}{r(t)}\,  \ll\,1,\end{equation}
and linearize with respect to $\alpha $. We take our initial conditions at $\tau=t=0$:
\begin{equation} \bx_0=
\begin{pmatrix}
-x_0\\b\\0
\end{pmatrix} \qquad\text{ and}\qquad
\bp_0=
\begin{pmatrix}
 p_0\\0\\0
\end{pmatrix}.
\end{equation}
To alleviate notations we will write from now on $\bx=(x_1,\,x_2,\,x_3)$ with lower indices.
Following tradition we consider the photon in the $x_1$-$x_2$ plane with energy $p_0>0$
 coming in from the left, $x_0>0$, with positive impact parameter $b$. We suppose $a\ll b$. Then we have to first order in $\alpha $:
 \begin{equation} \cE\sim(1-2\alpha _0) p_0,\qquad \,\frac{\bcL}{\cE}\, \sim\,-(1+4\alpha _0)\,b\begin{pmatrix}
0\\0\\1
\end{pmatrix} \ \text{ and}\ \lambda \sim \pm\,\frac{\cE}{r}\,  \sqrt{1-\lp\frac{\cL}{\cE}\rp^2\frac{1-8\alpha }{r^2}} . 
\end{equation}
The equations of motion (\ref{dxdtau}) become:
\begin{align}
\dot x_1&\sim-\cL(1-4\alpha )\,\frac{x_2}{r^2}\, +\lambda x_1,\\
\dot x_2&\sim+\cL(1-4\alpha )\,\frac{x_1}{r^2}\, +\lambda x_2,\\
\dot x_3&=0,
\end{align} 
implying
\begin{equation}
\dot\varphi=\,\frac{x_1\dot x_2-x_2\dot x_1}{r^2}\,  \sim -\,\frac{1-4\alpha }{r^2}\,  \cL
\end{equation}
Equation (\ref{drdtau}), $\dot r =\lambda r$, tells us that the distance of closest approach $r_p$ (`perihelion')  is reached when $\lambda $ vanishes. Therefore 
\begin{equation}\cL/\cE\sim b\sim (1+4\,\alpha _p)\,r_p,\end{equation}
and $r_p\sim b-4\,a+4\,ab/r_0$.

Our aim is to compute the scattering angle $\Delta \varphi $ for $x_0\rightarrow\infty$. As we have set the cosmological constant to zero, spacetime is flat far away from the mass and there coordinate and physical angles coincide. Denoting by $\varphi _p$ the angle of closest approach, we have $\Delta \varphi =\pi -2\varphi _p$. We can compute $\varphi _p$ by integrating
\begin{align}
 \,\frac{d\varphi }{dr}\,& =\,\frac{\dot \varphi }{\dot r}\,=\,\frac{\dot\varphi }{\lambda r}\,
\sim\mp
\,\frac{1-4\alpha }{r^2}\,\frac{\cL}{\cE}\,\lb1-\lp\frac{\cL}{\cE}\rp^2\frac{1-8\alpha }{r^2}\rb^{-1/2}\nonumber \\[1mm]
&\sim\,\mp \,\frac{1+4\alpha _p}{r_p}\,\frac{1-4\alpha }{r/r_p}\lb(r/r_p)^2-1\rb^{-1/2}\lp1-4\,\frac{\alpha -\alpha _p}{(r/r_p)^2-1}\rp \label{dphidr}
\end{align} 
between $r_0=\infty$ and $r_p$.   
In this interval both $r$ and $\varphi $ decrease and we must choose the positive signs in equation (\ref{dphidr}). Our initial angle is $\varphi _0=\pi $ and we obtain with $u\dpp=r/r_p$,
\begin{equation}
\pi -\varphi _p\sim (1+4\alpha _p)\int_{1}^{\infty}
\frac{1-4\alpha_p/u }{u}\lb u^2-1\rb^{-1/2}\lp1-4\alpha_p\,\frac{1/u -1}{u^2-1}\rp du\,=\,
\,\frac{\pi }{2}\, +4\,\frac{a}{r_p}\,. 
\end{equation}
Note the integrable singularity at the perihelion, $u=1$.
Finally, in linear approximation, the scattering angle takes its famous value: $\Delta \varphi \sim 4\,GM/r_p$.

We thus recover the known geodesic equations in the Schwarzschild metric, and the usual deflecting angle $\Delta \varphi$, from the Souriau-Saturnini formalism and putting $\bs = 0$. The resulting equations of motions (\ref{dxdtgeo}--\ref{dpdtgeo}) are first order equations, but are strictly equivalent to the second order geodesic equations. Now, the next step is to consider the spinning case, $\bs \neq 0$, hence considering the full equations of motions (\ref{dx_spin}--\ref{ds_spin}). This is done in the following sections.

\subsection{Numerical solutions}
\label{s:4}

Since solving the system of equations (\ref{dx_spin},\,\ref{dp_spin}) is not straightforward, we will use the help of numerical integration to propagate specific initial conditions. These numerical solutions will guide us towards perturbative ones.

The numerical integration meets the usual problem of
 accuracy errors when computing the difference of two almost identical numbers. It becomes relevant here because the present system of equations involves such computations, especially when conserved quantities are involved, \textit{e.g.} (\ref{xps}). This is why it is better to numerically solve all of the 9 differential equations (\ref{dx_spin}-\ref{ds_spin}), including those of the spin.

Even with such measures, integrating these equations over a long time can be tricky with Mathematica. The step algorithm seems overly cautious and is eager to stop the integration process due to stiffness problems, even though all quantities involved are well defined, finite, and smoothly evolving. We need to select the right precision parameters to keep the step algorithm from stopping the integration. Yet, this does not create instabilities in the trajectory of the simulation and we obtain very precise results.

It is convenient to take the initial conditions not at infinity but at perihelion $r_0=r_p$ of the geodesic trajectory of the spinless photon around the star located at the origin:
\begin{equation}
\bx_0 = \lp
\begin{array}{c}
r_0\\
0\\
0
\end{array}\rp,\qquad
\bp_0 = \lp
\begin{array}{c}
0\\
p_0\\
0
\end{array}\rp,\qquad
\bs_0 = \lp
\begin{array}{c}
0\\
s\\
s^\perp_0
\end{array}\rp.
\label{ic}
\end{equation}
Note that the first component of the initial transverse spin $\bs^\perp_0$ vanishes, because at perihelion ${d\bx}/{dt}|_0\cdot\bx_0 = 0$.

We use SI units here. The photon starts with a wavelength of $\lambda_0 = 600\,$nm and a helicity of $\chi = +1$, the star has a Schwarzschild radius of $a = 3\cdot10^3\,$m, and the initial distance from the center of the star to the perihelion is~$r_0 = 3\cdot10^5\,$m. The numerical integration runs from $0$ to $0.1\,$s. While we have $s = \hbar$ in the initial conditions (\ref{ic}), we will put $s_0^\perp = 0$ for the time being, due to trajectory instabilities when $s_0^\perp$ is close to $\hbar$. We will come back to the transverse spin in the perturbative analysis in the next section.

\begin{figure}
\centering
\includegraphics[scale=0.9]{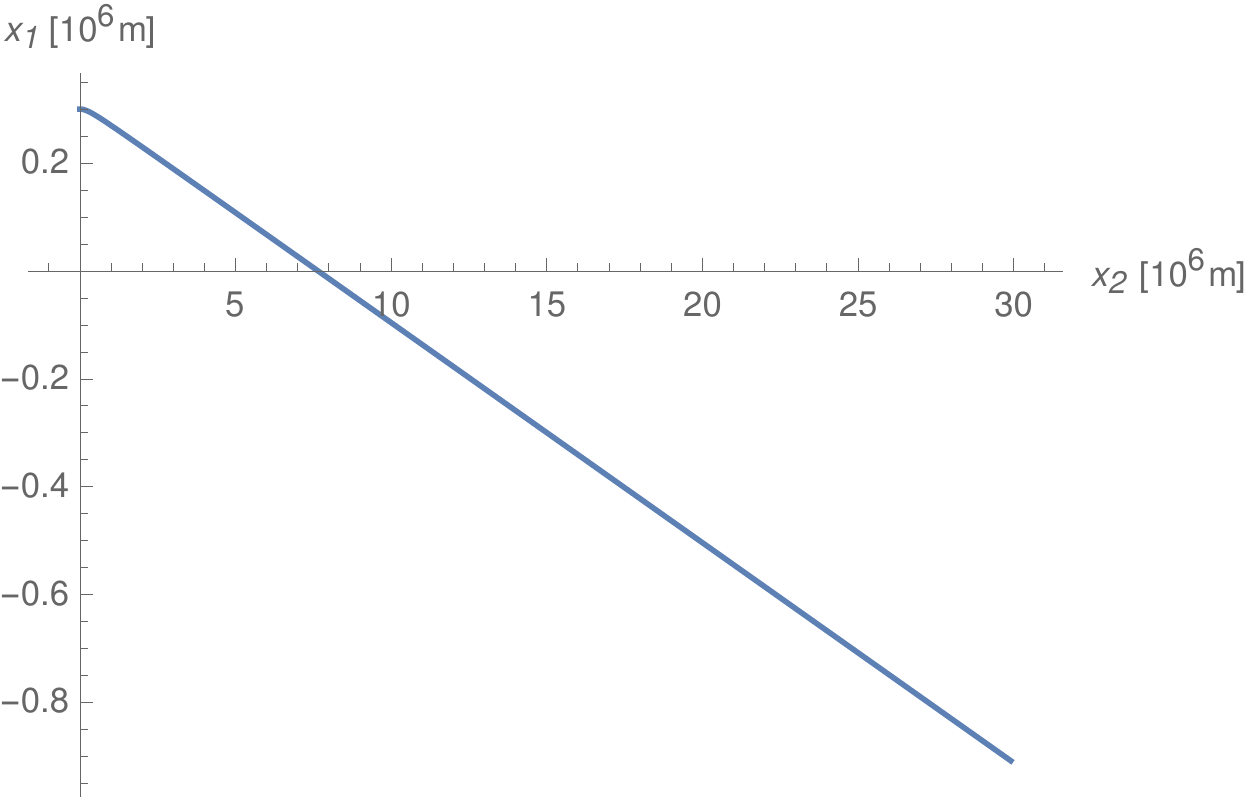}
\caption{Trajectory of the spinning photon in the geodesic plane. Visually, this trajectory is the same as that of the spinless photon.}
\label{fig_geod}
\end{figure}

\begin{figure}
\centering
\includegraphics[scale=0.9]{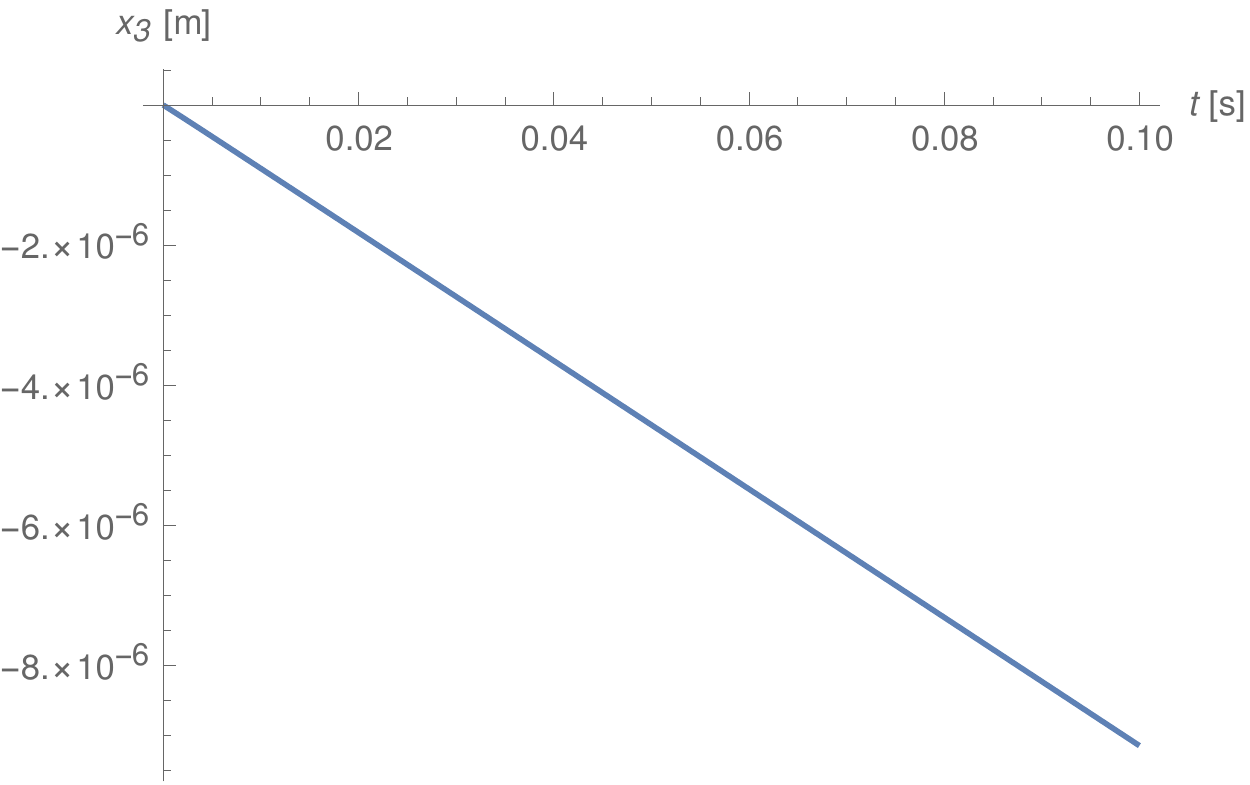}
\caption{Component $x_3$ of the trajectory of the photon as a function of time. The spinning photon leaves the geodesic plane, albeit with a very small angle.}
\label{fig_x3t}
\end{figure}

\begin{figure}
\centering
\includegraphics[scale=0.9]{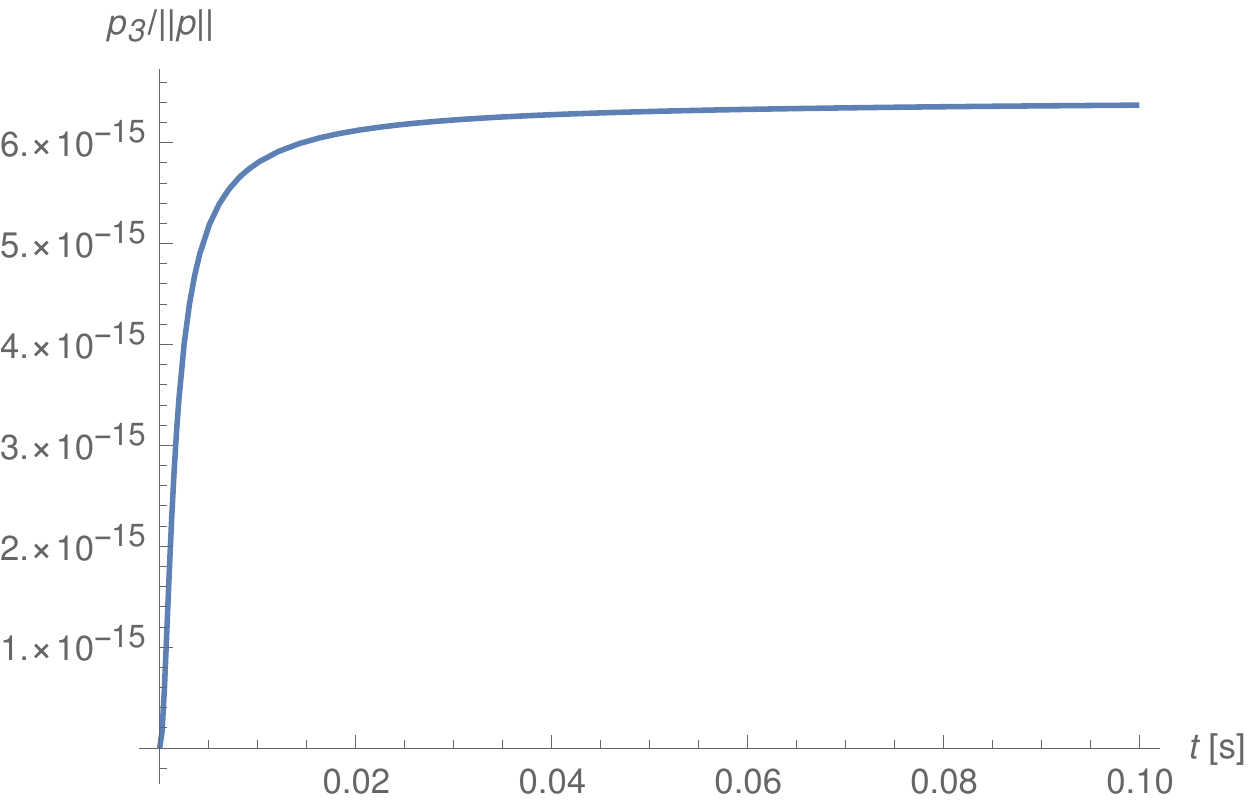}
\caption{Component $p_3$ of the momentum of the photon, normalized to the norm of the momentum, as a function of time. Just like with $x_3$, while the geodesic momentum is contained in the plane $(p_1, p_2)$, the momentum of the spinning photon has a component perpendicular to that plane. Notice that the sign of $p_3$ is opposite to that of $x_3$.}
\label{fig_p3t}
\end{figure}

Figure \ref{fig_geod} shows the trajectory of the spinning photon in the geodesic plane. This trajectory is almost identical to the null geodesic one. Indeed, the difference between the coordinates $x_1$ and $x_2$ of spinning and spinless photons is of the order of the nanometer at the end of the numerical integration. The main differences are the transverse components $x_3$ of the trajectory, and $p_3$ of the momentum, pictured in Figures \ref{fig_x3t} and \ref{fig_p3t} respectively. While the geodesic trajectory is contained within the plane $(x_1, x_2)$, the equations of motion (\ref{dx_spin}-\ref{ds_spin}) imply non-vanishing transverse components $x_3$ and $p_3$.

The angle $\beta$ of the trajectory going out of the plane is small, but constant. As shown here in figure \ref{fig_x3t}, it is about $\beta = - 6.3 \cdot 10^{-8}\, \arcsec$. The sign of the angle $\beta$ depends directly on the helicity $\chi$. Indeed when changing the helicity from $+1$ to $-1$, the amplitude of the angle remains the same, but the sign switches. We see from numerical integrations that the trajectories of two different helicity photons are  symmetric with respect to the null geodesic. The transverse momentum $p_3$ also shows the same behavior under helicity changes and its sign is again opposite to that of $x_3$.

In the next section, we will confirm and explain these results with a perturbative approach.

\subsection{Perturbative solutions}
\label{s:5}

We wish to compare the behavior of our system (\ref{dx_spin},\,\ref{dp_spin}) describing the trajectories of photons with their due spin to the behavior of null geodesics.

Now, define two \emph{constant} small parameters,
\begin{equation}
\label{params}
\alpha = \frac{a}{r_0} \qquad \& \qquad \epsilon = \frac{\hbar}{r_0\, p_0},
\end{equation}
where, for the sun, $\alpha$ is typically of the order of $10^{-6}$ and $\epsilon$ of the order of $10^{-16}$ for photons in the visible spectrum. A small $\epsilon$ corresponds to photons having a  wavelength much smaller than its distance to the star, which is a sensible hypothesis. Due to the particularities of this system of equations, namely $D$ (\ref{def_d}) being of order $\epsilon$, we must consider second order terms in $\epsilon$ to obtain the first order equations.  In $\alpha$, linear terms will be sufficient.

Let us redefine the spin by setting
\begin{equation}
s =\dpp \chi \hbar\qquad\&\qquad s^\perp_0=\dpp w\hbar,
\end{equation}
where $\chi = \pm 1$ is the helicity of the photon and $w$ is finite and dimensionless. 
We easily obtain the conserved quantities (\ref{cE}) and (\ref{cL}) from  the initial conditions (\ref{ic}),
\begin{equation}
\cE \sim (1-2 \alpha) \, p_0 \qquad \& \qquad \bcL \sim r_0\,p_0 \lp
\begin{array}{c}
0\\
(1-2\alpha) \, \chi\,\epsilon\\
(1+2\alpha)+(1-2 \alpha )\,w\,\epsilon
\end{array}\rp
\label{pert_qtt}
\end{equation}
We define the normalized quantities,
\begin{equation}
  \xl = \frac{\bx \cdot \bcL}{r\, \chi s}
  \qquad \& \qquad
  \xps = \frac{\bx \times \bp \cdot \bs}{r\, p\,\chi s}.
\end{equation}
We can then write the equations (\ref{dx_spin}-\ref{dp_spin}) as
\begin{align}
\frac{d\bx}{dt}\,& =\,
\frac{r^2(r-a)}{(r+a)^3\lp r\,\np-3(\bx\cdot\bp)\,\xl\rp} \lb r\, \bp-3\np \, \xl \,\bx+3\,\xps\, \bx\times\bp\rb \label{pert_dx},\\
\frac{d\bp}{dt}\,& =\,
\frac{2\,a}{(r+a)^4\lp r\,\np-3(\bx\cdot\bp)\,\xl\rp} \Big[ r(r-a)\lp(\bx\cdot\bp)-3 \,\frac{r^3}{(r+a)^3}\,  s \,\chi \, \xl \, \xps \rp\bp \nonumber \\
& \hspace{6.5cm} -r \np\lp(2r-a) \np-3(\bx\cdot\bp) \, \xl\rp\,\bx \nonumber \\
& \hspace{6.5cm} + 3 (r-a) \, \xps \, (\bx\cdot\bp) \,\bx\times\bp \, \Big]\label{pert_dp}.
\end{align}
Let us momentarily forget the physical aspect of this system and set $\epsilon = 0$ in (\ref{pert_qtt}). Then with the initial conditions (\ref{ic}) the differential equations (\ref{pert_dx}) and (\ref{pert_dp})  reduce to those of the null geodesics (\ref{dxdtgeo}) and (\ref{dpdtgeo}). Indeed, from (\ref{xps}), we have initially $\xps|_0 = 0$ and $\xl|_0 = 0$, reducing the initial system to the geodesic one. If we are on a geodesic trajectory, which is in the plane spanned by $\bx_0$ and $\bp_0$, then  $\xl=0$ and $\xps = 0$ continue to vanish due to geodesic conservation of angular momentum and the photon continues on the geodesic trajectory.

This heuristic argument and our numerical results in the last section motivate the ansatz
\begin{equation}
\bx \sim \lp\begin{array}{ccccc}
x_1 & + & \epsilon\, y_1 & + & \epsilon^2\, z_1\\
x_2 & + & \epsilon\, y_2 & + & \epsilon^2\, z_2\\
& & \epsilon\, y_3 & + & \epsilon^2\, z_3
\end{array}\rp
\quad \& \quad
\bp \sim \lp\begin{array}{ccccc}
p_1 & + & \epsilon\, q_1 & + & \epsilon^2\, u_1\\
p_2 & + & \epsilon\, q_2 & + & \epsilon^2\, u_2\\
& & \epsilon\, q_3 & + & \epsilon^2\, u_3
\end{array}\rp\,,
\label{ansatz}
\end{equation}
where $x_1,\, x_2,\, p_1,\, p_2$ solve the geodesic equations. Define $r_g = \sqrt{x_1^2+x_2^2}$ and similarly for $p_g$.
To leading order, we have
\begin{align}
\xl & = (1-2 \alpha) \, \frac{x_2}{r_g}+(1+2\alpha) \, \chi \frac{y_3}{r_g} + \cO(\epsilon), \label{pert_lx}\\
\xps & = \,\chi \,\frac{r_0p_0}{r_gp_g}\, \Bigg[
w+2\,\frac{a}{r_g}\,\frac{x_1 y_1 + x_2 y_2}{r_g^2}\, -\lp1 + 2\alpha + 2\frac{a}{r_g}\rp\,\frac{y_1p_2-y_2p_1+ x_1q_2-x_2q_1}{r_0p_0}\Bigg]
+ \cO(\epsilon). \label{pert_xps}
\end{align}

In order to recover the geodesics in the limit $\epsilon \rightarrow 0$, we thus need these two leading terms to be zero implying the initial transverse spin to vanish and some conditions on first order terms in $\epsilon$ that are valid at least to first order in $\alpha $:
\begin{align}
w&\sim 0,\label{w}\\
y_3 & \sim -\chi \,(1-4\alpha) x_2, \label{cond_y3}\\
x_1 \, y_1  + x_2 \, y_2& \sim 0,\\
y_1p_2-y_2p_1+ x_1q_2-x_2q_1 & \sim 0.
\label{constraints}
\end{align}

Plugging the ansatz (\ref{ansatz}) into the six scalar equations (\ref{pert_dx}) and (\ref{pert_dp}) we obtain twelve equations: six in $\epsilon^0$ and six in $\epsilon^1$. The six equations in $\epsilon^0$ are equivalent to the four equations (\ref{w}-\ref{constraints}). The six equations in $\epsilon^1$ yield:
\begin{equation}
y_1\sim y_2\sim q_1\sim q_2\sim { q_3\sim  \cO(\alpha )\qquad\qquad\text{ and}\qquad\qquad z_3\sim \cO(\alpha )}.
\end{equation}
At this point, we may even obtain the terms of order $\alpha \epsilon$ giving us constraints on $z_1$ and on the initial transverse spin and we end up with  $y_1 \sim y_2\sim q_1\sim q_2\sim \cO(\alpha^2)$ and 
\begin{align}
\epsilon \, y_3 & = - \epsilon \, \chi\lp(1-4\alpha)\,t-4 \, \alpha \, r_0 \, \ln\frac{t+\sqrt{r_0^2+t^2}}{r_0}\rp, \\
\epsilon \, q_3 & = 2 \, \epsilon \, \alpha \, \chi \, p_0 \left(1-\frac{r_0}{\sqrt{r_0^2+t^2}}\right),
\end{align}
and our perturbative solution reads
\begin{align}
\bx & = \lp\begin{array}{c}
r_0 + 4 \, \alpha \, r_0\, \left(1-\frac{\sqrt{r_0^2+t^2}}{r_0}\right) \\[4pt]
t - 4 \, \alpha \, r_0 \ln\frac{t+\sqrt{r_0^2+t^2}}{r_0} \\[4pt]
- \epsilon \, \chi\lp(1-4\alpha)\,t-4 \, \alpha \, r_0 \, \ln\frac{t+\sqrt{r_0^2+t^2}}{r_0}\rp
\end{array}\rp + \cO(\epsilon^2, \alpha^2), \\[2mm]
\bp & = \lp\begin{array}{c}
- 4 \, \alpha \, p_0\frac{t}{\sqrt{r_0^2+t^2}} \\[4pt]
p_0 - 2 \, \alpha \, p_0 \lp 1 - \frac{r_0}{\sqrt{r_0^2 + t^2}}\rp \\[4pt]
2 \, \epsilon \, \alpha \, \chi \, p_0 \left(1-\frac{r_0}{\sqrt{r_0^2+t^2}}\right)
\end{array}\rp+ \cO(\epsilon^2, \alpha^2).
\label{pert_sol}
\end{align}
Finally, using (\ref{bs}) and $\bs = \frac{\bp}{\np} s + \bs^\perp$, we obtain the perturbative solution for the transverse spin,
\begin{equation}
\bs^\perp = \chi \, \hbar \lp\begin{array}{c}
- \frac{t}{r_0}(1-4 \alpha)+4 \alpha \ln \frac{t+\sqrt{r_0^2+t^2}}{r_0} \\[4pt]
- \frac{4 \alpha t^2}{r_0 \sqrt{r_0^2+t^2}} \\[4pt]
0
\end{array}\rp + \cO(\epsilon^2, \alpha^2).
\label{pertspin}
\end{equation}

The most striking effect of the spin on the trajectory of the photon is that it leaves the geodesic plane, but its projection on this plane coincides up to order $\epsilon \alpha$ with the geodesic. The angle $\beta $ between the  trajectory and the geodesic plane is given from $\beta \sim d(\epsilon y_3)/dx_2$ at infinity, which is immediate with the help of (\ref{cond_y3}),
\begin{equation}
\beta \sim - (1-4 \alpha) \frac{\chi \, \lambda_0}{2 \pi \, r_0}
\label{angle_beta}
\end{equation}
with the definition (\ref{params}) for $\epsilon$ and where $\lambda_0$ is the wavelength of the photon at perihelion. Notice that this angle depends both on the helicity of the photon $\chi = \pm 1$ and on its wavelength. Photons of the two different helicities  follow  symmetric trajectories with respect to the geodesic and the dependence on $\lambda_0$  produces a rainbow effect. In the case of the sun, with $r_0$ its radius, this means that two photons starting at the perihelion with opposite helicity will have an offset given by $2 \beta = 5.7 \cdot 10^{-11}\, \arcsec$. If these two photons then travel to the Earth, the offset between them would be of the order of $41 \mathrm{\mu m}$ in perfect conditions. The angle $\beta$ has the curious property of not depending, at zeroth order in~$\alpha$, on the mass of the star. This seems to imply that this angle becomes a non zero constant as the mass of the star becomes arbitrarily small.  Let us note though, that the limit $\alpha \rightarrow 0$ is ill defined in the equations of motion and therefore in the perturbative solutions. Indeed, the first of the Souriau-Saturnini equations \eqref{ss_xdot} is independent of $a$ because both $R(S)(S)$ and $S R(S) P$ are proportional to $a$. The introduction of a cosmological constant will regularize this singularity, even at small scales, as we will see in a later section.

Also, we find no correction of order $\epsilon \, \alpha$ to the usual deviation angle $\Delta \varphi$ in the plane, computed in section \ref{sec_deltaphi}.

Note that the transverse component of the momentum quickly reaches its maximum at a distance of a few $r_0$, which is $\epsilon \, {q_3}_{max} = 2 \, \epsilon \, \alpha \, \chi \, p_0$. Since the angle $\beta$ comes from a spin-orbit-like effect of the star on the trajectory, we would expect it to only act close to the star. To avoid this problem, we define $\gamma $ to be the angle between the geodesic plane and the momentum carried by the spinning photon. We have:
\begin{equation}
\gamma \sim \chi \frac{a \, \lambda_0}{\pi \, r_0^2}.
\label{angle_gamma}
\end{equation}
This angle does depend on the mass of the star and is even smaller than $\beta$. For the sun we have $2 \gamma = 4.9 \cdot 10^{-16}\, \arcsec$.

Our perturbative results for $y_3$ and $q_3$  above match our numerical results with a relative error of about $10^{-9}$ and $10^{-4.5}$, respectively. The match is better for $y_3$ because it contains terms of order 1 and of order $\alpha $, while $q_3$ is of order $\alpha$.

\subsection{Remarks on pure Schwarzschild backgrounds}

For photons, quantum mechanics teaches us that the longitudinal component $s$ of the spin is $\pm \hbar$. This is in harmony with the conservation of $s$, which follows in general from the Souriau-Saturnini equations. Quantum mechanics also teaches us that the norm of the transverse spin $\nst$ is $ \hbar$. Two remarks arise from the present work. First, we saw in the radial case that the photon follows the null geodesic trajectory, and that the transverse spin undergoes the same evolution as the momentum: it is parallel transported. However, in our non-radial perturbative solution, equation (\ref{pertspin}), this norm vanishes at perihelion and then grows linearly with time~$t$ (to leading order). The linear growth implies that our perturbation theory breaks down for large times. This instability is absent in a generic Robertson-Walker metric where the norm of the transverse spin is proportional to the inverse Hubble parameter \cite{ChDTSRW}.

With its continuously varying transverse spin, the instability reminds us of the instability of the classical hydrogen atom and its continuously varying energy. Indeed, the equations we here use are purely classical. While the definition of the longitudinal spin comes from the co-adjoint representation of the Poincar\'e group \cite{Sou70}, what we call transverse spin here are the two additional degrees of freedom we obtain when considering dipole moments. It is not clear, from the geometrical derivation of these equations, if these two degrees of freedom are exactly akin to the transverse spin in quantum mechanics. A way to determine their exact meaning would be to derive the Souriau-Saturnini equations \eqref{ss_xdot}--\eqref{ss_sdot} from quantum mechanics, \`a la Eikonal.

Notice also that the out-of-plane momentum is in the opposite direction with respect to the offset. This means that the star is intrinsicly acting on the photon's position and momentum, \textit{i.e.} a spin-orbit effect. Yet, at large time $t$ in the perturbative solution, we see that the trajectory's offset keeps increasing linearly, while the momentum stays constant and in the opposite direction. We would expect, once we are sufficiently far away from the star, that the star loses grip on the photon. Since spacetime is flat far away, we  expect the photon's momentum to carry the trajectory, which is not what we see here. This is in line with the fact that we don't recover the equations of motion in flat spacetime in the limit $a \rightarrow 0$. 

For me, the most interesting features of birefringence in the Schwarzschild metric are the out-of-plane contributions to trajectory and momentum. First, we have the linearly growing offset -- given by an angle $2\beta $, equation (\ref{angle_beta}) -- between the trajectories of opposite polarisations. Then, the Souriau-Saturnini equations in the Schwarzschild metric become singular far away from the star, a singularity absent in the Kottler metric. Therefore we expect the offset induced by the angle $\gamma$ (\ref{angle_gamma}) to play a  more important role in observations. 
In any case, both angles, $\beta $ and $\gamma $ are wavelength dependent and the  offset  must feature a rainbow effect.

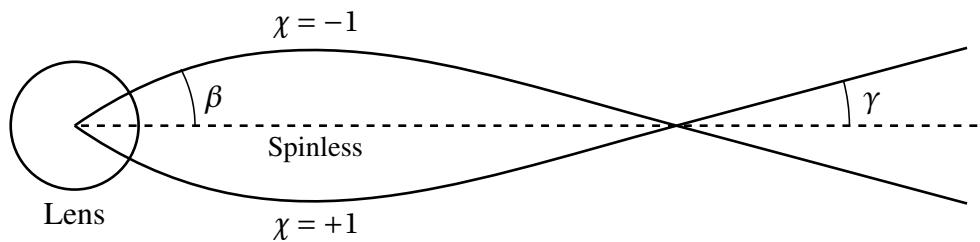
\begin{figure}[!ht]
\begin{tikzpicture}[line width=1pt,scale=1, every node/.style={transform shape}]
  \node [draw,circle,minimum size = 17mm, label=below:Lens] at (0,0) {};
  \draw [dashed] (12,0) coordinate (allright) -- node[pos=0.35,above,scale=1]{$\gamma$} (8,0) -- node[pos=0.8,below,scale=0.9]{Spinless} (2,0) -- node[pos=0.08,above,scale=1]{$\beta$} (0,0);
  \draw (8,0) -- +(-15:4);
  \draw (8,0) coordinate (center) -- +(15:4) coordinate (topleft);
  \draw (8,0) to[out=165,in=35] node[pos=0.58,above,scale=1]{$\chi = -1$} (0,0) coordinate (allleft);
  \draw (8,0) to[out=-165,in=-35] node[pos=0.58,below,scale=1]{$\chi = +1$} (0,0);
  \draw [line width=0.6pt] pic["",draw=black,-,angle radius=2.3cm] {angle=allright--center--topleft};
  \coordinate (abovecenter) at ($(0,0)+(28:2)$);
  \draw [line width=0.6pt] pic["",draw=black,-,angle radius=1.6cm] {angle=allright--allleft--abovecenter};
\end{tikzpicture}
\caption{Looks fishy}
\label{f:birefringence}
\end{figure} 

Let us see on the figure \ref{f:birefringence} how the trajectory would play out if we make the hypothesis that far away from the star the photon will start following its momentum. On this figure, we are looking at the system from ``above'', \ie we have in order, the star, then the photon with a starting point at the perihelion, then the observer, us. The usual null geodesic trajectory is the dashed line. We would expect the two trajectories of the $\chi = +1$ and $\chi = -1$ helicity photons to look like the plain lines. As explained in the previous sections, the star acts in a spin-orbit-like fashion on the photon when they are ``close'' to each other, by deviating the photon's trajectory in one way with an angle $\beta = -(1-4\alpha) \frac{\chi \lambda_0}{2\pi \, r_0}$, with $\alpha = a /r_0$, while changing the photon's momentum in the other direction. Thus, ``far'' from the star, where it loses its influence, one might want that the photon should start to follow its momentum, and curves back with an angle $\gamma = \chi \frac{\alpha \lambda_0}{\pi \, r_0}$. However, the distance at which this switch happens is not present in the theory with only the Schwarzschild metric, but introducing another length scale such as the cosmological constant may solve this.

\medskip

Let us add, finally, that gravitational birefringence has already been considered experimentally in 1974 \cite{Har74}, resulting in an upper bound for this effect in gravitational lensing of $2 \cdot 10^{-3}\, \arcsec$ for a wavelength of $3.9$~cm. For such a wavelength, we find $2 \beta = 4 \cdot 10^{-6} \, \arcsec$, and $2\gamma = 3\cdot 10^{-11}\, \arcsec$. The effects predicted here are thus weaker than the experimental precision at the time by a few orders of magnitude.

To date the highest attainable angular resolving powers in astronomy have been obtained by two mutually unrelated efforts \cite{HarwitPrivate}. The first is the GRAVITY interferometer at the European Southern Observatory, operating in the $2 \mu$m wavelength range and attaining angular resolving powers of $10 \cdot 10^{-6} \, \arcsec$ \cite{GRAVITY}. However with such wavelength, the angles predicted here are of the order of $10^{-10} \, \arcsec$ and $10^{-15} \, \arcsec$ for, respectively, $\beta$ and $\gamma$. The second is the globe-spanning consortium of telescopes going by the name of "The Event Horizon Telescope" \cite{EventHorizonTelescope}, which has attained an angular resolving power at $1.5$~mm wavelengths amounting to $20\cdot 10^{-6} \, \arcsec$. In this situation, the angles $\beta$ and $\gamma$ are, respectively, of the order of $10^{-7} \, \arcsec$ and $10^{-12} \, \arcsec$. While neither experiments have the necessary angular resolving power to observe any effect predicted in this section, the Event Horizon telescope is only 2 orders of magnitude away from the first angle $\beta$. 
 
\subsection{An attempt at improving the null infinity limit with the cosmological constant} 

Since our hypothesis to explain the discrepancy between the trajectory of the spinning photon and the direction of its momentum in the null infinity limit in the previous section is that the Souriau--Saturnini equations do not behave well in the Minkowski limit, we consider including the cosmological constant together with the Schwarzschild metric. Indeed, while the Souriau--Saturnini equations are ill-defined with the Minkowski metric, we have seen in section \ref{ss:desitter} that including the cosmological constant solves this ill-definition. We thus need here the co-called Kottler solution to Einstein's equations \cite{Kottler18}.

\subsubsection{Metric}
The Kottler metric, sometimes called the Schwarzschild-de Sitter metric, describes a spacetime containing a spherically symmetric mass and the cosmological constant. It can be written as
\begin{equation}
\label{g}
g = f(\rho,t) \, dt^2 - f(\rho,t)^{-1} \, d\rho^2 - \rho^2 \, d\Omega^2,
\end{equation}
with $f(\rho,t) = 1 - 2GM / \rho - \Lambda \, \rho^2 / 3$, and $2GM > \rho > \sqrt{3/\Lambda}$. We recover the usual Schwarzschild metric in the limit $\Lambda \rightarrow 0$.

As discussed in section \ref{ss:iso_coords}, to make computations easier, we need the metric written with isotropic coordinates. The form of the metric \eqref{g} is once again the one we have studied in \ref{ss:iso_coords}, with $\displaystyle C(\rho) = \sqrt{1 - 2GM / \rho - \Lambda \, \rho^2 / 3}$, and we must now solve the differential equation,
\begin{equation}
\label{decoords}
\frac{d\rho(r)}{dr} = \sqrt{1- \frac{4 a}{\rho(r)}-\frac{\Lambda}{3} \rho(r)^2} \; \frac{\rho(r)}{r},
\end{equation}
with $a = GM/2$ the Schwarzschild radius in isotropic coordinates. While we can solve this differential equation either in the pure Schwarzschild case ($\Lambda = 0$), see \eqref{schw_trsf}, or in the de Sitter case ($a = 0$), see section \ref{ss:iso_coords}, we do not know of an exact solution to the full equation \eqref{decoords}. Nevertheless, we can try to find a perturbative solution, based on two small quantities related to the Schwarzschild radius $a$ and the cosmological constant $\Lambda$. Define $\lambda := r^2 \Lambda$ and $\displaystyle \alpha := \frac{a}{r}$. Now, consider a solution of the differential equation \eqref{decoords} as an expansion in $\lambda$. We find,
\begin{equation}
\rho(r) = r \left( \left(1+\frac{a}{r}\right)^2 - \frac{\lambda}{12} \left(1+\frac{16 \, a}{r} + \cO\left(\alpha^2\right)\right) + \cO(\lambda^2) \right).
\end{equation}

Note that this is only valid for $\lambda = r^2 \Lambda < 1$. While this means we will not be able to properly study the equations at infinity, we can still study situations where $a \ll r \ll \sqrt{\Lambda^{-1}}$, which is a fairly large range for reasonable physical situations.

The leading term, without $\lambda$, is exact in $a/r$. note that we recover the Schwarzschild coordinate transformation law \eqref{schw_trsf} in the limit $\lambda \rightarrow 0$. The first term in $\lambda$ contains corrections up to first order in $a/r$. The Kottler metric in isotropic coordinates is then given by,
\begin{equation}
g = A^2 dt^2 - B^2 \Vert d\bx \Vert^2,
\end{equation}
 with the two functions $A$ and $B$,
\begin{align}
A & \vcentcolon=\,\frac{r-a}{r+a}-\frac{\lambda}{6}\left(1+ \frac{7\, a}{r}+\cO\left(\alpha^2\right)\right)+\cO(\lambda^2)\,, \\
B & \vcentcolon=\left(\frac{r+a}{r}\right)^2-\frac{\lambda}{12}\left(1+ \frac{16 \, a}{r}+\cO\left(\alpha^2\right)\right)+\cO(\lambda^2).
\end{align}

The Christoffel symbols are then computed to be,
\begin{align}
{\Gamma^j}_{ii}&=-{\Gamma^i}_{ji}=-{\Gamma^j}_{jj}=\frac{2a\, x^j}{r^2(r+a)} + \frac{\Lambda \, x^j}{6}\left(1+\frac{7a}{r} + \cO(\alpha^2) \right) + \cO(\lambda^2), \\
{\Gamma^j}_{44}&=\frac{2ar^3(r-a)\, x^j}{(r+a)^7} - \frac{\Lambda \, x^j}{3} \left(1-\frac{5 a}{2 r} + \cO(\alpha^2) \right) + \cO(\lambda^2), \\
{\Gamma^4}_{4j}&=\frac{2a\, x^j}{r\,(r+a)(r-a)} - \frac{\Lambda \, x^j}{3}\left(1+\frac{9 a}{2r} + \cO(\alpha^2) \right) + \cO(\lambda^2),
\label{Gammak}
\end{align}
for all $i\not=j=1,2,3$, no summation over repeated indices. 

For the Riemann tensor ${R^\mu}_{\nu\alpha\beta}=\partial_{\alpha}{\Gamma^\mu}_{\beta\nu}-\partial_{\beta}{\Gamma^\mu}_{\alpha\nu}+\cdots$ with $i,j$ and $k$ all different, we have, with $\sim$ denoting the same approximation scheme as above, \ie up to order $\lambda$ with a correction in $\alpha$,
\begin{align}
{R^i}_{jij}& \sim \,\frac{2a\,[2(x^k)^2-(x^i)^2-(x^j)^2]}{r^3(r+a)^2}+\frac{\Lambda}{3}\left(1+\frac{a[10(x^k)^2+7(x^i)^2+7(x^j)^2]}{2 r^3}\right)\,,\\
{R^4}_{i4i}& \sim \,\frac{2a\,[2(x^i)^2-(x^j)^2-(x^k)^2]}{r^3(r+a)^2}+\frac{\Lambda}{3}\left(1+\frac{a[10(x^i)^2+7(x^j)^2+7(x^k)^2]}{2 r^3}\right)\,,\\
{R^j}_{iki}& \sim -\,\frac{6a\,x^jx^k}{r^3(r+a)^2}-\frac{a \, \Lambda\,  x^j x^k}{2r^3}\,,\\
{R^4}_{i4j}& \sim \ \,\frac{6a\,x^ix^j}{r^3(r+a)^2}+\frac{a \, \Lambda\,  x^j x^k}{2r^3}\,.
\end{align}

\subsubsection{Momentum and spin}

Now that we have the metric and the derived quantities, we can define the 4-momentum of the photon. It is,
\begin{equation}
P=(P^\mu)\sim\left(
\begin{array}{c}
\displaystyle
\left(\frac{r^2}{(r+a)^2}+\frac{r^2\Lambda}{12}+\Lambda \, a \, r\right) \bp\\[10pt]
\displaystyle
\left(\frac{r+a}{r-a}+\frac{r^2 \Lambda}{6}+\frac{11 \Lambda \, a \, r}{6}\right) \np
\end{array}\right),
\label{Pkbis}
\end{equation}
with $\bp\in\bbR^3\setminus\{0\}$, the spatial linear momentum, and $\np\vcentcolon=\sqrt{\bp\cdot\bp}$. The 4-momentum is light-like, as usual, $P^2\sim 0$.

Accordingly the spin tensor is defined by,
\begin{equation}
S=({S^\mu }_\nu)\sim\left(
\begin{array}{cc}
j(\bs)&\displaystyle
-\frac{(\bs\times\bp)}{\np}\left(\frac{r^2(r-a)}{(r+a)^3}+\frac{r^2\Lambda}{12}\right)\\[6pt]
\displaystyle
-\frac{(\bs\times\bp)^T}{\np}\left(\frac{(r+a)^3}{r^2(r-a)}+\frac{\Lambda \,r(r+8a)}{12}\right)&0
\end{array}\right)
\label{Skbis}
\end{equation}
with the spin 
vector $\bs\in\bbR^3\setminus\{0\}$. In addition, we have the usual constraints, namely the Tulczyjew SSC $SP\sim 0$, and $-\half\Tr(S^2)\sim s^2$, with the conserved scalar spin $s\sim\frac{\bs\cdot\bp}{\np}$.

\subsubsection{Equations of motion}
Now that we have defined all the usual quantities appearing in the Souriau--Saturnini equations, let us compute the equation of motion $d\bx/dt$ for the trajectory. From the equation \eqref{ss_xdot}, we have $\displaystyle \frac{d\bx}{d\tau}=\boldsymbol{P} + 2 \frac{\boldsymbol{SR(S)P}}{R(S)(S)}$ and $\displaystyle \frac{dt}{d\tau}=P_4 + 2 \frac{SR(S)P_4}{R(S)(S)}$. Hence, assuming $R(S)(S) \neq 0$, we can directly write,
\begin{equation}
\frac{d\bx}{dt} = \frac{R(S)(S) \boldsymbol{P} + 2 \boldsymbol{SR(S)P}}{R(S)(S) P_4 + 2 SR(S)P_4}.
\end{equation}

The key to study the behavior of this equation is to expand, in terms of $\alpha$ and $\lambda$, the numerator and the denominator separately. We end up with,
\begin{equation}
\label{eom_kottler}
\frac{d\bx}{dt} = \frac{6a \frac{r^2(r-a)}{(r+a)^3}\Big(r^2 (\bs\cdot\bp) \bp - 3 \np^2 (\bs\cdot\bx) \bx + 3 [\bx\times\bp\cdot\bs] \bx\times\bp\Big)- \Lambda\, r^5 (\bs\cdot\bp) \bp + \cO(\lambda^2,\alpha \lambda)}{6 a \np \Big(r^2 (\bs\cdot\bp) - 3 (\bx\cdot\bp)(\bs\cdot\bx)\Big) - \Lambda \, r^5 (\bs\cdot\bp) \np+\cO(\lambda^2,\alpha \lambda)}.
\end{equation}

We distinguish three regimes in the above equation of motion: near the star where $1 \gg a/r \gg r^2 \Lambda$, far away from the star where $1 \gg r^2 \Lambda \gg a/r$, and a transition regime in between. In the first regime, near the star, the equation of motion reduces exactly to \eqref{dx_spin}, which is the equation we obtained for the pure Schwarzschild case. Now, in the second regime, far away from the star, the equation simplies considerably and we have,
\begin{equation}
\left.\frac{d\bx}{dt}\right|_{1 \gg r^2\Lambda \gg a/r} \sim \frac{\bp}{\np}.
\end{equation}

This is the qualitative behaviour we expected: near the star, the spin-orbit interaction of the star and the photon dominates, while sufficiently far away from the star, the photon follows the direction of its momentum. By comparing the contributions in the equation of motion \eqref{eom_kottler}, we see that this change of behaviour happens roughly at a distance of 
\begin{equation}
r_H = \sqrt[3]{\frac{a}{\Lambda}}.
\end{equation}

While considering the cosmological constant is certainly better than not, there is one shortfall: this distance $r_H$ seems too large. For example, for our sun this distance would be $r_H \approx 300$ light years. 

For the sack of completeness we also have,
\begin{equation}
\frac{d\bp}{dt} = \frac{N(p)}{D(p)},
\end{equation}
with
\begin{equation}
D(\bp) = 6 a \np \Big(r^2 (\bs\cdot\bp) - 3 (\bx\cdot\bp)(\bs\cdot\bx)\Big) - \Lambda \, r^5 (\bs\cdot\bp) \np+\cO(\Lambda^2),
\end{equation}
and
\begin{align}
N(\bp) = & \frac{12 a^2}{(r+a)^7} \Big[r^2(r-a)\Big(-3r(\bs\cdot\bx)[\bx\times\bp\cdot\bs]+(r+a)^3(\bs\cdot\bp)(\bx\cdot\bp)\Big) \bp \nonumber \\
& \hspace{1.5cm} - r(r+a)^3\Big(r(2r-a)(\bs\cdot\bp)-3(\bx\cdot\bp)(\bs\cdot\bx)\Big) \bx \nonumber \\
& \hspace{1.5cm} + 3(r+a)^3(r-a)[\bx\times\bp\cdot\bs] (\bx\cdot\bp) \bx\times\bp \Big] \nonumber \\
& + a \Lambda \Big[ -12 [\bx\times\bp\cdot\bs] (\bs\cdot\bx) \bp - 6 (\bs\cdot\bx)(\bx\cdot\bp) \np^2 \bx - (\bs\cdot\bp)(\bx\cdot\bp) r^2 \bp \nonumber \\
& \hspace{1.5cm}+ 5 (\bs\cdot\bp) \np^2 r^2 \bx + 3 [\bx\times\bp\cdot\bs] (\bx\cdot\bp) \bx\times\bp \Big] \nonumber \\
& - \frac{\Lambda^2}{6} r^5 (\bs\cdot\bp) \Big(\np^2 \bx + (\bx\cdot\bp) \bp\Big),
\end{align}
so that, much like for the first equation \eqref{eom_kottler}, in the regime near the star with $1 \gg a/r \gg r^2\Lambda$ the differential equation for the momentum above reduces to the one we found earlier in the pure Schwarzschild case \eqref{dp_spin}. In the case where we are far away from the star, the equation reduces to,
\begin{equation}
\left.\frac{d\bp}{dt}\right|_{1 \gg r^2\Lambda \gg a/r} \sim \frac{\Lambda}{6 \np} \left(\np^2 \bx + (\bx \cdot \bp) \bp\right),
\end{equation}
which is nothing more than the usual equation for the momentum in de Sitter spacetime \eqref{ds_dpdt}, at first order in $\Lambda$. 

Now, the cross term in $a \Lambda$ complicates the study of this equation of motion. 

Similarly, for the equation of motion for the spin vector, we have
\begin{equation}
\frac{d\bs}{dt} = \frac{N(\bs)}{D(\bs)},
\end{equation}
with, 

\begin{align}
N(\bs) = 
& \frac{6a}{(r+a)^4} \Big[3 (r - a) (r + a)^3 \Big[\big(2 \np^2 (\bs\cdot\bx) - (\bx\cdot\bp) (\bs\cdot\bp)\big) (\bx\times\bp) + \nonumber \\
& + \big(-r^2 \np^2 + (\bx\cdot\bp)^2\big) (\bs\times\bp) \Big] + 2 a (r - a) \Big[-r^2 (\bs\cdot\bp)^2 \bx - 3 [\bx\times\bp\cdot\bs]^2 \bx +  \nonumber \\
& \hspace{0.7cm} + r^2 (\bs\cdot\bx) (\bs\cdot\bp) \bp + 3 [\bx\times\bp\cdot\bs] (\bs\cdot\bx) (\bx\times\bp)\Big] \nonumber \\
& \hspace{0.7cm} + 2 a r \Big(r^2 (\bs\cdot\bp)-3(\bs\cdot\bp)(\bs\cdot\bx)\Big) \Big((\bs\cdot\bx) \bp - (\bx\cdot\bp) \bs)\Big)\Big] + \nonumber \\
& + \frac{a \Lambda}{2} \Big[9 [\bx\times\bp\cdot\bs] \np^2 r^2 \bx - 6 [\bx\times\bp\cdot\bs] (\bx\cdot\bp) (\bs\times\bx) - 6 [\bx\times\bp\cdot\bs] r^2 (\bs\times\bp) \nonumber \\
& \hspace{0.7cm}- 9 [\bx\times\bp\cdot\bs] (\bx\cdot\bp) r^2 \bp - 10 (\bs\cdot\bp) (\bs\cdot\bx) r^2 \bp + 2 (\bs\cdot\bp)^2 r^2 \bx + 8 (\bs\cdot\bp) r^2 (\bx\cdot\bp) \bs \nonumber \\
& \hspace{0.7cm} + 9 (\bs\cdot\bx) r^2 \np^2 (\bx\times\bp) -12 (\bs\cdot\bx) (\bx\cdot\bp)^2 \bs + 12 (\bs\cdot\bx)^2 (\bx\cdot\bp) \bp\Big] + \nonumber \\
& + \frac{\Lambda^2 r^5}{6} (\bs\cdot\bp) \Big[(\bs\cdot\bp) \bx + (\bs\cdot\bx) \bp - 2 (\bx\cdot\bp) \bs \Big],
\end{align}
and
\begin{equation}
D(\bs) = D(\bp') = 6 a \np \Big(r^2 (\bs\cdot\bp) - 3 (\bx\cdot\bp)(\bs\cdot\bx)\Big) - \Lambda \, r^5 (\bs\cdot\bp) \np+\cO(\Lambda^2).
\end{equation}

The same conclusions apply for this equation as for the equation on the momentum vector. Near the star, we recover the equation of motion \eqref{ds_spin} for a Schwarzschild background, far away we recover the equation of motion for a de Sitter background \eqref{ds_dsdt} at first order in $\Lambda$, and in the transition regime we have a cross term in $a \Lambda$ which complicates the study of this equation.

\subsection{Conclusions}

It seems that introducing the cosmological constant in the Schwarzschild background, hence considering the Kottler metric, helps solving a few oddities we encountered in the behaviour of a spinning photon in a Schwarzschild spacetime, however it raises some more.

Remember the two remarks we had. Firstly, the photon was not following its momentum, arbitrarily far away from the star. Secondly, the transverse spin grew unboundedly with time. Both points are addressed here since, while close to the star the equations of motion reduce to the ones we have found for the Schwarzschild case, at distances much larger than $r_H = \sqrt[3]{a/\Lambda}$, the equations of motion reduce to those of de Sitter. Hence, the photon will follow its momentum, and the transverse spin will be bonded. However, this is not sufficient, as $r_H$ is much too large for physical reasons. For the sun, this distance is of the order of 300 light years. While this is in a sense a much smaller distance than we could have expected with the cosmological constant, we find it hard to believe that the sun	has an effect on the trajectory of a photon 300 light years away from its center. Moreover, while the transverse spin is now bounded, it still grows to ridiculously high values before the de Sitter spacetime keeps it in check.

The previous comments are only qualitative, in the sense that we have presented no numerical simulation nor perturbative solutions. While qualitative comments are a good start, one would need to study the intermediate phase around $r_H$ where both the Schwarzschild and the de Sitter metrics have influence on the evolution of $\bp$ and $\bs$ to know if our hypothesis of the angle $\gamma$ is correct. However, the equations of motion in Kottler background are rather complicated, and due to the large value of $r_H$ compared to the Schwarzschild radius $a$, even numerical simulations are non trivial. 

\subsection{Comparison with existing literature}

The study presented here is not the only work which has been done to determine whether there is birefringence of light, or a Spin Hall Effect of Light, in a Schwarzschild spacetime. It is worth comparing the results of these different approaches.

The first two cases we can compare are when both use the Mathisson--Papapetrou--Dixon equations, but with different Spin Supplementary conditions. As we have discussed in the section \ref{s:sscs} about the different SSCs, the MPD equations together with the Mathisson--Pirani SSC lead to a photon following null geodesics \cite{Mas75,Duv78}, hence they do not predict a birefringence effect in a Schwarzschild spacetime. On the other hand, the MPD equations supplemented with the Tulczyjew SSC lead to the Souriau--Saturnini equations, which as we have seen in this section, or in \cite{Sat76,ChDLMTS}, do predict birefringence in such spacetime.

In \cite{FrolovS11}, the authors study standard (and modified) geometric optics with the (modified) eikonal equation, starting from Maxwell's equations in stationary spacetimes. They first note that standard geometric optics lead, at any order in the expansion, to null geodesics. They then justify modifying the eikonal equation such that helicity effects are not visible locally, which agrees with standard geometric optics, but compound to a non zero effect at ``large distances''. In the end, they find a deviation with respect to null geodesics in Kerr spacetime, however this deviation vanishes in the limit of a Schwarzschild spacetime.

Next, in \cite{Gos06}, a massless particule of spin one is described by the Bargmann--Wigner equations \cite{BargmannW48}, and the Hamiltonian associated to these equations. After diagonalizing the Hamiltonian at first order in $\hbar$, they find that the Berry curvature, associated to the Berry phase \cite{Berry84} of the photon's momentum, couples to the helicity of the photon, thus introducing an anomalous velocity term in the equations of motion of a photon. The authors find a deviation with respect to the null geodesic in a Schwarzschild spacetime, out of the geodesic plane, exactly equal to the angle $\gamma$ \eqref{angle_gamma} we found with the Souriau--Saturnini equations.\footnote{Note that there is a typo in~\cite{Gos06} for the expression of the birefringence angle (last unlabelled equation of p.~5). While it is written in the article that the birefringence angle adds to the existing lensing angle, their equations of motion do predict that this angle yields a trajectory out of the geodesic plane.} However, this may just be a coincidence, as there are not many possibilities to construct an angle with the physical quantities of the problem.

Finally, a recent work \cite{OanceaJDRPA20} carried out a Wentzel--Kramers--Brillouin (WKB) analysis of Maxwell's equations in curved spacetime. The authors have three main hypothesis. First, the vector potential satisfies the Lorenz gauge, then that the initial phase gradient be future-oriented and null, and finally the beam have initially circular polarization. In the end, they find an anomalous velocity which relates to Berry curvature, which is reminiscent of \cite{Gos06}. They perform a numerical simulation comparing their equation of motion with the equation of motion found in \cite{Gos06}. Though they do not give a deviation angle, they say that the two equations of motion they compare give the same trajectory. Hence, at infinity, they should have the same expression of the angle as that of \cite{Gos06}, which is the expression of our angle $\gamma$ \eqref{angle_gamma}. It is also worth mentioning that in their figure, we see that at first the photon is deviated in one direction out of the geodesic plane, and after a little while the photon is pulled back and deviates to the opposite direction. This is reminiscent of our two angles, $\beta$ and $\gamma$, which seem to do just that. However, the comparison is not direct as their photon starts far away from the star, then does a fly-by, while our photon starts at perihelium. Additionally, the distance at which this ``pull-back'' happens looks much more reasonable that we have with the help of the cosmological constant, though they do not give any measure of this. 

\section{A photon in a gravitational wave background}
\label{ss:gw}

\subsection{Introduction}

Gravitational wave detection in interferometers such as the Laser Interferometer Gravita\-tional-Wave Observatory (LIGO) and the Virgo observatory involves laser beams travelling through a gravitational field perturbed by a gravitational wave inhomogeneity. The wave profile is reconstructed from the difference of time of flight of the laser light in two perpendicular linear arms. Presently, the time of flight is computed by treating the beam as a collection of photons, with each photon moving on a geodesic in a given (gravitational wave) background. However, as we have seen all along this chapter, geodesics are only followed by {\em spinless} particles. In the present section, we thus try to include the photons' spin into its equations of motion and check whether it could lead to a measurable effect.

Note that the Tulczyjew SSC has already been used in the problem of massive spinning-particle motion in an exact gravitational wave solution \cite{Moh01}. In \cite{ObuSilAleTer17}, classical as well as quantum massive fermions were studied, with application to a gravitational-wave background (among others).

\subsection{The Souriau--Saturnini equations in a GW background}

While the Souriau--Saturnini equations \eqref{ss_xdot}--\eqref{ss_sdot} work rather well in a Robertson-Walker background, see section \ref{ss:flrw}, or in the proximity of a star, see section \ref{ss:schw}, they break down when the curvature of the gravitational background vanishes. This is due to the lonely term $R(S)(S)$ in the denominator of (\ref{ss_xdot}). When the curvature vanishes, the equations become those of a plane wave traveling at the speed of light. Indeed, massless and chargeless particles cannot be localized in flat spacetime with this approach. It becomes a problem for a metric of gravitational waves, as they are usually computed as a perturbation around flat spacetime.

Let us consider similar equations to those of Souriau--Saturnini, but this time, for massive particles, where $P^2 = m^2 \neq 0$, and still adopt the Tulczyjew constraint $SP = 0$. we have the similar equations \cite{Kun72,Obu11},
\begin{align}
\dot{X} & = P - \,\frac{2 \, S R(S) P}{4 \, P^2 - R(S)(S)}, \label{xdot_massive}\\
\dot{P} & = - \half R(S) \dot{X},\label{pdotnew}\\
\dot{S} & = P\overline{\dX}-\dX\barP.\label{sdotnew}
\end{align}

Notice that we recover the Souriau--Saturnini equations in the limit $P^2 \rightarrow 0$, which is not, \textit{a priori}, trivial. For example, this would not be the same had we considered the Mathisson--Pirani constraint.

Now, for massive particles, the denominator of (\ref{xdot_massive}) behaves in a nicer way. When the Riemann tensor goes to zero, or when $m^2 \gg R(S)(S)$, we recover the usual geodesic equation. To be sure that the denominator does not vanish in the massive case, we should have $4 m^2 > R(S)(S)$. We thus have a lower bound on the mass of the test particle. With $f$ the frequency of the gravitational wave and $c$ the speed of light, that requirement becomes
\begin{equation}
m^2 > \frac{\epsilon \,\pi^2 \, f^2 \, \hbar^2}{c^4}
\end{equation}
Note that this depends on the amplitude $\epsilon$ of the gravitational waves. As this amplitude goes to zero, the mass restriction reduces to $m > 0$. In the case of gravitational wave detections, the frequency of gravitational waves is typically around $f = 50$Hz, and the amplitude around $\epsilon = 10^{-20}$. This gives
\begin{equation}
\label{m_constraint}
m > 10^{-59}\,\mathrm{ kg},
\end{equation}
to have a consistent set of equations describing a massive particle with spin in a typical background with gravitational waves.

The main idea to compute the time delay due to the photon's spin in a background of gravitational waves is to only compute the effect in the direction defined by the momentum. Indeed, the photon goes back and forth in one direction of propagation, so here we are not interested in the full trajectory in space of the photon/particle.  Therefore, to compute the delay, we can compute the effect of spin on a massive particle, though with a mass much smaller than its momentum. Since we only compute the time delay in the direction defined by the momentum, and since (\ref{xdot_massive}) reduces to (\ref{ss_xdot}) in the limit $P^2 \rightarrow 0$, the mass will drop out of the equations when compared to the momentum, thus giving us the expected time of flight delay for a photon.

Notice that, in any case, the best experimental measurements on the mass of a photon give us an upper bound for the mass of about $10^{-50}\,$kg to $10^{-54}\,$kg depending on the type of measurements and assumptions \cite{Wil71,Ryu07}. These upper bounds are a few orders of magnitude higher than the constraint on the mass of the photon (\ref{m_constraint}) in the massive equations.

\subsection{Equations of motion for the ultrarelativistic photon}
\label{s:eom}
Using Cartesian coordinates $(x^1,x^2,x^3,t)$, we linearize the gravitational field equations with the metric, 
\begin{equation}
\label{metric_gw}
g_{\mu\nu}= \eta_{\mu\nu} + \epsilon \, h_{\mu\nu} + \cO(\epsilon^2)
\end{equation}
where $(\eta_{\mu\nu}) = \diag(-1,-1,-1,1)$ is the flat Minkowski metric, $h_{\mu\nu}$ the linear deviation of the metric to flat spacetime, and $\epsilon \ll 1$ a small parameter encoding the amplitude of the gravitational wave.

Linearizing the Einstein field equations in $\epsilon$, and considering a gravitational wave propagating in the direction of the $z$ axis, leads to the well-known solution for the perturbation~$h_{\mu\nu}$,
\begin{equation}
(h_{\mu\nu}) = 
\left(
\begin{array}{cccc}
f_+(t-x_3) & f_\times(t-x_3) & 0 & 0 \\
f_\times(t-x_3) & -f_+(t-x_3) & 0 & 0 \\
0 & 0 & 0 & 0 \\
0 & 0 & 0 & 0
\end{array}
\right)
\end{equation}
with $f_+$ and $f_\times$ two functions describing the two polarization states of the gravitational waves.

For concreteness, take $f_+(t-x_3) = \cos(\omega(t-x_3))$ and $f_\times(t-x_3) = 0$ with $c = 1$. The linearized metric thus takes the form,
\begin{equation}
\label{g_e}
(g_{\mu\nu}) = 
\left(
\begin{array}{cccc}
-1 + \epsilon \cos(\omega(t-x_3)) & 0 & 0 & 0 \\
0 & -1 - \epsilon \cos(\omega(t-x_3)) & 0 & 0 \\
0 & 0 & -1 & 0 \\
0 & 0 & 0 & 1
\end{array}
\right)+ \cO(\epsilon^2)
\end{equation}

Up to linear order in $\epsilon$, we have ${R^3}_{1 3 1} = -{R^3}_{1 4 1} = - {R^3}_{2 3 2} = {R^3}_{2 4 2} = {R^4}_{1 3 1} = - {R^4}_{1 4 1} = - {R^4}_{2 3 2} = {R^4}_{2 4 2} = - \half \omega^2 \epsilon \cos(\omega(t-x_3))$. 

Now, to alleviate notations, we write $k \equiv \cos(\omega(t-x_3))$. The conditions $P^2 = m^2$, and to recover the usual four-momentum $P$ in the limit $\epsilon \rightarrow 0$, dictate the expression,
\begin{equation}
(P^\mu) = \left(
\begin{array}{c}
\displaystyle p_1 \left(1 + \frac{\epsilon}{2} k\right) \\[1.3ex]
\displaystyle p_2 \left(1 - \frac{\epsilon}{2} k\right) \\[1.3ex]
\displaystyle p_3 \\[1.3ex]
\sqrt{m^2 + \np^2}
\end{array} 
\right)+ \cO(\epsilon^2),
\end{equation}
where the $p_i = p_i(t), i = 1, 2, 3$ are the unknown components of the 3-momentum, and with $\np^2 = p_1^2 + p_2^2 + p_3^2$.

Likewise, the spin tensor is defined by its constraints. To linear order in $\epsilon$ we have, with $s_i = s_i(t)$ understood,
\begin{equation}
\label{def_s}
({S^\mu}_\nu)=
\left(
\begin{array}{cccc}
0 & -s_3\left(1 + \epsilon k\right) & s_2\left(1 + \frac{\epsilon}{2} k\right) & \frac{(p_2 s_3 - p_3 s_2)}{\sqrt{m^2+\np^2}}\left(1 + \frac{\epsilon}{2} k\right) \\
s_3\left(1 - \epsilon k\right) & 0 & -s_1\left(1 - \frac{\epsilon}{2} k\right) & \frac{(p_3 s_1 - p_2 s_3)}{\sqrt{m^2+\np^2}} \left(1 - \frac{\epsilon}{2} k\right) \\
-s_2\left(1 - \frac{\epsilon}{2} k\right) & s_1\left(1 + \frac{\epsilon}{2} k\right) & 0 & \frac{(p_1 s_2 - p_2 s_1)}{\sqrt{m^2+\np^2}} \\
\frac{(p_2 s_3 - p_3 s_2)}{\sqrt{m^2+\np^2}}\left(1 - \frac{\epsilon}{2} k\right) & \frac{(p_3 s_1 - p_2 s_3)}{\sqrt{m^2+\np^2}}\left(1 + \frac{\epsilon}{2} k\right) & \frac{(p_1 s_2 - p_2 s_1)}{\sqrt{m^2+\np^2}} & 0 
\end{array}
\right)
\end{equation}
such that $S$ is skew-symmetric, and still up to linear order,
\begin{equation}
SP = 0 \qquad \mathrm{and} \qquad -\half \Tr(S^2) = j^2
\end{equation}
with
\begin{equation}
j^2 = \frac{(\bs\cdot\bp)^2+m^2 \ns^2}{\np^2+m^2}.
\end{equation}
Note that in the limit $m \rightarrow 0$ in the above relation, we recover the square of the scalar spin, or longitudinal spin, of a massless particle. In other words, in the massless case, the longitudinal spin is the projection of the spin vector along the direction of the momentum.

Next, we have,
\begin{equation}
\Pf(R(S)) = \cO(\epsilon^2).
\end{equation}

See Appendix \ref{AppendixA} for the expressions of $R(S)(S)$ and $S\,R(S)\,P$.

We then have the equations of motion for the position of the massive particle (\ref{xdot_massive}),
\begin{equation}
\dot{X} = P - \,\frac{2 \, S R(S) P}{4 \, P^2 - R(S)(S)},
\end{equation}

So, we get the equations of motion on 3d-space, with respect to the time coordinate $t$, in the 3+1 splitting $(\bx,t)$, as
\begin{equation}
\frac{d \bx}{dt} = \frac{\left(2 m^2-\half R(S)(S)\right) {\bP}-\boldsymbol{SR(S)P}}{\left(2 m^2-\half R(S)(S)\right) P_4-SR(S)P_4}
\end{equation}

At this point, the mass terms allow us to take the limit $\epsilon \rightarrow 0$. From \eqref{pdotnew} and \eqref{sdotnew}, which we can rewrite as equations for $d\bp/dt$ and $d\bs/dt$ with the $3+1$ split, we see that $d\bp/dt \sim d\bs/dt \sim \cO(\epsilon)$. Hence, if we take the following initial conditions for the photon,
\begin{equation}
\bx_0 = \left(\begin{array}{c}
0 \\
0 \\
0
\end{array}
\right), \quad \bp_0 = \left(\begin{array}{c}
0 \\
{p_2}_0 \\
0
\end{array}
\right), \quad \bs_0 = \left(\begin{array}{c}
{s_1}_0 \\
{s_2}_0 \\
{s_3}_0
\end{array}\right),
\end{equation}
we have the following momentum and spin, $\bp(t) = \bp_0 + \epsilon \, \bq(t) + \cO(\epsilon^2)$ and $\bs(t) = \bs_0 + \epsilon \, \bsigma(t) + \cO(\epsilon^2)$. Since we only want the equation of motion $dx_2/dt$ at linear order in $\epsilon$, it is sufficient to have $\bp(t)$ and $\bs(t)$ at the zeroth order in $\epsilon$. Indeed, as we will see below, contributions in $\bq(t)$ and $\bsigma(t)$ vanish after the ultrarelativistic limit.

Thus, for the velocity in the direction we are interested in, at first order in $\epsilon$, we have,
\begin{align}
\frac{dx_2}{dt} = & \frac{{p_2}_0}{\sqrt{m^2+{p_2}_0^2}} + \epsilon \frac{m^2 q_2(t)}{(m^2+{p_2}_0^2)^{3/2}} + \nonumber \\
& - \frac{\epsilon}{2} \, \frac{{p_2}_0(m^2+{p_2}_0^2) + \omega^2 \Big({p_2}_0({s_1}_0^2-{s_3}_0^2) - \sqrt{m^2+{p_2}_0^2} {s_2}_0 {s_3}_0 \Big)}{(m^2+{p_2}_0^2)^{3/2}} \, \cos(\omega(t-x_3)) + \nonumber \\
& + \cO(\epsilon^2)
\end{align}

We might be interested here in the behaviour of the function $q_2(t)$. From \eqref{pdotnew} and the $3+1$ split, we get, 
\begin{equation}
\frac{d q_2(t)}{dt} = \frac{\epsilon}{2} \omega \big({p_2}_0 \sin(\omega(t-x_3)) - {s_1}_0 \omega \cos(\omega(t-x_3))\big).
\end{equation}
The important take away here is that $q_2(t)$ does not contain any mass term. Thus, when ${p_2}_0^2 \gg m^2$, we have
\begin{equation}
\label{dxdt_ur}
\frac{d x_2}{dt} = 1 - \frac{\epsilon}{2} \cos(\omega(t-x_3)) - \frac{\epsilon}{2} \frac{\lambda_\gamma^2}{\lambda_{\mathrm{GW}}^2} \frac{\left({s_1}_0^2-{s_3}_0^2-{s_2}_0{s_3}_0\right)}{\hbar^2} \cos(\omega(t-x_3)) + \cO(\epsilon^2)
\end{equation}
with $\lambda_\gamma$ the wavelength associated to the photon, and $\lambda_{\mathrm{GW}} = 2\pi/\omega$ is the wavelength of the gravitational wave. With values taken from LIGO/Virgo, $\lambda_\gamma = 1064$nm, 
\begin{equation*}
\frac{\epsilon}{2} \frac{\lambda_\gamma^2}{\lambda_{\mathrm{GW}}^2} \sim 10^{-46}.
\end{equation*}

This means that geodesic effects of order $\epsilon^2 \sim 10^{-40}$ would be seen before observing any spin effect in LIGO/Virgo type detectors.

The effect is maximum when photons are polarized such that $\bs = (0, \hbar, \hbar)$, at least in the classical limit. In that case, the measured time delay is decreased from~$\Delta\tau$ to
\begin{equation}
\widetilde{\Delta\tau} = \Delta\tau\left(1 - 2 \frac{\lambda_\gamma^2}{\lambda_{\mathrm{GW}}^2}\right)
\end{equation}

A corollary is that two photons of different polarization will have different times of flight. Thus, a beam made up of photons of random polarization will introduce a noise due to spin curvature effects. A way to eliminate this noise is to polarize the beams of light before sending them into the arms. However, the amplitude of the noise created by this birefringence is of the relative order of $10^{-46}$ in LIGO/Virgo, which is much below the current sensitivity in LIGO and Virgo experiments.

\subsection{Conclusions}
\label{s:conclusions}

To take into consideration the possible effects of the photon's spin on its trajectory in curved space, we used the Mathisson-Papapetrou-Dixon equations for spinning test particles, together with two possible supplementary conditions for photons, by Frenkel-Pirani, or by Tulczyjew. While for a massive spinning body, such as a spinning star, the choice of SSC does not seem to have much practical impact on the observable trajectory (unless the angular momentum of the body is extremely large \cite{Semerak99}), this choice has potentially visible consequences for elementary particles.

The Frenkel-Pirani SSC for a massless particle leads to a trajectory along a null geodesic, regardless of the gravitational background. In that case, there would be no change to the geodesic trajectory of photons in a background of gravitational waves.

The Tulczyjew SSC for a massless particle predicts a very small effect due to the polarization of the light on its trajectory. Since the massive equations with this condition lead to the massless equations in the limit $m \rightarrow 0$, and because of the instability of the localization of the test particle in the equations near zero curvature, the photon is treated in this paper as an ultrarelativistic massive particle. This mass, which can be both large compared to the spin-curvature coupling term $R(S)(S)$ and extremely small compared to the momentum of the photon, allows for convenient limits to be taken in the equations. The geodesic equations in a gravitational wave background are recovered, together with a new term depending on the spin polarization of the photon. This means that with this supplementary condition, the time of flight of a photon in a detector depends on its polarization state. This dependence is, however, many order of magnitudes lower than the first order effects of gravitational waves on the time of flight. But, if we achieve that kind of precision, polarizing the laser beam in a specific way would be an easy way to reduce the noise introduced by birefringence. With enough precision, this could even potentially be a way to discriminate between the two possible Spin Supplementary Conditions.

\section{Final remarks}
\label{ss:conclusions}

The Souriau--Saturnini equations \eqref{ss_xdot}--\eqref{ss_sdot} describe the trajectory of a photon, while taking its spin into account. We have seen four examples of application of these equations, namely a spinning photon in de Sitter spacetime, see section \ref{ss:desitter}, in a FLRW spacetime, see section \ref{ss:flrw}, in a Schwarzschild spacetime in section \ref{ss:schw}, and in a gravitational wave background in \ref{ss:gw}. Thanks to these four examples, we can try to extract some key points about the application of these equations.

From the de Sitter example, we saw a good coordinate system to compute the Souriau--Saturnini equations, which are isotropic coordinates. They are suitable do to the amount of Euclidean scalar products and vector products that typically arise in this system of equations. We also saw that in de Sitter spacetime, the Souriau--Saturnini equations reduce to the light-like geodesic equation, together with an equation for the evolution of the transverse spin. While the Souriau--Saturnini equations are ill-defined in Minkowski spacetime, there is no ambiguity on the localization of the photon as long as the cosmological constant is non zero.

Then, remember we mentioned that, after deriving the Souriau--Saturnini equations in \ref{ss:eom2}, we cannot recover the light-like geodesic equation from the Souriau--Saturnini equations in the limit of vanishing $\hbar$, or the vanishing limit of the longitudinal spin $s \rightarrow 0$. However, we have seen in the example of a photon in FLRW spacetime that a key ingredient to solve the Souriau--Saturnini equations perturbatively was to consider the transverse spin $s^\bot$ of the photon. While it is not possible to take the limit where the longitudinal spin goes to zero, the transverse spin is not constrained by the Souriau--Saturnini equations. As it turns out, the limit $s^\bot \rightarrow 0$ in a FLRW spacetime corresponds to light-like geodesics, as computed in \cite{ChDTSRW}. 

The example of a Schwarzschild background in \ref{ss:schw} teaches us that this trick about the transverse spin seems like an isolated case. Indeed, the limit $s^\bot \rightarrow 0$ does not lead to light-like geodesic equations. Even worse, the transverse spin needs to be somewhat fine-tuned for the equations of motions to make sense. In this study, the only way we recovered a light-like geodesic was in the case of radial motion, regardless of the photon's spin state, but our efforts to approximate a solution around the radial case were in vain. However, we saw that two small parameters helped us derive a perturbative solution. The first parameter shows how small the Schwarzschild radius of the star is compared to the distance of the photon at perihelion, and the second one shows how small $\hbar$ is compared to the same distance multiplied by the momentum of the photon. However, even with these two small quantities, one needs to be careful in the derivation of the perturbative solution. 

Then we have the study of a spinning photon in a gravitational wave background. We have solved this system in a yet again completely different way. Note that here, since the metric itself is a perturbative approximation to the Einstein equations, we did not need a special coordinate system. However, to avoid the Souriau--Saturnini equations to be ill-defined in the limit of vanishing curvature, we had to introduce a dummy mass for the photon. We have also given arguments why the introduction of this mass is not a problem in practice to compute the time of flight of a massless photon in this situation.

To summarize, we have here four studies solving the Souriau--Saturnini equation in different contexts, and four different ways of doing so. One constant we have is the use of isotropic coordinates, though only to leading order in the case of gravitational waves. However, this coordinate system might prove problematic in more complex studies, for example with a Kerr metric. Another constant we observe, and which could have been expected, is that any deviation of the trajectory of the spinning photon with respect to the light-like geodesic depends on the wavelength of the photon. In the limit where the photon has infinite energy, this deviation disappears.

However, this system of equations is not without problems, as we have seen. The Schwarzschild example is the most problematic. Indeed, we have seen that the norm of the transverse spin grows unbounded with time, and that the photon's trajectory does not follow the direction of its momentum at infinity, where spacetime is asymptotically flat, and hence where we could expect the trajectory to reduce to a light-like geodesic. These may be an artefact of the fact that the Souriau--Saturnini equations are a completely classical set of equations that describe the motion of a single photon. Even though $\hbar$ is a constant appearing in the system, no quantum effects are taken into consideration. 

Given these problems, and the possibility that they arise from ignoring quantum - or wave- effects, it is legitimate to wonder whether some cases, for example a Schwarzschild background, should be studied with a more complete theory. Efforts in this direction exist. For example in \cite{Gos06,OanceaJDRPA20} semi-classical equations of motions are derived.

Speaking of a more complete theory, a question arises naturally: could this classical theory be elevated to a quantum theory with the help of Geometric Quantization. Souriau showed in \cite{Sou74} how to write the MPD equations with Tulczyjew SSC for a massive and charged particle of spin $1/2$ in both gravitational and electromagnetic background in a symplectic framework, and that it can be pre-quantized. However, in 1974, at his time of writing, he found that the polarization step was inaccessible. Since then, new techniques in Geometric Quantization have developed, and we wonder if they could deal with such a system.

%

	\newpage
	\newcommand\doi[1]{\href{http://dx.doi.org/#1}{DOI: #1}}
	\addcontentsline{toc}{chapter}{Bibliography}
    \bibliographystyle{mystyle}
    \bibliography{mybib}
	\newpage

	\setcounter{chapter}{1}
	\setcounter{section}{0}
	\renewcommand{\thesection}{\Alph{section}}
	
	\chapter*{Appendix}
	\addcontentsline{toc}{chapter}{Appendix}

\section{Computational details}

\label{AppendixA}
From the expression of the Riemann tensor, of the spin tensor (\ref{def_s}), with the shorthand $k \equiv \cos(\omega(t-x_3))$, we get,

\begin{equation}
\begin{split}
R(S)(S) = \frac{2 \omega^2 \epsilon k}{m^2+\np^2}& \Big[
2 (p_1 s_1 - p_2 s_2) s_3 \left(p_3 - \sqrt{m^2+\np^2}\right) - \left(p_1^2-p_2^2\right)s_3^2 + \\
& - \left(s_1^2 - s_2^2\right) \left(p_3 \left(p_3 - 2 \sqrt{m^2+\np^2}\right)+\left(m^2+\np^2\right)\right)
\Big]+ \cO(\epsilon^2).
\end{split}
\end{equation}

Similarly, we obtain, with $SR(S)P^\mu = {R^\mu}_{\nu\lambda\sigma} P^\nu S^{\lambda\sigma}$,

\begin{equation}
SR(S)P = \left(\begin{array}{c}
SR(S)P_1 \\
SR(S)P_2 \\
SR(S)P_3 \\
SR(S)P_4
\end{array}\right),
\end{equation}
with,

\begin{align}
SR(S)P_1 = & K \Bigg(s_3 \left(m^2+\np^2\right) \left(\sqrt{m^2+\np^2}-p_3\right) \left(s_1 \left(\sqrt{m^2+p^2}-p_3\right)+p_1 s_3\right) + \nonumber \\
& -\left(s_2 \sqrt{m^2+\np^2} \left(\sqrt{m^2+\np^2}-p_3 \right)+p_2 s_3 \right) \times \\
& \times \left(\left(\sqrt{m^2+\np^2}-p_3\right) \left(p_2 s_1+p_1 s_2\right)+2 p_1 p_2 s_3\right)\Bigg) + \cO(\epsilon^2), \nonumber \\
SR(S)P_2 = & K \Bigg(s_3 \left(m^2+\np^2\right) \left(p_3-\sqrt{m^2+\np^2}\right) \left(s_2 \left(\sqrt{m^2+\np^2}-p_3\right)+p_2 s_3\right) + \nonumber \\
& +\left(s_1 \sqrt{m^2+\np^2} \left(\sqrt{m^2+\np^2}-p_3 \right)+p_1 s_3 \right) \times \\
& \times \left(\left(\sqrt{m^2+\np^2}-p_3\right) \left(p_2 s_1+p_1 s_2\right)+2 p_1 p_2 s_3\right)\Bigg) + \cO(\epsilon^2), \nonumber \\
SR(S)P_3 = & K \sqrt{m^2+\np^2} \Bigg(\left(s_2^2-s_1^2\right) \sqrt{m^2+\np^2} \left(\sqrt{m^2+\np^2}-p_3\right)^2 +\nonumber \\
& + \left(\sqrt{m^2+\np^2}-p_3\right) \left(-s_3 \sqrt{m^2+vp^2} \left(p_1 s_1-p_2 s_2\right)+p_2^2 s_1^2-p_1^2 s_2^2\right)+\\
& + 2 p_1 p_2 s_3 \left(p_2 s_1-p_1 s_2\right) \Bigg) + \cO(\epsilon^2), \nonumber \\
SR(S)P_4 = & K \sqrt{m^2+\np^2} \Bigg(2 p_3^2 \left(\sqrt{m^2+\np^2}-p_3\right) \left(s_1^2-s_2^2\right)+s_3 \left(p_1^3 s_1-p_2^3 s_2\right) + \nonumber \\
& +3 s_3 p_1 p_2 \left(p_2 s_1-p_1   s_2\right)+\left(m^2+3 p_3^2-3 \sqrt{m^2+\np^2} p_3\right) s_3 \left(p_1 s_1-p_2 s_2\right) + \\
& -\left(\sqrt{m^2+\np^2}-2 p_3\right) \left(p_1^2 s_2^2-p_2^2   s_1^2\right)+\left(\sqrt{m^2+\np^2}-p_3\right) s_3^2 \left(p_1^2-p_2^2\right) + \nonumber \\
& -p_3 \left(p_1^2 s_1^2-p_2^2 s_2^2\right)- m^2 p_3 \left(s_1^2-s_2^2\right)\Bigg) + \cO(\epsilon^2), \nonumber 
\end{align}
and,
\begin{equation}
K = \frac{\omega^2 \epsilon \cos (\omega (t-x_3)) }{\left(m^2+\np^2\right)^{3/2}}.
\end{equation}

\end{document}